\documentclass[review]{elsarticle}
\usepackage[margin=1.2in]{geometry}
\usepackage[english]{babel}
\usepackage[utf8]{inputenc} % Required for inputting international characters
\usepackage[T1]{fontenc} % Output font encoding for international characters
\RequirePackage{times} % favourite for note.

\usepackage{siunitx} % typesets numbers with units very nicely
\usepackage[italicdiff]{physics}
\usepackage{amsfonts}
\usepackage{subfigure}
\usepackage{amsmath}
\usepackage{amssymb}
\usepackage{graphicx,epstopdf}
\usepackage{appendix}
\usepackage{xspace}
\usepackage{array}
\usepackage{lineno}
\usepackage{hyperref}
\usepackage{color}
\usepackage{xcolor}
\usepackage{my_symbols}
\usepackage{CJK}
\usepackage{bm}
\usepackage{verbatim}
\usepackage{lipsum}
\usepackage[normalem]{ulem}

\RequirePackage{mdframed} % Required for creating the theorem, definition, exercise and corollary boxes
{\par}

\newcounter{mycomment}

\newcommand\rmv{\bgroup\markoverwith {\textcolor{red}{\rule[0.5ex]{2pt}{0.4pt}}}\ULon}
\usepackage{graphicx}% Include figure files
\usepackage{dcolumn}% Align table columns on decimal point
\usepackage{bm}% bold math
\usepackage{bbold}% bbold package
\usepackage{amsfonts,color}
\usepackage{amsmath}
\usepackage{hyperref}%reference doi
\newcounter{lastnote}

\begin{document}

%\linenumbers

\title{Magnetic texture based magnonics}
\date{\today}

\author[1,2]{Haiming Yu}
\ead{haiming.yu@buaa.edu.cn}
\author[3,4]{Jiang Xiao}
\ead{xiaojiang@fudan.edu.cn}
\author[5]{Helmut Schultheiss}
\ead{h.schultheiss@hzdr.de}

\address[1]{Fert Beijing Institute, School of Microelectronics, Beijing Advanced Innovation Center for Big Data and Brain Computing, Beihang University, 100191 Beijing, China}
\address[2]{Shenzhen Institute for Quantum Science and Engineering (SIQSE), and Department of Physics, Southern University of Science and Technology (SUSTech), 518055 Shenzhen, China}
\address[3]{Department of Physics and State Key Laboratory of Surface Physics, Fudan University, Shanghai 200433, China}
\address[4]{Institute for Nanoelectronics Devices and Quantum Computing, Fudan University, Shanghai 200433, China}
\address[5]{Helmholtz-Zentrum Dresden–Rossendorf, Institute of Ion Beam Physics and Materials Research, Bautzner Landstraße 400, 01328 Dresden, Germany}

\date{\today}

\begin{abstract}
The spontaneous magnetic orders arising in ferro-, ferri- and antiferromagnets stem from various magnetic interactions. Depending on the interplay and competition among the Heisenberg exchange interaction, Dzyaloshinskii-Moriya exchange interaction, magnetic dipolar interaction and crystal anisotropies, a great variety of magnetic textures may be stabilized, such as magnetic domain walls, vortices, Skyrmions and spiral helical structures. While each of these spin textures responds to external forces in a specific manner with characteristic resonance frequencies, they also interact with magnons, the fundamental collective excitation of the magnetic order, which can propagate in magnetic materials free of charge transport and therefore with low energy dissipation. Recent theories and experiments found that the interplay between spin waves and magnetic textures is particularly interesting and rich in physics. In this review, we introduce and discuss the theoretical framework of magnons living on a magnetic texture background, as well as recent experimental progress in the manipulation of magnons via magnetic textures. The flexibility and reconfigurability of magnetic textures are discussed regarding the potential for applications in information processing schemes based on magnons.
\end{abstract}

\maketitle

\newpage
\tableofcontents


\newpage

% !TEX root = review_magnonics_in_texture.tex

Spin textures refer to magnetic ground states with a non-collinear alignment of magnetic moments. Prominent examples range from magnetic domain walls and vortices to more complex structures in helical or chiral material systems where the interplay of the symmetric Heisenberg exchange interaction and the asymmetric Dzyaloshinskii-Moriya interaction leads to the formation of spin spirals and Skyrmions. Despite their complexity these special spin configurations exhibit extraordinary stability which is often referred to as protected by topology and they offer a richness of dynamic excitation. Moreover, spin textures are present in all types of magnetically ordered materials namely ferro-, ferri-, and antiferro-magnetic systems where the energies of magnetic excitation span several orders of magnitude ranging from GHz to THz frequencies.

Spin waves, or magnons as their quanta, are the fundamental excitation of magnetically ordered systems. The concept of magnons was first introduced by Bloch in 1930~\cite{bloch_zur_1930} in order to understand the temperature dependence of ferromagnets and led later to the concept of ferromagnetic resonance developed by Charles Kittel~\cite{Kittel1948}. Driven by advances in computational power and micromagnetic simulations as well as the development of novel experimental methods with unprecedented spatial resolution and frequency range, the magnon eigenmodes of spin textures become accessible even for complex systems and open new routes for fundamental and application-oriented research in condensed matter physics.

The research field of Magnonics, or Magnon Spintronics~\cite{Kru2010,Len2011,Stamps:2014en,ChumakNP15,Yu_Xiao_Pirro_2018,Demi_Review_2020} targets the transport of spin angular momentum via magnons. Besides the fundamental questions arising in the field of magnonics and the interaction with charge based spin transport in the field of Spintronics, research on magnons is also driven by the desire to find alternatives for standard CMOS computing technology to pave the road for a post-Moore technology~\cite{Khi2010,Chu2017,Csa2017}. Using magnons as carriers of information is a promising alternative for state-of-the-art information technology which does not rely on charge transport with the inevitable waste heat problems due to Ohmic losses. Moreover, magnons would allow for combining the non-volatile storage capabilities of magnetic materials with high-frequency data processing in the THz regime. Thus, Magnonics could finally fuse memory and computation in a single hardware unit opening the bottleneck of modern CPU and memory architectures. In this review, we give an overview of current research on magnonics based on spin textures.  Using spin textures for magnonics offers unique opportunities for the generation and manipulation of magnons at the nanoscale, with the benefit of the non-volatile and still reprogrammable character of complex spin structures which go far beyond the possibilities offered by  homogeneous magnetic systems or artificially patterned structures.~\cite{Kra2014,Neusser2009,Adeyeye_2010,Tacchi_BLS_2017}.

\section{Theoretical background of magnetic texture based magnonics}\label{sec:intro}

In this section we introduce the theory for understanding the main energy contributions relevant for the formation of complex spin textures in ferro- and antiferromagnetic materials. The second spatial derivatives of the free energy distribution in magnetic systems not only result in an inhomogeneous landscape of the internal magnetic field, but are also the origin of the torques acting on magnetic moments which are responsible for the richness of spin textures and their robustness as well as the complexity of dynamic magnetic phenomena. Regarding the dynamics it is important to distinguish three fundamentally different types of excitation:
\begin{enumerate}
\item There are magnons, which are the collective excitation of a spin system~\cite{demokritov_bose-einstein_2006}. The magnon frequencies typically adjust to the local strength of the total internal magnetic field, but due to their collective nature they also sense the magnetic field environment on dimensions comparable to their coherence length. In some sense the magnetic field inside a magnetic material can be considered as an effective potential for magnons similar to the electric potential in the case of electrons. Hence, inhomogeneities of the magnetic field, in particular local extrema, cause spatial confinement of the magnon wavefunction resulting in a discretisation of the magnon energies. As a consequence, spin textures such as domain walls and skyrmions cause bound magnon states localized inside the texture and they cause scattering of magnons impinging from the outside.
\item Each type of spin texture also has a characteristic, dynamic response to time dependent torques (anything that causes a time dependence of the second derivative of the free energy of the magnetic system). Typically, the excitation of spin textures can be considered as a motion of the magnetic ground state itself with a resonance frequency smaller than the magnon eigenfrequencies. Prominent examples are the resonant oscillations of magnetic domain walls around a pinning center or effective magnetic field~\cite{saitoh_current-induced_2004} at tens of MHz or the gyroscopic motion of a vortex core at hundreds of MHz~\cite{chou_direct_2007}.
\item Interactions between the first two scenarios exist, where magnons can be generated by moving spin textures or where spin textures can be moved by impinging magnons. Intensive studies of these phenomena recently led to the discussion of hybridized excitation in magnetic systems, where magnons and the motion of spin textures cannot be separated from each other~\cite{Wintz2016} and form an entirely new form of excitation.
\end{enumerate}

The goal of this first section is to introduce step by step the background information needed for a deeper understanding of these three different scenarios of magnetization dynamics, taking into account also the differences in ferromagnetic and antiferromagnetic systems.

\begin{figure}[t]
  \centering
\includegraphics[width=0.85\textwidth,trim=0 0 0 0,clip]{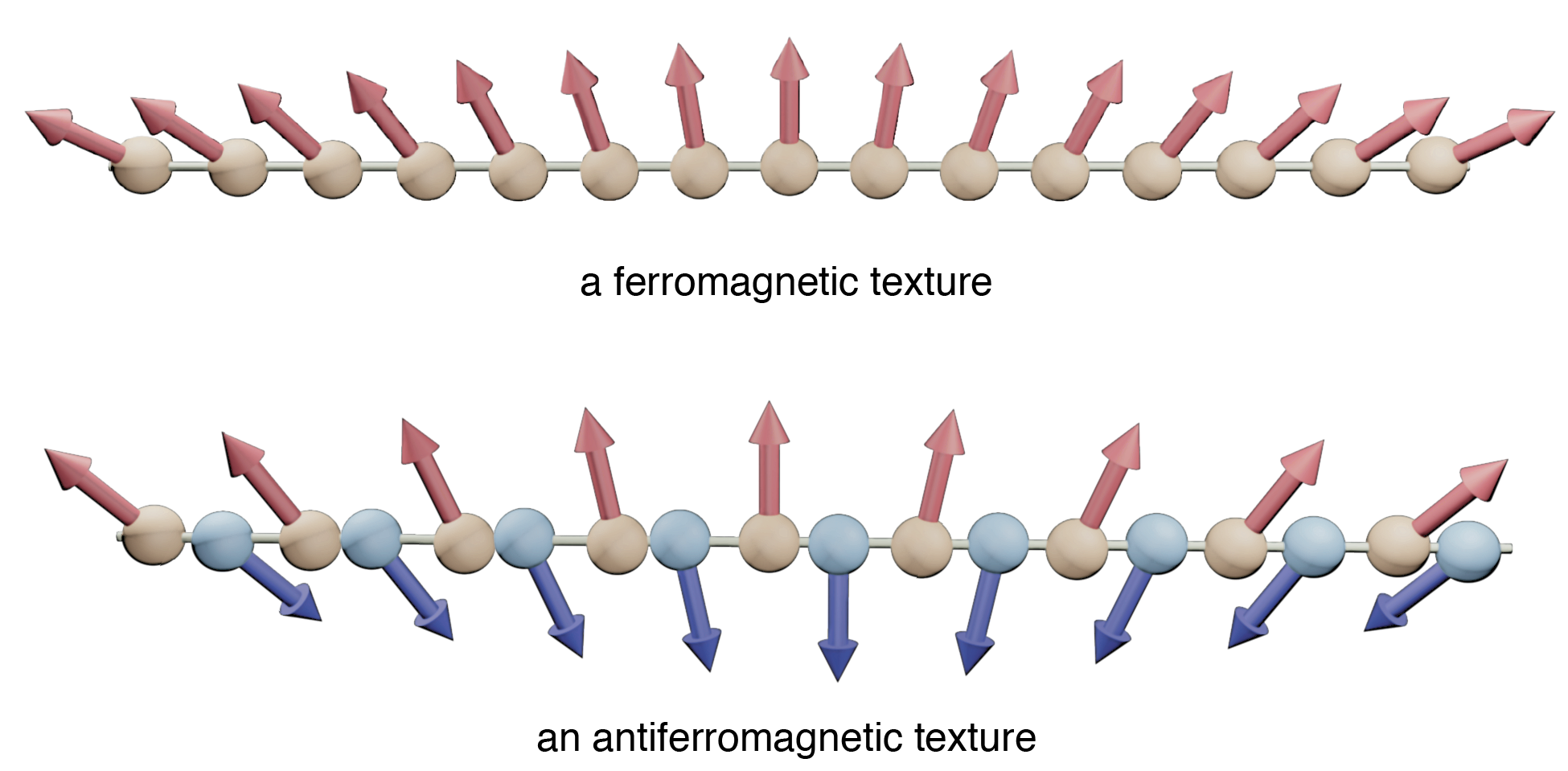}
\caption{Illustrations of one-dimensional ferromagnetic (top) and antiferromagnetic (bottom) textures.}
\label{fig:FM-AFM}
\end{figure}

In a solid crystal, for a spin $\bS_i$ residing on an atomic site of index $i$, the associated magnetic moment is $\bM_i = \gamma\bS_i$ with gyromagnetic ratio $\gamma$. Due to the exchange interaction, the neighboring spins may favor parallel or antiparallel alignment. The former results in a ferromagnet, and the latter in an antiferromagnet. A ferromagnet can be represented by a single continuous order parameter $\bM(\br)$ when the variation of $\bM_i$ is tiny over one lattice constant. In bipartite antiferromagnets, however, two order parameters $\bM^{(1)}(\br)$ and $\bM^{(2)}(\br)$ are required, representing the magnetization from two magnetic sublattices (see \Figure{fig:FM-AFM}) pointing in opposite directions. One may also use the sum and difference of $\bM^{(1),(2)}$ as two new order parameters: The net magnetization $\bM = (\bM^{(1)} + \bM^{(2)})/2$ and the N\'{e}el vector $\bN = (\bM^{(1)} - \bM^{(2)})/2$. Because of this simple sign change in the exchange interaction, ferromagnets and antiferromagnets are drastically different, in terms of both static configurations and their dynamic excitation. In fact, there exist also antiferromagnets with three sublattices~\cite{Ros2015,diaz_topological_2019} that can form \eg fractional antiferromagnetic skyrmions~\cite{Gao2020}. Nevertheless, in this review, the bipartite antiferromagnets remain our main focus. Before discussing the various types of spin textures in (anti)ferromagnets, we shall first introduce the energy contributions of ferromagnets and antiferromagnets separately, which determine their static and dynamic properties.

%-----------------------------------
%-----------------------------------

\subsection{Magnetic energetics}\label{sec:free energy}

The total energy of a magnetic system includes various contributions, such as the Zeeman energy in the presence of an external magnetic field, shape anisotropy caused by dipolar fields, crystalline anisotropy energy originated from the magnetocrystalline effect, the Heisenberg exchange energy due to the quantum Pauli exclusion principle, as well as the antisymmetric exchange energy (or the Dzyaloshinskii-Moriya interaction) in systems without spatial inversion symmetry. This subsection gives a brief overview of these different energy contributions in both ferromagnetic and antiferromagnetic systems.

\subsubsection{Free energy for ferromagnets}\label{sec:free energy ferromagnets}

{\bf Zeeman energy} -  The energy of a ferromagnetic body in the presence of a uniform external field $\bH_\ssf{ext}$ is called the Zeeman energy:
%--------------
\begin{equation}
  \label{eqn:EZeeman}
  E_\ssf{Zeeman} = -\mu_0 M_s \int \mb(\br) \cdot \bH_\ssf{ext}  \dd[3]{\br},
\end{equation}
%--------------
where the unit vector $\mb(\br) \equiv \bM(\br)/M_s$ represents the magnetization direction at $\br$ with $M_s$ the saturation magnetization, and $\mu_0$ is the vacuum permeability. To minimize the Zeeman energy, the magnetization tends to align with the external magnetic field.

{\bf Crystalline anisotropy} - In crystalline solids, the electronic orbitals reflect the symmetry defined by the crystal structure, which results in non-spherical electron orbitals. The energy of such orbital depends on the orientation of the orbital with respect to the crystal structure. Through the spin-orbit interaction, the electron spin, thus the magnetization, (dis)favors the parallel alignment with certain axes. Consequently, the energy of a ferromagnet is anisotropic with respect to the direction of the magnetization. This magnetization-orientation dependent energy contribution is called the magnetocrystalline anisotropy energy. \cite{Stancil} The simplest type is the uniaxial anisotropy with one single special axis $\hbu$, for which the anisotropic energy can be expanded in even orders of the magnetization projection on $\hbu$:

%--------------
\begin{equation}
  \label{eqn:Eani}
  E_\ssf{ani} = - M_s\int \qty[K_{u1} \qty(\mb\cdot \hbu)^2 + K_{u2} \qty(\mb\cdot \hbu)^4 + O(\mb^6)] \dd[3]{\br},
\end{equation}
%--------------
where $K_{u1, u2}$ are the anisotropy constants. When $K_{ui}$ is positive, $\hbu$ is called the easy axis. When $K_{ui}$ is negative, $\hbu$ is called the hard axis and the plane perpendicular to $\hbu$ is called the easy plane. To minimize the anisotropy energy, the magnetization tends to align along the easy axis or in the easy plane, depending on the sign of $K_{ui}$. The $K_{ui}$ of the crystal structure has further consequences, \eg the qunenching of the orbital momentum. A good introduction is given in Ref.~\cite{Daal1990}.

In certain materials, there could be more than one easy axis. For example, materials with a cubic lattice such as iron have three easy axes $\be_i$ that are mutually orthogonal: $\be_i\cdot\be_j = \delta_{ij}$. The energy for such cubic anisotropy can be written as an expansion in terms of the magnetization components along these three axes:
%--------------
\begin{equation}
  E_\ssf{ani} = - M_s\int \qty[K_{c1} \qty(m_1^2m_2^2 + m_2^2m_3^2 + m_3^2m_1^2) + K_{c2} m_1^2m_2^2m_3^2 + O(\mb^8)] \dd[3]{\br},
\end{equation}
%--------------
where $m_i = \mb\cdot\be_i$.

Another type of magnetic anisotropy occurs due to material surfaces or interfaces. Because of the reduced number of neighbors for atoms at surfaces or interfaces, the magnetization at the surface experiences different anisotropy energies from those in the bulk, and the anisotropy axis is the surface normal. Such an anisotropy can also be either positive or negative, resulting in an easy axis surface anisotropy or easy plane surface anisotropy. The easy axis surface anisotropy is the origin of the perpendicular magnetic anisotropy (PMA)~\cite{Guoqiang2015PMA,RMP_PMA_2017,Avci2017,Lat2018,Sou2018,Rana:2019ce,Marko:2019bm,Chen2019PMA,X_Qiu2020}.

%-----------------------------------
\begin{figure}[t]
  \centering
\includegraphics[width=0.98\textwidth]{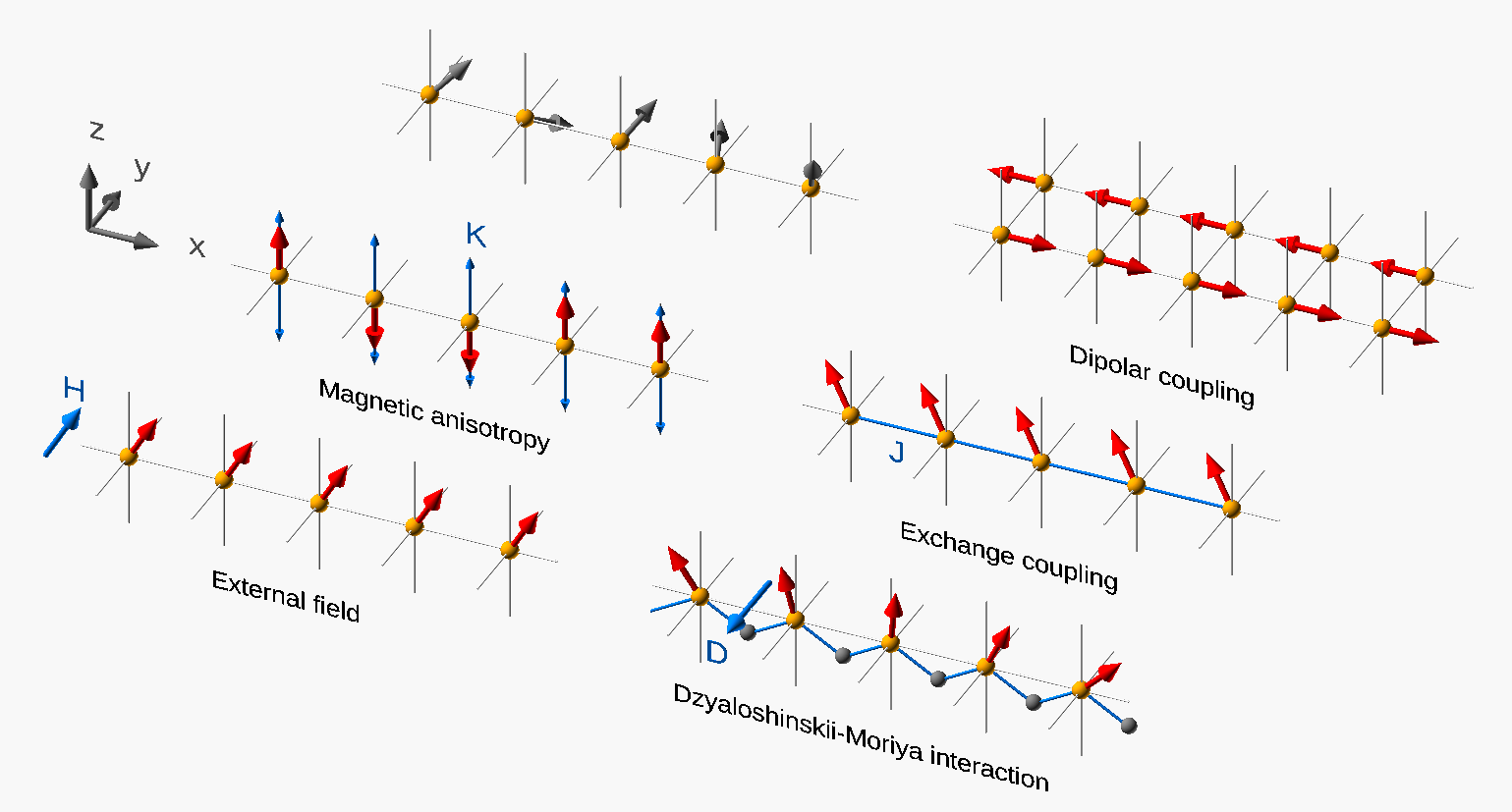}
\caption{Top left: the random magnetization distribution. The uniaxial magnetic anisotropy along $\hbz$, caused by the underlying lattice structure, aligns magnetization in $\pm\hbz$. The Zeeman energy due to the external field $\bH$ aligns all magnetic moments along the field direction. The dipolar coupling among the magnetic moments gives rise to the demagnetization field which creates magnetic domains to minimize the stray field outside the bulk. The Heisenberg exchange coupling tends to remove all inhomogeneity. DMI: the symmetry breaking caused by an extra atom gives rise to the anti-symmetric exchange coupling, with which the magnetization tends to twist about the DM vector.}
\label{fig:FME}
\end{figure}
%-----------------------------------

{\bf Demagnetization} - The demagnetization energy, or magnetostatic energy, is due to the nonlocal dipolar interaction between magnetic moments. The effect of this energy contribution usually leads to a demagnetization field $\bH_\ssf{dem}$ which tends to suppress the overall magnetization of a finite system. This field can be derived using the magnetostatic approximation in Maxwell's equations where all fields are static and the electric component is vanishing:
%--------------
\begin{equation}
  \label{eqn:BH}
  \nabla\cdot\bB = 0 \qand
  \nabla\times\bH_\ssf{dem} = 0
  \qwith \bB = \mu_0\qty(\bH_\ssf{dem} + \bM).
\end{equation}
%--------------
The curl-less demagnetization field can be expressed as the gradient of a scalar potential: $\bH_\ssf{dem} = - \nabla u$. Then \Eq{eqn:BH} becomes
%--------------
\begin{equation}
  \label{eqn:Mu}
  \nabla^2 u = \nabla\cdot\bM,
\end{equation}
%--------------
where the divergence of $\bM$ on the right hand side can be understood as a magnetic charge. The energy associated with the demagnetization field is given by
%--------------
\begin{equation}
  \label{eqn:Edem}
  E_\ssf{dem} = - {\mu_0\ov 2} \int \bM \cdot \bH_\ssf{dem}(\br) \dd[3]{\br},
\end{equation}
%--------------
where the factor $1/2$ accounts for double counting. It is worth to mention that $E_\ssf{dem}$ involves integration over the entire sample twice, because the demagnetization field $\bH_\ssf{dem}$ itself involves integration of the dipolar fields from the whole sample. Therefore, the demagnetization field is a non-local phenomenon, which makes it computationally very expensive, and it is long-range compared to the exchange energies.

{\bf Heisenberg exchange interaction} - Since electrons are fermions, the full wave function for a system of many electrons has to be antisymmetric. The Coulomb potential energy between any two electrons depends on their spatial separation. Due to the Pauli exclusion principle, parallel spins have the tendency to "repel" each other in space, thus having a larger average spatial separation than antiparallel spins. Consequently, the state of parallel spins has a lower Coulomb energy than that of antiparallel spins. This energy difference is the exchange energy.
This exchange effect can be captured by the Heisenberg model for two spins at site $i$ and $j$:

%--------------
\begin{equation}
  E_\ssf{ex}^{ij} = - J_{ij} \bS_i \cdot \bS_j,
\end{equation}
%--------------
where $J_{ij}$ is the exchange integral, or the exchange energy difference for the spins at the two sites with parallel and antiparallel spin alignment. For cubic or isotropic lattice structures, the coupling constant $J_{ij} = J$, and the total exchange energy is $E_\ssf{ex} = - J \sum_{\avg{ij}} \bS_i \cdot \bS_j$. In the continuum limit, it becomes~\cite{kittel_physical_1949}
%--------------
\begin{equation}
  \label{eqn:Eex}
  E_\ssf{ex} = M_s\int \half A \qty[\nabla\mb(\br)]^2 \dd[3]{\br},
\end{equation}
%--------------
where $A \propto J$ is the exchange constant.
It is evident that to minimize the exchange energy, the magnetic moments tend to be collinear over space such that its gradient vanishes.

{\bf Dzyaloshinskii-Moriya interaction (DMI)} - The exchange effect discussed above is symmetric, \ie the exchange energy remains the same if the two spins are swapped ($\bS_i \leftrightarrow \bS_j$). Dzyaloshinskii~\cite{dzyaloshinsky_thermodynamic_1958} and Moriya~\cite{Moriya1960} predicted that there exists an asymmetric exchange coupling in noncentrosymmetric materials with broken inversion symmetry, and the associated energy changes sign when the two spins are swapped and can be expressed as
%--------------
\begin{equation}
  E_\ssf{DM}^{ij} = \bD_{ij}\cdot\qty(\bS_i\times\bS_j).
\end{equation}
%--------------
where $\bD_{ij}$ is the DM vector, whose direction can be determined using the Moriya rules. There are two basic types of inversion symmetry breaking~\cite{fert_skyrmions_2013}, \ie the bulk type inversion symmetry breaking and the interfacial type inversion symmetry breaking. For the bulk type~\cite{Tok2010}, typically found in materials with B20 structure (such as MnSi, FeCoSi, and FeGe), the DM vector points in the direction connecting site $i$ and $j$, \ie $\bD \parallel \br_{ij} \equiv \br_i - \br_j$. For the interfacial type, mostly investigated in ferromagnetic/heavy metal bilayers (such as Mn/W, Fe/Ir, and Fe/W), the DM vector points in the direction perpendicular to the interface normal and $\br_{ij}$, \ie $\bD \parallel  \bn \times \br_{ij} $. In the continuum limit, the DMI energy contribution is written as
\cite{Abert2019,Moon2013}
%--------------
\begin{subequations}
  \label{eqn:EDMI}
\begin{align}
\mbox{for bulk type DMI:}\quad
E_\ssf{DM} &= - M_s\int \dd[3]{\br} D~\mb\cdot\qty(\nabla\times \mb),
\\
\mbox{for interfacial type DMI:}\quad
E_\ssf{DM} &= -M_s\int \dd[3]{\br} D~\qty[\mb\cdot\nabla(\bn\cdot \mb)-(\nabla\cdot\mb)(\bn\cdot\mb)],
\end{align}
\end{subequations}
%--------------
here $D \propto \abs{\bD}$ is the strength of the DMI,
%\jx{Double check $D$ expression.}
which can be experimentally characterized by spin-wave nonreciprocity \cite{Moon2013, Cor2013} as elaborated in Section \ref{sec:DMI_spin_wave_nonreciprocity}.

{\bf Total Energy} - When keeping the uniaxial anisotropy along $\hbu$ and assuming the bulk type DMI, the total free energy (from here on we use the bulk type DMI in all cases unless specified otherwise) is given by
%--------------
\begin{align}
  \label{eqn:FFM}
  F_\ssf{FM} &= E_\ssf{Zeeman} + E_\ssf{ani} + E_\ssf{dem} + E_\ssf{ex} + E_\ssf{DM} \nn
  &= M_s \int \dd[3]{\br}\qty[
  - \mu_0\mb\cdot\qty(\bH_\ssf{ext} + \half \bH_\ssf{dem})
  - \half K\qty(\mb\cdot\hbu)^2
  + \half A \qty(\nabla\mb)^2
  - D \mb\cdot\qty(\nabla\times \mb) ],
\end{align}
%--------------
where we have assumed the simplest uniaxial anisotropy with $K = K_{u1}$ and ignored all higher orders like $K_{u2}$.

\subsubsection{Free energy for antiferromagnets}

In antiferromagnets, the magnetizations from the two magnetic sublattices point in opposite directions, and consequently the dipolar fields from all magnetic moments compensate each other. Therefore, there is no demagnetization field in antiferromagnets. Apart from this, the energetics for an antiferromagnet is quite similar to that of a ferromanget, except that there are now two antiferromagnetically coupled magnetic sublattices. Due to this difference, the free energy of an antiferromagnet can be written as
%--------------
\begin{align}
  \label{eqn:FAFM1}
  F_\ssf{AFM}\qty[\mb_1(\br),\mb_2(\br)]
  &= M_s\sum_{i=1,2}\int \dd[3]{\br}\qty[
  -\mu_0\mb_i\cdot\bH_\ssf{ext}
  -\half K\qty(\mb_i\cdot\hbu)^2
  +\half A \qty(\nabla\mb_i)^2
  -D \mb_i\cdot\qty(\nabla\times\mb_i) ] \nn
  &+J \int \dd[3]{\br}\mb_1 \cdot\mb_2
  +\bd\cdot \int \dd[3]{\br}\mb_1 \times\mb_2,
\end{align}
%--------------
where $A$ and $J$ are the inhomogeneous and homogeneous Heisenberg exchange coupling constants, and $D$ and $\bd$ are the inhomogeneous and homogeneous DMI constants, respectively \cite{bogdanov_magnetic_2002}. Here the inhomogeneous and homogenous coupling refer to the intra- and inter-sublattice coupling, respectively. Using the net magnetization $\mb = (\mb_1 + \mb_2)/2$ (here $\mb$ is not a unit vector) and the N\'{e}el vector $\bn = (\mb_1 -\mb_2)/2$ ($\mb\cdot\bn = 0$ and $\abs{\mb} \ll \abs{\bn}$ and $\bn$ is approximately a unit vector when $\mb_1$ and $\mb_2$ are close to antiparallel), and ignoring the terms involving the derivatives in $\mb$, the antiferromagnetic free energy \Eq{eqn:FAFM1} can be rewritten as~\cite{landau_statistical_1980}
%--------------
\begin{align}
  \label{eqn:FAFM2}
  F_\ssf{AFM}\qty[\mb(\br),\bn(\br)]
  &= M_s\int \dd[3]{\br}\qty[
  -2\mu_0\mb\cdot\bH_\ssf{ext}
  - K\qty(\bn\cdot\hbu)^2 + J \mb^2 + A \qty(\nabla\bn)^2] \nn
  &- M_s\int \dd[3]{\br} 2\qty[
  D\bn\cdot\qty(\nabla\times\bn)
  +\bd\cdot\qty(\mb\times\bn)].
\end{align}
%--------------

\subsection{Static magnetic textures}

%-----------------------------------
\begin{figure}[t]
\centering
\includegraphics[width=\textwidth]{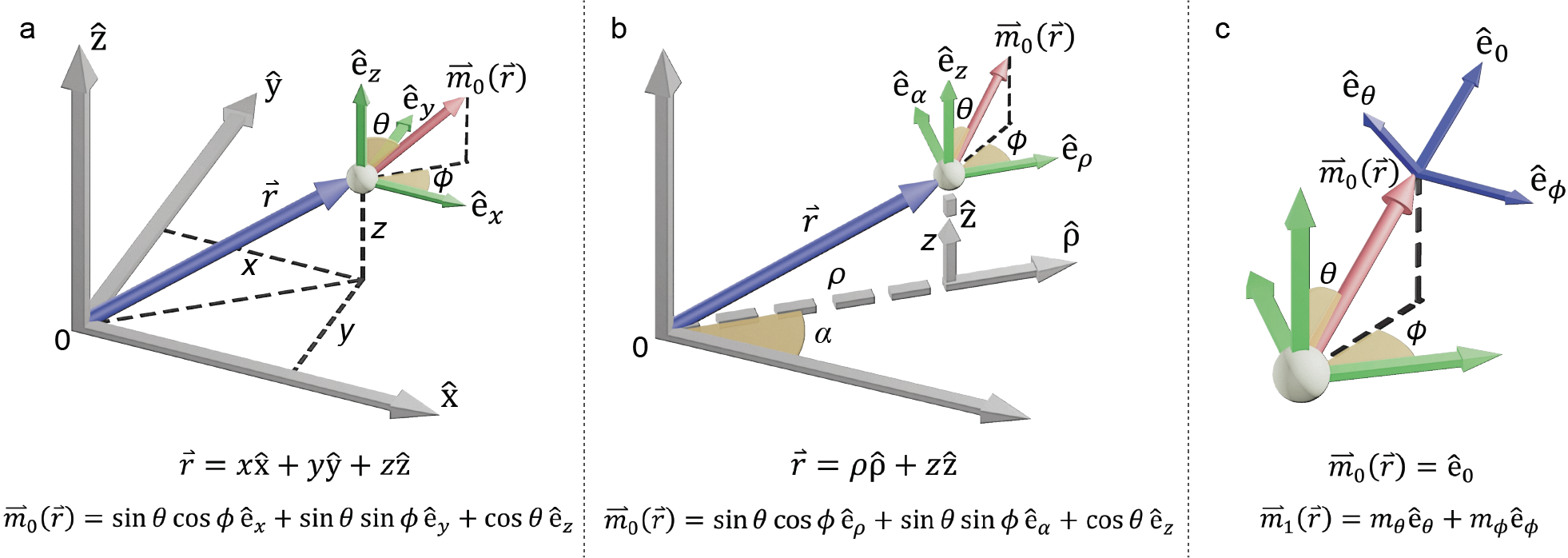}
\caption{The definition of coordinate systems: (a) the Cartesian coordinates for the real position space ($\hbx,\hby,\hbz$) and spherical coordinates for spin space with polar $\theta$ and azimuthal $\phi$ angle defined with respect to the $\hbe_z$ and $\hbe_x$ axis. (b) the Cylindrical coordinates for the real position space and spherical coordinates for spin space with polar $\theta$ and azimuthal $\phi$ angle defined with respect to the $\hbe_z$ and $\hbe_\rho$ axis. (c) The local frame for magnetization excitation $\mb_1(\br,t) = m_\theta(\br,t)\hbe_\theta + m_\phi(\br,t)\hbe_\phi$ and the static background $\mb_0(\br) = \hbe_0$.}
\label{fig:coordinates}
\end{figure}
%-----------------------------------

\begin{figure}[t]
\centering
\includegraphics[width=\textwidth]{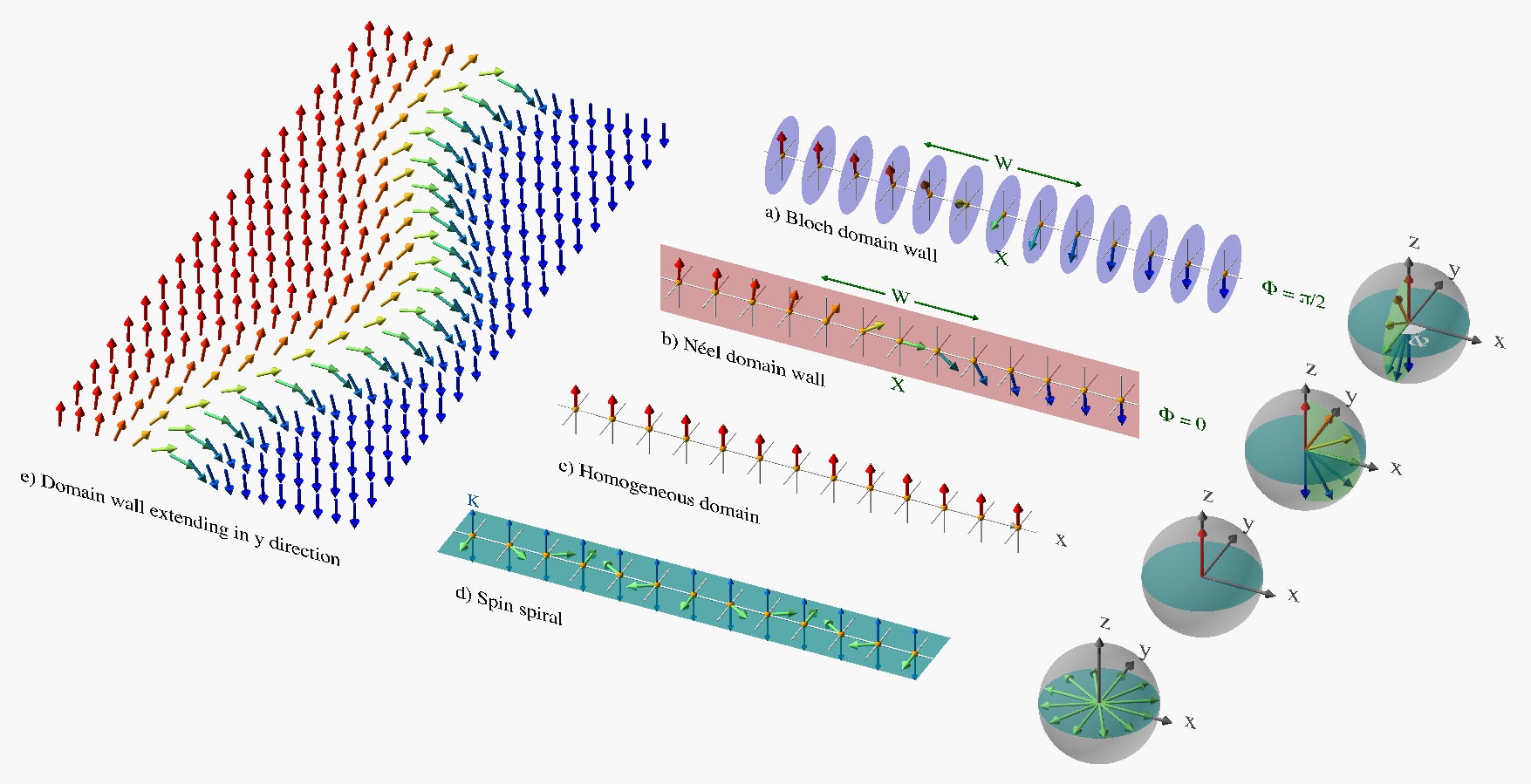}
\caption{In a homogeneous magnetic domain, all moments point in the same direction along the magnetic anisotropy axis ($\hbe_z$). For a \ang{180} N\'{e}el type magnetic domain wall, the magnetization rotates from north pole ($+\hbz$) to south pole ($-\hbz$) via a longitude line with $\Phi = 0$ (in the $\hbx$-$\hbz$ plane), and $\Phi = \pi/2$ (in the $\hby$-$\hbz$ plane) for a Bloch type domain wall. Here $X, W$ denote the position and width of the domain wall. And in a spin spiral, the magnetization rotates constantly within an easy plane. The top right inset shows a curved domain wall in a 2-dimensional plane.}
\label{fig:FM_texture}
\end{figure}

Because of the competing energy contributions outlined above, the total energy landscape typically has multiple local minima for specific directions of $\mb_0(\br)$, resulting in the richness of magnetic textures.
As shown in \Figure{fig:coordinates}(a,b), the position vector $\br$ can be expressed in either Cartesian or Cylindrical coordinates, and the magnetization vector $\mb(\br)$ can be expressed in polar coordinates with the polar $\theta(\br)$ and azimuthal $\phi(\br)$ angle defined as associated with different coordiantes:
%--------------
\begin{subequations}
\begin{align}
  \br = x\hbx + y \hby + z \hbz &\qand
  \mb_0(\br) = \sin\theta(\br)\cos\phi(\br)\hbe_x + \sin\theta(\br)\sin\phi(\br) \hbe_y + \cos\theta(\br)\hbe_z, \\
  \br = \rho\hbrho + z \hbz &\qand
 \mb_0(\br) = \sin\theta(\br)\cos\phi(\br)\hbe_\rho + \sin\theta(\br)\sin\phi(\br) \hbe_\alpha + \cos\theta(\br)\hbe_z,
\end{align}
\end{subequations}
%--------------
These two angles uniquely define the magnetic texture in space. When considering only the magnetic anisotropy and magnetic exchange energies, the lowest energy of such magnetic system is realized as a homogeneous magnetic domain, where all magnetic moments are pointing in the same direction along the magnetic anisotropy axis as shown in \Figure{fig:FM_texture}. For samples with a finite size, the shape anisotropy and the energy of the dipolar magnetic field generated outside the magnetic volume also play an important role in the formation of magnetic textures~\cite{Wegrowe1999,Ralph2000,Beach2005,Parkin2008,Jung2008,Sato:2019cn}. Only very limited forms of non-trivial (inhomogeneous) texture are stable as listed below.

%-----------------------------------
%-----------------------------------

{\bf Magnetic domain wall} - The simplest non-trivial magnetic texture is a magnetic domain wall. Due to the dipolar interaction, a ferromagnet of finite size tends to break into multiple domains, where the magnetization points in different directions in each domain~\cite{hubert_magnetic_1998}. The transition region, called the magnetic domain wall, forms at the boundaries between two different magnetic domains. The magnetization direction continuously changes from one domain to another across the domain wall, as shown in \Figure{fig:FM_texture}(a,b). The length scale of such a domain wall, or the domain wall width, is determined by the competition between the magnetic anisotropy energy and the magnetic exchange energy. The exact form of the domain wall texture can be calculated using the free energy expression in \Eq{eqn:FFM}, where we only keep the uniaxial anisotropy (along $\hbe_z$), exchange, and DMI for simplicity.

We assume a \ang{180}-domain wall along the $x$ direction with $\theta(\br) = \theta(x), \phi(\br) = \phi(x)$ and $\theta(x\ra -\infty) = 0, \theta(x\ra + \infty) = \pi$.
In the absence of DMI ($D = 0$), one type of texture minimizing the free energy corresponds to
%--------------
\begin{equation}
  \label{eqn:DWWalker}
  \theta(x) = 2\arctan\qty[\exp\qty(\frac{x-X}{W})] \qand
  \phi(x) = \Phi = \const
\end{equation}
%--------------
where $W = 2\pi\sqrt{A/K}$ is the domain wall width. The domain wall profile given by \Eq{eqn:DWWalker} is called the Walker profile~\cite{lilley_lxxi_1950}.
When  $\Phi = \pi/2$ or $3\pi/2$, it is the Bloch type domain wall as shown in \Figure{fig:FM_texture}(a), where the magnetization rotates in the $y$-$z$ plane. When $\Phi = 0$ or $\pi$, it is the N\'{e}el type domain wall as shown in \Figure{fig:FM_texture}(b), where the magnetization rotates in the $x$-$z$ plane or the rotation axis ($\hbe_y$) is perpendicular to the domain wall direction ($\hbe_x$). In general, there can be a domain wall of mixed type for $\Phi$ equals to another constant value. In the presence of bulk (interfacial) type DMI ($D \neq 0$), only one solution among the \Eq{eqn:DWWalker} survives with $\Phi = \pi/2$ ($\Phi = 0$), \ie the Bloch (N\'{e}el) type domain wall, and the domain wall also has a topological chiral character~\cite{schoenherr_topological_2018, chen_novel_2013, chen_tailoring_2013}.

The above described domain wall is point-like object in a one-dimensional magnetic chain. In a two dimensional magnetic thin film, however, the magnetic domain wall extends in the other dimension as shown in \Figure{fig:FM_texture}(e), and becomes a string-like structure.

In antiferromagnets, domain walls have a very similar structure as in the ferromagnetic case, but with two magnetic sublattices whose magnetization point in approximately opposite directions.

{\bf Magnetic spin spiral} - Periodic magnetic structures like a spin spiral may also be stabilized. One particular case is in an easy-plane magnet as shown in  \Figure{fig:FM_texture}, where a spin spiral with a constant pitch may be stabilized in a one-dimensional magnet with hard-axis along $\hbz$ (and $x$-$y$ plane is the easy plane), when the magnetic moments at the two ends are pinned externally (along $-\hby$ in this case). Such an easy-plane magnet can also be realized qualitatively with the help of DMI, because the DM vector represents an effective hard-axis, therefore the plane perpendicular to the DM vector is an effective easy-plane \cite{dzyaloshinskii_theory_nodate,bode_chiral_2007-2}.

Thus, spin spiral structures can also be formed in easy-axis magnets if the DMI is strong enough to overcome the easy-axis anisotropy.
It is also possible to realize spin spirals in systems without DMI in rutile type crystals such as MnO$_2$ \cite{yoshimori_new_1959}.

%-----------------------------------
\begin{figure}[t]
  \centering
\includegraphics[width=\textwidth]{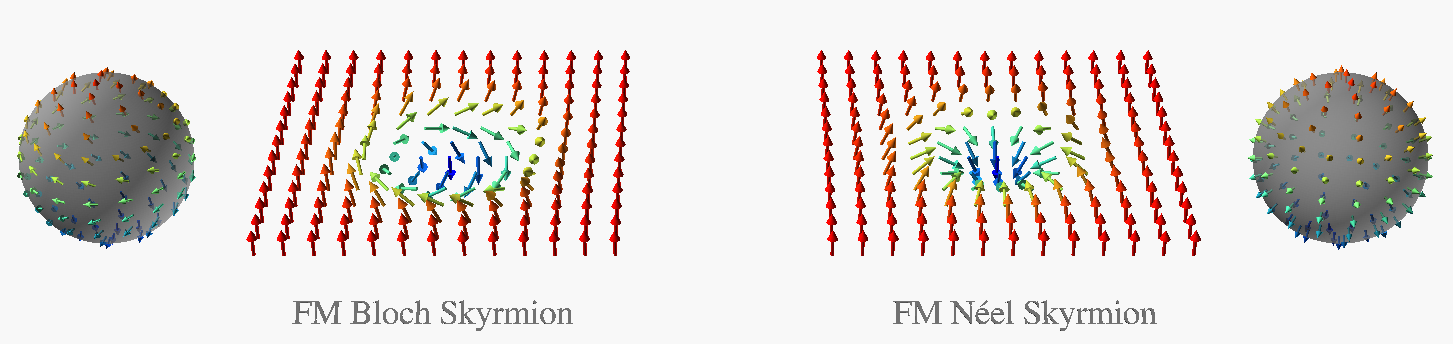}
\caption{The Bloch and N\'{e}el type magnetic Skyrmions.}
\label{fig:2d_texture}
\end{figure}
%-----------------------------------

{\bf Magnetic Skyrmion} - Another type of stable magnetic texture formed in two dimensional films is the magnetic Skyrmion, where the magnetization rotates by \ang{180} from its core to its perimeter, as shown in the right panel of \Figure{fig:2d_texture}. Such structures can be stabilized by dipolar interaction or DMI. The spin texture has a rotational symmetry around the core of the Skyrmion, for which we adopt cylindrical coordinates for the magnetization at $\br = (\rho, \alpha, z)$:
%--------------
\begin{equation}
 \mb_0(\br) = \sin\theta(\br)\cos\phi(\br)~\hbe_\rho + \sin\theta(\br)\sin\phi(\br)~\hbe_\alpha + \cos\theta(\br)~\hbe_z,
\end{equation}
%--------------
where $\theta(\br) = \theta(\rho), \phi(\br) = \phi(\rho)$ are only a function of the radius measured from the Skyrmion core. The minimization of the free energy yields the possible Skyrmion-type solutions with $\theta(\rho), \phi(\rho)$ satisfying:
%--------------
\begin{equation}
  \label{eqn:fSk}
  0 = \rho B\sin\theta + \rho K\sin2\theta + 2 A \qty(-\rho\theta''-\theta'+\frac{\sin2\theta}{2\rho})
  \pm D \sin^2\theta
\end{equation}
%--------------
and  $\phi = \pm {\pi\ov 2}$~\cite{XSWang2018} for Bloch Skyrmions~
\footnote{If the interfacial type DMI is used, the stable structure would be the N\'{e}el type Skyrmion with $\phi = 0, \pi$.} with the boundary condition as $\theta(\rho=0) = 0$ and $\theta(\rho\ra\infty) = \pi$. Similar to the magnetic domain walls, there are also two types of Skyrmions, the N\'{e}el type and Bloch type as shown in \Figure{fig:FM_texture}, which can be stabilized by the interfacial type DMI and bulk type DMI, respectively. Here we only describe the simplest model. Notably, there are exceptions~\cite{kezsmarki_ne-type_2015} where N\'{e}el-type Skyrmions can be stablized in non-chiral but polar crystals with $C_{nv}$ symmetry. However, here we only describe the simplest model and more sophisticated cases as Ref.~\cite{kezsmarki_ne-type_2015} are beyond our simple theoretical consideration.

Magnetic Skyrmions can also arrange in a hexagonal pattern forming a Skyrmion lattice~\cite{Muh2009}. In bulk, Skyrmions can extend in the depth direction forming Skyrmion tubes or magnetic bobbers when the texture terminates in the bulk at a chiral Bloch point~\cite{zheng_experimental_2018}.

\begin{figure}[t]
  \centering
\includegraphics[height=6.0cm]{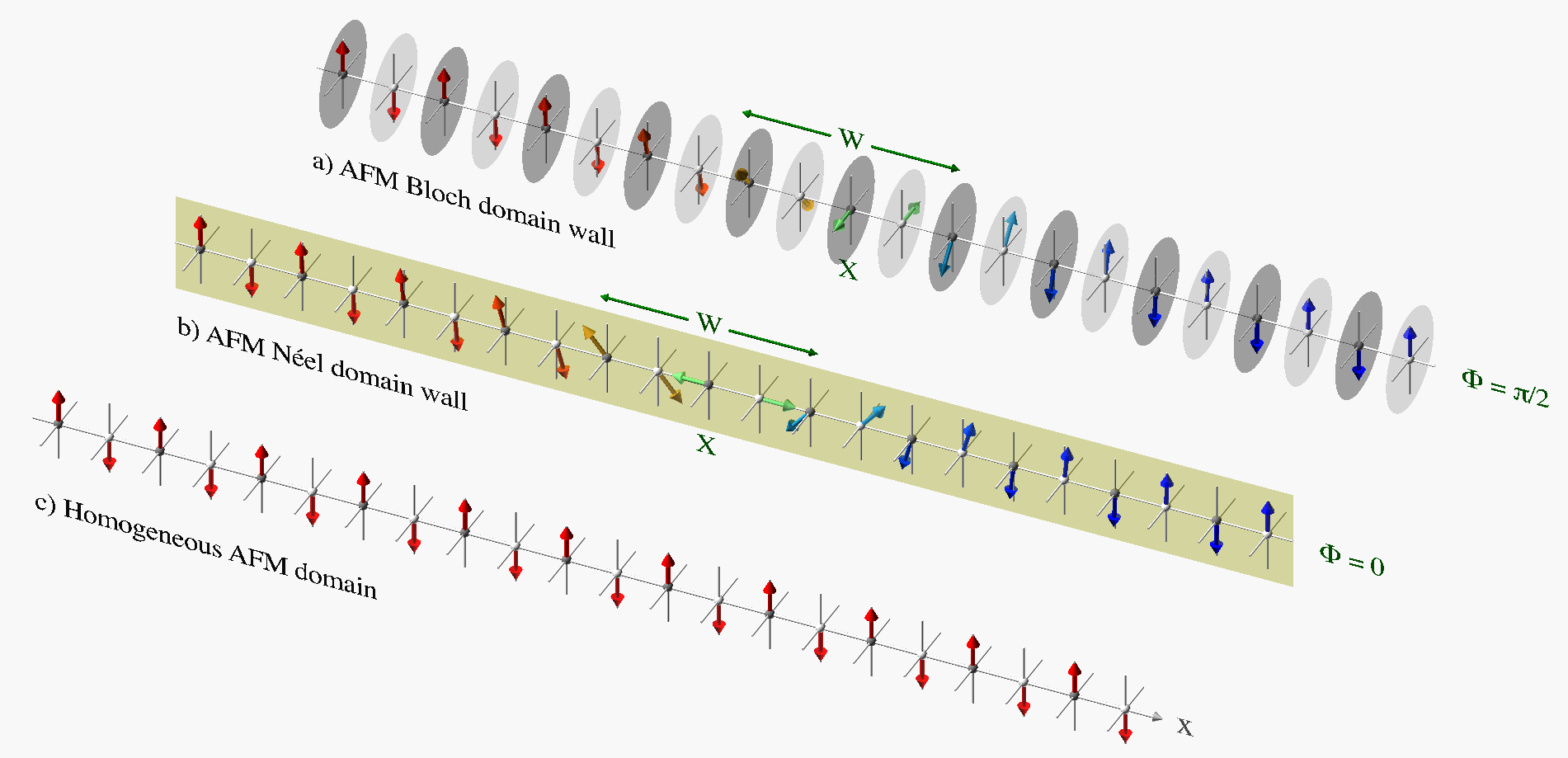}
\hspace{0.2cm}
\includegraphics[height=6.0cm]{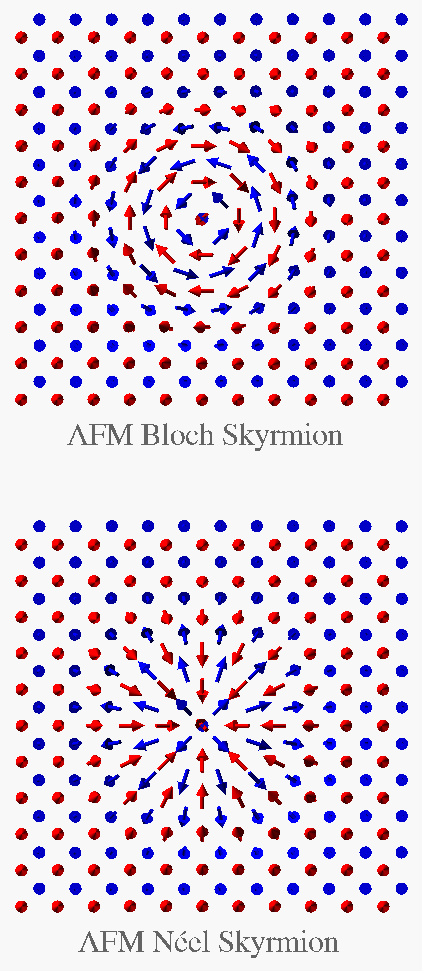}
\caption{The Bloch and N\'{e}el type domain wall (left) and Skyrmion (right, top view) textures in antiferromagnets. The magnetic moments from two magnetic sublattices are represented by the full color and half-transparent color arrows.}
\label{fig:AFM_texture}
\end{figure}

{\bf Antiferromagnetic domain wall and Skyrmion} - In antiferromagnets (AFM), magnetic textures such as domain walls or Skyrmions (see \Figure{fig:AFM_texture}) can also be stabilized. Most of the descriptions of AFM textures are the same as those in ferromagnetic textures. One major simplification in an AFM texture is that the demagnetization field can be ignored. For the static uniform AFM, magnetization from two sublattices compensate each other completely. However, in static AFM textures, there can be a non-vanishing intrinsic magnetization \cite{tveten_intrinsic_2016}.

%-----------------------------------
%-----------------------------------

\subsection{Landau-Lifshitz-Gilbert phenomenology for magnetization dynamics}\label{sssec:LLG}

The minimization of the free energy determines the static magnetization profile in a system. Upon the static profile, the magnetization may develop dynamic motion. The magnetization dynamics can be captured by the Landau-Lifshitz-Gilbert phenomenology as described below.

{\bf Ferromagnet} - In ferromagnets, the equation of motion for $\mb(\br,t)$ is the Landau-Lifshitz-Gilbert (LLG) equation:
\cite{bloch_zur_1932,landau_3_1992,gilbert_phenomenological_2004}

%--------------
\begin{equation}
  \label{eqn:LLG}
  \dv{\mb}{t} = -\gamma \mb \times\bB_\ssf{eff} + \alpha \mb\times\dv{\mb}{t},
\end{equation}
%--------------
where $\gamma$ is the gyromagnetic ratio, $\alpha$ is the phenomenological Gilbert damping constant, and the effective magnetic field $\bB_\ssf{eff}$ can be obtained by the functional derivative of the free energy \Eq{eqn:FFM} with respect to the magnetization within a small volume $\Delta V$:

%--------------
\begin{equation}
  \label{eqn:Heff}
  \bB_\ssf{eff}(\mb) = - {1\ov M_s \Delta V}\frac{\delta F}{\delta \mb}
  = \mu_0\bH_\ssf{ext} + K\qty(\mb\cdot\hbu)\hbu + A \nabla^2\mb + D\nabla\times\mb,
\end{equation}
The LLG equation \Eq{eqn:LLG} describes the time evolution of the angular momentum $\bL = -\bM/\gamma$ in the presence of a torque given by $\bT = \bM\times\bH_\ssf{eff}$, where the minus sign is due to the fact that the electron has a negative charge such that its magnetic moment points in the opposite direction of the associated angular momentum. Spin waves in ferromagnets with zero wave vector do not invoke exchange interaction such that their frequencies are determined mainly by the anisotropy and Zeeman energy and are typically in the gigahertz (GHz) range.

{\bf Antiferromagnet} - In antiferromagnets, where the magnetization for each magnetic sublattice is denoted by $\mb_{1,2}(\br,t)$,  the equation of motion can be phenomenologically described by two coupled Landau-Lifshitz-Gilbert (LLG) equations: \cite{keffer_theory_1952}
%--------------
\begin{equation}
  \label{eqn:LLG-AFM}
  \dv{\mb_i}{t} = -\gamma \mb_i \times\bB_\ssf{eff}^i + \alpha \mb_i\times\dv{\mb_i}{t}
  \qfor i = 1, 2,
\end{equation}
%--------------
where an effective field on the sublattice $\mb_i$ can be described as
%--------------
\begin{equation}
  \label{eqn:Heffi}
  \bB_\ssf{eff}^i = - {1\ov M_s\Delta V}{\delta F_\ssf{AF}\ov\delta \mb_i}
  = \mu_0\bH_\ssf{ext}
  + K\qty(\mb_i\cdot\hbu)\hbu + A \nabla^2\mb_i + D\nabla\times\mb_i
  - J \mb_{\bar{i}} + \bd \times\mb_{\bar{i}},
\end{equation}
%--------------
with $\bar{1} \equiv 2$ and $\bar{2} \equiv 1$.
The LLG equations \Eq{eqn:LLG-AFM} are coupled because the effective field for $\mb_1$ contains $\mb_2$ and vice versa. The frequencies of spin waves in antiferromagnets are typically much higher compared to their ferromagnetic counterpart because the inter-sublattice exchange interaction is involved even for spin waves with zero wavevector ($k=0$), which increases the spin-wave frequencies by a factor of $\sqrt{J/K}$ into the terahertz (THz) regime.

\Eq{eqn:LLG-AFM} can be rewritten in terms of the net magnetization $\mb \equiv (\mb_1 + \mb_2)/2$,
and the N\'{e}el vector $\bn \equiv (\mb_1 - \mb_2)/2$.
\footnote{For small amplitude excitation, $\abs{\bn} \simeq 1$ can be regarded as a unit vector.}
With the constraint that $\mb\cdot\bn = 0$, the net magnetization $\mb = \dot{\bn}\times\bn/J$ becomes a slave quantity of the N\'{e}el order $\bn$, whose dynamics is governed by
%--------------
\begin{equation}
  \label{eqn:LLG-n}
    J^{-1}\bn \times \qty(\ddot{\bn}\times\bn) = \bn \times \qty(\gamma \bB_\bn \times \bn)
    - \alpha \bn \times \qty(\dot{\bn}\times\bn)
\end{equation}
%--------------
where $\bH_\bn$ is the effective field for the stagger order $\bn$:
%--------------
\begin{equation}
  \label{eqn:Hn}
  \bB_\bn = - {1\ov M_s\Delta V}{\delta F_\ssf{AFM}\ov \delta \bn}
  = \half\qty(\bB_\ssf{eff}^1 - \bB_\ssf{eff}^2)
  = K (\bn\cdot\hbu)\hbu + A \nabla^2\bn + D \nabla\times\bn - J \bn + \bd \times \bn.
%  = \hbe_\beta \hH_{\beta\gamma} n_\gamma.
\end{equation}
%--------------

{\bf Ferrimagnet} - As seen in \Eqs{eqn:LLG}{eqn:LLG-n}, the dynamics for the ferromagnet and antiferromagnet are the first order and second order in time derivative, respectively. In ferrimagnets, which possess both magnetic order and N\'{e}el order, the dynamics contains both first and second order time derivatives~\cite{Serga2010,Barker2016,Nambu2020}. The low-energy dynamics of the collinear anti-ferromagnet can be described using the net magnetization vector $\bn$ as
\cite{oh_coherent_2017,kim_self-focusing_2017,kim_tunable_2019}

%--------------
\begin{equation}
  \label{eqn:LLG-ferri}
  s\dot{\bn} + \rho_{0} \bn\times\ddot{\bn}
  = \bn \times \bB_n - \alpha \bn \times\dot{\bn} ,
\end{equation}
%--------------
where $s$ is the net spin density along the $\bn$, $\rho_{0}$ is the inertia associated with the dynamics of $\bn$. This equation nicely reduces to the LLG equation \Eq{eqn:LLG} for ferromagnets when $\rho_{0} = 0$, and to \Eq{eqn:LLG-n} for antiferromagnets when $s = 0$.

{\bf Linearized LLG} - Due to the complexity of the magnetic free energy in the presence of spin textures an analytic solution of the LLG equation is often impossible. Thanks to the increase of computational power and GPU accelerated parallelized algorithms the most straightforward approach to solve the LLG equation \Eq{eqn:LLG} or \Eq{eqn:LLG-AFM} for the study of spin waves in complex spin textures is by micromagnetic simulations (using the packages such as OOMMF or MuMax, \etc). Because the effective field $\bB_\ssf{eff}(\mb)$ is a function of $\mb$, the LLG equation is a non-linear equation. A fully analytic treatment of the non-linear phenomena~\cite{Demi_Nonlinear2001,Wu_Nonlinear2007,Slavin_Nonlinear2009,Kam_Nonlinear2011} requires complicated perturbation theory and is described elsewhere~\cite{gurevich_magnetization_1996}. However, when one is interested in small amplitude excitation, the nonlinearity can be neglected and the equation of motion becomes linear and the analytic solution is less tedious.

A standard approach is to separate the static and dynamic components of the order parameter.
In ferromagnets, the magnetic order is separated as: $\mb(\br,t) = \mb_0(\br) + \mb_1(\br,t)$, where the dynamic component is called the spin wave and is usually assumed to be much smaller than the static component: $\mb_1 \ll \mb_0$. To guarantee the magnitude of the magnetization to be constant, we also require the dynamic component to be transverse to the static component:  $\mb_1 \perp \mb_0$.
In antiferromagnets, we distinguish the dynamic excitation of $\bn$ from its static background as: $\bn(\br,t) = \bn_0(\br) + \bn_1(\br,t)$ with $\bn_1 \ll \bn_0$ and $\bn_1 \perp \bn_0$. In doing so, the linearization of LLG equations is obtained by keeping only the zeroth and first order terms of $\mb_1$ or $\bn_1$ in \Eq{eqn:LLG} or \Eq{eqn:LLG-n}.

\subsection{Spin wave}\label{sec:spin_wave}

Although the spin wave dynamics can be well described by the Landau-Lifshitz-Gilbert phenomenology described above, the physical phenomenon may be drastically different when comes to different magnetic systems. In this subsection, we present the spin wave behaviors in several typical magnetic systems: the mostly widely encountered case of a ferromagnetic thin film with homogenous equilbirium magnetization, and the various ferro- or antiferro-magnetic textures as the central topic of this review.

\subsubsection{Spin waves in ferromagnetic thin films}\label{Sec:spin waves in ferromagnetic thin films}

In the context of magnonics, the most studied sample geometry is an in-plane magnetized thin film, where the static magnetization $\mb_0$ points uniformly in the film plane. For such a system, there are typically two regimes for spin wave dynamics~\cite{Kal1986}: the exchange regime for spin waves with short wavelengths where the exchange interaction dominates, and the dipolar (demagnetization field) regime for spin waves with long wavelengths where the dipolar interaction dominates. For thin films with the thickness of tens or hundreds of nanometers, the boundary between the exchange and dipolar regimes is approximately $kd \sim 1$~\cite{Len2011}, where $k$ is the in-plane wave vector and $d$ is the film thickness, \ie the dipolar (exchange) interaction dominates when the wavelength is much longer (shorter) than the film thickness. Experimentally, the excitation of dipolar spin waves is straightforward, typically via a microwave antenna, such as a stripline~\cite{Demi2009} or a coplanar waveguide~\cite{Yu2014}. However, exchange spin waves are difficult to excite because of their short (sub-micrometer) wavelengths. Some possible exciting methods using, \eg a sub-micrometer grating~\cite{Yu2016,Liu2018} or via a magnetic vortex~\cite{Wintz2016,Die2019} have been demonstrated. The experimental techniques used to excite and detect exchange spin waves are introduced in Sections~\ref{sec:PSWS} and \ref{sec:TR-STXM}.

\begin{figure}[t]
  \centering
\includegraphics[height=6.5cm]{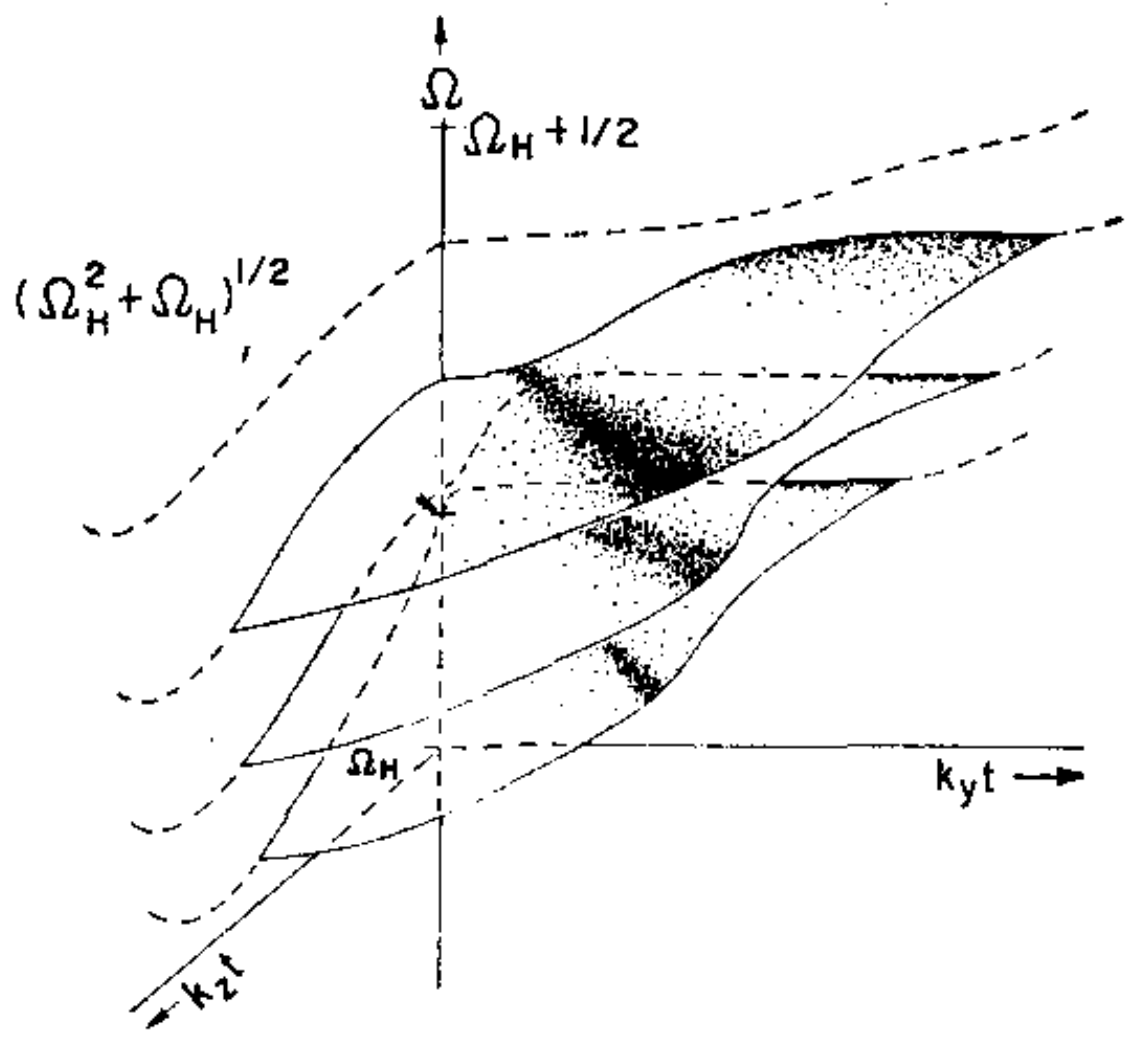}
\includegraphics[height=6.5cm]{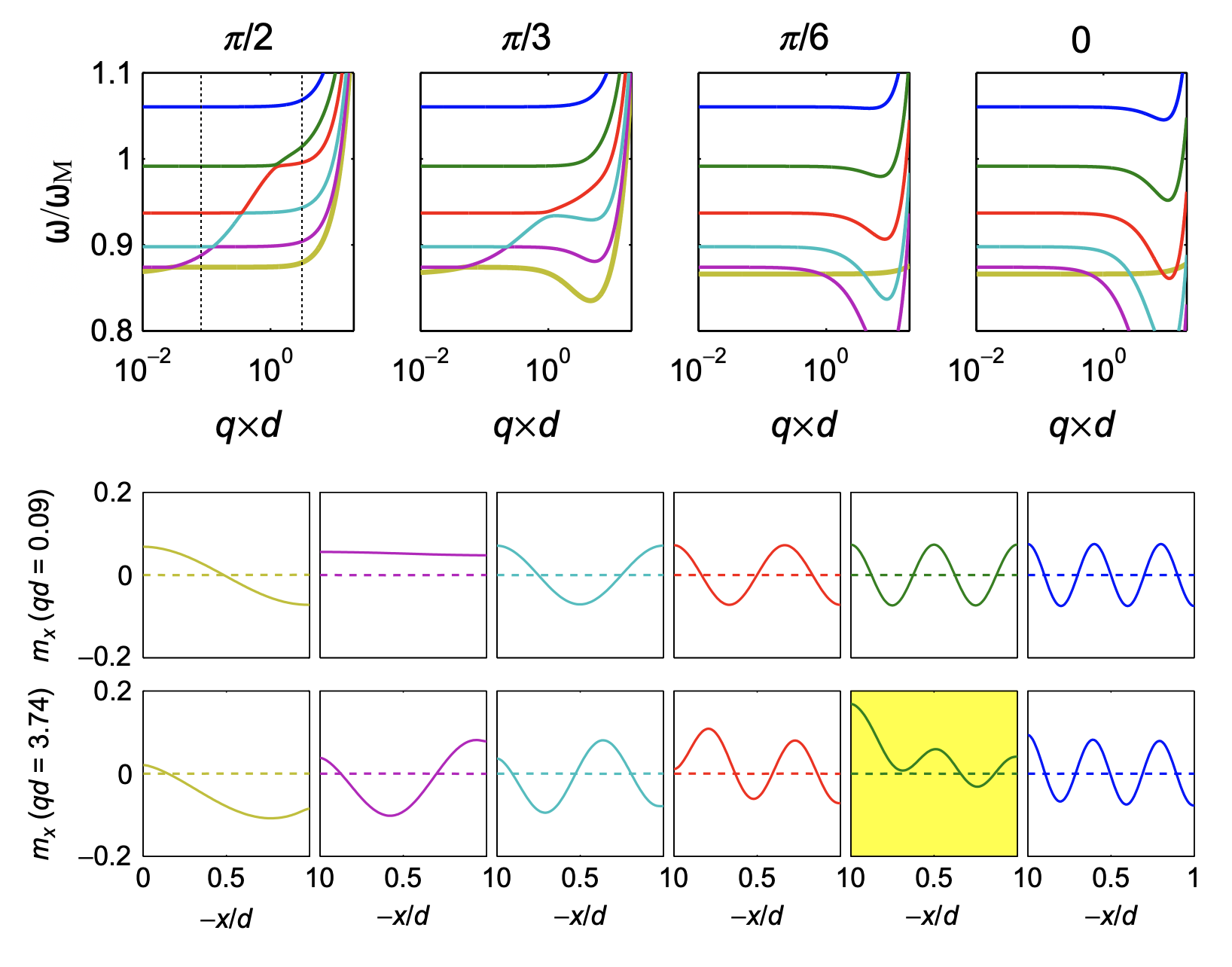}
\caption{Left: the schematic view of the spin-wave dispersion manifold in a magnetic thin film with magnetization pointing along the $\hbz$ direction. Reproduced from Ref.~\cite{de_wames_dipole-exchange_1970}. Top right: the slicing cut of the manifold for various orientation of magnetization in a magnetic thin film of thickness $d = 0.61~\mu$m (the angle is between the magnetization and wavevector $\bk$). Bottom right: the spin-wave profile in the thickness direction for $kd=0.09$ and $3.74$ respectively. Reproduced from Ref.~\cite{wu_book_2013}}.
\label{fig:dsw}
\end{figure}

The spin-wave dispersions in both regimes are well understood~\cite{eshbach_surface_1960,damon_magnetostatic_1961,camley_surface_1978,de_wames_dipole-exchange_1970,Kal1986,Hil1990,hurben_theory_1995}. Here we only briefly discuss the spin-wave modes in magnetic thin films, a more thorough discussion on this matter can be found elsewhere~\cite{Stancil,gurevich_magnetization_1996,xiao_chapter_2013}.
In the exchange regime, the spin-wave dispersion is relatively simple as $\omega(\bk) = \omega_0 + \gamma A \abs{\bk}^2$, where $\omega_0$ is the spin-wave gap \ie the ferromagnetic resonance (FMR) frequency with $k=0$, $A$ is the exchange constant, and $\bk = (k_y, k_z)$ is the in-plane wavevector of the mode.
In the dipolar regime, the spin-wave dispersion is complicated because of the anisotropic nature of the dipolar (demagnetization) fields, depending on the angle $\theta$ between the wavevector and the magnetization. The typical spin-wave dispersion in a ferromagnetic thin film is presented in \Figure{fig:dsw}(left), clearly showing that spin waves propagating along $z$ and $y$ are drastically different. Slicings of these dispersions along different propagating directions are shown in \Figure{fig:dsw}(top right). When the wavevector is parallel to the magnetization ($\theta = 0$), the spin wave modes are the backward volume magnetostatic waves (BVMSWs), with different branches corresponding to the standing waves in the film thickness direction. When the wavevector is perpendicular to the magnetization ($\theta = \pi/2$), the spin wave modes include the forward volume magnetostatic waves (FVMSWs) and magnetostatic surface waves (MSSWs), also known as the Damon-Eshbach modes~\cite{eshbach_surface_1960}.

The typical spin wave profiles in the film thickness direction are shown in \Figure{fig:dsw}(bottom right) for $kd = 0.09,~3.74$, and the second and the fifth panel in each row are the surface spin wave modes for the corresponding wavevectors. The magnetostatic surface waves owe their name to the fact that their profile decays exponentially along the film thickness with a maximum at the film surface. There are two interesting peculiarities worth mentioning: Firstly, Damon-Eshbach modes are non-reciprocal, i.e. upon inversion of the wavevector the profile of the spin wave is also inverted and the mode is propagating at the opposite surface. Secondly, Damon-Eshbach modes exist only in a narrow angular range around the direction perpendicular to the magnetization, there is no surface spin wave mode propagating along the magnetization direction ($\theta = 0$).

%-----------------------------------
%-----------------------------------

\subsubsection{Spin waves in ferromagnetic textures}\label{sec:spin_wave_ferromagnetic_textures}

Let the full spatial-temporal distribution of the ferromagnetic magnetization be expressed as
\footnote{This frame for the magnetization usually coincides with the frame of the position in real space: $\br = x\hbe_x + y\hbe_y + z\hbe_z$.}
%--------------
\begin{equation}
  \label{eqn:mxyz}
  \mb(\br, t) = m_x(\br,t) \hbe_x + m_y(\br,t) \hbe_y + m_z(\br,t) \hbe_z.
\end{equation}
%--------------
Since we are mainly interested in the dynamic excitation relative to the static texture $\mb_0(\br)$, it is more convenient to define $\mb_0(\br)$ as the local spin quantization axis. Therefore, in a local coordinate system
%--------------
\begin{equation}
  \label{eqn:mrtp}
  \mb(\br, t) = m_0(\br) \hbe_0(\br) + m_\theta(\br,t) \hbe_\theta(\br) + m_\phi(\br,t) \hbe_\phi(\br),
\end{equation}
%--------------
where $\hbe_0(\br) = \mb_0(\br)$ is chosen pointing in the same direction as the local static magnetization, and $\hbe_{\theta, \phi}$ are the two transverse directions with respect to $\hbe_0$ as shown in \Figure{fig:coordinates}(c). In the varying local frame, the magnetization includes the static background $\mb_0(\br)$ and the dynamic part $\mb_1(\br,t)$:
%--------------
\begin{equation}
  \label{eqn:m0m1}
  \mb(\br, t) = \mb_0(\br) + \mb_1(\br,t) \qwith
  \mb_0(\br) = \hbe_0(\br) \qand
  \mb_1(\br,t) = m_\theta(\br,t) \hbe_\theta(\br) + m_\phi(\br,t) \hbe_\phi(\br).
\end{equation}
%--------------
We would like to express the LLG equation in terms of the dynamic components $m_\theta, m_\phi$, which can also be combined into a complex function as $\psi(\br,t) \equiv m_\theta(\br,t) - i m_\phi(\br,t)$. In some typical cases as shown below, the linearized LLG equation can be recast approximately into an effective Schr\"{o}dinger equation as~\cite{Yan2011,Lan2015}
%--------------
\begin{equation}
  \label{eqn:LLGSch}
  i\hbar(1+i\alpha)\pdv{\psi}{t} =
  \qty[ {1\ov 2\mu}\qty(-i\hbar\nabla - \bA)^2 + V]\psi,
\end{equation}
%--------------
where $\mu = \hbar/2\gamma A$ is the effective mass, and $V(\br)$ and $\bA(\br)$ are the constant effective scalar and vector potentials that contain the information about the static magnetic texture $\mb_0(\br)$.
The non-vanishing Gilbert damping parameter $\alpha$ on the left hand side of \Eq{eqn:LLGSch} indicates that there is no probability conservation because of dissipation. Therefore,  spin-wave propagation upon some magnetic texture, as governed by the effective Schr\"{o}dinger equation in \Eq{eqn:LLGSch}, can be understood as an electron moving in certain scalar and vector potentials, that are determined by the magnetic texture.

{\bf Homogeneous magnetic domain} - The simplest case possible is the homogeneous case with the uniaxial axis along $\hbu\parallel \hbz$ and the external field $\bH_\ssf{ext} = H\hbz$ also points in the $\hbz$ direction, then the static magnetization at all locations points in the $\hbz$ direction: $\mb_0(\br) = \hbz$ with $\theta_0 = \phi_0 = 0$. In this case,
%--------------
\begin{equation}
  V = \gamma \hbar \qty(K + H + {D^2\ov 4A})  \qand
  \bA = {\hbar\ov 2A}D\mb_0.
\end{equation}
%--------------
Here, the demagnetization field is ignored for simplicity, and therefore the shape (either film or bulk) is not taken into account at this point. When $\alpha = 0$, the solutions to \Eq{eqn:LLGSch} are simply plane waves
%--------------
\begin{equation}
  \psi(\br) = m_\theta - i m_\phi = C e^{i(\omega t - \bk\cdot\br)}
%  \qwith \omega = \gamma \qty(K + H + {D^2\ov 4A}) + \gamma A \abs{\bk - {D\ov 2A}\mb_0}^2,
  \qwith \omega = \gamma \qty(K + H + {D^2\ov 2A}) + \gamma A \abs{\bk}^2 - \gamma D \bk\cdot\mb_0,
\end{equation}
%--------------
where the second expression is the quadratic spin wave dispersion $\omega(\bk)$ for a homogeneous magnetic domain in bulk. Due to DMI, spin waves of a certain frequency have different wave vector when propagating parallel or anti-parallel to $\mb_0$. As expected for a simple ferromagnet, the spin wave solution above is of right-circular polarization, where the $m_\theta$ and $m_\phi$ components are of the same amplitude but out of phase by $\pi/2$.
For the more general case where the external magnetic field misaligns with the anisotropy axis, the LLG equation does not reduce so nicely to a Schr\"{o}dinger-like equation, but involves both $\psi$ and $\psi^*$, meaning that both right- and left-circular polarization are involved, and the magnetization precession becomes elliptical. A prominent example is an in-plane magnetized thin film, where the precession is strongly elliptical due to the shape anisotropy.

\begin{figure}[t]
\includegraphics[width=\textwidth]{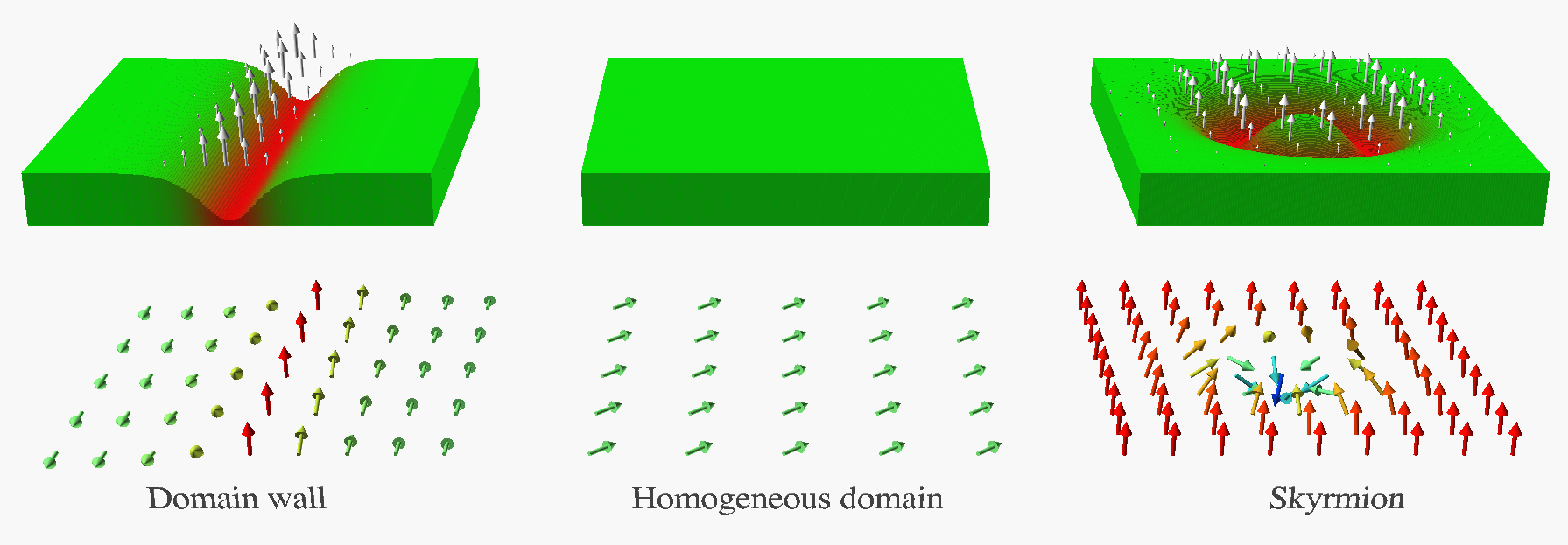}
\caption{The effective scalar potential and effective magnetic field for the domain wall, homogeneous domain and Skyrmion.
}
\label{fig:VB_Texture}
\end{figure}

{\bf Magnetic domain wall} - For simplicity, let's consider a Bloch type magnetic domain wall in the absence of an external magnetic field and dipolar field ($\bH_\ssf{ext} = \bH_\ssf{dem} = 0$), assuming that the domain wall is along the $x$ direction and the magnetization rotates from $+\hbz$ to $-\hbz$ (see \Figure{fig:FM_texture}). In this scenario, $\mb_0(\br) = (0,\sin\theta_0(x),\cos\theta_0(x))$ with $\phi_0(\br) = \pi/2, \theta_0(-\infty) = 0$ and $\theta_0(+\infty) = \pi$. The stability of the domain wall texture requires:
%--------------
\begin{equation}
  \label{eqn:dw_stable}
  K \sin\theta_0\cos\theta_0 - A\theta''_0 = 0
  \qRa \sqrt{A/K}~\theta'_0 = \sin\theta_0,
\end{equation}
%--------------
which gives the same Walker domain wall profile as in \Eq{eqn:DWWalker}, even in the presence of the DMI. \footnote{It should be noted that the domain wall profile may not be the simple Walker profile for other types of domain walls shown in \Figure{fig:FM_texture}.}
Under the approximation that $D/w\ll K$ for wide domain walls or weak DMI, \Eq{eqn:LLG} reduces again to an effective Schr\"{o}dinger equation as \Eq{eqn:LLGSch}, but with spatially varying effective scalar and vector potentials:
%--------------
\begin{equation}
  V(x) = \gamma \hbar \qty[K\cos2\theta_0(x) + {D^2\ov 4A}]  \qand
  \bA(x) = {\hbar\ov 2A}D\mb_0(x).
\end{equation}
%--------------
As shown in the left panel of \Figure{fig:VB_Texture}, the scalar potential has a well defined minimum across the domain wall, forming a potential well. The curl of the effective vector potential gives the effective magnetic field:
%\jx{Check if the magnetic field can be in plane.}
%--------------
\begin{equation}
  \bB(x) = \nabla\times\bA(x) = {\hbar\ov 2A}D\mb_0(x)\theta'_0(x),
\end{equation}
%--------------
which points in the same direction as the magnetization, but its magnitude is proportional to the magnetization direction gradient, thus concentrates in the domain wall region. The perpendicular component of the effective magnetic field for a \ang{180} Bloch domain wall is plotted as arrows in \Figure{fig:VB_Texture}. The spin wave scattering behavior by a domain wall can thus be understood as an "electron" moving in a potential landscape and magnetic field as shown in the left panel of~\Figure{fig:VB_Texture}. This non-trivial effective potential well along the domain wall gives rise to possible bound spin-wave states beneath the bulk spin-wave gap localized near the domain wall~\cite{Gar2015,Lan2015,Wagner2016,Wintz2019,SF_Lee2020}. The effective magnetic field due to the DMI leads to a chiral behavior of such bound spin-wave states when propagating along the domain wall~\cite{Lan2015}.

%-----------------------------------
%-----------------------------------

{\bf Magnetic Skyrmion} - A more complicated example is the magnetic Skyrmion, where the magnetization profile has rotational symmetry around the Skyrmion core. Because of this symmetry, it is more convenient to use cylindrical coordinates (see \Figure{fig:coordinates}(b)) 
The rotational symmetry of a Skyrmion means that the polar and azimuthal angle of the magnetization $\theta(\br) = \theta(r)$ and $\phi(\br) = \phi(r)$ only depend on the radius $r = \abs{\br}$.

We consider a Skyrmion with
%--------------
\begin{equation}
  \mb_0(\rho, \alpha, z) = \sin\theta_0(r)\cos\phi_0(r)\hbe_\rho + \sin\theta_0(r)\sin\phi_0(r)\hbe_\alpha + \cos\theta_0(r) \hbe_z
\end{equation}
%--------------
where $\theta_0(r=0) = 0$ and $\theta_0(r=+\infty) = \pi$. Assuming the external field points in the $\hbe_z$ direction, then the Skyrmion-like texture stabilizes for a Bloch type Skyrmion with $\phi_0(r) = \pi/2, \phi'_0 = \phi''_0 = 0$ and $\theta_0(r)$ is determined by
\footnote{$\phi_0 = -\pi/2$ also satisfies the first equation, but this solution corresponds to a unstable solution.}
%--------------
\begin{equation}
  0 = H\sin\theta_0 + K \sin\theta_0\cos\theta_0
  + A \qty({\sin\theta_0\cos\theta_0\ov\rho^2} - {1\ov\rho}\theta'_0 - \theta''_0)
  + D {\sin^2\theta_0 \ov\rho}.
\end{equation}
%--------------
The LLG equation \Eq{eqn:LLG} reduces approximately to an effective Schr\"{o}dinger equation as \Eq{eqn:LLGSch}, but with spatially varying effective scalar and vector potentials:
%--------------
\begin{subequations}
  \begin{align}
  V(\br) &= \gamma \hbar \qty[H\cos\theta_0(\rho) + \qty(K+{A\ov\rho^2})\cos2\theta_0(\rho) + {D\sin2\theta_0(\rho)\ov \rho}]  + O(D^2), \\
  \bA(\br) &= {\hbar\ov 2A}\qty[D\mb_0(\rho)+{2A\cos\theta_0(\rho)\ov\rho}\hbe_\alpha].
  \end{align}
\end{subequations}
%--------------
The scalar potential is a ring shaped potential well around the Skyrmion core as shown in the right panel of \Figure{fig:VB_Texture}. Again the curl of the effective vector potential gives an effective magnetic field:
%--------------
\begin{equation}
  \bB(\br) = \nabla\times\bA(\br) = {\hbar\ov 2A}
  \qty[D\theta'_0\mb_0(\br) +  (D-2A\theta'_0){\sin\theta_0\ov\rho}\hbz].
%  = {\hbar\ov 2A} \qty{
%  \qty[ {\sin\theta_0\ov\rho} (D-2A\theta'_0) + D\cos\theta_0 \theta'_0]\hbz
%  + D\sin\theta_0\theta'_0 \hbe_\alpha}.
\end{equation}
%--------------
whose $\hbz$ component is shown in \Figure{fig:VB_Texture}.
This effective field can originate from both the DMI and the spatial curvature of the magnetization profile. With these effective potentials and magnetic field, it is relatively straightforward to picture the spin wave behaviour near a Skyrmion. First of all, the ring-shaped potential well can confine some bound spin-wave states, just like the bound states confined by the domain wall above. The frequencies of these bound states are below the bulk spin wave gap. In fact, the excitation of these bound spin wave states corresponds to the various breathing modes of the Skyrmion. The effective magnetic field produced by the Skymion can also deflect spin waves impinging on the Skyrmion from the outside, giving rise to the magnon Hall effect~\cite{Onose2010}.
\subsubsection{Spin wave polarization}\label{sssec:sw_pol}

%-----------------------------------
\begin{figure}[t]
  \centering
\includegraphics[width=0.98\textwidth]{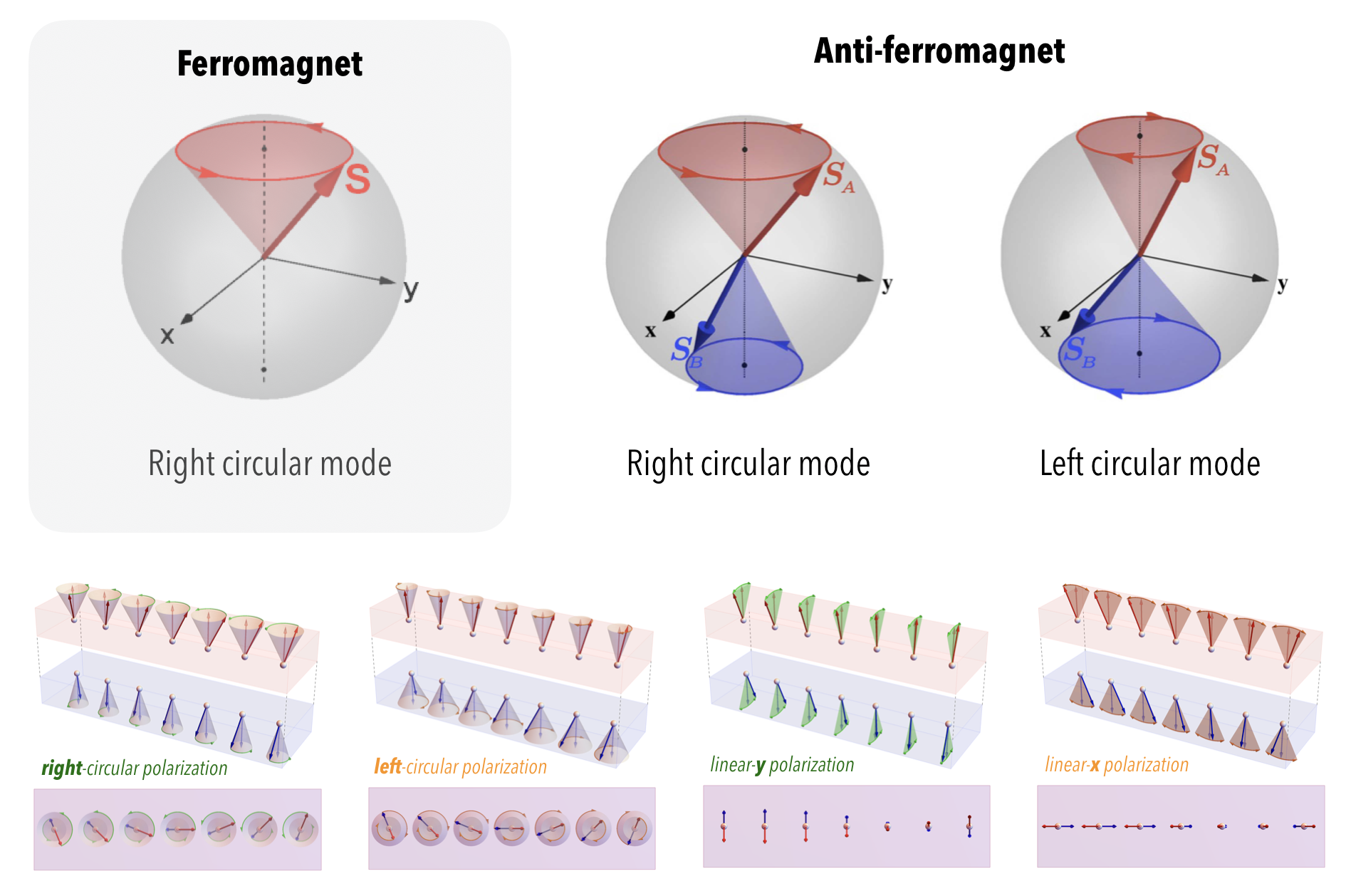}
\caption{Spin wave polarization in FM and AFM. Adapted from Ref.~\cite{Cheng2016}. In FM, only right circular spin waves exist. In AFM, both right and left circular spin waves exist and they are degenerate, therefore they can be combined into arbitrary polarization including linear polarization. The bottom panels depict spin waves with some typical polarization in a SyAF.}
\label{fig:sw_pol}
\end{figure}
%-----------------------------------

In ferromagnets (FM), the net magnetization breaks the time reversal symmetry, and hence there exists only the right-circularly polarized spin-wave mode for ferromagnets with an uniaxial anisotropy, as shown in the left panel in \Figure{fig:sw_pol}. In a more general case, the ferromagnetic spin waves may not be exactly circularly polarized, and may become elliptically polarized when additional interactions are considered. For example, for a thin film with its thickness being confined in the $y$ direction (using the coordinates defined in \Figure{fig:sw_pol}), the additional shape anisotropy tends to confine the magnetization within the $x$-$z$ plane, which squeezes the spin-wave mode~\cite{Kam2017} into an elliptically polarized mode elongated in $x$ direction. Nevertheless, the elliptical polarized mode still possesses the right-circular character.

However, in antiferromagnets (AFM), the time reversal symmetry is preserved\footnote{The time reversal symmetry is preserved in combination with a spatial translation.}, and there exist two degenerate spin-wave modes in the absence of magnetic fields with circular polarization, \ie left- and right-circular modes as shown in the right panel in \Figure{fig:sw_pol}. Because of the degeneracy, these two circular modes can be linearly combined into arbitrary polarizations, including the linearly polarized modes as shown in \Figure{fig:sw_pol}. Therefore, in antiferromagnets, the polarization degree of freedom of spin waves becomes quite similar to that of photons. The spin-wave polarization in antiferromagnets is conserved only in the absence of interactions that can couple left and right circular modes. These spin non-conserving interactions include: an external magnetic field along a direction different from the polarization axis, the dipolar interaction between the two magnetic sublattices~\cite{Kam2017,shen_magnon_2020}, DMI, or additional anisotropy along a different axis, all of which can couple the left- and right- circular modes~\cite{Kawa2019}, and thus the spin wave polarization is no longer a conserved quantity.

In ferri-magnets~\cite{Serga2010,Woo2018,Velez2019,Ding2019,Lukas2019}, the two magnetic sublattices are not identical, and thus there is a non-zero net magnetic moment. Therefore, the left- and right-circularly polarized spin-wave eigenmodes are no longer degenerate. Consequently, the two opposite circular types of polarization progress at different paces, which indicates that the linearly polarized mode will not keep its linear polarization state during propagation, but will rotate.

It is possible to detect the spin-wave polarization electrically via the spin pumping effect~\cite{tserkovnyak_enhanced_2002}. In ferromagnets, a spin current pumped by spin waves can be detected via the inverse spin Hall effect~\cite{saitoh_conversion_2006} or via the spin Seebeck effect~\cite{uchida_observation_2008,uchida_spin_2010,jaworski_observation_2010,xiao_theory_2010,adachi_linear-response_2011}. Because the spin-wave polarization is fixed by the magnetization direction, the spin current caused by spin pumping and the resulting inverse spin Hall voltage have a fixed sign depending on the magnetization direction. In antiferromagnets, although the magnetizations from the two magnetic sublattices compensate each other, the pumped spin currents from each sublattice actually add up~\cite{cheng_spin_2014}. Therefore, antiferromagnetic spin waves are able to pump spin currents into an adjacent normal metal, but the sign of the spin current depends on the polarization of the antiferromagnetic spin waves. Spin pumping from antiferromagnets has been observed experimentally by Li~\textit{et al.}~\cite{li_spin_2020}. Therefore, it is possible to detect the polarization of antiferromagnetic spin waves via the sign of the inverse spin Hall voltage. However, in the absence of external magnetic fields, the spin Seebeck effect in antiferromagnets would vanish because of the cancellation between the two types of degenerate spin wave polarization~\cite{ohnuma_spin_2013}.

\subsubsection{Spin waves in antiferromagnetic textures}
We define in a similar manner spherical coordinates for the staggered, antiferromagnetic order such that
%--------------
\begin{equation}
  \label{eqn:n0n1}
  \bn_0(\br) = \hbe_\rho(\br) \qand
  \bn_1(\br,t) = n_\theta(\br,t) \hbe_\theta(\br) + n_\phi(\br,t) \hbe_\phi(\br).
\end{equation}
%--------------
The static solution $\bn_0(\br)$ to \Eq{eqn:LLG-n} requires that $\bH_n\parallel \bn_0$ everywhere. The dynamic equations to the linear order of $n_{\theta,\phi}$ from \Eq{eqn:LLG-n} can be rewritten into a matrix form as

%--------------
\begin{equation}
  \label{eqn:LLGntp}
  \qty({1\ov J}\pdv[2]{t} - \alpha \pdv{t}) \mqty(n_\theta \\ n_\phi)
  = \qty[ -A \nabla^2 + \hV(\br) + \hW^j(\br) \partial_j]
  \mqty(n_\theta \\ n_\phi),
\end{equation}
%--------------
where $\hV, \hW$ are $2\times 2$ (in general non-Hermitian) matrices, containing the information about the magnetic texture in background depends on the detailed texture in consideration. If we define
%--------------
\begin{equation}
  \label{eqn:psipm}
  \mqty(\psi_- \\ \psi_+)
  = {1\ov\sqrt{2}}\mqty(1 & -i \\ 1 & i) \mqty(n_\theta \\ n_\phi)
  \equiv \hU \mqty(n_\theta \\ n_\phi),
\end{equation}
%--------------
with $\psi_+ = \psi_-^*$,
then \Eq{eqn:LLG-n} can be reformulated as:
%--------------
\begin{equation}
  \label{eqn:LLG-n-psi}
  \qty({1\ov J}\pdv[2]{t} - \alpha \pdv{t}) \mqty(\psi_-(\br,t) \\ \psi_+(\br,t))
  = \qty[ -A \nabla^2 - \hU \hV \hU^\dagger - \hU \hW^j \hU^\dagger \partial_j ]
  \mqty(\psi_-(\br,t) \\ \psi_+(\br,t)) .
\end{equation}
%--------------
For the simplest case of a homogeneous antiferromagnetic domain
%\jx{Double check}
%--------------
\begin{equation}
  \hU\hV\hU^\dagger = K + J \qand
  \hU\hat{\bW}\hU^\dagger = -iD \mqty(1 & 0 \\ 0 & -1) \bn_0.
\end{equation}
%--------------
$\hV$ and $\hW$ have more complicate form for inhomogeneous magnetic textures such as domain wall or Skyrmions.
Both forms \Eq{eqn:LLGntp} and \Eq{eqn:LLG-n-psi} can be useful. When the spin wave eigenstate possesses circular polarization, it is more convenient to use \Eq{eqn:LLG-n-psi}. if the spin wave eigenstate is of linear polarization, it is more favorable to use \Eq{eqn:LLGntp}.

{\bf Homogeneous antiferromagnet} - For a homogenous antiferromagnet with uniaxial anisotropy pointing along $\hbu = \hbz$, the static $\bn_0(\br) = \hbz$, assuming that the homogeneous DMI vanishes with $\bd = 0$, \Eq{eqn:LLG-n-psi} reduces to a Klein-Gordon-like equation:
\footnote{Since both $\psi_-$ and $\psi_+$ satisfy the same equation as \Eq{eqn:LLG-KG}, if $\psi_-$ the right-handed circular polarization is a solution, then its conjugate $\psi_+ = \psi_-^*$ representing the left-handed circular polarization is also a solution.}
%--------------
\begin{equation}
  \label{eqn:LLG-KG}
  \qty({1\ov J}\pdv[2]{t} - \alpha \pdv{t}) \psi_-(\br,t)
  = \qty[{1\ov 2\mu}\qty(-i\hbar\nabla-\bA)^2 + V] \psi_-(\br,t).
\end{equation}
%--------------
where
%--------------
\begin{equation}
  V = \gamma\hbar\qty(K + J + {D^2\ov 4A}) \qand
  \bA = {\hbar D\ov 2A}\bn_0.
\end{equation}
%--------------
When $\alpha = 0$, the solutions to \Eq{eqn:LLG-KG} are plane waves \cite{keffer_theory_1952}
%--------------
\begin{equation}
  \psi_-(\br) = n_\theta - i n_\phi = C e^{i(\omega t - \bk\cdot\br)}
  \qwith \omega = \pm\sqrt{\gamma \qty(K + H + {D^2\ov 2A}) + \gamma A \abs{\bk}^2 - \gamma D \bk\cdot\mb_0}.
\end{equation}
%--------------
where the $\pm$ indicates that the AFM spin waves can be either right- or left-circularly polarized.

{\bf 1D antiferromagnetic domain wall} - Let $\hbu = \hbz$ and neglecting $\bd = 0$, a static antiferromagnetic domain wall from $+\hbz$ to $-\hbz$ can be stabilized with $\bn_0(\br) = (0,\sin\theta_0(x),\cos\theta_0(x))$ with $\phi_0(\br) = \pi/2$, $\theta_0(-\infty) = 0$ and $\theta_0(+\infty) = \pi$. The stability of the antiferromagnetic domain wall is exactly the same as that in the ferromagnetic case as in \Eq{eqn:dw_stable}.
When DMI is present, it is more convenient to use the $n_{\theta, \phi}$ representation in \Eq{eqn:LLGntp} as
\footnote{Because we consider the special case of a 1D domain wall along $x$, and $\bn_0$ is in the $y$-$z$ plane, the vector potential (or the $\hW^j_{\theta\phi}$ term) here does not play any role.}
%--------------
\begin{equation}
  \label{eqn:LLGnAFdw}
  \qty({1\ov J}\pdv[2]{t} - \alpha \pdv{t}) \mqty(n_\theta \\ n_\phi)
  = \qty[-{\hbar^2\ov 2\mu} \partial_x^2 + K\cos2\theta_0
  + {D\ov w}\sin\theta_0\mqty(0 & 0\\ 0 & 1)
  ] \mqty(n_\theta \\ n_\phi).
\end{equation}
%--------------
This equation indicates that the equation of motion for $n_\theta$ and $n_\phi$ components are decoupled and can be solved separately as two independent linearly polarized spin wave solutions to the Klein-Gordon-like equations above. More importantly, the scattering potential due to the domain wall differs for the two types of linear polarization because of the DMI. This feature enables the antiferromagnetic domain wall to function as a spin-wave polarizer which will be further discussed later~\cite{Lan2017}.

{\bf 2D antiferromagnetic Skyrmion} - In principle, it is straightforward to write down similar equations as \Eqs{eqn:LLGntp}{eqn:LLG-n-psi} for antiferromagnetic Skyrmions~\cite{Ros2015,diaz_topological_2019,Gao2020}, but with much more tedious algebra. Therefore, we skip the discussion on this issue.

\subsubsection{Spin waves in ferrimagnetic textures}

When two sublattices described above in the antiferromagnetic case become non-identical, the system turns ferrimagnetic. Based on the equation of motion \Eq{eqn:LLG-ferri} for the net magnetic vector $\bn$ in a ferrimagnet, Kim~\textit{et al.} derived an equation of motion for high energy magnons in the presence of non-trivial static ferrimagnetic textures~\cite{kim_tunable_2019}:
%--------------
\begin{equation}
  \label{eqn:LLGnFerri}
  -q\hbar s\qty(i\hbar\partial_t - q\phi) \psi_q
  +\rho\qty(i\hbar\partial_t - q\phi)^2 \psi_q
  = A \qty(-i\hbar \nabla - q\bA)^2 \psi_q(\br,t),
\end{equation}
%--------------
where $\psi_q$ defined in \Eq{eqn:psipm} with $q = \pm$ represent the left- or right-circularly polarized magnons, and $\phi, \bA$ represent the texture- and DMI-induced scalar and vector potentials. This equation can be regarded as a combination of the effective Schr\"{o}dinger equation for the ferromagnetic case and the effective Klein-Gordon equation for the antiferromagnetic case described above.

\subsection{Dynamics of magnetic textures driven by spin waves}\label{sec:dynamics of magnetic textures}

The previous subsection shows that a static magnetic texture can influence spin wave behavior in a similar fashion to an electron moving in certain gauge potentials depending on the exact texture. The reversed process is also true that the magnetic texture may respond to spin waves and starts to move in space or deform its shape. This latter part has no direct correspondence to the electronic analog, because the real electronic potentials are usually fixed and cannot change. An exception is the electromigration effect where the electronic potential can be changed due to the gradual motion of the ions caused by the momentum transfer between the electrons and diffusing metal atoms \cite{pierce_electromigration_1997}.

While, in contrast, the magnetic texture that gives rise to the effective potentials for the spin wave is not a rigid object and therefore can be modified. The analysis of magnetic texture in response to external driving forces usually adopts the collective coordinates method that has been used by Thiele~\cite{Thie1973}, \ie assuming the texture is rigid and can only translate in real space or rotate uniformly in spin space, therefore the dynamics of the magnetic texture can be captured using two collective coordinates that characterize the central position of the texture and the overall orientation of the texture.

\subsubsection{Spin wave driven ferromagnetic domain wall motion}

Under external driving forces such as the magnetic field, electric current~\cite{Jam2010}, coherent spin waves \cite{Thi2002,Han2009,Wie2010,Yan2011,Seo2011,Wang2012,GWang2012,Kim2012,Moon2013,Wang2013,Zhang2014,Hata2014,Wang2015,Yang2015} or thermal magnons~\cite{Nowak2011,Jiang2013,Sch2014,Yan2015,Sel2016,More2017},
a magnetic domain wall can be regarded as a rigid body with its overall shape undistorted, and therefore the dynamics of a magnetic domain wall can be described by its collective coordinates: domain wall central position $X$, domain wall width $W$, and tilting angle $\Phi$ from the magnetization rotation plane (see \Figure{fig:FM_texture}). For an extended domain wall in a two-dimensional film, these collective coordinates are a function of an additional dimension, therefore effectively becoming a domain wall string~\cite{rodrigues_effective_2018}. According to Noether's theorem, the angular momentum of a ferromagnetic domain wall corresponds to a linear motion, while the linear momentum of the wall corresponds to a rotation of its magnetization tilt~\cite{Yan2013}. The former part can be easily understood, since the moving domain wall can be equivalently viewed as expanding one domain while shrinking the other, thus changing the total angular momentum (magnetization) of the domains (not of the domain wall).

For a simple one-dimensional domain wall as in \Figure{fig:FM_texture}, when driven by external forces, its dynamics can be described using the collective coordinates model (or sometime referred as $q$-$\phi$ model)~\cite{Sch1974,Thi2002},
%--------------
\begin{subequations}
  \label{eqn:FMDW_qphi}
  \begin{align}
  \dot{X}/W - \alpha \dot{\Phi} &=J_{\text{angular}}\left(\mbox{angular momentum transfer rate}\right), \\
  \dot{\Phi} + \alpha \dot{X}/W &=J_{\text{linear}}\left(\mbox{linear momentum transfer rate}\right),
  \end{align}
\end{subequations}
%--------------
where the angular/linear momentum transfer can be driven by various means, including external field, electric/spin current, or magnons. Notice that, resulting from the first derivative of the LLG equation, the above equations of motion for ferromagnetic domain walls are also in the first derivative, which means that the ferromagnetic domain wall has no inertia.

Typically, the transmission of spin waves or electron spins transfers angular momentum to the domain wall, whereas the reflection of spin waves or electron spins transfers linear momentum to the domain wall. Therefore, in order to know the domain wall dynamics, it is sufficient to study the transmission/reflection behavior of spin waves at the domain wall. For spin waves, the angular momentum transfer due to transmission is determined by the number of magnons passing through: each magnon transfers $2\times\hbar = 2\hbar$ angular momentum (for an \ang{180} domain wall). Therefore, the total angular momentum transfer rate or the angular momentum current $J_\ssf{angular}$ is simply proportional to the total number of magnons being transmitted or proportional to the square of the amplitude of propagating spin waves $\rho^2$ and the transmission probability $T$:
%--------------
\begin{equation}
  \label{eqn:swA}
  J_\ssf{angular} =  {M_s\ov \gamma \hbar}\int \dd{\omega} T(\omega)\rho^2(\omega)\hbar v_g(\omega),
\end{equation}
%--------------
where the prefactor ${M_s\ov \gamma \hbar}$ can be considered as the saturated spin density, and $v_g(\omega)$ stands for the spin-wave group velocity. The consequence of the angular momentum transfer is that the domain wall is always pulled towards the source of spin waves. Since spin waves also carry linear momentum, they can transfer linear momentum to a domain wall via spin-wave reflection from the domain wall. The linear momentum transfer rate is proportional to the number of reflected magnons, and each magnon transfers $2\hbar k$ linear momentum to the domain wall. Therefore, the total linear momentum transfer rate or the linear momentum current is proportional to the square of the spin-wave amplitude $\rho^2$ and the reflection probability $R$:
%--------------
\begin{equation}
  \label{eqn:swL}
  J_\ssf{linear} =  {M_s\ov \gamma\hbar}\int \dd{\omega} R(\omega)\rho^2(\omega) 2\hbar k(\omega) v_g(\omega).
\end{equation}
%--------------
By plugging \Eqs{eqn:swA}{eqn:swL} into \Eq{eqn:FMDW_qphi}, one may find how a ferromagnetic domain wall responds to spin waves, providing that the transmission and reflection probabilities are calculated beforehand based on the method in the previous section. The magnon-induced friction of domain walls has also been studied theoretically by Kim~\textit{et al.}~\cite{Kim2018}.

Notably, the mutual interaction between spin waves and domain walls has been recently demonstrated in experiments by Han~\textit{et al.}~\cite{Han2019}, which will be further discussed in Section~\ref{sec:Domain wall driven by spin waves}.

\subsubsection{Spin wave driven antiferromagnetic domain wall motion}

In a similar rigid model, the dynamics of an antiferromagnetic domain wall are also captured by the time evolution of its collective coordinates, \ie the position $X$, width $W$, and tilting angle $\Phi$, and the equations of motion can be expressed as
%--------------
\begin{subequations}
  \label{eqn:AFDW_qphi}
  \begin{align}
  M\ddot{X} + \alpha \dot{X}/W &=J_{\text{angular}}\left(\mbox{angular momentum transfer rate}\right), \\
  I\ddot{\Phi} + \alpha W^2 \dot{\Phi} &=J_{\text{linear}}\left(\mbox{linear momentum transfer rate}\right),
  \end{align}
\end{subequations}
%--------------
where $M = 2/WJ$ and $I = MW^2$ are the effective mass and moment of inertia of the antiferromagnetic domain wall. By comparing with \Eqs{eqn:FMDW_qphi}{eqn:AFDW_qphi},
one sees two main qualitative differences between the ferromagnetic and antiferromagnetic domain wall motion. One is that the association between the linear/angular momentum transfer with the linear/angular motion of the domain wall is opposite for ferro- and antiferromagnetic cases. The second is that the equation of motion for the antiferromagnet is in the second time derivative, thus $M$ and $I$ in \Eq{eqn:AFDW_qphi} are the effective mass and momentum of inertia of an antiferromagnetic domain wall. More theoretical discussions can be found in Refs.~\cite{Lan2017,yu_polarization-selective_2018}. The angular and linear momentum transfer between propagating spin waves and the antiferromagnetic domain walls have the same form as in the ferromagnetic case in \Eqs{eqn:swA}{eqn:swL}.

\subsubsection{Spin wave driven ferromagnetic Skyrmion motion}

For a typical Skyrmion formed on a two-dimensional ferromagnetic thin film, the dynamics of a ferromagnetic Skyrmion can be describe by the Thiele equation~\cite{Thie1973,Zar2018}
%--------------
\begin{equation}
  \label{eqn:theile}
  M\ddot{\bX} + Q \dot{\bX} \times \hbz + \eta \dot{\bX} =J_{\text{linear}}\left(\mbox{linear momentum transfer rate}\right),
\end{equation}
%--------------
where the terms on the left-hand side are the inertial force, Magnus force and the friction force acting on the Skyrmion. $M$ is the effective mass of the Skyrmion, $Q$ is the topological charge (the number of times that the order parameter wraps around the unit sphere) of the Skyrmion, and $\eta$ describes the frictional force caused by the Gilbert damping. Recently, further theoretical developments of the Skyrmion Thiele equation~\cite{Zar2018}, spin-wave-driven Skyrmion Hall effect~\cite{Woo2018}, vortices drag~\cite{Rei2018} and coupled breathing modes in 1D-Skyrmion lattice~\cite{Kim2018a} have also been introduced and discussed.

\section{Experimental techniques for studying spin waves in magnetic textures}

To study spin-wave dynamics in magnetic textures, various experimental tools can be utilized. In this section, we primarily introduce six different experimental techniques for investigating spin waves in magnetic textures, which are successively: Brillouin light scattering (BLS) spectroscopy based on inelastic scattering of photons and magnons (Section~\ref{sec:BLS}); Propagating spin wave spectroscopy (PSWS) using integrated microwave antennas to excite and detect spin waves based on a vector network analyzer (Section~\ref{sec:PSWS}); Time resolved magneto-optical Kerr effect (MOKE) by analyzing the temporal change of the reflected light polarization (Section~\ref{sec:MOKE}); Time-resolved scanning transmission X-ray microscopy (TR-STXM) capable of studying short-wavelength spin waves (Section~\ref{sec:TR-STXM}); Detection of magnon transport by inverse spin Hall effect (ISHE) for DC detection of spin waves in magnetic textures (Section~\ref{sec:ISHE}); Nitrogen-vacancy (NV) centers and similar spin systems by probing the static and dynamic stray field of magnetic textures with ultra-high sensitivity (Section~\ref{sec:NV centers}). 

Apart from these techniques elaborated in the following subsections, there are certainly other techniques used to study the dynamics of magnetic textures, \eg ferromagnetic resonance force microscopy~\cite{Chia2012,Volo2018} and X-ray ferromagnetic resonance~\cite{Hicken2020}. In addition, there are several versatile techniques to characterize static magnetic textures, \eg magnetic force microscopy (MFM)~\cite{Hart2003,Kimura2012,JX_Zhang2018,Claudia_Parkin_2020}, Lorentz transmission electron microscopy (Lorentz TEM)~\cite{Tok2010,H_Du2018,W_Jiang2019,Back2020} and spin-polarized low-energy electron microscopy (SPLEEM)~\cite{chen_novel_2013,chen_tailoring_2013,Y_Wu2019}.

\subsection{Brillouin light scattering spectroscopy}\label{sec:BLS}
\begin{figure}[b]
\begin{center}
\includegraphics[width=16cm]{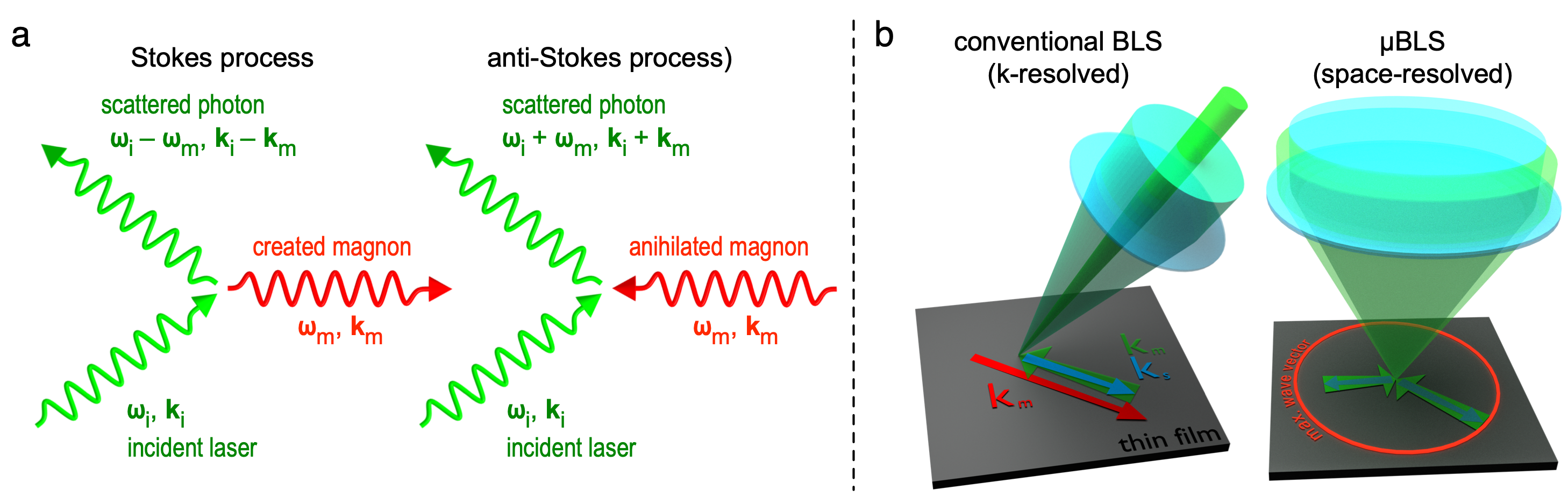}
\caption{(a) Schematics of the magnon-photon interaction during the Brillouin light scattering process.  (b) Illustrations of the scattering geometries for conventional, wavevector-resolved BLS (left) and a BLS microscope (right).}
\label{fig:BLS_1}
\end{center}
\end{figure}

Brillouin light scattering (BLS) is an optical method for the detection of magnons in thin films based on the inelastic scattering of photons and magnons. The first reports were based on magnons in anti-ferromagnetic materials in the THz frequency range~\cite{Fle1966} and were soon extended to ferro- and ferrimagnetic materials with frequencies in the GHz range~\cite{San1973}. The fundamentals of the photon-magnon interaction can be understood from Fig.~\ref{fig:BLS_1}(a). The photons from a continuous wave laser scatter inelastically with magnons under the conservation of energy (given by the frequency $\omega$ and momentum $\vec{k}$. In the case that a magnon is created (Stokes process), the frequency of the scattered photons is reduced. If a magnon is annihilated (anti-Stokes process), the frequency of the scattered photons is increased. Depending on the type of magnons, this frequency shift can be rather small and in order to separate the inelastically scattered photons from the elastically reflected photons (Rayleigh peak) an interferometer with a frequency resolution of a few tens of MHz and a contrast better than $10^{10}$ is needed. Most of the interferometers used for BLS on magnetic excitation are based on the six-pass Tandem-Fabry-Perot interferometer developed by J.R.~Sandercock. Details on the state of the art version of this apparatus can be found at \url{www.tablestable.com}. Suitable laser sources are continuous wave lasers with a linewidth smaller than 1~MHz, a wavelength in the visible range (depending on the magnetic material) and power up to 100~mW.
%-----------------------------------
%-----------------------------------

There are two basic operation modes of a BLS experiment, which are both relevant in the context of texture based magnonics: the traditional backscattering BLS geometry and the BLS microscope geometry which is also referred to as microfocus BLS or just $\mu$BLS. Traditional BLS offers wavevector resolution with limited spatial resolution in the order of tens of micrometers, whereas $\mu$BLS offers high spatial resolution down to the diffraction limit of light by sacrificing wavevector resolution. The difference of the scattering geometry for both modes of operation is sketched in Fig.~\ref{fig:BLS_1}(b): For traditional BLS a narrow laser beam with a diameter of a few millimeters is focused using a lens with a long focal length. Hence, the effective numerical aperture is rather small and the spot size on the sample is much larger than the diffraction limit of light, however, the scattering geometry, \ie the angle of incidence of the laser with respect to the sample, is well defined. In the backscattering geometry the photons scattered inelastically by magnons are collected by the same lens used for focusing the incident laser beam and, thus, only the light scattered by magnons with a well defined wavevector $\vec{k}_\mathrm{s}$ is possible. This wavevector of the detected magnons can be changed by varying the sample orientation with respect to the incident laser beam and therefore allows for measuring the magnon dispersion. Further details on the capabilities of traditional BLS can be found in the review by Demokritov~\textit{et al.}~\cite{Demi_Nonlinear2001}. For $\mu$BLS a microscope lense with a large numerical aperture (small focal length) is used to focus the incident laser on the sample to a spot size down to the diffraction limit. Since the inelastically scattered photons are also collected by this microscope lens, magnons with wave vectors in all directions and magnitudes are collected, where the maximum absolute wave vector is limited by the wavelength of the laser and the numerical aperture of the microscope lens (red circle in Fig.~\ref{fig:BLS_1}(b)). Please note that even though BLS and $\mu$BLS deliver a signal proportional to the magnon intensity measured directly in the frequency domain, it is possible for both modes of operation to reconstruct the magnon's phase and spatio-temporal evolution using additional equipment as described in details in the review by Sebastian and coworkers~\cite{Seb2015} and the references therein.

%-----------------------------------
\begin{figure}[t]
\begin{center}
\includegraphics[width=16cm]{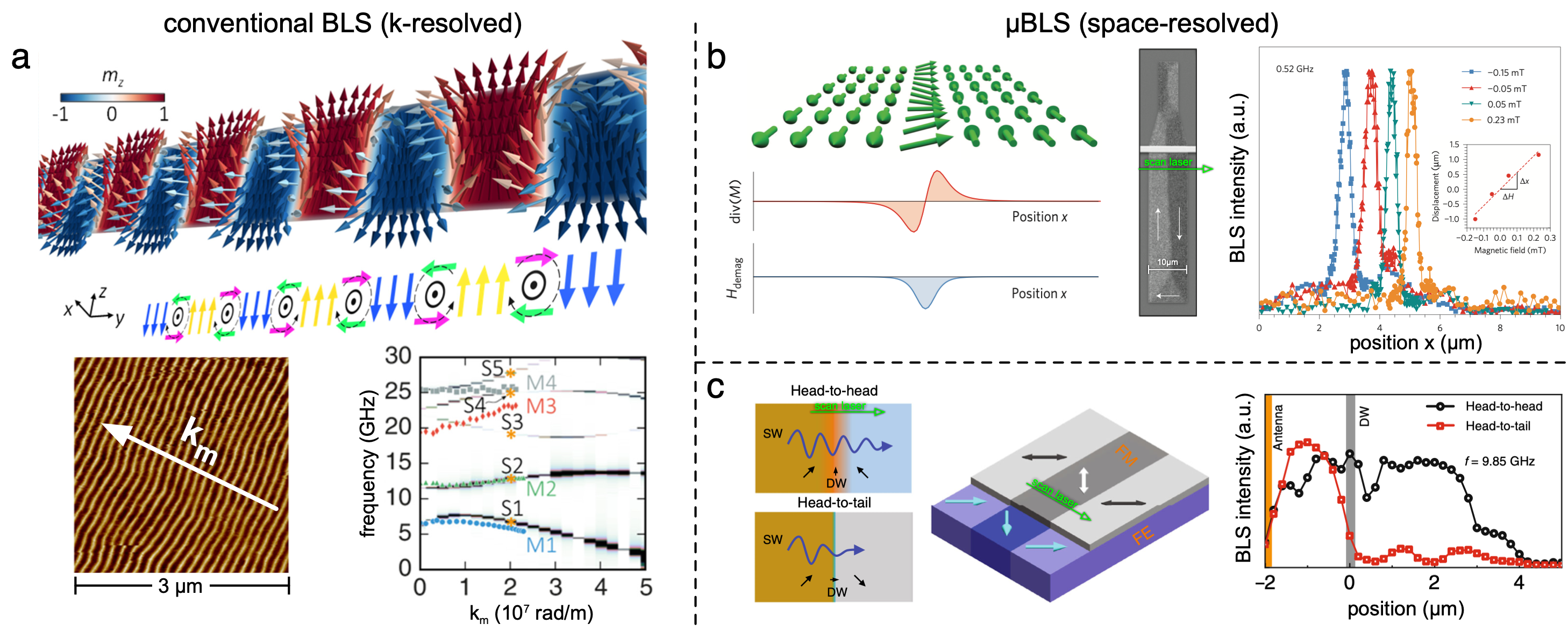}
\caption{(a) Example for traditional BLS on a line domain pattern in a Co/Pd multilayer, which creates a texture based magnonic crystal. Top shows a micromagnetic simulation of the magnetic configuration, bottom left the corresponding MFM image after demagnetization with an in-plane magnetic field and bottom right shows the magnon dispersion measured by BLS for wave vectors $\vec{k}_\mathrm{m}$ in the direction perpendicular to the domain walls as indicated in the MFM image.  Images rearranged and reproduced from Ref.~\cite{Ban2017}. (b) Example for $\mu$BLS measurements of magnons driven by microwaves channeled inside a nano-sized 180 degree N\'eel domain wall. Left shows a schematic of the magnetic configuration with the magnetic volume charges and demagnetizing field calculated using micromagnetics; Middle displays the magnetic domain pattern of a bottle shaped Permalloy structure imaged by wide-field Kerr microscopy; Right shows $\mu$BLS intensity profiles when scanning the laser across the domain wall for an excitation at 500~MHz for various magnetic fields applied in the direction of the domain wall. Images rearranged and reproduced from Ref.~\cite{Wagner2016}. (c) Examples for $\mu$BLS measurements of magnons propagating through 90 degree domain walls, which are pinned by an underlying ferro-electric material. Images rearranged and reproduced from Ref.~\cite{Hama2018}.}
\label{fig:BLS_2}
\end{center}
\end{figure}
%-----------------------------------

The characteristic length-scale in magnetic textures due to the interplay of different exchange energies, anisotropies and dipolar interactions is typically smaller than one micrometer. Nevertheless, standard BLS with its spatial resolution of several tens of micrometers is still a powerful tool for investigating magnons in non-collinear spin textures, if long-range order of the texture is given within the area of the probing laser spot, because it is probing magnons directly in $k$-space. A recent example for such an experiment is shown in Fig.~\ref{fig:BLS_2}(a). Banerjee and coworkers~\cite{Ban2017} studied magnons in a Co/Pd multilayer. This material has a strong perpendicular magnetic anisotropy (c.f. Section~\ref{sec:free energy ferromagnets}), meaning the magnetization wants to align perpendicular to the film surface. However, due to minimization of dipolar energy the magnetization breaks down into very narrow magnetic domains, which can be oriented by an in-plane magnetic field and act as a texture based magnonic crystal. If the periodicity and alignment of the domain pattern is well defined within the area of the probing laser spot, the wave vector resolution of standard BLS allows for observing the formation of magnonic band gaps and quantification of the frequency splitting induced by the periodic change of the magnetization direction. Shown is only the dispersion for wave vectors perpendicular to the domain walls, the dispersion for wave vectors along the magnetization can simply be measured by rotating the sample around the film normal by 90 degree.

If a regular periodicity is not given over distances of a few tens of micrometers, the standard BLS approach will not yield well defined resonance peaks. And if single textures like isolated domain walls are of interest, the overall signal strength will drop because most of the incident photons will not see the area of interest and therefore do not scatter inelastically. Thus, $\mu$BLS can be employed to focus the incident laser to a spot with a diameter of about 300-400~nanometers. This approach is sensitive enough to measure the thermal magnon spectrum even in scattering volumes as small as transverse domain walls in magnetic nanowires~\cite{San2008} or in rings with a width of a few hundreds of nanometers magnetized in the onion texture~\cite{Sch2008}. Although measuring the thermal magnons has the advantage that the full eigenmode spectrum can be observed without any symmetry breaking due to the antenna geometry, the excitation of magnons by microwave magnetic fields generated by different antenna geometries can boost the signal and is commonly used to study magnon transport phenomena. Two examples for $\mu$BLS measurements on magnetic textures are summarized in Fig.~\ref{fig:BLS_2}(b,c): The upper, right panel (b) shows the detection of magnons propagating inside a 180 degree N\'eel domain wall in a 40~nm thick Permalloy element~\cite{Wagner2016}. Even though this domain wall has a width of approximately 40~nm, which also sets the confinement length of the rf-driven magnons perpendicular to the transport direction, the signal strength allows for an unambiguous determination of the position of the channeled magnons. Naturally, the peak width in spatially resolved scans is limited at the lower end due to the size of the focusing laser spot, nevertheless, the signal from the inelastic scattering of magnons confined to such a small area can be detected, and the peak-position of magnons in a single texture can be determined with far greater precision than the diffraction limit of light. The second example is shown in panel (c) where the transmission of rf-driven magnons was investigated upon propagation through different types of 90 degree N\'eel domain walls~\cite{Hama2018}. Using $\mu$BLS, H{\"a}m{\"a}l{\"a}inen and coworkers~\cite{Hama2018} could demonstrate that the transmission of magnons through domain walls will strongly depend on the domain wall configurations, \eg head-to-head or head-to-tail. This reprogrammable spin-wave propagation over domain walls will be further discussed in Section~\ref{sec:reprogrammable SW with DW}. Other examples for spatially mapping of magnons in spin textures using $\mu$BLS can be found in Refs.~\cite{Per2005,Vogt2012,Vogt2014,Pirro2015,Alb2017,Sch2017,SF_Lee2020}.

One final remark regarding BLS: There is a common misunderstanding in the wavevector range of magnons for which standard BLS and $\mu$BLS is applicable. Since the detection is based on the inelastic scattering under the assumption of conservation of momentum, a fundamental maximum wave vector of magnons is assumed upon which photons can scatter. While this is true for the interaction of photons with plane-wave magnons this argument does not hold for spin waves confined on length scales comparable to their wavelength. Once such confinement is present, the magnons form standing waves. In this case the magnon's wave function cannot be described by a single wave vector but by a continuum of wave vector in reciprocal space, which allows for inelastic scattering with photons with smaller wave vectors.

\subsection{Propagating spin wave spectroscopy}\label{sec:PSWS}
\begin{figure}[t]
\begin{center}
\includegraphics[width=16cm]{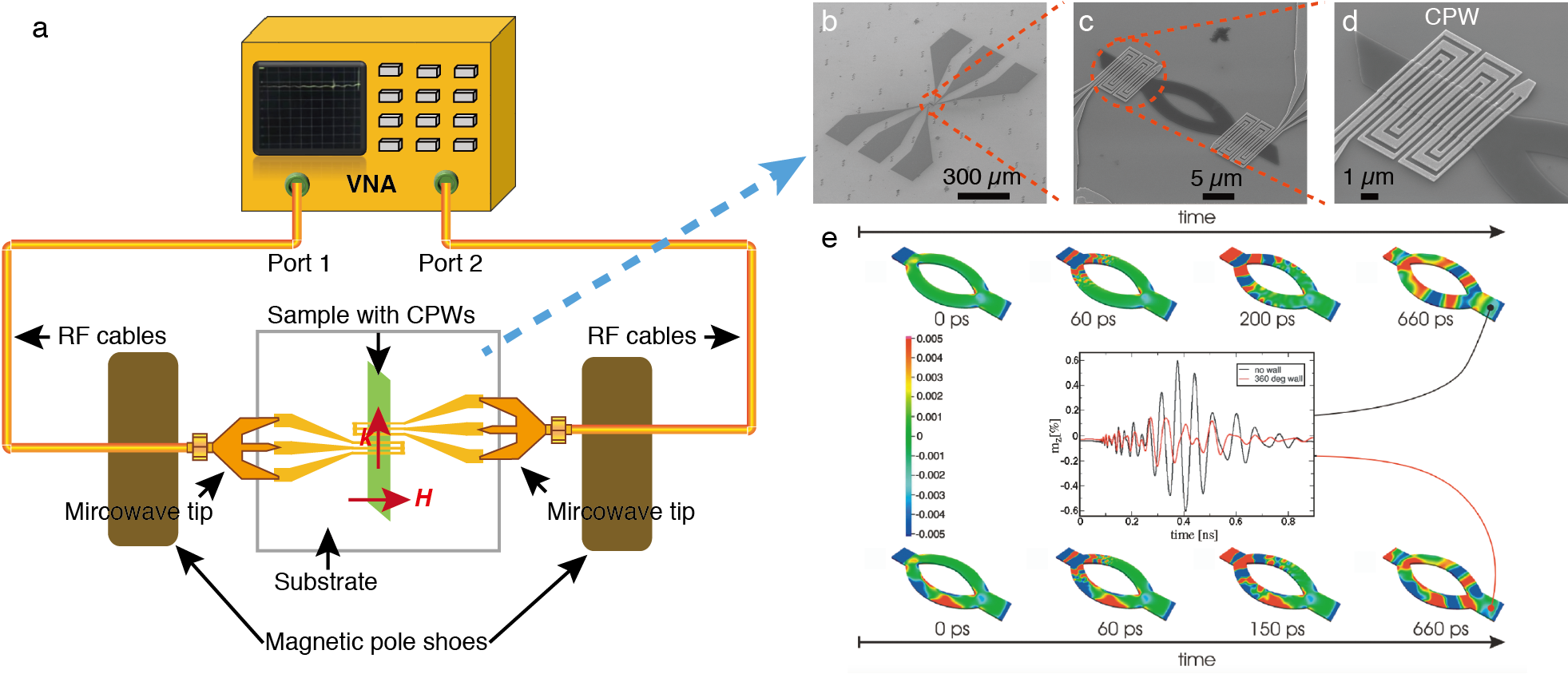}
\caption{(a) An illustrative diagram of the propagating spin-wave spectroscopy measurement setup based on a VNA.  (b-d) Scanning electron microscope images on a spin-wave interferometer device with integrated coplanar waveguides. (b) A full scale image of the device with large pads designed for microwave probes. (c) A blow-up image of the spin-wave interferometer device. (d) A high-resolution SEM image of the integrated coplanar waveguide (CPW). SEM images in courtesy to Dr. Chuanpu Liu at Peking university. (e) Micromagnetic simulation results on spin wave phase shift induced by domain walls. Reproduced from Ref.~\cite{Her2004}}
\label{fig:PSWS_1}
\end{center}
\end{figure}

\begin{figure}[t]
\centering
\includegraphics[width=16cm]{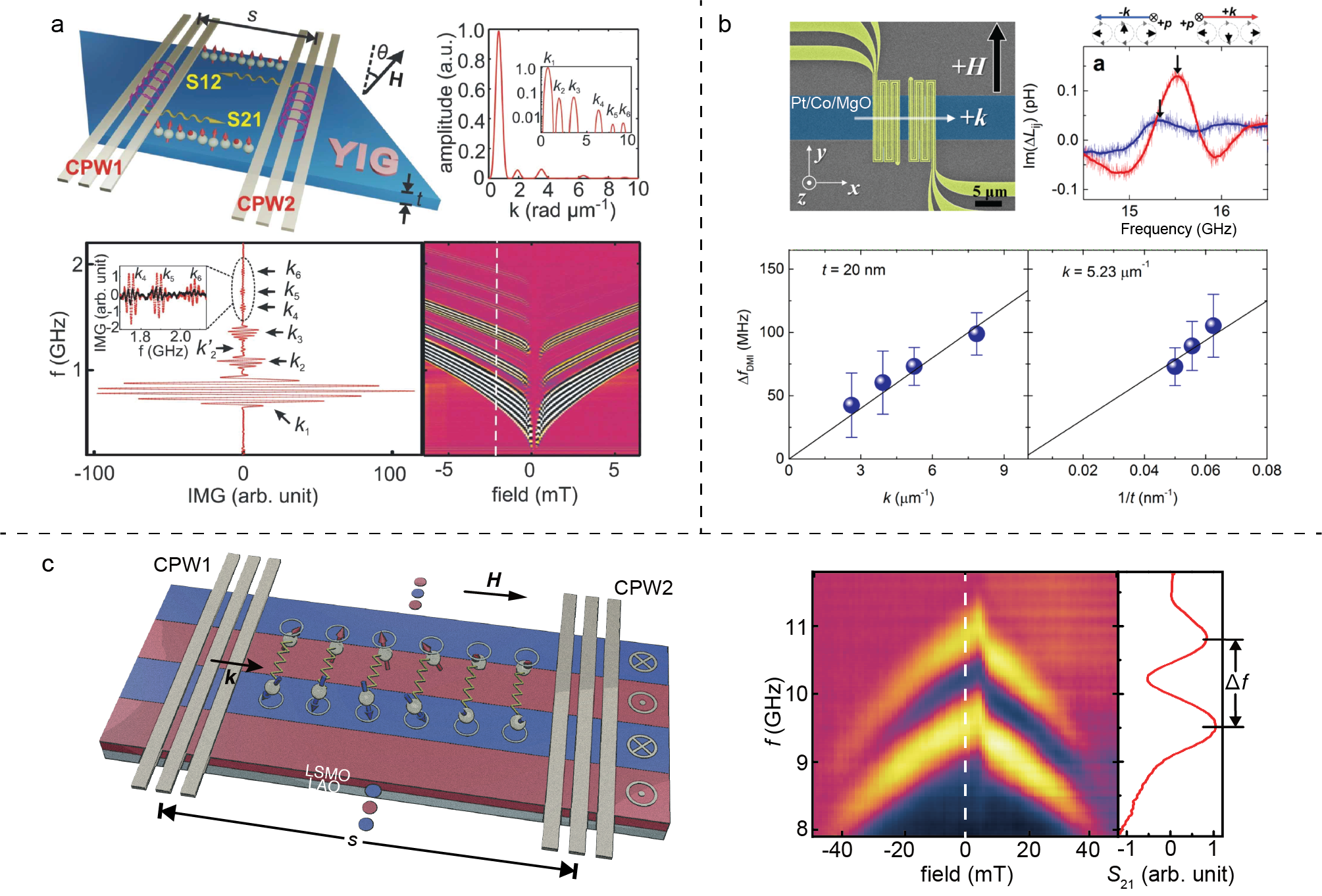}
\caption{(a) An example of PSWS measurements on a 20 nm-thick YIG film, where two CPWs are integrated on the film. The Fourier transformation shows multiple high-order excitation modes besides the main excitation $k_1$ mode. The propagating spin-wave spectra $S_{12}$, \ie spin waves excited by CPW2 and detected by CPW1, are measured with an applied field parallel to the CPW signal lines. Multiple transmission modes are observed corresponding to the CPW excitation. Adapted from Ref.~\cite{Yu2014}. (b) PSWS used to characterize DMI in Pt/Co/MgO multilayers. The DMI-induced frequency shift $\Delta f_{\text{DMI}}$ scales up with an increasing spin-wave wavevector $k$. The thickness dependence of $\Delta f_{\text{DMI}}$ indicates that DMI is of an interfacial type~\cite{Moon2013}. Reproduced from Ref.~\cite{Lee2016}. (c) An illustrative diagram for the spin-wave propagation in periodical domain stripes on LSMO thin films. Spin-wave transmission spectra $S_{21}$ measured by the PSWS with the field swept parallel to the spin-wave wavevector $k$. The lineplot is extracted at zero field indicated by the white dashed line. The peak-to-peak frequency span $\Delta f$ is extracted to calculate the group velocity based on Eq.~(\ref{vg1}). Reproduced from Ref.~\cite{Liu2019}.}
\label{fig:PSWS_2}
\end{figure}

The propagating spin-wave spectroscopy (PSWS)~\cite{Vla2008}, also sometimes referred to all-electrical spin-wave spectroscopy (AESWS)~\cite{Neu2010}, uses microwave techniques to excite and detect spin wave propagation based on a vector network analyzer (VNA). A schematic diagram of the measurement setup is shown in Fig.~\ref{fig:PSWS_1}(a). Microwave antennas, \eg coplanar waveguides (CPWs)~\cite{CPW1969} are integrated on top of magnetic thin-film structures under investigation for example as shown in \Figure{fig:PSWS_1}(b-d) (scanning electron microscopy (SEM) images courtesy of Dr. Chuanpu Liu at Peking University). The spin-wave wavevector $k$ distribution is determined by the Fourier transformation of the CPW spatial excitation profile~\cite{Vla2010}. A typical example is shown in Fig.~\ref{fig:PSWS_2}(a) where spin-wave transmission of multiple high-order excitations are resolved from the PSWS experiments in the high-quality yttrium iron garnet (YIG) thin films with excellent damping properties~\cite{Liu2014,Yu2014}. The $S_{21}$ ($S_{12}$) parameter of the VNA measures spin waves propagating from CPW1 (CPW2) to CPW2 (CPW1). As an example, \Figure{fig:PSWS_1}(b-d) show SEM images taken of an actual magnonic device with two integrated CPWs on each side of a ring structure made from a YIG thin film~\cite{Che2016}. The spin-wave Mach-Zehnder-type interferometer~\cite{Kos2005,Lee_Kim2008,Rousseau2015} shown in \Figure{fig:PSWS_1}(c) is designed for studying the spin-wave phase shift induced by a domain wall~\cite{Her2004}.

%-----------------------------------
%-----------------------------------

\indent The PSWS technique has recently been utilized to study the Dzyalonshinskii-Moriya interaction (DMI), which gives rise to magnetic textures such as Skyrmions~\cite{Muh2009,Tok2010,Sam2013}. Lee~\textit{et al.}~\cite{Lee2016} studied the interfacial DMI in Pt/Co/MgO mulitilayers as shown in Fig.~\ref{fig:PSWS_2}(b). Theoretical studies~\cite{Moon2013} imply that the interfacial DMI gives rise to a frequency nonreciprocity between counter-propagating spin waves, \ie $S_{21}$ ($+k$) and $S_{12}$ ($-k$) exhibit a frequency shift $\Delta f_{\text{DMI}}$ in a Damon-Eshbach (DE) configuration~\cite{DE1961}. Such frequency nonreciprocity is used to characterize the interfacial DMI with both PSWS~\cite{Lee2016} and BLS techniques~\cite{Nem2015,Li2017} (c.f. Section~\ref{sec:DMI_spin_wave_nonreciprocity}). In the PSWS measurements by Lee~\textit{et al.}~\cite{Lee2016}, only two modes are observed due to severe damping of the material system. However, by varying the integrated CPW design, a few different $k$ values can be achieved and the corresponding DMI-induced frequency shifts $\Delta f_{\text{DMI}}$ are obtained, and thereby the DMI constant can be extracted from the linear fit such as in Fig.~\ref{fig:PSWS_2}(b). The thickness dependence of $\Delta f_{\text{DMI}}$ indicates that the DMI is of an interfacial type. The interfacial DMI-induced spin-wave frequency nonreciprocity increases with thinner films, and is sizable for ultra-thin films with thickness of typically a few nanometers. The amplitude nonreciprocity for DE spin waves~\cite{DE1961,Demi2009,Saitoh2019} increases with thicker films, and becomes sizable for films with thickness of typically tens of nanometers or more.

%-----------------------------------
%-----------------------------------

\indent In comparison with the BLS technique (c.f. Section~\ref{sec:BLS}), the PSWS shows both advantages and disadvantages. Although the BLS is clearly more powerful in the $k$-dependent and spatial resolved measurements, the PSWS shows advantages in the frequency domain. By integrating nanoscale antennas such as nano-striplines~\cite{Ciu2016,Tobias2017}, broadband spin-wave excitation and detection can be easily achieved with fine frequency resolution~\cite{Neu2010}. The PSWS can also be used as a tool to characterize the spin-wave group velocity $v_{\text{g}}$~\cite{Vla2008,Neu2010,Yu2012,H_Wang2020} based on
\begin{equation}\label{vg1}
v_{\text{g}}=\frac{d\omega}{dk}=\frac{2\pi\Delta f}{2\pi/s}=\Delta f\cdot s,
\end{equation}
where $s$ is the propagation distance and $\Delta f$ is the peak-to-peak frequency span in the transmission spectra indicating a phase change of 2$\pi$ as indicated in Fig.~\ref{fig:PSWS_2}(c), where propagating spin waves in periodic domain stripes are characterized in La$_{0.67}$Sr$_{0.33}$MnO$_{3}$ (LSMO) thin films~\cite{Liu2019}. Taking the propagation distance $s=2~\mu$m and $\Delta f\approx$ 1.3 GHz, the group velocity $v_{\text{g}}$ can be estimated to be 2.6~km/s.

Lucassen~\textit{et al.}~\cite{Luc2019} has recently made some endeavors to further optimize the CPW design in the PSWS, particularly in service for experimentally extracting the DMI constant in ultrathin films~\cite{Luc2020,H_Wang2020}. Very recently, Che~\textit{et al.}~\cite{Che2020} have demonstrated the efficient excitation and detection of exchange magnons with wavelengths down to 100 nm by using ferromagnetic CPWs (mCPWs). In spite of the convenience in frequency sweeping, to vary the spin-wave wavevector $k$ in PSWS, several micro-structured or nano-structured antennas need to be integrated with well calibrated impedance matching. The effort to fabricate various integrated antennas for PSWS is quite demanding in comparison with the BLS~\cite{Nem2013} and the time resolved magneto-optical Kerr effect~\cite{Back1999}, where the spin-wave wavevectors can be tuned more easily.

\subsection{Time resolved magneto-optical Kerr effect}\label{sec:MOKE}
The magneto-optical Kerr effect (MOKE) describes the change of the polarization of light upon reflection from a magnetic surface, where the actual change of the polarization state (rotation of the polarization and/or change of ellipticity) depends on the direction of the magnetization with respect to the plane of incidence and the original polarization direction of the light impinging on the magnetic surface~\cite{Hubert_1998}. Since there are numerous different geometries for conducting MOKE experiments which can be selected depending on the type of material (magnetization in the sample plane or magnetized along the surface normal due to perpendicular magnetic anisotropies), we do not report detail here but refer to the comprehensive overview given in the recent review article by J. McCord~\cite{McCord2015} and the references therein.

\begin{figure}[t]
\begin{center}
\includegraphics[width=16cm]{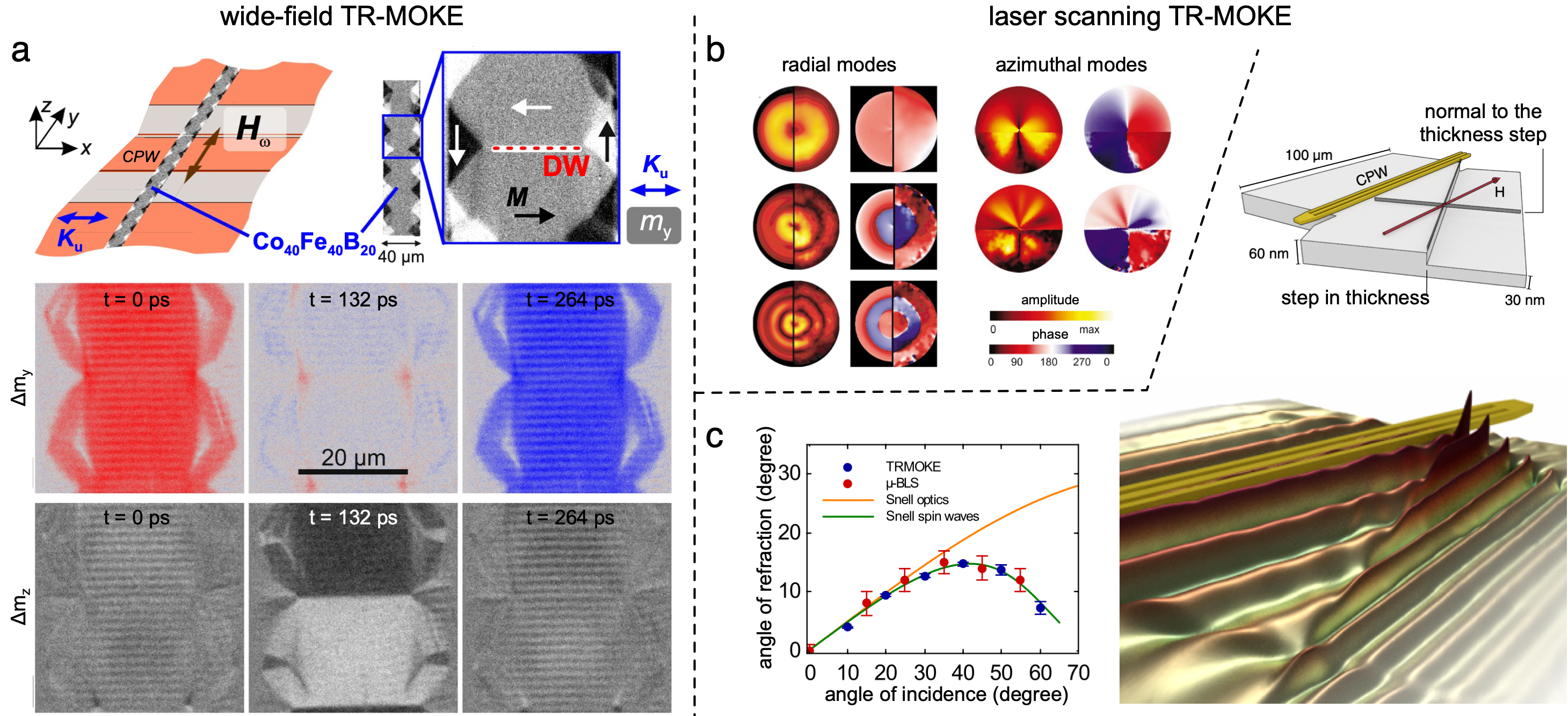}
\caption{(a) Examples for wide-field TR-MOKE on magnons emitted by domain walls. A CoFeB film was patterned into a 40~$\mu$m wide stripe with its short axis parallel to the uniaxial anisotropy. The sample was placed on top of a CPW for excitation with an oscillating magnetic field. The differential dynamic magnetic response in the lower intensity graphs clearly show periodic wavefront oriented parallel to the domain walls. Images rearranged and reproduced from Ref.~\cite{Holl2018}. (b) Examples for laser-scanning based TR-MOKE on a 15~nm-thick Py disc with a diameter of 6~$\mu$m magnetized in a vortex state. Magnons were driven by a circular microwave antenna patterned around the disc. Shown is the magnon's amplitude and phase for several different eigenmodes, where the left part of the disc is the result from micromagnetic simulations and the right part is from TR-MOKE. Images rearranged and reproduced from Refs.~\cite{Buess2005,Buess2004}. (c) Examples for laser-scanning TR-MOKE on propagating magnons driven by microwaves applied to a coplanar waveguide antenna. The Py element has a step in thickness from 60~nm to 30~nm where an anomalous refraction of magnons could be observed. Images rearranged and reproduced from Ref.~\cite{Stig2016}.}
\label{fig:MOKE_1}
\end{center}
\end{figure}

Since a magnon is locally changing the direction of the magnetization at GHz frequencies due to the collective precession of the magnetic moments, magnons can be detected using time-resolved magneto-optical Kerr effect (TR-MOKE) by analysing the temporal change of the polarization of the reflected light. In most scenarios this is done using pulsed lasers, which allow for a stroboscopic probe of the magnetization with a controllable time or phase delay with respect to the excitation source of the magnons. This excitation can be either all-optical using a femtosecond laser pulse in a pump-probe experiment for the excitation of magnons~\cite{Kampen2002}, by converting a fast pump laser pulse into a magnetic field using a photoconductive switch~\cite{Hie1997,Ger2002} or by synchronizing the pulsed laser to a signal generator or arbitrary waveform generator connected to a microwave antenna~\cite{Per2005,Buess2005,Neu2006,Au2011,Dav2015}. All the experiments in the references listed here so far were based on a laser-scanning approach, where the spatial profile of magnons is acquired by moving the sample with respect to a focused laser. In a recent work Holl\"ander and coworkers~\cite{Holl2018} could observe magnons emitted from magnetic domain walls~\cite{mozooni_direct_2015} using a wide-field Kerr microscopy, as shown in Fig.~\ref{fig:MOKE_1}(a) using a picosecond laser system synchronized to a CCD camera as described in detail in Ref.~\cite{McCord2015}. Two examples for TR-MOKE based on the laser-scanning approach are shown in Fig.~\ref{fig:MOKE_1}(b,c). Both experiments are based on microwave excitation of magnons and detection of the dynamic magnetic response by a fs-laser synchronized and phase locked to the signal generator driving the magnons. One advantage of measuring magnons using TR-MOKE is the direct access to the magnon's phase. This directly allows for distinguishing quantized magnons, as shown by Buess and coworkers ~\cite{Buess2005,Buess2004} for a permalloy (Py) disc magnetized in the vortex state (Fig.~\ref{fig:MOKE_1}(b)) from propagating magnons. This is shown, for example, in Fig.~\ref{fig:MOKE_1}(c), where magnons propagating through a thickness step in a Py element are refracted~\cite{Stig2016}.

%-----------------------------------
%-----------------------------------

Noteworthy are two recent advances of laser-scanning based MOKE: First, a novel heterodyne magneto-optical microwave microscope (H-MOMM) was developed by Nembach and coworkers~\cite{Nem2013} by mixing two frequency detuned CW lasers for detecting the polarization changes using a heterodyne mixing scheme, which aimed to enhance the sensitivity to measure even single nanomagnets with size below the diffraction limit; Second, Woltersdorf and coworkers developed a super-Nyquist sampling MOKE (SNS-MOKE) which removes the frequency resolution limit in the standard TR-MOKE geometries~\cite{Farle2013,Dre2018}.

It is often assumed that TR-MOKE and $\mu$BLS (c.f. Section~\ref{sec:BLS}) are complementary in the simple sense that TR-MOKE is operating in the time-domain and $\mu$BLS in the frequency domain. The difference between both methods is even more fading in experiments, where the modulation of the polarization is detected using a fast photodiode with a bandwidth of several tens of GHz connected to a vector network analyzer. A laser-scanning microscope using this detection scheme was recently described as microfocused frequency-resolved magneto-optic Kerr effect ($\mu$FR-MOKE) by Liensberger~\textit{et al.}~\cite{Lie2019}, where also a comparison is made to the signal of propagating magnons detected by means of $\mu$BLS. There have been earlier reports comparing TR-MOKE and $\mu$BLS on geometries theoretically by Hamrle~\textit{et al.}~\cite{Ham2010} and experimentally, where magnons are driven by a microwave antenna \cite{Per2005,Stig2016,Stig2018,Flac2019}. While both methods yield similar results, there is a fundamental difference. In standard TR-MOKE the reflected light is analyzed, whereas in $\mu$BLS the detected light, which is diffracted by an effective optical grating, may be generated even by a single magnon. The consequence is that using $\mu$BLS even thermally excited magnons or magnon auto-oscillations driven by spin currents can be detected, because no fixed phase correlation is required in contrast to the stroboscopic approach of TR-MOKE. This is of special importance if nonlinear effects such as magnon-magnon scattering create non-coherent magnons~\cite{Sch2009,Sch2012,Sch2019}, which are not detectable using a time-domain stroboscopic method. On the other hand, the advantage of TR-MOKE on the other side is the direct access to the magnon phase over the entire frequency range. Furthermore, since TR-MOKE is sensitive only to magnons coherent to the excitation source, a direct comparison of TR-MOKE and $\mu$BLS data allows to separate the signal originating from coherent and non-coherent magnons as shown in injection-locked, constriction-based spin Hall nano-oscillators by Hache and coworkers~\cite{Hache2019}.

\subsection{Time resolved scanning transmission X-ray microscopy}\label{sec:TR-STXM}

\indent Spin waves have become promising as information carriers for nanoscale spintronic devices for low-power consumption computing, and hence nanomagnonics~\cite{Grundler2016,Wagner2016,Ade2016} arises as a modern or future version of conventional magnonics~\cite{Kru2010}. One major challenge in nanomagnonics is to be able to excite and detect short-wavelength spin waves that enter the exchange-dominated regime~\cite{Kal1986,Len2011}. Much research effort has been made to be able to efficiently excite coherent short-wavelength spin waves with techniques such as waveguide tapering ($\lambda\approx 900$ nm)~\cite{Demi2011} or with thickness steps ($\lambda< 1~\mu$m)~\cite{Ono2018}, spin-transfer torque (STT) oscillators ($\lambda\approx 74$ nm)~\cite{Demi2010,Madami2011,Ura2014,Ake2018}, a spin-wave grating coupler ($\lambda\approx 90$ nm)~\cite{Yu2013,Yu2016,Yu2017}, intrinsic excitation from spin textures, \eg magnetic domain walls ($\lambda\approx 100-1600$ nm)~\cite{Hama2017}, magnetic vortices ($\lambda\approx 125$ nm)~\cite{Wintz2016} and ferromagnetic coplanar waveguides (mCPWs)~\cite{Che2020}. Up to now, coherent excitation of spin waves with the shortest wavelength have been achieved by Liu~\textit{et al.}~\cite{Liu2018} ($\lambda\approx 50$ nm). However, to be able to detect these short-wavelength spin waves is challenging. Micro-focused BLS excels in $k$ dependence and spatial resolution, but has a wavelength detection limit down to about 200 nm~\cite{Madami2011}. The PSWS can technically detect spin waves with wavelengths even below 50 nm~\cite{Liu2018} as long as the frequency limit of the VNA is not reached (\eg with $\lambda\approx 50$~nm, $f\approx 30$~GHz considering spin waves in a YIG thin film), but cannot achieve spatial resolution.

\begin{figure}[t]
\centering
\includegraphics[width=13cm]{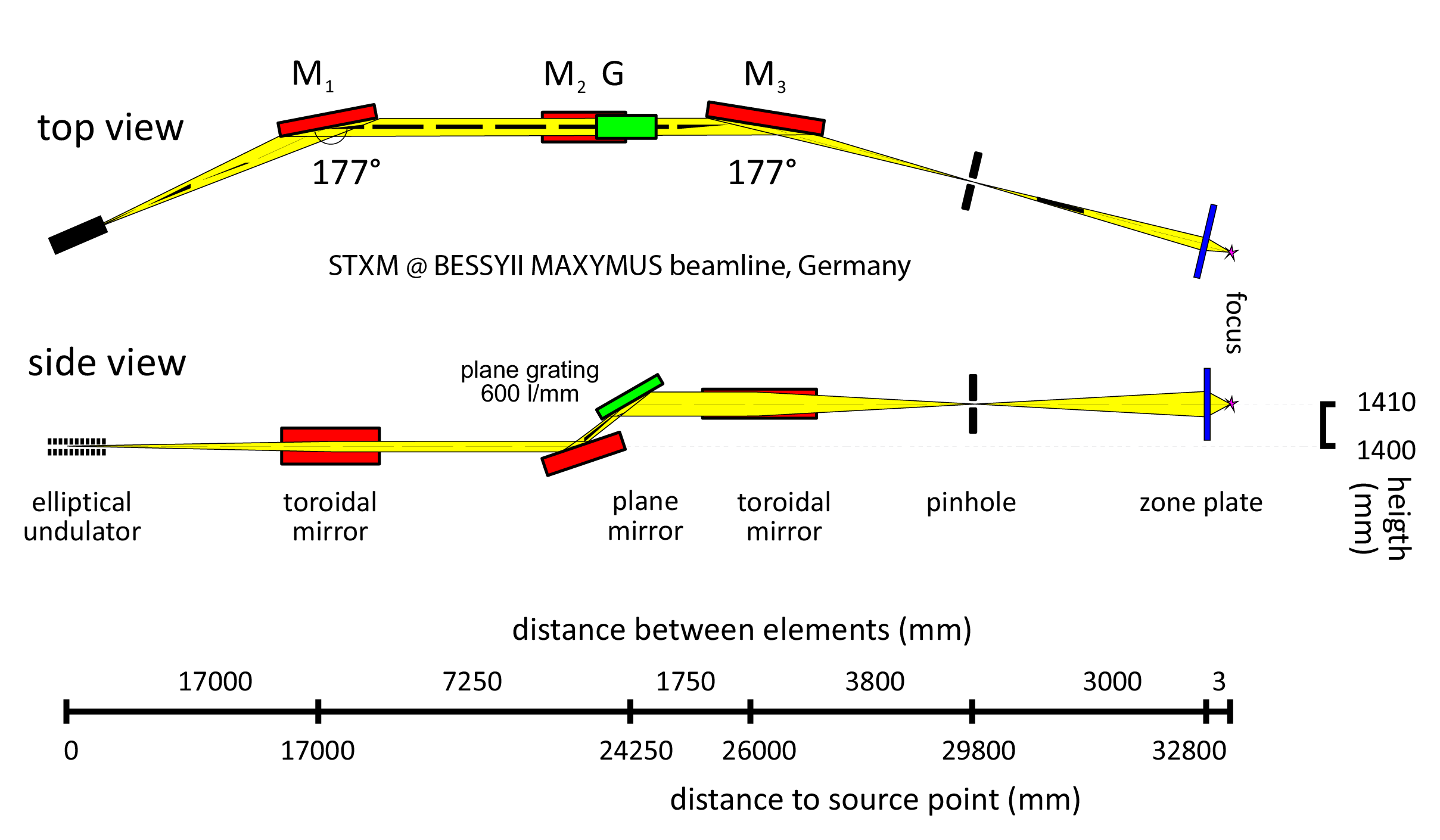}
\caption{Schematic for the optical design of the BESSY\uppercase\expandafter{\romannumeral2} MAXYMUS (MAgnetic X-raY Microscope with UHV Spectroscopy) beamline at Helmholtz Zentrum Berlin, as an example of the facility where TR-STXM is conducted. The image courtesy of Helmholtz Zentrum Berlin~\cite{MAXYMUS}. }
\label{fig:STXM}
\end{figure}

\indent At this point, the time resolved scanning transmission X-ray microscopy (TR-STXM) plays a critical role in detecting sub-100 nm wavelength spin waves with temporal and spatial resolution. Wintz~\textit{et al.}~\cite{Wintz2016} have utilized the BESSY\uppercase\expandafter{\romannumeral2} MAXYMUS beamline~\cite{MAXYMUS} (Fig.~\ref{fig:STXM}) to conduct TR-STXM and thereby have successfully resolved the short-wavelength propagating spin waves emitted from the magnetic vortex cores of NiFe/Ru/Co multilayers. The short-wavelength spin waves are emitted from the vortex cores and propagate outwards. The propagating spin wave patterns are resolved at three different frequencies of 1 GHz, 2 GHz and 4 GHz. A decrease of wavelength of the emitted spin waves from the vortex core is observed with the shortest wavelength for 4 GHz mode being 125 nm. The spin-wave dispersion shows a deviation from the linear relation which indicates the exchange regime is reached where spin-wave dispersion follows a $k^{2}$ dependence~\cite{Kal1986,Len2011}.

%-----------------------------------
%-----------------------------------

Dieterle~\textit{et al.}~\cite{Die2019} have used TR-STXM technique to study the short-wavelength spin waves in a permalloy disk with thickness $d=80$ nm. The first-order perpendicular standing spin waves (PSSW)~\cite{Kittel1948,Kli2018,Qin2018} with an additional finite in-plane wave vector are found to be coherently emitted from a vortex core. The dispersion relations for such heterosymmetric spin waves (c.f. Section~\ref{Sec:spin waves in ferromagnetic thin films}) considering both an in-plane wave vector $k$ and an out-of-plane wave vector $n\pi/d$ can be written as
\begin{equation}\label{eq:hetero}
f=\frac{\gamma}{2\pi}\sqrt{\left[Ak^2+A\left(\frac{n\pi}{d}\right)^2\right] \left[Ak^2+A\left(\frac{n\pi}{d}\right)^2+\mu_{0}M_{s}\right]},
\end{equation}
where $\gamma$ is the gyromagnetic ratio, $A$ is the exchange constant, $M_s$ is the saturation magnetization, and $n$ is the PSSW mode number with the first order being 1~\cite{Kittel1948,Stancil}. In-plane propagating first-order PSSWs with a wavelength down to 80 nm are observed at a frequency of 10 GHz~\cite{Die2019} resulting from a rather flat dispersion relation (Eq.~(\ref{eq:hetero})) for the first-order PSSWs with additional wave vector contribution of $\frac{n\pi}{d}$. At the same frequency, the uniform modes have much longer wavelengths ($\lambda>1~\mu$m). Very recently, Pile~\textit{et al.}~\cite{Pile2020} used the TR-STXM technique to probe spin waves in confined permalloy microstructures with non-standing characteristics.

\begin{figure}[t]
\centering
\includegraphics[width=13cm]{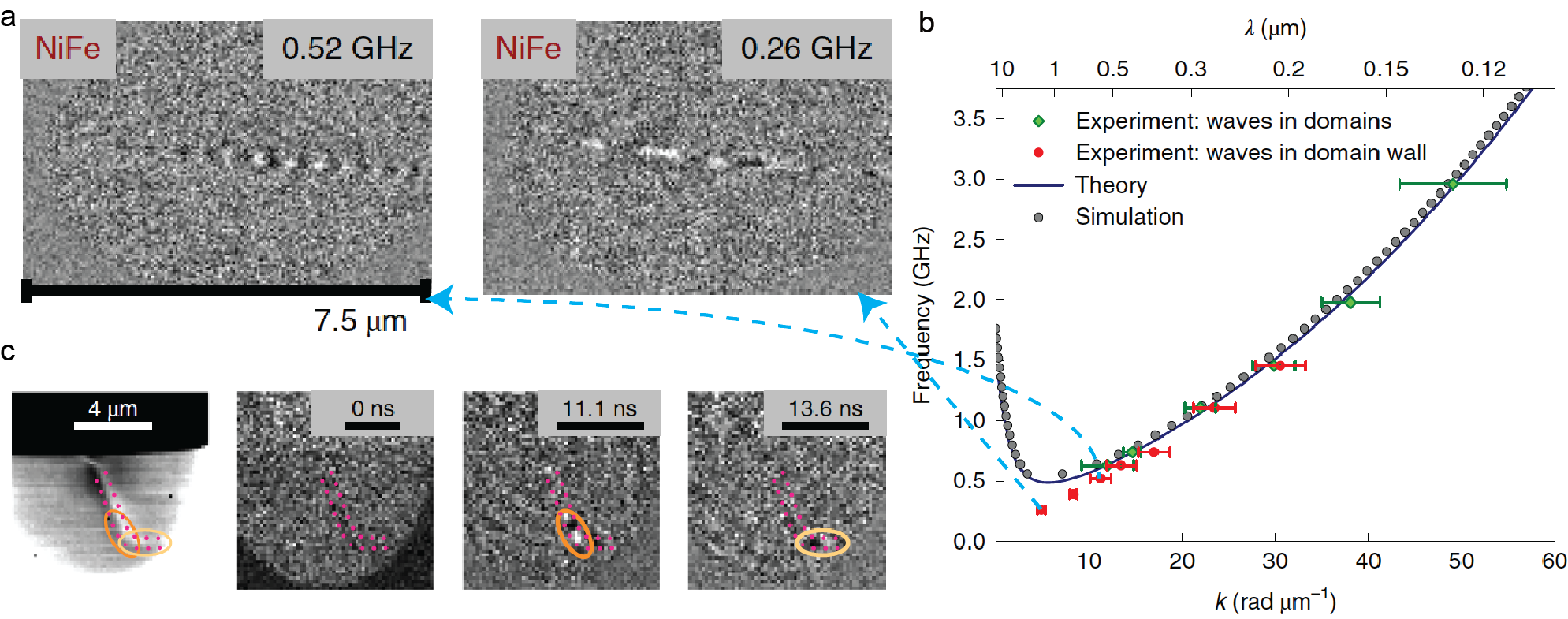}
\caption{(a) The TR-STXM characterization of the propagating spin waves confined in a domain wall excited at 0.52 and 0.26 GHz as indicated, with respectively different wavelengths. The contrast represents the out-of-plane component of the NiFe layer magnetization. (b) Spin-wave dispersion relations. The red dots (green diamonds) are data points extracted from the TR-STXM experimental results for spin waves in the domain walls (domains). The plane wave dispersions (blue continuous line) and the micromagnetic simulations (grey dots) are in good agreement. (c) The experimental demonstration of the spin wave propagation along a domain-wall-based waveguide around a corner. Images from left to right are, successively, an out-of-plane magnetic contrast image of a domain wall, TR-STXM snapshots of a spin wave packet with a magnetic field pulse at the excitation, after 11.1~ns and 13.6~ns. Reproduced from Ref.~\cite{Wintz2019}.}
\label{fig:STXM2}
\end{figure}

Spin waves propagating inside domain walls were first discovered by $\mu$BLS~\cite{Wagner2016} (c.f. Section~\ref{sec:BLS}), but using TR-STXM has recently allowed for measuring the dispersion~\cite{Wintz2019}. Figure~\ref{fig:STXM2}(a) shows the TR-STXM images for spin wave propagation in a magnetic domain wall nanochannel~\cite{Wagner2016} at two different frequencies of 0.52 GHz and 0.26 GHz. Different wavelengths of propagating spin waves are resolved. These results contribute as two data points in the spin wave dispersion relations shown in Fig.~\ref{fig:STXM2}(b), where the red dots are data points for spin waves in domain walls and the green diamonds are data for spin waves in domains. At low frequencies, only the spin waves in domain walls are allowed due to the low internal magnetic field inside the domain wall, and therefore the spin waves are guided in the domain wall nanochannels. Interestingly, using TR-STXM, Sluka~\textit{et al.}~\cite{Wintz2019} demonstrated that the spin waves can turn around a corner, if the domain wall nanochannel is curved. Thanks to the time-resolve functionality of the STXM, one may observe how the spin-wave packet stimulated by a magnetic pulse propagates after several nanoseconds (Fig.~\ref{fig:STXM2}(c)). After some 13.6 ns, the spin-wave packet around the corner is observed within the domain-wall nanochannel. Spin waves are observed to turn a corner~\cite{Vogt2012,Vogt2014,Vogel:2018gr} following the domain-wall track.

%-----------------------------------
%-----------------------------------

Further progress based on TR-STXM is shown by Albisetti~\textit{et al.}~\cite{Alb2019} combined with another novel technique of thermally assisted magnetic scanning probe lithography (tam-SPL)~\cite{Alb2016}. The tam-SPL can realize reconfigurable domain patterns in ferromagentic thin films~\cite{Alb2016,Alb2018}, which allows for artificially creating domain wall nanochannels for spin waves to propagate through. Very recently, Albisetti~\textit{et al.}~\cite{Alb2019} used tam-SPL to realize coherent spin wave interference and focusing based on 
reconfigurable domain patterns in synthetic antiferromagnets~\cite{Gru1981,duine_synthetic_2018,Shiota2020} and detected this using TR-STXM. Moreover, a strong spin-wave nonreciprocity~\cite{Di2015,Kwo2016,Chen2019} was also studied both with experiments and micromagnetic simulations.

In recognition of the versatility and advantages of TR-STXM for characterizing short-wavelength spin waves with high spatial and temporal resolution, there are some limitations, \eg the facilities are obviously not of table-top type~\cite{MAXYMUS} and more importantly, the samples must be prepared on substrates suitable for X-ray studies such as silicon nitride membranes, which constrain the measurements on high-quality magnetic oxide materials such as yttrium iron garnet (YIG), that become amorphous~\cite{Wes2017,Gomez2020} when prepared on membrane substrates. It is noteworthy that some recent progress in studying spin waves in YIG using TR-STXM has been made via, \eg transferring a lamella of YIG thin film from a gadolinium gallium garnet (GGG) substrate to a silicon nitride membrane~\cite{Forster2019} or etching away the underlying GGG substrate~\cite{Forster2019b}. 

\subsection{Detection of magnon transport by inverse spin Hall effect}\label{sec:ISHE}

%-----------------------------------
\begin{figure}[t]
\centering
\includegraphics[width=13cm]{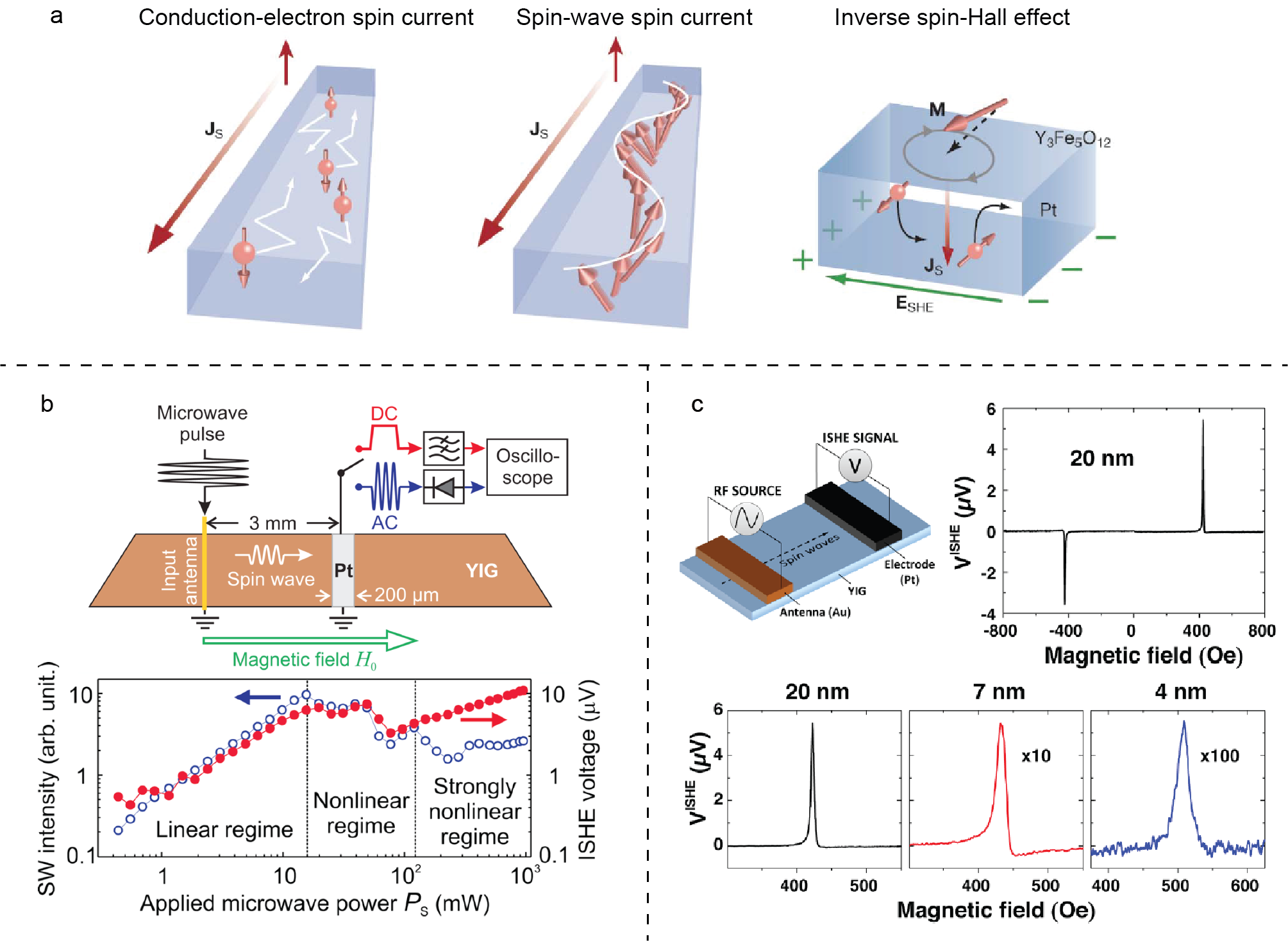}
\caption{(a) Conceptual illustrations for the conduction-electron spin current, spin-wave spin current and inverse spin Hall effect. Reproduced from Ref.~\cite{Saitoh2010}. (b) Spin waves excited by a microwave pulse and detected by ISHE with a Pt bar 3 mm apart from the excitation antenna. Both induced AC and ISHE DC voltage signals are measured as a function of the input microwave power. The linear, nonlinear, and strongly nonlinear regimes are indicated. Reproduced from Ref.~\cite{Chu2012}. (c) ISHE detection of propagating spin waves in ultrathin YIG films. The results on 20 nm-thick YIG film exhibits a sharp and narrow linewidth which yields an ultra-low damping of about 2.3$\times 10^{-4}$. Reproduced from Ref.~\cite{Madjid2014}.}
\label{fig:ISHE}
\end{figure}
%-----------------------------------

The inverse spin Hall effect (ISHE)~\cite{Sai2006,Val2006,Kim2007,Hoffmann2015,W_Han2019,Roma2020} can convert a spin current into an electric voltage in a heavy metal layer, usually a Pt bar. Propagating spin waves can be considered as a spin current conveyed by magnons~\cite{Saitoh2010,Maekawa_Spin_Current_2017,W_Han2020}. At the ferromagnet/heavy metal interface, spin waves are converted by the spin pumping effect~\cite{tserkovnyak_enhanced_2002,San2008,Heinrich2011,Y_Sun2013,Weiler2013} into a pure spin current in the heavy metal, which can be detected by a transverse voltage induced by the ISHE~\cite{Mosendz2010PRB,Mosendz2010PRL,Ando2011}. Figure~\ref{fig:ISHE}(a) shows conceptual illustrations of the conduction-electron spin current, spin-wave spin current and inverse spin-Hall effect~\cite{Saitoh2010}. Spin waves excited by spin-transfer torque (STT) or spin-orbit torque (SOT) can be detected by the inverse spin Hall effect~\cite{Saitoh2010,Chu2012,Col2016}. Kajiwara~\textit{et al.}~\cite{Saitoh2010} applied a large DC current into a Pt bar on one end of the YIG film and excited spin waves by STT which were detected on the other end of the YIG film as a voltage response when the current overcomes the magnetic damping above a critical value. The detected voltage is extremely small, down to a level of 1 nV. Chumak~\textit{et al.}~\cite{Chu2012} used a conventional stripline antenna to coherently excite spin waves at GHz frequencies (similar to the excitation technique used in the PSWS, c.f. Section~\ref{sec:PSWS}) in a 2.1~$\mu$m-thick YIG film detected both electromagnetically induced AC and ISHE-induced DC signals with a Pt strip 3~mm away from the antenna as shown in Fig.~\ref{fig:ISHE}(b). The power dependence measurements revealed instructive information that the linear regime holds until approximately 16 mW ($\sim$ 12~dBm). Thanks to recent advances in the fabrication of ultra-thin YIG films~\cite{Liu2014,Chang2014,Madjid2014,Yu2014}, ISHE detection of spin waves in nanometer-thick YIG/Pt system has been demonstrated~\cite{Madjid2014} with YIG thicknesses from 20 nm down to 4 nm. The linewidths extracted from the field-dependent ISHE voltage increase dramatically when the film thickness goes below 20 nm, which indicates an increase in magnetic damping (Fig.~\ref{fig:ISHE}(c)).

%-----------------------------------
\begin{figure}[t]
\centering
\includegraphics[width=13cm]{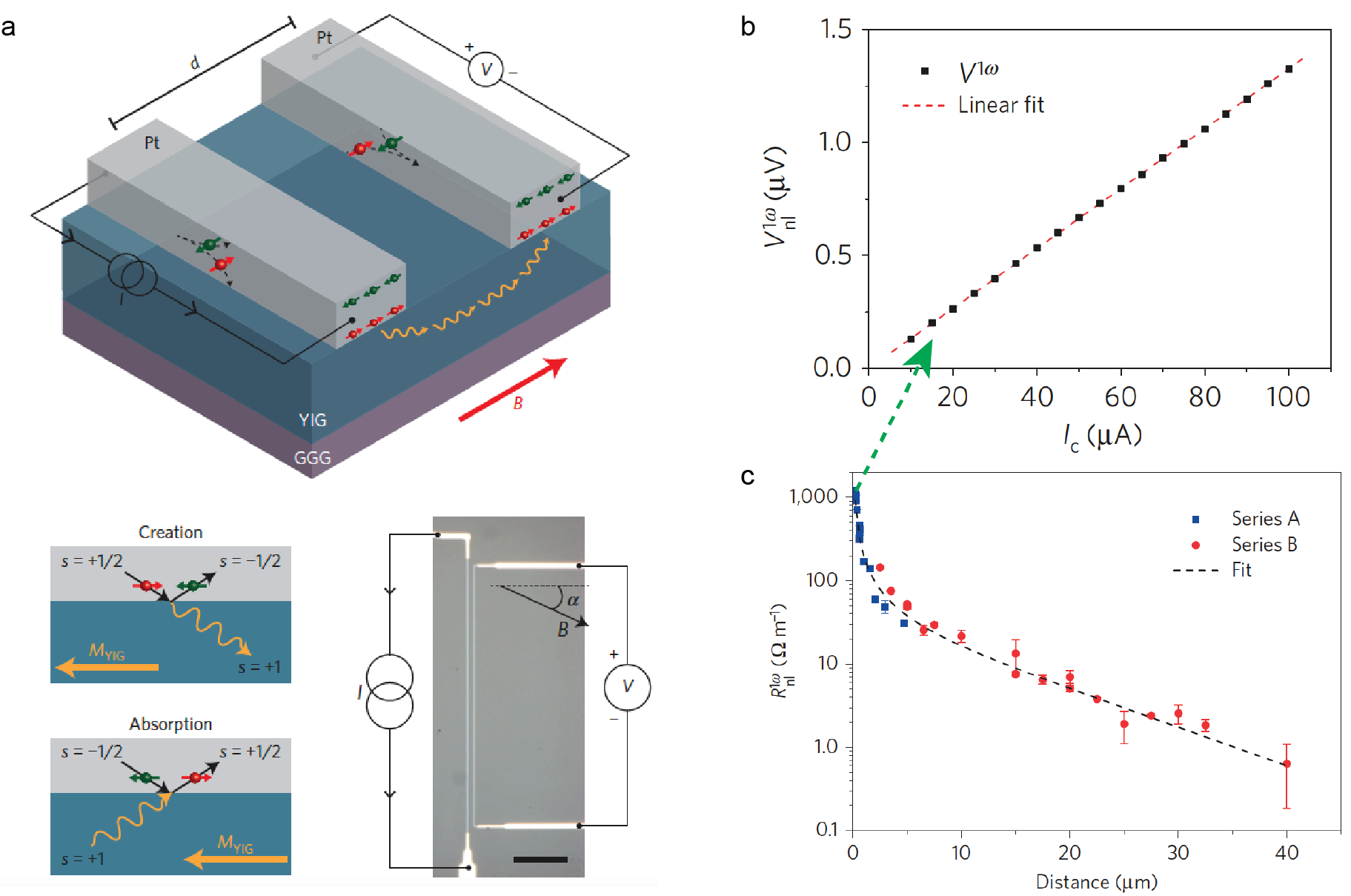}
\caption{(a) A schematic diagram of the creation and detection of magnons with Pt strips on top of a YIG film grown on a GGG substrate. A conduction electron scatters off and transfers angular momentum to YIG and creates magnons and the reverse process occurs for the absorption. The actual device is shown in the scanning electron microscope (SEM) image, where an electric current is injected into one Pt strip and detected in the second one with a lock-in amplifier. (b) The non-local voltage $V^{1\omega}_{\text{nl}}$ measured as a function of the injected current $I_{\text{C}}$ at a lock-in frequency of 5.939 Hz.The distance between the injector and detector for this investigated device is 200 nm. By a linear fit of the data, the non-local magnon resistivity $R^{1\omega}_{\text{nl}}$ can be extracted. The results measured on this device contribute as one data point in (c) indicated by a green arrow. (c) The non-local resistivity $R^{1\omega}_{\text{nl}}$ measured as a function of the magnon transport distance. Blue squares and red dots represent data points from two series of devices. The dashed line is a fit based on Eq.~(\ref{eq:nonlocal}), where a decay length of 9.4$\pm0.4~\mu$m is estimated. Reproduced from Ref.~\cite{Cor2015}.}
\label{fig:ISHE2}
\end{figure}
%-----------------------------------

Cornelissen~\textit{et al.}~\cite{Cor2015} have used two Pt strips to inject and detect spin-wave spin current based on the spin Hall effect and inverse spin Hall effect. The studied device is illustrated in Fig.~\ref{fig:ISHE2}(a). An AC current ($I_{\text{C}}$) is injected into one Pt strip for the generation of spin current and the nonlocal voltage at the first harmonic $V^{1\omega}_{\text{nl}}$ is measured on another Pt strip by a lock-in amplifier at a frequency of 5.939~Hz. The experimental results are shown in Fig.~\ref{fig:ISHE2}(b) for the device with an injector-detector distance of 200~nm, where a linear dependence of $V^{1\omega}_{\text{nl}}$ as a function of $I_{\text{C}}$, and thereby a non-local magnon resistance $R^{1\omega}_{\text{nl}}$, can be extracted from a linear fitting of the data. This extracted $R^{1\omega}_{\text{nl}}$ contributes to one of the blue squares in Fig.~\ref{fig:ISHE2}(c), where more data points are extracted from devices with different magnon transport distances $d$. With a two-parameter fitting function of
\begin{equation}\label{eq:nonlocal}
R^{1\omega}_{\text{nl}}=\frac{C}{\lambda}\frac{\text{exp}\left(d/\lambda\right)}{1-\text{exp}\left(2d/\lambda\right)},
\end{equation}
where $C$ is a distance-independent prefactor, one may extract the magnon spin diffusion length $\lambda$ to be around 9.4 $\mu$m at room temperature. Even in amorphous YIG, Wesenberg~\textit{et al.}~\cite{Wes2017} have demonstrated long-distance spin transport up to 100 $\mu$m. This is considered to result from strong local exchange interaction, which does not require long range order. In general, domain structures do not often form in YIG thin films. However, Mendil~\textit{et al.}~\cite{Men2019} recently observed zigzag magnetic domains in ultrathin YIG/Pt bilayers by X-ray photoelectron emission microscopy, which offers possibilities to study spin-wave propagation in domain structures in YIG/Pt bilayers by the ISHE. Notably, in addition to the ISHE there are other mechanisms involved in the electrical detection of magnetization dynamics, such as the spin rectification effect~\cite{Bai2013,C_Hu2016} and bolometric effect~\cite{Gui2007,Hahn2013}. 

\subsection{Nitrogen-vacancy centers and similar single spin systems}\label{sec:NV centers}

%-----------------------------------
\begin{figure}[t]
\centering
\includegraphics[width=14cm]{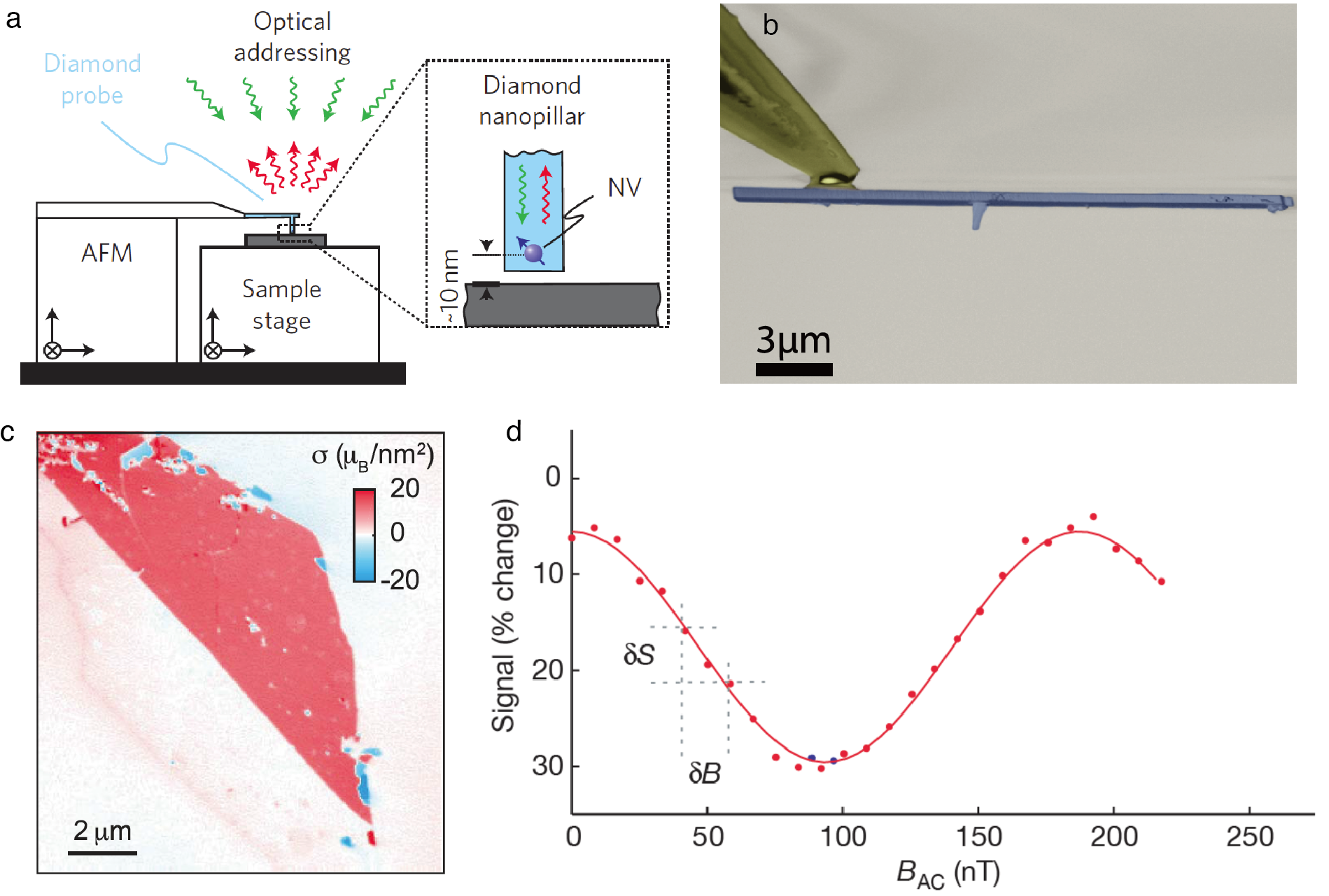}
\caption{(a) A schematic diagram of the experimental setup for NV center magnetometry, consisting of a combined optical and atomic
force microscope. A 532 nm laser (green arrows) is used to address the scanning NV center through its red fluorescence (red arrows). The scanning NV center resides in a diamond nanopillar (inset) and its proximity to the sample is maintained by means of atomic force microscope feedback. Reproduced from Ref.~\cite{Mal2012}. (b) An SEM image of the NV center probe attached to the end of a quartz tip. The image  is adapted from Ref.~\cite{App2016}. (c) Map of magnetization probed by an NV center scanning in few-layer flakes of CrI$_3$. Reproduced from Ref.~\cite{Thi2019}. (d) Experimental results for AC magnetometry. $\delta$B/$\delta$S determines the NV's sensitivity to changes in B$_{\text{AC}}$. This figure is adapted from Ref.~\cite{Maz2008}.}
\label{fig:NV1}
\end{figure}
%-----------------------------------

\begin{figure}[t]
\centering
\includegraphics[width=15cm]{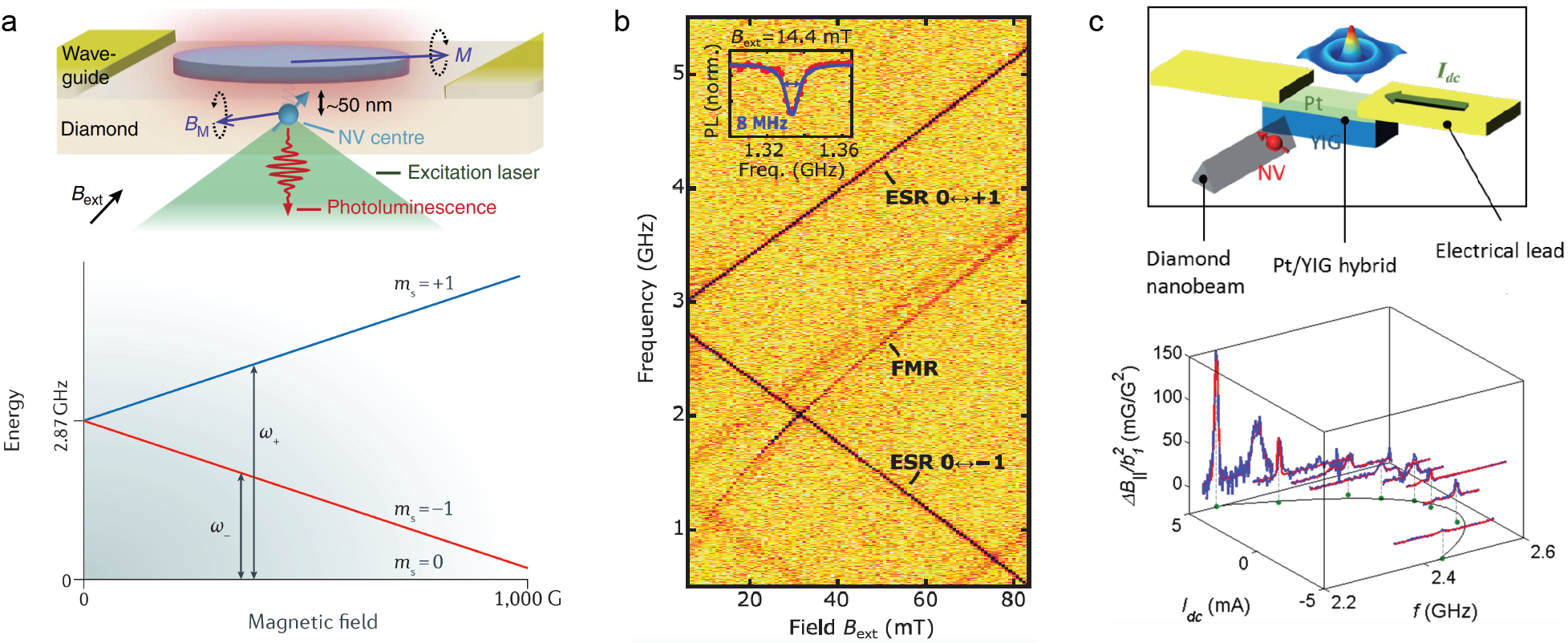}
\caption{(a) Probing spin waves using NV magnetometry on a permalloy microdisc (diameter 6 $\mu$m, thickness 30 nm) fabricated on top of a diamond surface. The ESR frequencies show Zeeman splitting under an external magnetic field applied parallel to the NV axis. Reproduced from Refs.~\cite{Sar2015,Cas2018}. (b) Normalized photoluminescence (PL) as a function of the external magnetic field B$_{\text{ext}}$ and the frequency of the injected microwave. Inset: A lineplot at $B_{\text{ext}}=14.4$~mT showing the 8-MHz linewidth. Reproduced from Ref.~\cite{Du2017}. (c) A sketch of spin-orbit torque oscillation induced by a DC current in a Pt bar and detected by NV magnetometry. The NV-measured stray fields generated by spin-wave resonance in YIG as a function of DC current in Pt strips. The stray fields are found to be dramatically enhanced when the DC current in Pt exceeds a critical value of about 4.5~mA. Reproduced from Ref.~\cite{Zhang2019}.}
\label{fig:NV2}
\end{figure}

When it comes to sensing the stray field of magnetic textures, nitrogen-vacancy (NV) center magnetometry can provide superbly high spatial resolution in imaging the magnetic textures in both static and dynamic approaches. The electron spin of the diamond nitrogen-vacancy (NV) defect can act as a magnetic field sensor which can be initialized optically and read out by spin-dependent photoluminescence as illustrated in Fig.~\ref{fig:NV2}(a). The spin is pumped into the $\ket{0}$ state by optical excitation, the excitation states of $\ket{\pm 1}$ can decay through metastable singlet states. Microwave excitation is able to manipulate the ground-state spin. The spin state can be detected through the emitted fluorescence. By applying an external magnetic field along the NV axis, the energy levels of the NV spin or the electron spin resonance (ESR) frequencies will exhibit a Zeeman splitting, which is particularly important for dynamic NV detection of spin waves at GHz frequencies. A more detailed description of the theoretical and experimental principles of the NV technique can be found in Ref.~\cite{Cas2018} and the references therein. As demonstrated by Maze~\textit{et al.}~\cite{Maz2008}, NV centers can probe extremely small magnetic field changes down to several nT. In addition, NV centers~\cite{Dohe2013} can realize ultralong spin coherence times at room temperature~\cite{Bala2009}. Figure~\ref{fig:NV1}(a) shows a sketch of the experimental setup of the NV center magnetometry as an example, where the NV center is manipulated by an atomic force microscope tip on top of the sample~\cite{Mal2012}. The key component of the entire setup is essentially the NV center probe, which is fabricated by nanolithography. A NV center probe recently developed by Malentinsky's group in Basel is shown in Fig.~\ref{fig:NV1}(b) as an example~\cite{App2016}. A recent experimental work of probing magnetism in two-dimensional van der Waals crystals CrI$_{3}$~\cite{Thi2019} has demonstrated its high sensitivity to extremely small stray fields and also the ultrahigh spatial resolution down to approximately 50~nm, which is mainly determined by the distance between the NV center and the sample surface.

%-----------------------------------
%-----------------------------------

The NV center magnetometry has recently been proven to be powerful for resolving static magnetic textures on the nanoscale such as magnetic domain structures~\cite{App2019}, magnetic domain wall structures~\cite{Tet2015}, magnetic Skyrmions~\cite{Jen2019} and antiferromagnetic chiral textures~\cite{Chau2019,Hay2020}. In addition, NV center magnetometry can also be used to study spin dynamics in magnetic textures at GHz frequencies~\cite{Cas2018}. As such, it becomes an exceptional tool for probing spin waves at the nanoscale with high sensitivity and therefore provides more possibilities for magnetic texture based magnonics. The NV center itself has an intrinsic electron spin resonance (ESR) frequency in the vicinity of 3~GHz. The ESR frequencies of NV centers can vary slightly from one NV center to another and are dependent on the magnetic field as shown in Fig.~\ref{fig:NV2}(a). Such resonant frequencies of NV centers are in the same range of the ferromagnetic resonance (FMR) frequencies of many ferromagnetic materials. For example, a permalloy (Py) microdisc has been chosen by {van der Sar}~\textit{et al.}~\cite{Sar2015} to study spin waves using NV magnetometry. The ESR of the NV center is excited by optical pumping and matches the frequency of the FMR in the Py disc excited by a coplanar waveguide as illustrated in Fig.~\ref{fig:NV2}(a). Apart from probing spin waves in ferromagnetic metals, Du~\textit{et al.}~\cite{Du2017} have recently demonstrated using NV magnetometry to probe spin waves in magnetic insulator YIG. The spin waves are excited by a gold stripline on top of the YIG film. A Pt bar is integrated on top of YIG film a few micrometers away from the gold stripline to vary the magnon chemical potential via the spin Hall effect~\cite{Liu2012}. The chemical potential change as a function of the current density in the Pt strip is probed by the NV magnetometry~\cite{Du2020}. The NV photoluminescence (PL) is highly sensitive to magnetic fields at the ESR frequencies~\cite{Sar2015}. Figure~\ref{fig:NV2}(b) shows the normalized PL spectra measured as a function of the external magnetic field B$_{\text{ext}}$ and the microwave frequency, where a sharp linewidth of 8 MHz is detected for the FMR of the YIG at 14.4~mT. When the current in the Pt bar exceeds a critical value, the associated spin orbit torque can overcome the damping of the YIG film and generate self-oscillations, and therefore a spin-orbit torque oscillation is realized~\cite{Col2016}. Very recently, the spin-orbit torque induced spin-wave resonance in YIG film was studied by Zhang~\textit{et al.}~\cite{Zhang2019} using NV magnetometry. When the DC current injected into the Pt strip is tuned up gradually and reaches a critical value of about 4.5~mA, the self oscillation occurs due to the spin-orbit torque effect and consequently the stray fields generated by the magnons in YIG are drastically enhanced, which has been observed experimentally using the NV magnetometry as shown in Fig.~\ref{fig:NV2}(c).

In general, FMR can be considered as a $k=0$ spin-wave mode~\cite{Stancil}. In recognition of its superb sensitivity and exceptionally high resolution, the challenge remains for NV magnetometry of how to use it to probe short wavelength spin waves (or exchange spin waves)~\cite{Kal1986,Yu2016,Wintz2016,Liu2018}. One severe technical challenge is that the resonance frequency of NV centers are constrained around 2 or 3 GHz regime, whereas the exchange spin waves are normally at 20 to 30 GHz~\cite{Liu2018} or at even higher frequencies. While one tunes the NV center frequency up with increasing magnetic field, the spin wave frequency increases too. However, spin waves in magnetic textures such as Skyrmions~\cite{Dirk2015} including artificial Skyrmions~\cite{Ding2013,Dustin2015} usually show relatively low resonance frequencies~\cite{Dirk2015} because of a completely different dispersion relation in comparison with the saturated ground state of the thin film. Therefore, the short-wavelength spin waves in magnetic Skyrmions may be detected by NV magnetometry with high field sensitivity and high spatial resolution.

% !TEX root = review_magnonics_in_texture.tex
\section{Spin waves in magnetic textures in ferromagnetic thin films}

\subsection{Spin wave propagation in ferromagnetic domain textures}

\subsubsection{Spin waves propagating inside nanochannels formed by domain-walls}\label{sec:SW in domain wall}

\begin{figure}[t]
\centering
\includegraphics[width=14cm]{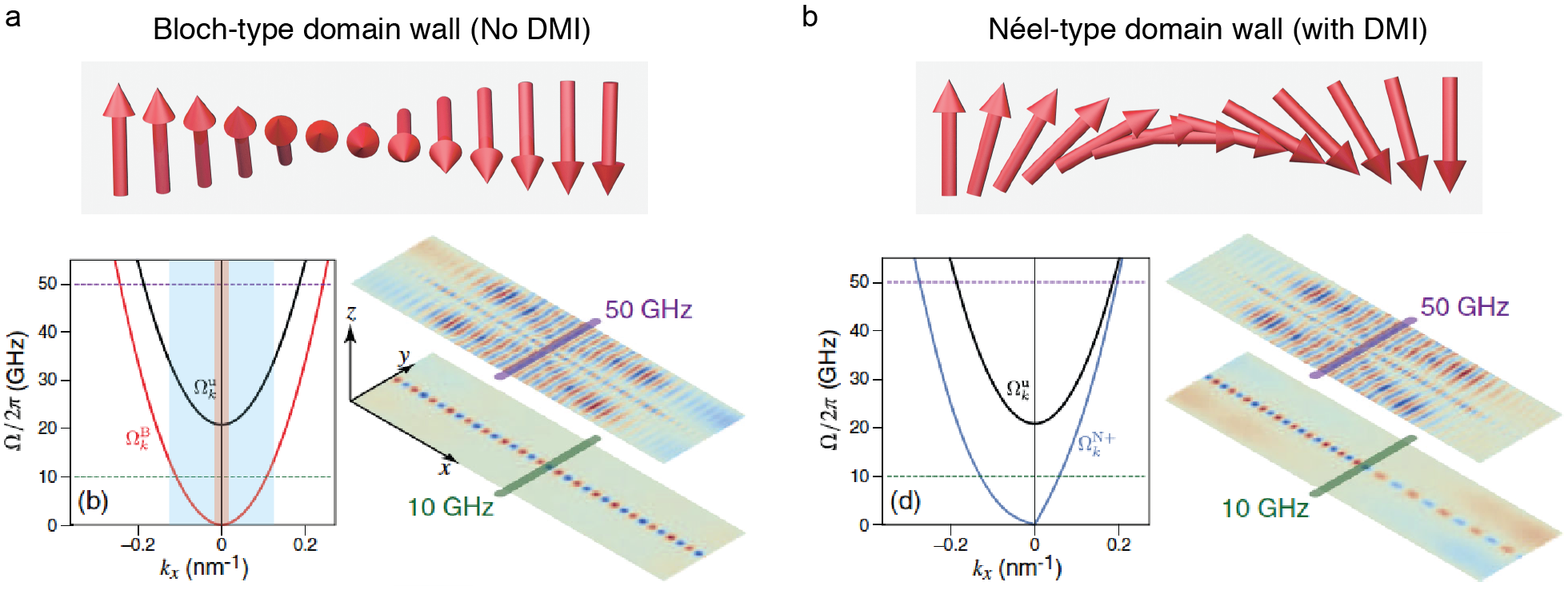}
\caption{(a) Spin wave propagation in Bloch-type domain wall texture at different frequencies of 10 GHz and 50 GHz studied by micromagnetic simulations. The red and black curves indicate spin-wave dispersions in the wall and the domain, respectively. (b) Spin wave propagation in N\'eel-type domain wall texture with DMI at 10 GHz and 50 GHz studied by micromagnetic simulations. The blue curve shows an asymmetric spin-wave dispersion for spin waves propagating in the domain wall in the presence of a DMI of $D=1.5$ mJ/m$^2$. A clear spin-wave nonreciprocity is observed. Illustration for domain walls in courtesy to Dr. Hanchen Wang. Figures reproduced from Ref.~\cite{Gar2015}.}
\label{fig:nanochannel}
\end{figure}

The pioneering work of Winter~\cite{Win1961} predicted that a Bloch-type domain wall can host a special wall-bound spin-wave mode, which is allowed to propagate along the domain wall texture. It has been theoretically found that quantized spin-wave modes can be formed in ferromagnetic and antiferromagnetic domain structures~\cite{wieser_quantized_2009}. Garcia-Sanchez~\textit{et al.}~\cite{Gar2015} predicted theoretically and in micromagnetic simulations using $\mu$max~\cite{umax} that spin waves can propagate through a nanochannel created by a magnetic domain wall in an ultrathin film having 1~nm thickness with strong perpendicular magnetic anisotropy (PMA). The fact that for a certain frequency range spin waves are confined within the domain wall is due to the reduction of the total internal magnetic field within the texture. The rotation of the magnetization across the domain wall creates magnetic volume charges resulting in a large dipolar magnetic field. This field is oriented anti-parallel to the magnetization and thus reduces the internal magnetic field with respect to the surrounding, uniformly magnetized domains. As a consequence, a spin-wave band for propagation along the domain walls is formed with a frequency range below the spin-wave band in the magnetic domains as shown in Fig.~\ref{fig:nanochannel}. The width of the domain wall nanochannel is approximately 18 nm as given by $\sqrt{A/K_{0}}$~\cite{Gar2015}, where $A$ is the exchange constant and $K_{0}$ is the effective perpendicular magnetic anisotropy. In the case of a N\'eel domain wall, when a Dzyaloshinskii-Moriya interaction (DMI) of $D=1.5$ mJ/m$^2$ is introduced into the simulation, a spin-wave frequency nonreciprocity is found in the domain-wall nanochannel (Fig.~\ref{fig:nanochannel}(b)). Lan~\textit{et al.}~\cite{Lan2015} have used this spin wave non-reciprocity induced by DMI to design a spin-wave diode that is promising for magnonic applications in low-power consumption nanoscale devices. The new paradigm of channeling spin waves in domain wall textures as proposed by Garcia-Sanchez~\textit{et al.}~\cite{Gar2015} is an important theoretical idea for the development of magnetic texture-based magnonics.

%-----------------------------------
%-----------------------------------

\begin{figure}[t]
\centering
\includegraphics[width=14cm]{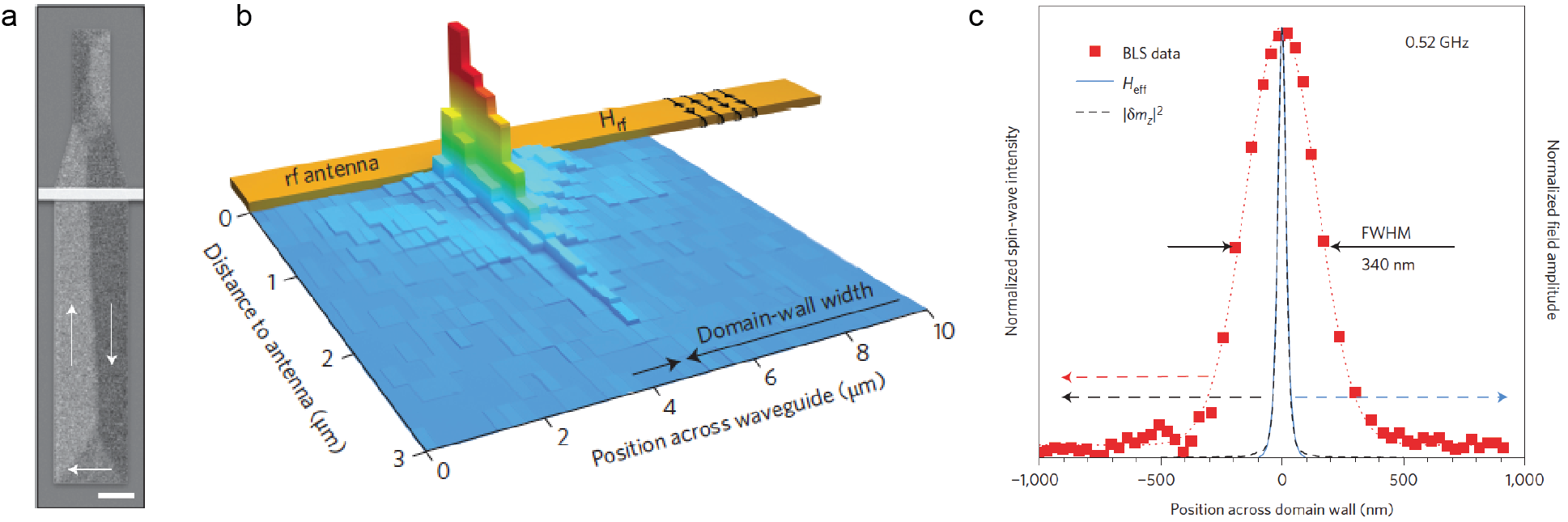}
\caption{(a) A Kerr microscopy image characterizing the domain pattern in a Py bar structure with an integrated microwave stripline. The arrows indicate the magnetization orientation. The scale bar is 5 $\mu$m. (b) Spin-wave spectra measured by BLS microscopy plotted as a two-dimensional intensity distribution of spin waves propagating along the nanochannel formed by the domain wall in the centre of the waveguide. (c) Spin-wave intensity across the domain wall width determined from experiment (squares) and simulation (dashed
line) for 0.52 GHz. The FWHM of 340 nm is on the same order as the focal spot size of the BLS microscope, which suggests that lateral confinement of spin waves in domain wall is even smaller. The simulation results shows a strong confinement of 40 nm for the spin waves propagating within the domain-wall nanochannel.  Figures reproduced from Ref.~\cite{Wagner2016}.}
\label{fig:nanochannel2}
\end{figure}

Remarkably, about the same time as when Garcia-Sanchez~\textit{et al.}~\cite{Gar2015} theoretically predicted a confined spin-wave propagation in domain walls, Wagner~\textit{et al.}~\cite{Wagner2016} demonstrated in experiments that spin waves propagate in a domain wall nanochannel in a permalloy film using Brillouin light scattering (c.f. Section~\ref{sec:BLS}). The experiment~\cite{Wagner2016} revealed the same physics discussed by Garcia-Sanchez~\textit{et al.}~\cite{Gar2015}, but was conducted independently. In the experimental work, Wagner and coworkers~\cite{Wagner2016} did not use thin films with PMA as studied in the theoretical work~\cite{Gar2015}, but instead the domain wall was created between two in-plane magnetized domains as shown in Fig.~\ref{fig:nanochannel2}. Experimental observations in films with PMA and iDMI are still lacking due to the high magnetic damping in materials with large spin-orbit interaction~\cite{Lat2018} and the low spin-wave group velocity resulting from the flat spin-wave dispersion typical in ultrathin films. Recent developments on doped yttrium iron garnet (YIG) thin films with PMA and low-damping may provide opportunities to overcome this challenge~\cite{Chen2019PMA,Sou2018,Deb2019}. Nevertheless, magnetic domain structures do not form in these low-damping garnet films as easily as in other PMA iron garnet films with much higher damping, \eg Ce-doped YIG ($\alpha\sim 0.048$)~\cite{Keh2015,Bilei2019} and TmIG ($\alpha\sim 0.013$)~\cite{Wu2018,Beach2019,Velez2019,Ding2019,Shao2019,F_Yang2020,Beach2020}. The ultralow damping CoFe~\cite{Sch2016,A_Lee2017,Flac2019,Y_Li2019,J_Du2020} and magnetic Heusler thin films~\cite{Fel2009,Miz2009,Seb2012,Bai2012,Seb2013,Loong2014,Qiao2016,Tobias2017,Ludbrook2017,S_Wang2018,Z_Zhang2019} may also provide excellent platforms for all-metal magnonic devices based on spin textures. Moreover, in ultrathin films with thicknesses around 1 nm, the signal is so small as to be undetectable by most spin-wave probing techniques. The second major challenge is to be able excite exchange spin waves with wavelength down to 60 nm as studied by Garcia-Sanchez~\textit{et al.}~\cite{Gar2015} with simulations. As discussed earlier in Section~\ref{sec:TR-STXM}, excitation of short-wavelength spin waves is in fact not trivial by any means. With the method proposed by Garcia-Sanchez~\textit{et al.}~\cite{Gar2015} as depicted in Fig.~\ref{fig:nanochannel}, one would need to fabricate ultra-narrow microwave striplines in order to be able to excite efficiently the high $k$ mode in the film. The current record for the narrowest stripline for spin-wave excitation is about 125 nm wide fabricated by Ciubotaru~\textit{et al.}~\cite{Ciu2016}, which generated the highest detected wavevector of $k=17$ rad/$\mu$m (wavelength $\lambda\sim$ 370 nm), \ie 0.017~rad/nm being merely at the very beginning of the spin-wave dispersion shown as the orange region in Fig.~\ref{fig:nanochannel}(a). By far, the shortest wavelength that can be coherently excited is approximately 50 nm, which yields a large wavevector up to 0.13 rad/nm~\cite{Liu2018}, approaching the $k$ vector studied in the simulation as indicated by the blue region in Fig.~\ref{fig:nanochannel}(a). However, in order to excite such a short wavelength, additional magnetic nanostructures need to be integrated on the 1 nm film, and the domain texture will be strongly influenced.

In the low $k$ regime, where most of the experimental works are relevant, the contribution of the dipole-dipole interaction cannot be neglected and actually plays a dominant role in the dispersion relation~\cite{Kal1986}. Conceptually in line with Garcia-Sanchez~\textit{et al.}~\cite{Gar2015}, Wagner~\textit{et al.}~\cite{Wagner2016} have demonstrated experimentally spin-wave propagation in a domain-wall nanochannel observed by Brillouin light scattering microscopy ($\mu$BLS) as shown in Fig.~\ref{fig:nanochannel2}. The device under investigation is a Ni$_{81}$Fe$_ {19}$ (permalloy, Py) bar structure, where the magnetic domain pattern is stabilized with one end being 5 $\mu$m wide and the other end being 10 $\mu$m wide. The magnetization is in the film plane as shown in the magneto-optical Kerr microscopy image (Fig.~\ref{fig:nanochannel2}(a)), which is the main difference compared to the PMA configuration proposed by Garcia-Sanchez~\textit{et al.}~\cite{Gar2015}. A Landau-like domain pattern is formed with a 180 degree N\'eel wall in the center separating two domains with opposite magnetization. A microwave stripline is then integrated on top of the Py structure perpendicular to the domain wall orientation as shown in Fig.~\ref{fig:nanochannel2}(a). The spin-wave intensity measured in the Py domain wall by $\mu$BLS (see Section~\ref{sec:BLS}) is shown in the 2D plot of Fig.~\ref{fig:nanochannel2}(b), where the propagating spin waves excited by the stripline antenna are found to be confined in the domain wall. The spatial distribution of the detected signal across the domain wall is shown in Fig.~\ref{fig:nanochannel2}(c). The full width at half maximum (FWHM) is found to be 340 nm which is very close to the BLS spot size. This means the real width of the spin-wave nanochannel created by the domain wall texture is below 100 nm in agreement with micromagnetic simulations revealing a domain-wall width and the width of the confined spin-wave beam to be about 40 nm. Notably, the spin-wave nanochannel created by the domain wall can be steered by an applied field, demonstrated in the $\mu$BLS measurements (see Fig.~\ref{fig:BLS_2}(b)).

Recently, Sluka~\textit{et al.}~\cite{Wintz2019} observed spin-wave propagation along domain walls within a Py microdisc by time-resolved scanning transmission X-ray microscope (TR-STXM) as presented in Section~\ref{sec:TR-STXM}. The spin waves follow the domain wall direction even when the domain wall turns at an angle~\cite{Wintz2019} as shown in Fig.~\ref{fig:STXM2}. A recent work done with micromagnetic simulations on spin wave propagation in a Bloch-type domain wall by Henry~\textit{et al.}~\cite{Hen2019} with micromagnetic simulations showed a unidirectional spin-wave channeling as a result of dynamic dipolar interactions~\cite{Hil1990,Pig2012}. Based on the simulation results of Henry~\textit{et al.}~\cite{Hen2019}, in stripe domain structures~\cite{Liu2019,Tac2014} spin waves propagate in opposite directions in neighbouring domain walls. Recent progress on the development of thermally assisted magnetic scanning probe lithography (tam-SPL)~\cite{Alb2016,Alb2018,Alb2019} for patterning magnetic domain textures has offered more opportunities and versatility in designing nanomagnonic circuits based on reconfigurable domain-wall textures~\cite{Alb2018a}.

%-----------------------------------
%-----------------------------------

\subsubsection{Spin wave propagation across domain walls}\label{sec:reprogrammable SW with DW}

The theory developed by Thiele~\cite{Thie1973} demonstrated that spin waves can propagate through a Bloch-type domain wall without any reflection. Although the transmitted spin wave amplitude is sustained, its phase, however, can be changed as studied theoretically by Bayer~\textit{et al.}~\cite{Bay2005} as
\begin{equation}\label{eq:phase_DW}
\Delta\phi(k_{\text{in}})=\int_{-\infty}^{+\infty}\left(k(x)-k_{\text{in}}\right)dx,
\end{equation}
where $k_{\text{in}}$ is the wavevector of the spin wave impinging on the domain wall, $x$ is the position in the direction of propagation perpendicular to the domain wall. The propagating spin waves tend to maintain their frequency while passing through the domain wall texture, which results in an adjustment of the wavevector $k(x)$ to conform with the dispersion relations along the domain wall width and thereby a phase shift is generated between the inbound and outbound spin waves across the Bloch-type domain wall. The interactions between spin waves with a N\'eel type domain wall~\cite{macke_transmission_2010,borys_spinwave_2016}, on the other hand, have been investigated by Chang~\textit{et al.}~\cite{Cha2018} in micromagnetic simulations.
The domain wall motion induced by a magnon-driven spin-transfer torque has been studied theoretically by Yan~\textit{et al.}~\cite{Yan2011}. Micromagnetic simulations~\cite{Han2009,Kim2012,Wang2012} have also been conducted to study the domain wall motion induced by propagating spin waves. Experiments by Woo~\textit{et al.}~\cite{Woo2017} have demonstrated domain wall depinning assisted by spin-wave pulses. Theoretically, Le Maho~\textit{et al.}~\cite{Mah2009} studied the spin-wave contributions to current-induced domain wall dynamics. In addition to exciting magnons by microwave pulses, a temperature gradient can also generate a directional flow of magnons which would exert a thermal spin-transfer torque (TST)~\cite{Hatami2007,Yu2010,Slon2010} on magnetic domain walls~\cite{Nowak2011}, even in magnetic insulators~\cite{Jiang2013}. The TST eventually leads to domain-wall motion if the temperature gradient reaches a critical value, \eg approximately 50 K/cm, which has been demonstrated experimentally by Jiang~\textit{et al.}~\cite{Jiang2013}. In general, to move domain walls purely by spin waves is rather challenging and requires considerably large spin-wave amplitudes or thermal gradient, which will be further discussed in Section~\ref{sec:Domain wall driven by spin waves}.

%-----------------------------------
\begin{figure}[t]
\centering
\includegraphics[width=14cm]{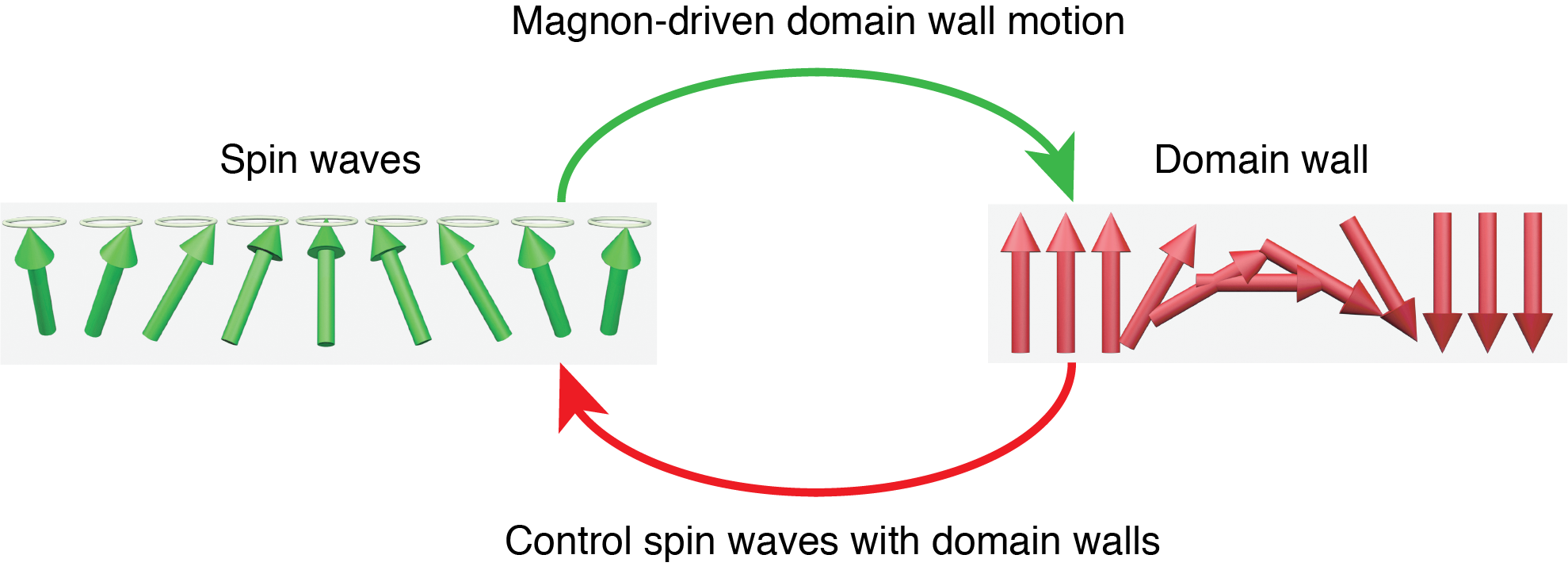}
\caption{The interaction between spin waves and a domain wall. Spin waves with sufficient power flow can move domain walls. Inversely, one can control spin wave propagation with domain walls.}
\label{fig:across_DW}
\end{figure}
%-----------------------------------

\begin{figure}[t]
\centering
\includegraphics[width=14cm]{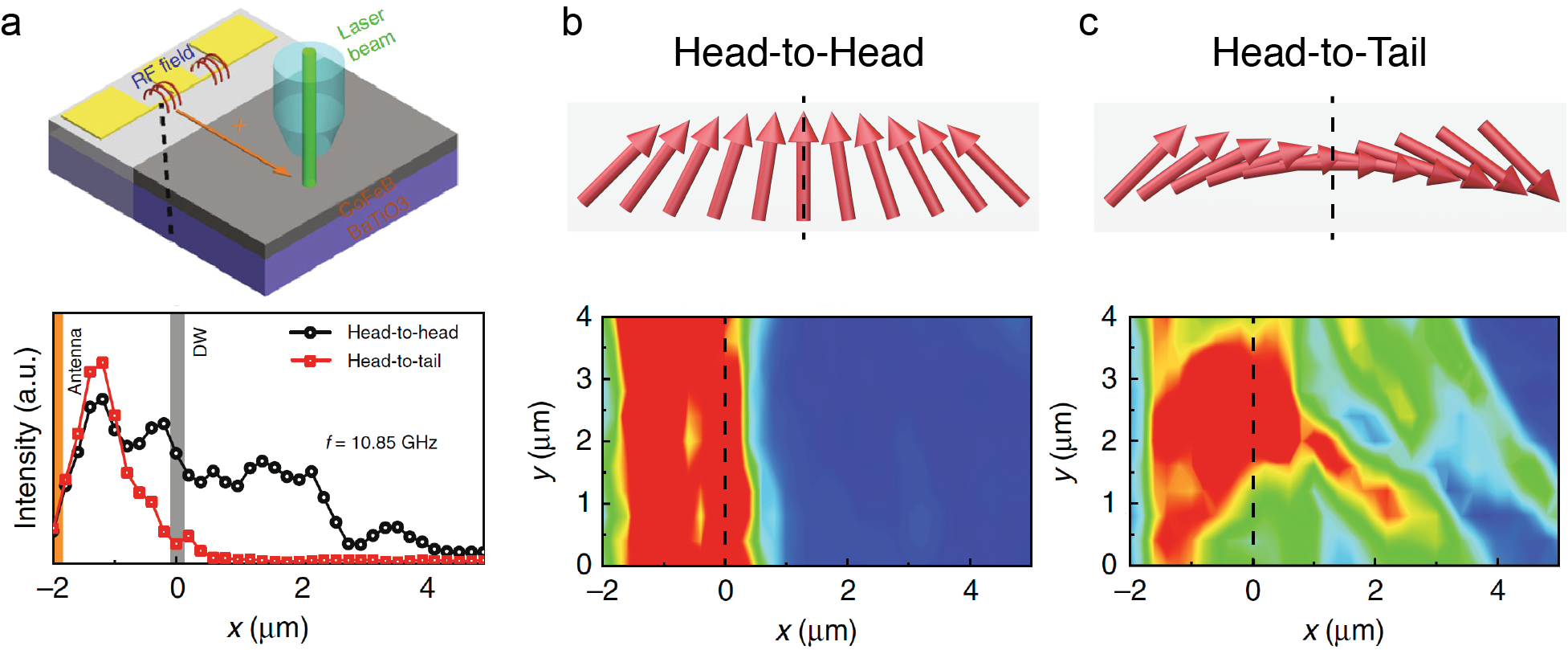}
\caption{(a) Spin waves propagating across domain walls detected by $\mu$BLS in a CoFeB/BaTiO$_{3}$ bilayer system. Spin waves are excited by a stripline antenna at 10.85~GHz and travel across different types of domain walls resulting in different amplitudes. Two dimensional spin-wave intensity maps are presented in (b) Head-to-Head and (c) Head-to-Tail domain walls. Reproduced from Ref.~\cite{Hama2018}.}
\label{fig:across_DW2}
\end{figure}

However, the reverse process, \ie the influence of domain walls on propagating spin waves (Fig.~\ref{fig:across_DW}), can be highly effective, which provides possibilities to control spin wave propagation via domain walls. H\"am\"al\"ainen~\textit{et al.}~\cite{Hama2018} demonstrated experimentally and with simulations that different domain wall textures can affect spin-wave propagation. 90$^{\circ}$ N\'eel type domain walls were formed in CoFeB imprinted from ferroelectric domain stripes in BaTiO$_{3}$ substrates~\cite{Lah2011}. Two types of 90$^{\circ}$ N\'eel domain wall textures can be formed with head-to-head (tail-to-tail) and head-to-tail configurations as shown in Fig.~\ref{fig:across_DW2}. In their experiments, spin waves were excited by a stripline antenna, propagate across a domain wall, and were detected by Brillouin light scattering microscopy ($\mu$BLS) as illustrated in Fig.~\ref{fig:across_DW2}(a). Spin waves can propagate across the Head-to-Head domain wall while maintaining their amplitude, whereas spin waves propagating across the Head-to-Tail domain wall are strongly suppressed. The spatial distribution of the detected BLS signals is shown in Fig.~\ref{fig:across_DW2}(b) and (c) for spin waves propagating across Head-to-Head and Head-to-Tail, respectively. Such a distinctive behavior of spin waves propagating across different domain wall textures provides possibilities to create a planar magnetic spin-wave valve where one can control spin-wave propagation by switching the magnetization of the middle domain sandwiched between two pinned domains. This magnetic spin-wave valve proposed by H\"am\"al\"ainen~\textit{et al.}~\cite{Hama2018} can be considered a planar version of the magnon valves~\cite{Han2018,Cor2018,Cra2018} in a vertical configuration. In vertical magnon valves~\cite{Han2018,Cor2018,Cra2018}, magnon transport perpendicular to the film plane can be allowed or prohibited by switching the magnetization of two separated YIG layers between parallel and antiparallel configurations. In the planar magnon valve~\cite{Hama2018}, one can control spin waves either to propagate through the domain walls or to be blocked by the domain walls depending on the domain wall configuration of either Head-to-Head/Tail-to-Tail or Head-to-Tail, which is tuneable by switching the middle domain orientation. Furthermore, the annihilation~\cite{Gho2017} or collision~\cite{Ram2014} of domain walls can emit spin waves. The domain wall textures can also reduce the propagation length of antiferromagnetic spin waves due to scattering at domain walls~\cite{ross_propagation_2020}.

%-----------------------------------
%-----------------------------------

\subsubsection{Spin wave phase shifts induced by domain walls}
\begin{figure}[t]
\centering
\includegraphics[width=16cm]{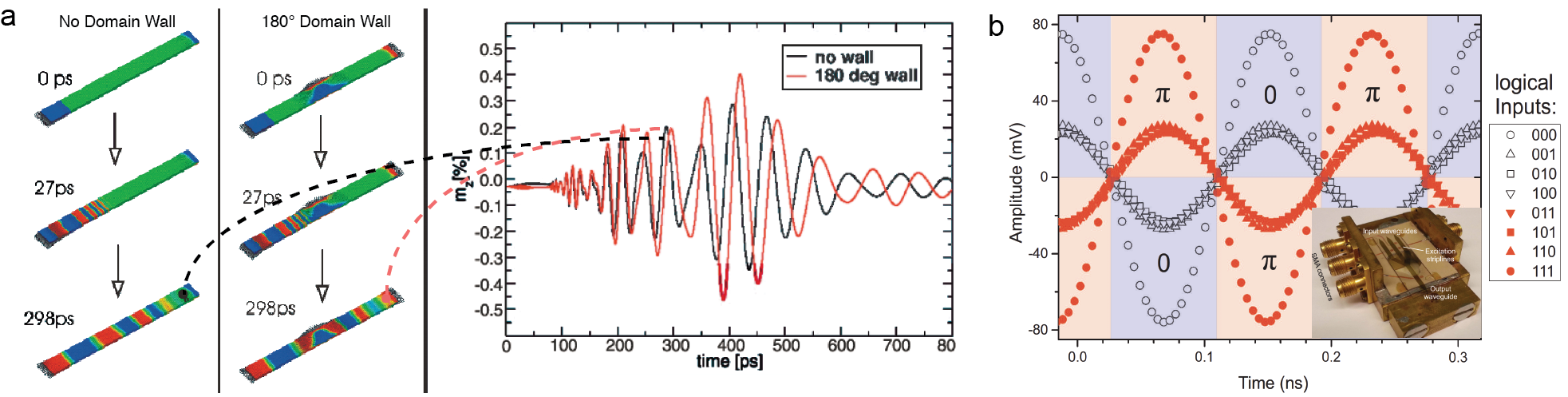}
\caption{(a) Simulation results of spin-wave propagation through a 180$^\circ$ domain wall in comparison with no domain wall. The time-domain signals are plotted for 180$^\circ$ domain wall (red curve) and no wall (black curve), where a clear phase shift is observed. Reproduced from Ref.~\cite{Her2004}. (b) An experimental prototype of a spin-wave logic majority gate (inset). The output signals are measured by an oscilloscope with a high sampling rate, which allows direct mapping of the spin-wave amplitude. The logic inputs are shown on the right-hand side. Reproduced from Ref.~\cite{Fis2017}.}
\label{fig:phase_shift}
\end{figure}

Spin waves propagating across domain walls will not only have a change in amplitude, but can also experience a change in phase as described in Eq.~(\ref{eq:phase_DW}). This domain-wall-induced phase shift of spin waves was first investigated using micromagnetic simulations by Hertel~\textit{et al.}~\cite{Her2004}. Their results are shown in Fig.~\ref{fig:phase_shift}(a) where a clear phase shift is manifested by comparing the spin wave propagation with and without a domain wall. To be able to control the spin-wave phase is of great importance in designing a spin-wave logic device based on spin-wave interference~\cite{Fis2017} (Fig.~\ref{fig:phase_shift}(b)). In principle, the idea is to deploy constructive and destructive spin-wave interference to implement spin-wave computing with phase shifts being $\Delta \phi=\pi$ and $\Delta \phi=0$, respectively~\cite{Khi2005,Kos2005}. The challenge, however, lies in how to effectively generate a phase shift of spin waves, where domain walls can be of great use. Spin waves propagating through a domain wall need to maintain the same frequency. However, the alternating magnetic fields inside domain walls force spin waves to adiabatically adjust their wavevector $k$. In other word, $k(x)$ changes due to an alternating $H(x)$ in domain walls and hence a phase shift $\Delta \phi$ is accumulated in the process of spin wave propagation through domain walls as described in Eq.~(\ref{eq:phase_DW}). Based on this concept, several spin-wave logic devices are proposed~\cite{Khi2010,Csa2017}.

%-----------------------------------
%-----------------------------------

Interestingly, it has recently been demonstrated analytically and with micromagnetic simulations~\cite{Bui2016} that different types of domain walls, \eg Bloch walls and N\'eel walls, can generate different spin-wave phase shifts. Buijnsters~\textit{et al.}~\cite{Bui2016} predicted that the phase shift of spin waves propagating through a domain wall is chirality dependent, which originates from the interfacial Dzyaloshinskii-Moriya interaction and is interpreted as a geometric phase. 

\subsubsection{Domain wall motion driven by spin wave emission}\label{sec:Domain wall driven by spin waves}

%-----------------------------------
\begin{figure}[t]
\centering
\includegraphics[width=16cm]{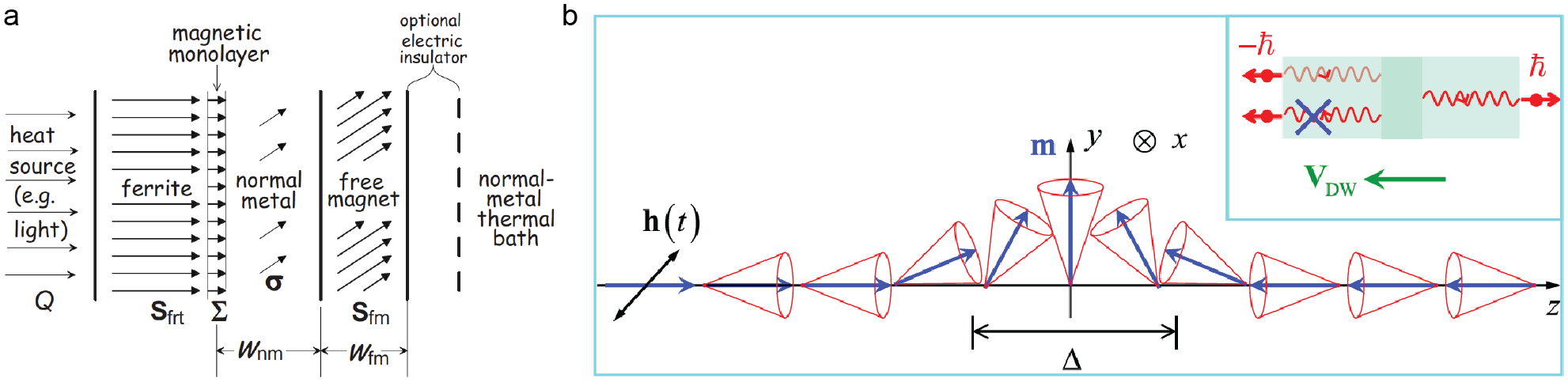}
\caption{(a) A magnetic multilayer structure proposed by Slonczewski~\cite{Slon2010} to study magnonic STT. Reproduced from Ref.~\cite{Slon2010}. (b) A sketch for describing the mechanism of magnon-driven domain wall motion. Inset shows the spin angular momentum transfered from the propagating spin waves to the domain wall. Reproduced from Ref.~\cite{Fis2017}.}
\label{fig:DW_motion1}
\end{figure}
%-----------------------------------

It has been discussed in previous sections that domain walls can influence spin wave propagation by creating a phase shift and modulating the amplitude, which also depends on different domain wall textures. As illustrated in Fig.~\ref{fig:across_DW}, spin waves can, in turn, act on domain walls, and when the strength of the spin waves is sufficiently large, domain walls can be moved as predicted first by Yan~\textit{et al.}~\cite{Yan2011}. It is well known that a spin-polarized electric current can exert a spin-transfer torque (STT)~\cite{Slon1996,Ber1996} on the magnetization. When the current density exceeds a critical value, typically of the order of $10^{7}$A/cm$^2$ (this value may vary with different materials depending on the spin polarization at the Fermi energy $E_{\text{F}}$), the magnetization can be switched by the STT~\cite{Bra2012}. Domain wall motion driven by STT~\cite{Shi2011} is the key to realization of the magnetic domain-wall racetrack memory~\cite{Parkin2008,Parkin2015}. After the theoretical prediction of STT in 1996~\cite{Slon1996}, Slonczewski once again predicted in 2010~\cite{Slon2010} a magnonic spin torque that is mediated by a spin-wave spin current, rather than a spin-polarized electric current. Figure~\ref{fig:DW_motion1}(a) depicts the multilayer system as proposed by Slonczewski for studying the magnonic spin torque. Le Maho~\textit{et al.}~\cite{Mah2009} have discussed theoretically the spin-wave contributions to the current-induced domain wall dynamics. Only very recently, Wang~\textit{et al.}~\cite{YWang2019} have demonstrated experimentally the magnetization switching by magnon-mediated spin torque in a heterostructure of Bi$_2$Se$_3$/antiferromagnetic insulator NiO/ferromagnetic metal Py. Just as conventional STT can drive domain-wall motion~\cite{Yama2004,Haya2006}, Yan~\textit{et al.}~\cite{Yan2011} predicted theoretically that the magnonic spin torque can also drive domain-wall motion as sketched in Fig.~\ref{fig:DW_motion1}(b). 

\begin{figure}[t]
\centering
\includegraphics[width=16cm]{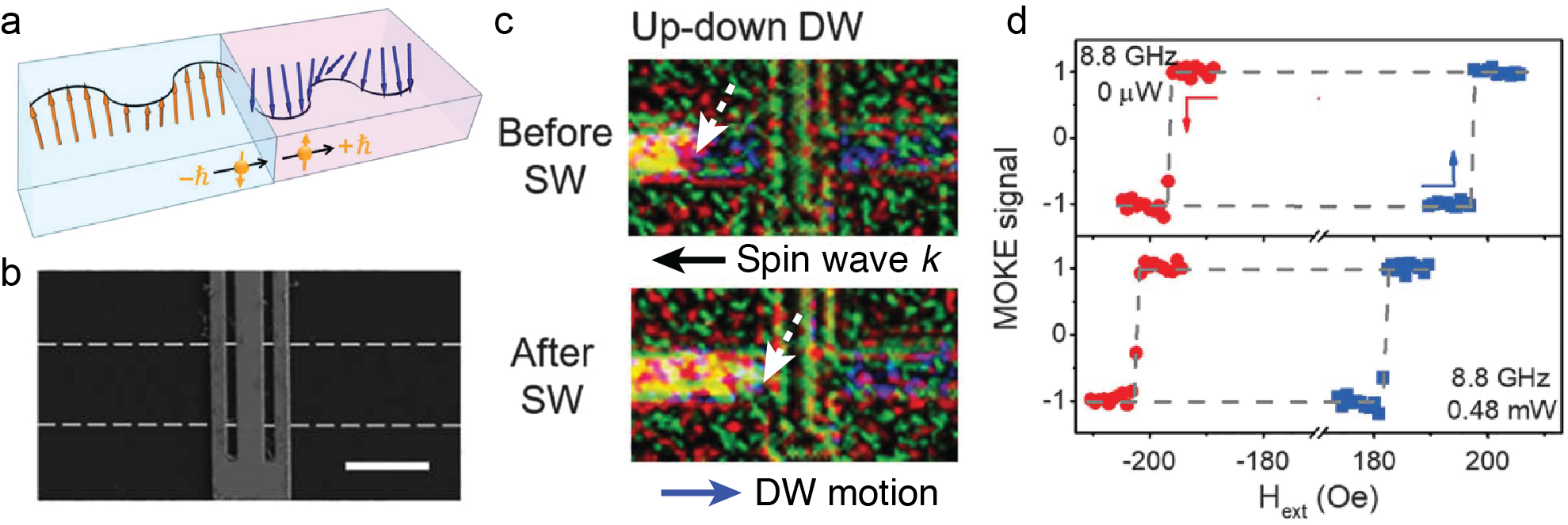}
\caption{(a) A schematic diagram of spin-wave propagation across a domain wall in a magnetic thin film with perpendicular magnetic anisotropy. Spin angular momentum is transferred from spin waves to the domain wall. (b) An SEM image of the investigated device where a CPW is integrated on a magnetic channel with a domain wall. The scale bar is 4 $\mu$m. (c) MOKE images showing the change of domain-wall position (white arrows) with spin waves injected from the CPW. The domain-wall motion direction is opposite to the spin-wave wavevector $k$. (d) The change of the domain-wall depinning field observed by MOKE with an injection of spin waves at the spin-wave resonance frequency of 8.8 GHz. Reproduced from Ref.~\cite{Han2019}.}
\label{fig:DW_motion2}
\end{figure}

Very recently, Han and coworkers~\cite{Han2019} have managed to experimentally observe the domain-wall motion induced by coherent spin waves in cobalt/nickel multilayer films using PSWS (c.f. Section~\ref{sec:PSWS}) and MOKE (c.f. Section~\ref{sec:MOKE}). The Co/Ni multilayer films exhibit strong perpendicular magnetic anisotropy (PMA). A coplanar waveguide (CPW) is patterned on top of the Co/Ni multilayer channel where microwaves can be injected. The MOKE microscopy images of the domain structures are presented in Fig.~\ref{fig:DW_motion2}(c), where an up-down domain wall is nucleated by external magnetic field pulses in the boundary of the light and dark regions. After applying spin waves, the domain wall moves close to the microwave antenna, indicating that the moving direction of the domain wall is against the spin wave transport direction, which is in excellent agreement with the theoretical prediction by Yan~\textit{et al.}~\cite{Yan2011}. As discussed in Section~\ref{sec:dynamics of magnetic textures}, the Noether theorem suggests that the angular momentum of the domain wall corresponds to a linear motion and the linear momentum of the domain wall corresponds to a rotation of its magnetization tilting~\cite{Yan2013}. The equation of motion and the frequency dependence of the domain wall motion driven by magnons are described by Rising$\aa$rd~\textit{et al.}~\cite{Rising2017}. In addition, the magnonic spin torque is experimentally determined by the switching phase diagram of the domain wall by applying both spin waves and an external magnetic field. The depinning field shifts when applying spin waves at the resonance frequency, indicating spin waves assist the domain wall to move close to the microwave antenna, while resisting moving it in the opposite direction. This observed direction of the domain-wall motion is different from the field-induced domain-wall motion under the assistance of transient spin waves by Woo~\textit{et al.}~\cite{Woo2017}, where spin waves assist domain walls to move along their propagation direction. With x-ray magnetic circular dichroism photoelectron emission microscopy, the magnetic domain walls were found to be steered by an ultrashort pulsed laser~\cite{Sho2019}, which was attributed to the torque exerted by thermally-induced magnons. Domain wall motion driven by thermally-induced magnons has also been studied~\cite{Jiang2013,Yan2015,Mazo2016,More2017}. Recently, Kim~\textit{et al.}~\cite{Kim2019} theoretically demonstrated unidirectional magnon-driven domain wall motion in the presence of inversion symmetry breaking. The unidirectionality lies at the heart of spin logic devices~\cite{Lan2015,bracher_creation_2017,Chen2019}, for example, spin diodes. When the intrinsic asymmetry, namely the DMI, is introduced to the magnetic/nonmagnetic bilayer systems, a unidirectional domain-wall motion can be induced by spin waves with both $+k$ and $-k$ wave vectors. In antiferromagnets, chiral domain walls can be controlled using spin-wave with opposite helicity as demonstrated theoretically by Qaiumzadeh and coworkers~\cite{qaiumzadeh_controlling_2017}, which is further discussed in Section~\ref{sec:AFM domain wall motion}.

%-----------------------------------
%-----------------------------------

\subsection{Helimagnon and Skyrmion excitations}

\subsubsection{Helimagnon resonances in magnonic crystals}
\begin{figure}[t]
\centering
\includegraphics[width=15cm]{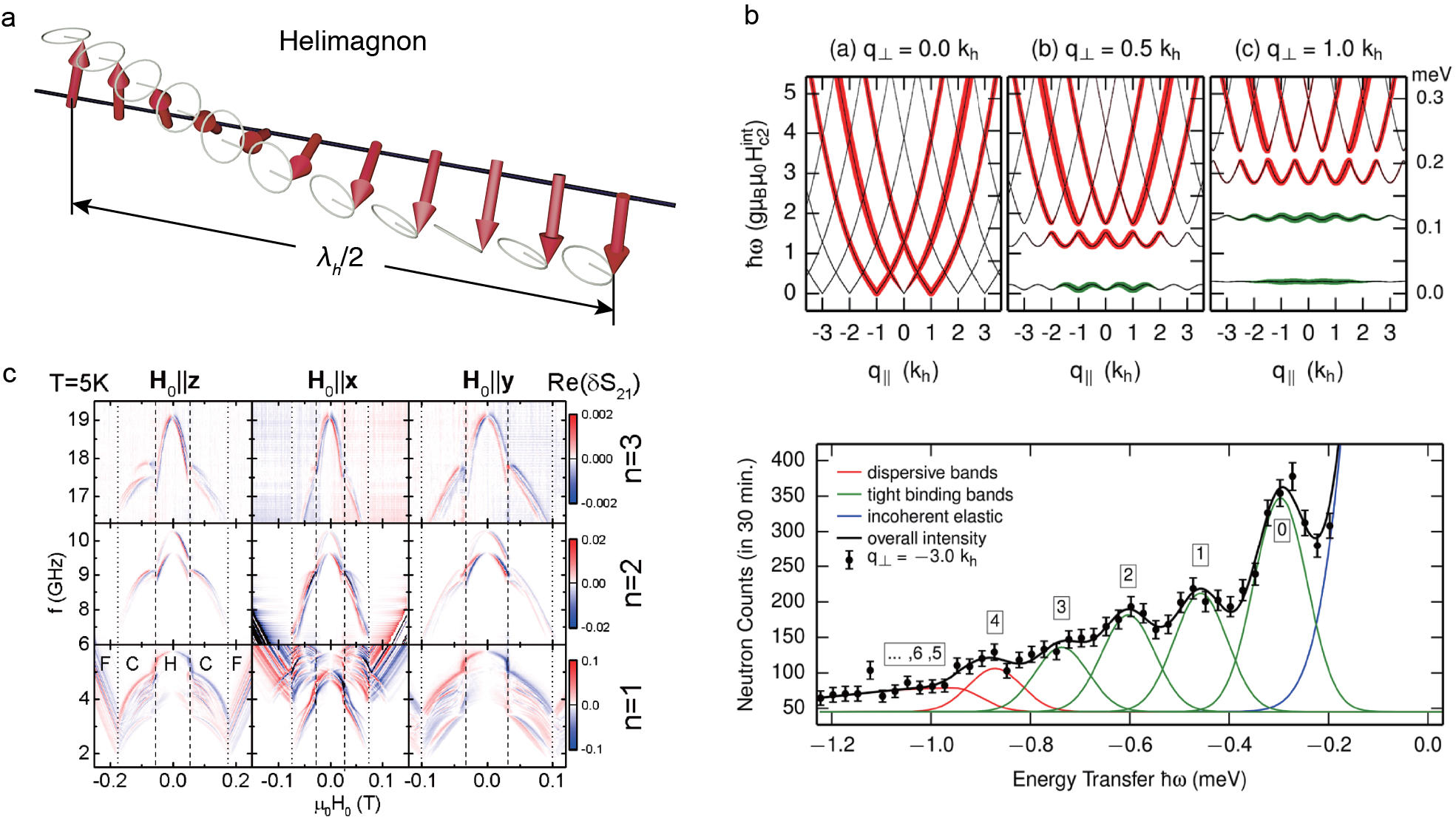}
\caption{(a) A sketch of the low-energy spin wave excitation in a helimagnet with a momentum $q$ parallel to the helix vector. Reillustrated according to Ref.~\cite{Jan2010}. $\lambda_{h}$ indicates the helix pitch length. (b) Universal helimagnon spectrum for various values of perpendicular momentum $q\bot$. Inelastic neutron scattering experiments using neutrons with fixed incident energy $E=5.04$ meV were conducted and five helimagnon bands are observed. Reproduced from Ref.~\cite{Kug2015}. (c) Color-coded VNA spectra $S_{21}$ recorded as a function of the applied magnetic field at 5 K for three different directions. Reproduced from Ref.~\cite{Wei2017}.}
\label{fig:Helimagnon}
\end{figure}

Chiral magnets with inversion symmetry breaking can host complex magnetic textures, such as helimagnetic spirals and Skyrmions. The magnetic helical state can be formed naturally by the interplay of the exchange energy $J$ and the Dzyaloshinskii-Moriya interaction (DMI) $D$ (c.f. Section~\ref{sec:free energy} for theoretical descriptions). The collective magnetic excitation of helimagnons, with pitch length $\lambda_{h}=2\pi/k$ (Fig.~\ref{fig:Helimagnon}(a)), is promising for investigating magnonic crystals beyond the limit of nanolithography processing~\cite{Jan2010,Kug2015}. The length scale of $\lambda_{h}$ is typically 10 nm to 100 nm resulting from the competition of the ferromagnetic-exchange interaction and DMI, \eg 14 nm for the B20 compound MnSi~\cite{Jan2010} and 60 nm for Cu$_{2}$OSeO$_{3}$~\cite{Wei2017}.

%-----------------------------------
%-----------------------------------

Sketches of spin waves with the wavelength $\lambda_{\rm{ex}}$ in the field polarized state and the helical state are shown in Fig.~\ref{fig:Helimagnon}(a). In the field polarized state, the spins are aligned in the same direction and spin waves propagate with wave vector $k$. In the helimagnet with the momentum $q$ parallel to the helix propagation vector $k_{h}$, two principle scenarios can exist, namely, 1) the wavelength $\lambda_{\rm{ex}}$ is smaller than the helix pitch $\lambda_{h}$. 2) the wavelength $\lambda_{\rm{ex}}$ is much larger than the helix pitch $\lambda_{h}$ as illustrated in Fig.~\ref{fig:Helimagnon}(a). Inelastic neutron scattering was used to observe the helimagnon bands, formed by multiple Umklapp scattering in MnSi at zero field~\cite{Jan2010}. The measured spectra with the magnetic Bragg satellites $k_1$ to $k_4$ correspond to the complex superpositions of four chiral domains. The obtained dispersions using a universal Ginzburg-Landau theory are shown in Fig.~\ref{fig:Helimagnon}(b). Ignoring the spiral distortion and the dipolar interaction in the magnon spectrum, the magnon dispersion with the helical wave vector $nQ$ can be described by~\cite{Wei2017}:
\begin{equation}\label{Heli_dispersion}
    \hbar \omega_n=|n|\frac{g\mu_{B}B_{c2}}{1+N\chi}\sqrt{n^2+(1+\chi)\rm{sin}^2\Theta},
\end{equation}
where $\mu_{B}$ is the Bohr magneton, $B_{c2}$ is the critical field of the phase transitions from the conical to field polarized phase, $N$ is the demagnetization factor, $\chi=\mu_0\frac{JM_{\rm{S}}^2}{D^2}$, $\mu_{0}$ is the permeability and $\rm{cos}\Theta=\mu_0H_0/B_{c2}$ is the conical angle. An energy scan is also shown in Fig. \ref{fig:Helimagnon}(b) measured at 20 K where five bands are clearly resolved. The elastic peaks appear because of the incoherent scattering.
Furthermore, microwave-based broadband spectroscopy measurements (similar to the PSWS technique described in Section~\ref{sec:PSWS}) at $Q=0$ were performed on helimagnets~\cite{Wei2017,Dirk2015}. The dipolar interaction and the cubic anisotropy are taken into consideration for the magnon energy to precisely describe the helimagnon spectra. The transmission spectra measured on a Cu$_2$OSeO$_3$ single crystal at 5~K are shown in Fig.~\ref{fig:Helimagnon}(c). Helical, conical and field polarized phases can be distinguished by multiple resonances. In both conical and helical phases, the higher order $n=2$ and $n=3$ modes are observed due to the multi-domain state.

\subsubsection{Skyrmion excitation and Skyrmion motion driven by magnons}

%-----------------------------------
\begin{figure}[t]
\centering
\includegraphics[width=16cm]{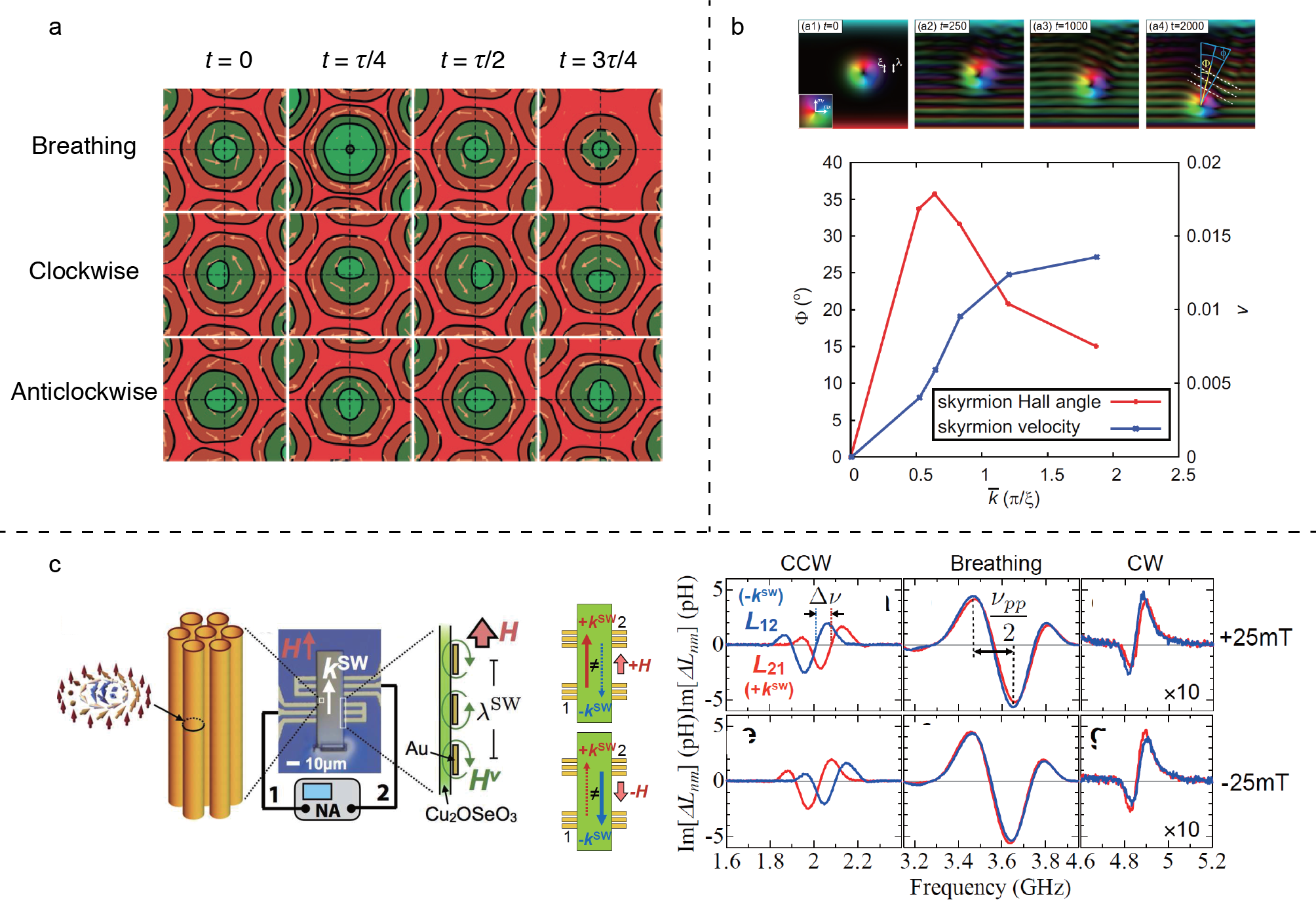}
\caption{(a) Illustrations of the real space Skyrmion resonance modes: clockwise mode, counterclockwise mode and breathing mode. Reproduced from Ref.~\cite{Dirk2015}. (b) Snapshots of the Skyrmion-magnon scattering processes with $k \xi \simeq 1.87 \pi$. The Skyrmion Hall angle and velocity are also shown for different $k$ values. Reproduced from Ref.~\cite{Iwa2014}. (c) Sketch of Skyrmion strings in a Cu$_2$OSeO$_3$ lamella. The all-electrical measurement technique based on the VNA is shown and the spin wave nonreciprocity can be observed in the measured spectra in the CCW mode. Reproduced from Ref.~\cite{Seki2019}.}
\label{fig:Skyrmion}
\end{figure}
%-----------------------------------

Magnetic Skyrmions are topologically protected spin textures, hosting promising applications in memory and logic devices~\cite{Nag2013,liu_skyrmion_2015,ye-hua_dynamics_2015,Fert2017}. Skyrmions can be formed by DMI, and have been observed in intrinsic chiral magnets~\cite{Muh2009,Tok2010,schulz_emergent_2012,garst_collective_2017,C_Jin2017,H_Du2018} with bulk DMI and magnetic multilayers~\cite{Jiang2015,Sou2017} with interfacial DMI. Typical Skyrmions have a size below 100 nm, determined by the interplay between the DMI and Heisenberg exchange interaction. Spin-polarized charge currents are used to move and manipulated Skyrmions in magnetic systems and Skyrmion dynamics is a significant issue for building Skyrmion based devices~\cite{Woo2016}. Spin waves, propagating without Joule heating, are desirable for Skyrmion motion with low-energy consumption.

A theoretical study of the spin wave excitation in a Skyrmion crystal of insulating magnets was proposed by Mochizuki~\cite{Moc2012}, and several spin-wave resonances near 1 GHz for in-plane oscillating fields were observed. The distribution of the out-of-plane spin components circulate around the Skyrmion core, shown in Fig.~\ref{fig:Skyrmion}(a). The directions of these two circulations are opposed, resulting in clockwise (CW) and counterclockwise (CCW) rotations. In addition, a breathing mode~\cite{Kim2014} was observed with an out-of-plane oscillating magnetic field. The resonance frequencies of these Skyrmion modes are different and have been experimentally observed in chiral magnets~\cite{Oka2013,Dirk2015}. Schwarze~\textit{et al.}~\cite{Dirk2015} have performed all-electrical spectroscopy in metallic, semiconducting and insulating chiral magnets MnSi, Fe$_{1{-}x}$Co$_x$Si and Cu$_2$OSeO$_3$, respectively, using conventional microwave antennas. Among those materials, Cu$_2$OSeO$_3$ shows small damping of $1\times10^{-4}$ at 5 K~\cite{Sta2017}, which is suitable for exploring Skyrmion based devices for microwave frequency applications. Recently, Satywali~\textit{et al.} found a gyrotropic resonance of N\'eel Skyrmions in Ir/Fe/Co/Pt multilayers~\cite{Sat2018}, which might open the door for Skyrmion nanoscale devices.

Spin waves are capable of driving Skyrmion motion. Theoretical studies have demonstrated the elementary process, involving spin waves and a single Skyrmion~\cite{Iwa2014}. The Landau-Lifshitz-Gilbert (LLG) equation was numerically solved and micromagnetic simulations were conducted. Fig. \ref{fig:Skyrmion}(b) shows snapshots of the scattering processes $k\xi \simeq 1.87\pi$ where $k$ is the spin-wave wavevector and $\xi$ is the Skyrmion size. The incident spin waves are scattered by the Skyrmions as determined by the Skyrmion Hall angle, which is more efficient for larger $k$. The Skyrmion is moved but not diminished by spin waves, due to the topological protection. The Skyrmion Hall angle and Skyrmion velocity as a function of $k$ are shown in Fig.~\ref{fig:Skyrmion}(b). Meanwhile, Skyrmion motion driven by a temperature gradient and the resulting transport of thermal magnons was observed~\cite{Lin2014,Moc2014}. In the presence of a thermal gradient, Skyrmions can move from the cold to the hot side due to the magnonic spin transfer torque similar to the thermal gradient induced domain wall motion~\cite{Jiang2013,Sho2019} (c.f. Section~\ref{sec:Domain wall driven by spin waves}) and induced by thermal spin torque~\cite{Hatami2007,Yu2010}. Skyrmion motion in nanowires driven by magnons is theoretically studied~\cite{X_Zhang2017}. Different geometries are proposed to build Skyrmion circuits based on spin waves~\cite{Zhou2015,li_dynamics_2018}. Skyrmion lattices or Skyrmion crystals~\cite{Petrova2011,F_Ma2015,F_Ma2015b,roldan-molina_topological_2016} provide a special platform for propagating spin waves, \eg spin wave refraction was studied numerically in arrays of Skyrmions~\cite{moon_control_2016}.

So far, an experimental realization of Skyrmion motion driven by spin waves~\cite{ding_motion_2015} is still missing. Very recently, Seki~\textit{et al.}~\cite{Seki2019} demonstrated propagating spin excitation along Skyrmion strings, shown in Fig.~\ref{fig:Skyrmion}(c). A Skyrmion can extend into the third dimension, forming a string structure, which contains uniform two-dimensional Skyrmions along the string direction. The PSWS (c.f. Section~\ref{sec:PSWS}) was used to study spin-wave propagation along the Skyrmion strings in Cu$_2$OSeO$_3$. A spin-wave nonreciprocity, associated with group velocities and decay lengths, was observed and these spin waves can propagate over a distance exceeding $10^3$ times the Skyrmion diameter. Such experimental observations are important steps towards unidirectional information transfer along topologically protected Skyrmion strings.

\subsubsection{Propagating spin waves as a probe of the Dzyaloshinskii-Moriya interaction}\label{sec:DMI_spin_wave_nonreciprocity}

%-----------------------------------
\begin{figure}[t]
\centering
\includegraphics[width=16cm]{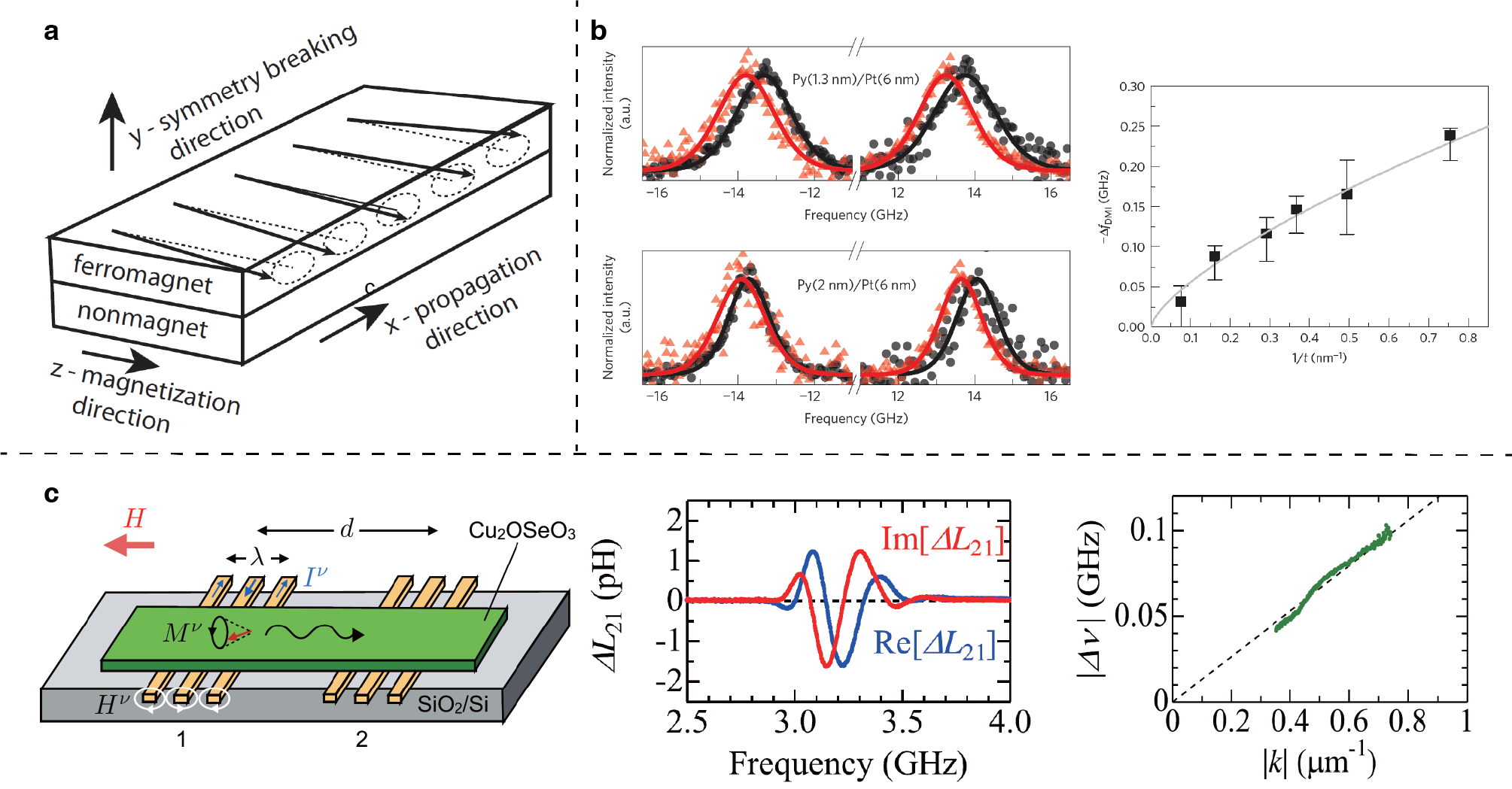}
\caption{(a) Sketch of spin waves propagating at the ferromagnet-nonmagnet interface with wave vector $k\parallel x$ and magnetization $M\parallel z$. Reproduced from Ref.~\cite{Moon2013}. (b) BLS spectra measured at $k=16.7$ $\mu$m$^{-1}$ for two opposite magnetization polarities (circles and triangles) on two samples with different permalloy thicknesses. The frequency difference of the counter-propagating spin waves as a function of $1/t$ is shown. Reproduced from Ref.~\cite{Nem2015}. A sketch of all-electrical spin-wave measurements of the bulk DMI in Cu$_2$OSeO$_3$ is shown in (c). The external field is parallel to the wave vector $k$. The frequency difference of the counter-propagating spin waves as a function of $k$ is plotted. Reproduced from Ref.~\cite{Seki2016}.}
\label{fig:spin_wave_DMI}
\end{figure}
%-----------------------------------

The interfacial DMI originates from spin-orbit coupling and inversion symmetry breaking~\cite{Moon2013}. In magnetic thin films with interfacial DMI, N\'eel type domain walls with well defined chirality form and give rise to a spin-orbit torque (SOT) induced fast domain wall motion~\cite{Emo2013}. Recently, a SOT-driven magnetic domain-wall logic~\cite{Z_Luo2020} was reported with coupled magnetic domains induced by interfacial DMI~\cite{Heide2008,Fran2014,Z_luo2019}. The determination of DMI is crucial for understanding the physics of SOT systems and might contribute to chiral spin texture based spintronic devices. The interfacial DMI can generate an asymmetric dispersion for spin waves in magnetic systems, following the equation~\cite{Moon2013}:

\begin{equation}\label{dispersion2}
\omega=\gamma\sqrt{\left(\mu_{0}H+\frac{1}{4}\mu_{0}M_{s}+Ak^2\right)\left(\mu_{0}H+\frac{3}{4}\mu_{0}M_{s}+Ak^2\right)-\left(\frac{1}{4}\mu_{0}M_{s}\right)^2e^{-4|k|d}\left(1+2e^{2|k|d}\right)}+pD^*k,
\end{equation}

where $H$ is the applied field, $A$ is the exchange constant, $p=\pm1$, $D^*=2D/\mu_0M_{s}$ and $D$ is the DMI constant. Figure~\ref{fig:spin_wave_DMI}(a) shows a sketch of spin-wave propagation in the $k\perp H$ configuration at the ferromagnet-nonmagnet interface with wavevector $k\parallel \pm x$ and magnetization $M\parallel z$. Due to the interfacial DMI, spin waves propagating in opposite directions but perpendicular to the magnetization direction have different dispersion relations. The frequency shift for counter propagating spin waves $\Delta f=\gamma \mu_0 D^* |k|\pi$ in the presence of interfacial DMI is proportional to the wavevector $k$.  Spin-wave nonreciprocity induced by interfacial DMI in magnonic waveguides was studied by Ma~\textit{et al.}~\cite{F_Ma2014} with micromagnetic simulations.

Different experimental methods have been used to obtain the spin wave dispersion relation and extract the interfacial DMI value, including BLS (c.f. Section~\ref{sec:BLS})~\cite{Nem2015,Yang2015a,Li2017,HYang2020}, highly resolved spin-polarized electron energy loss (SPEEL) spectra~\cite{Zak2010}, propagating spin-wave spectrascopy (c.f. Section~\ref{sec:PSWS})~\cite{Lee2016,Luc2019,H_Wang2020} and time-resolved scanning Kerr microscopy (c.f. Section~\ref{sec:MOKE})~\cite{Kor2015}. Interfacial DMI originating from heavy metal/ferromagnet interfaces have been characterized by BLS spectroscopy~\cite{Nem2015}. Stokes and anti-Stokes BLS spectra of Py/Pt films with different permalloy thicknesses are shown in Fig.~\ref{fig:spin_wave_DMI}(b). For a wavevector of $|k|=16.7$ $\mu$m$^{-1}$ a $D=0.15$ mJ/m$^2$ and $D=0.10$ mJ/m$^2$ could be quantified for samples with permalloy thicknesses of 1.3 nm and 2.0 nm, respectively. Reference samples without the heavy metal layer were measured and indicated the absence of interfacial DMI. The frequency shift between counter-propagating spin waves observed in BLS measurements is shown as a function of $1/t$, where the DMI-induced frequency shift increases for thinner samples. This thickness dependence is in stark contrast to that of spin-wave frequency nonreciprocity induced by asymmetric interfacial anisotropy~\cite{Gla2016,Cos2018} and graded magnetization~\cite{Gal2019b}. Very recently, Wang~\textit{et al.}~\cite{H_Wang2020} reported an all-garnet interfacial DMI~\cite{Beach2019,Velez2019,Ding2019,F_Yang2020,Y_Jiang2020,Beach2020,F_Yang2020b} in ultrathin yttrium iron garnet films with a low Gilbert damping, which may contribute to build future functional chiral magnonic devices~\cite{J_Chen2020}. Metamaterials with a periodic DMI were proposed theoretically by Gallardo~\textit{et al.}~\cite{Gal2019} for the implementation of magnonic crystals, which show flat bands and indirect band gaps, as revealed by micromagnetic simulations.

Bulk DMI can also influence spin-wave spectra in chiral magnets. Cort\'es-Ortu$\rm{\tilde{n}}$o~\textit{et al.} derived a spin wave dispersion for a bulk DMI system~\cite{Cor2013}. Similar to an interfacial DMI system, bulk DMI generates an asymmetric spin-wave dispersion relation in the backward-volume spin-wave configuration where the wavevector $k$ is parallel to the magnetization. The magnitude of the frequency shift can be defined as: $\Delta f=\frac{4\gamma DM_{\rm{S}}|k|}{2\pi}$. Takagi~\textit{et al.}~\cite{Tak2017} studied the propagation character of spin waves in chiral magnets FeGe and Co-Zn-Mn alloys. The bulk DMI values were experimentally evaluated using the PSWS technique (c.f. Section~\ref{sec:PSWS})~\cite{Seki2016}. Furthermore, periodical inhomegeneous DMI~\cite{Lee2017} can result in spin-wave band gaps~\cite{Gal2019} in a chiral magnonic crystal~\cite{L_Zhang2013,Shindou:2013cp,K_Chang2018}.

\subsection{Spin waves in magnetic vortices}
\subsubsection{Magnetic vortex cores as tunable spin-wave emitters}

\begin{figure}[htbp]
\vskip -10pt
\begin{center}
\includegraphics[width=1\textwidth]{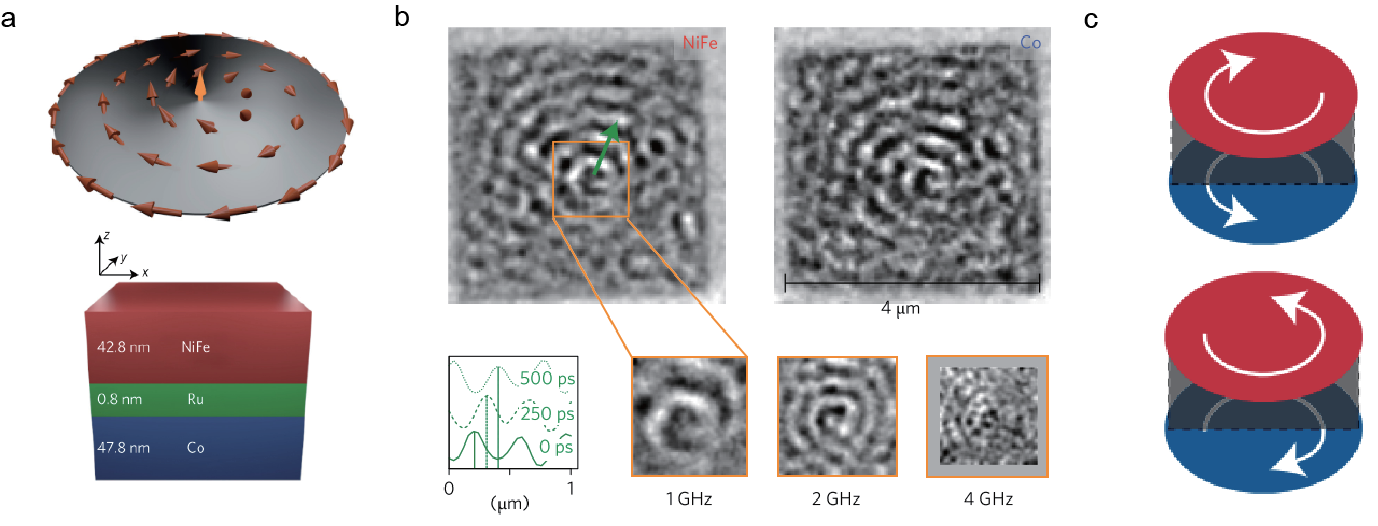}
\caption{(a) A sketch of a vortex with continuous planar magnetization circulation ($c = -1$, anticlockwise) and perpendicular core ($p = +1$, pointing upwards), and a sketch of a trilayer stack consisting of Co/Ru/NiFe layers. (b) Time-resolved scanning transmission X-ray microscopy images of spin-wave emission from magnetic vortex cores driven by different frequencies in Co and NiFe layers (c) The upper image corresponds to type 1 ($c_{\rm 1(Co)} =+1$, $c_{\rm 2(NiFe)} = -1$), and the below image corresponds to type 2 ($c_{\rm 1(Co)} =-1$, $c_{\rm 2(NiFe)} = +1$). Reproduced from Ref.~\cite{Wintz2016}.}
\label{fig1}
\end{center}
\vskip -10pt
\end{figure}

Magnetic vortices as non-trivial topological structures are promising candidates for future nonvolatile data storage devices~\cite{Bussmann99}. A vortex consists of a spot of perpendicular magnetization at the center and planar circulations, which can be characterized by the binary core polarity $p=\pm1$ ($p=+1$ represents that the core magnetization is pointing upwards) and the circulation $c=\pm1$ ($c=+1$ represents that the static magnetization in the vortex plane is an counterclockwise rotation) as shown in Fig.~\ref{fig1}(a)~\cite{Tretiakov07}. Two ferromagnetic layers with strong antiferromagnetic interlayer exchange coupling~\cite{hillebrands_spin-wave_1990,Kli2018,Qin2018,Chen2018,Y_Fan2020,Y_Li2020} have been theoretically predicted to form a coupled magnetic vortex pair with different circulations~\cite{Grunberg86}.

Recently, a magnetic vortex core was used by Wintz $et$ $al.$~\cite{Wintz2016} for the excitation of short-wavelength propagating spin waves, which is difficult to generate but plays a key role in nanoscale spintronic signal-processing devices~\cite{Demi2010}. The traditional methods suffer from technological restrictions in fabricating nanoscale transducers, matching their impedance, low excitation efficiency, non-locality and poor control of the nanoscale spin waves~\cite{Stancil}. In their work, the magnetic vortex pair was created in a Co(47.8 nm)/Ru(0.8 nm)/Ni$_{81}$Fe$_{19}$(42.8 nm) trilayer stack (Fig.~\ref{fig1}(a)). An in-plane sinusoidal magnetic field was used to excite the vortex pair to generate propagating nanoscale spin waves, which were observed by the time-resolved scanning transmission X-ray microscopy (TR-STXM, c.f. Section~\ref{sec:TR-STXM}). The tunability of the wavelength was achieved by varying the driving frequency as shown in Fig.~\ref{fig1}(b).

By studying the vortex properties with analytic calculations, they found that the dipole-exchange spin waves excited by a vortex core exhibit a linear dispersion with no frequency gap

\begin{equation}\label{dispersion1}
    \omega=V(\phi_k)k
\end{equation}

where $\phi_k$ is the angle between the wave vector $k$ and the in-plane direction of the magnetization in the ferromagnetic layers. The velocity $V(\phi_k)$ can be calculated based on

\begin{equation}
\begin{aligned}
\phi_k=\omega_{\rm M}d\left[\sqrt{\frac{\lambda_{\rm ex}^2}{d^2}+\frac{1}{12}\left[4+6\frac{d_{\rm s}}{d}+F\left(\frac{\lambda_{\rm ex}}{d}\right)\right]}\sin^2\phi_k-\frac{1}{2}\sin\phi_k\right]
\end{aligned}
\label{dispersion2}
\end{equation}

where $\omega_{\rm M}=\gamma\mu_0M_{\rm S}$, $\lambda_{\rm ex}$ is the exchange length, $d$ is the thickness of the ferromagnetic flims, $d_{\rm s}$ is the separation between the films, and the function $F(u)$ is given by

\begin{equation}
\begin{aligned}
F(u)=1-12u^2+24u^3\tanh(\frac{1}{2u})
\end{aligned}
\label{vg}
\end{equation}

Oscillating magnetic fields drive the gyration of the vortex core around its equilibrium position~\cite{Gus2008,Choe2004,lee_radiation_2005,park_interactions_2005,Vansteenkiste09}. Even if the driving frequency is far off the resonance of the gyroscopic motion of the vortex core but matches the frequencies of radial and azimuthal dipolar spin-wave modes~\cite{Buess2004,Per2005,Kammerer11}, the motion of the vortex core couples to the spin wave eigenmodes of the disk and causes spin-wave emission.

\subsubsection{Short-wavelength spin waves in magnetic vortices}

%-----------------------------------
\begin{figure}[t]
\centering
\includegraphics[width=16cm]{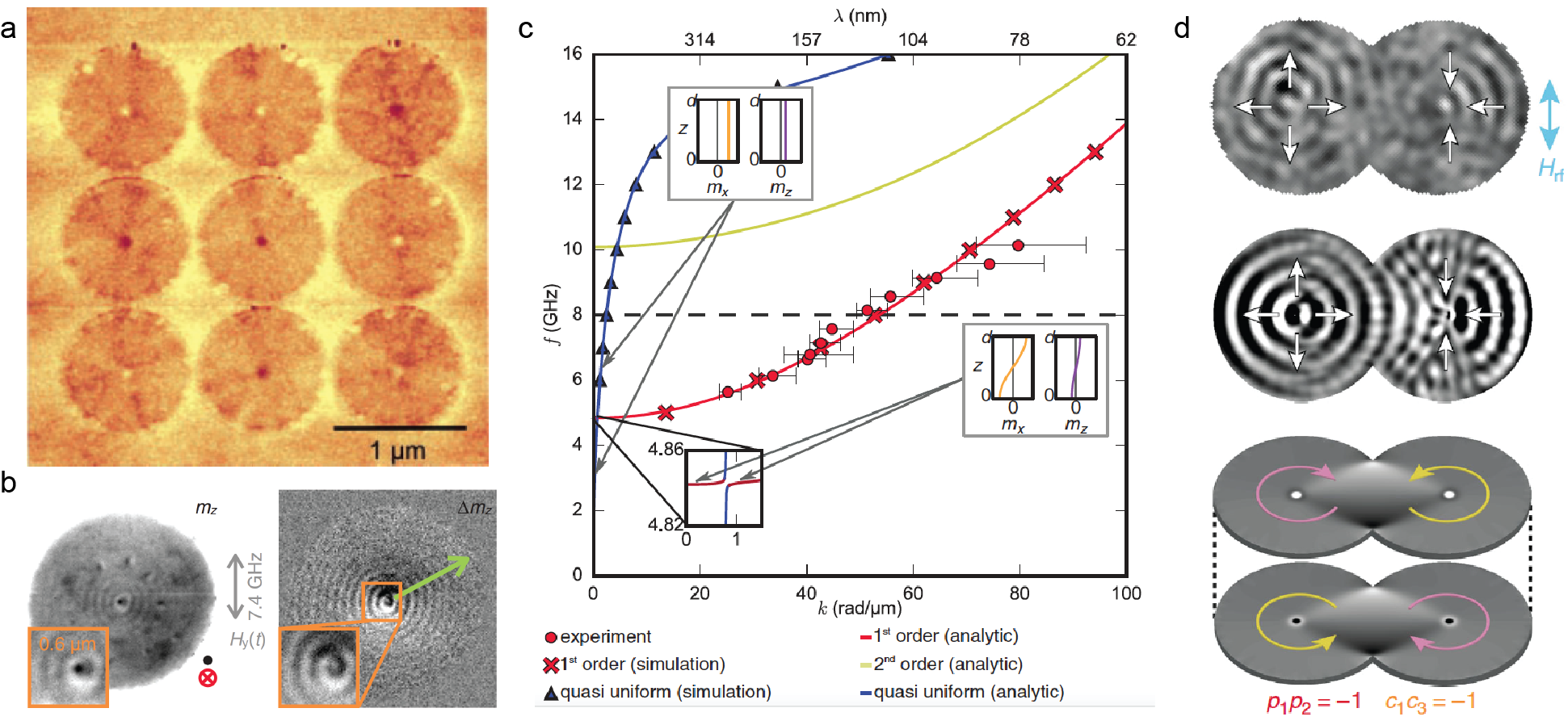}
\caption{(a) An MFM image of a magnetic vortex with different core magnetization at its center. The permalloy dots are 1~$\mu$m in diameter and 50~nm in thickness. Reproduced from Ref.~\cite{Shin2000}. (b) TR-STXM snapshot image of short-wavelength spin waves in a magnetic vortex with out-of-plane sensitivity. The RF field is applied in plane at 7.4 GHz. Reproduced from Ref.~\cite{Die2019}. (c) Spin wave interference and diffraction in two partially overlapping disks. The disks are 800~nm in diameter consisting of two exchange-coupled permalloy layers that are each 50 nm in thickness. From top to bottom are successively the TR-STXM snapshot image, micromagnetic simulation results of dynamic spin-wave modes and simulated static magnetic vortex state of the bilayer system. Reproduced from Ref.~\cite{Beh2018}.}
\label{fig:vortex2}
\end{figure}
%-----------------------------------

There have been intensive efforts to study spin-wave modes in ferromagnetic nanodot arrays, for example Co~\cite{Due2011} or Py~\cite{Ding2014,Lup2015} nanodot arrays, in the framework of magnonic crystals~\cite{Kra2014,Chu2017,Gub_book} with artificial band gaps.
Ferromagnetic nanodots can sustain a magnetic vortex state with a magnetic vortex core at the center, which can be imaged by magnetic force microscopy~\cite{Shin2000}, for example in Fig.~\ref{fig:vortex2}(a). The dynamics of magnetic vortices and their core reversal processes can be imaged by time-resolved magnetic Kerr optical microscope (TR-MOKE)~\cite{Back1999,Acre2000} (c.f. Section~\ref{sec:MOKE}). Recently, propagating spin waves in magnetic vortices were imaged with high spatial and temporal resolution using TR-STXM ~\cite{Wae2006} (c.f. Section~\ref{sec:TR-STXM}). Figure~\ref{fig:vortex2}(b) shows a snapshot image of TR-STXM on a permalloy disk with a diameter of 3~$\mu$m and a thickness of 80 nm, where short-wavelength spin waves with $\lambda=140$ nm are clearly observed. A quasi-uniform dipole-exchange spin-wave mode would reach a frequency of up to about 15 GHz with $\lambda=140$ nm based on the theory of Kalinikos and Slavin~\cite{Kal1986} (blue dispersion curve in Fig.~\ref{fig:vortex2}(c)). The observation from TR-STXM can be explained by considering a first-order heterosymmetric mode as described in Section~\ref{sec:TR-STXM} (red dispersion curve in Fig.~\ref{fig:vortex2}(c)), where the out-of-plane spin wave profile resembles that of perpendicular standing spin waves~\cite{Stancil} in the film thickness direction. The heterosymmetric mode is free of surface boundary conditions and therefore yields a dispersion relation as shown in Eq.~(\ref{eq:hetero}). The red and yellow curves in Fig.~\ref{fig:vortex2}(c) are the calculated dispersion for the first-order and second-order heterosymmetric modes, respectively, based on Eq.~(\ref{eq:hetero}). At 8 GHz, the wavevector $k$ for the first-order heterosymmetric mode is substantially larger than that of the quasi-uniform mode. In other words, at the same frequency, one can observe spin waves with much shorter wavelengths with first-order heterosymmetric modes, which explains why the TR-STXM observes ultra-short-wavelength spin waves at much lower frequencies than expected. This also offers an important route towards exciting spin waves with wavelengths even shorter than 50 nm~\cite{Liu2018} within the frequency range of a conventional vector network analyzer (up to 150 GHz). Remarkably, the spin wave dispersion at about 4.84 GHz (the inset of Fig.~\ref{fig:vortex2}(c)) exhibits an anticrossing behavior when the quasi-uniform mode (blue curve) and first-order heterosymmetric mode (red curve) intersect. Very recently, such anticrossing or hybridized modes of dipole-exchange spin waves were systematically studied by Tacchi~\textit{et al.}~\cite{Tac2019} in a 78~nm-thick nitrogen-implanted iron (Fe-Ni) film with BLS measurements (c.f. Section~\ref{sec:BLS}) and micromagnetic simulations. If the film thickness is small, \eg 20~nm~\cite{Liu2014,Yu2014}, the starting frequency of the first-order perpendicular standing spin wave is about 22~GHz, much higher than Damon-Eshbach type spin waves and therefore no mode hybridization can be observed. In this case, the scattering between the chiral magnetostatic surface mode and volume standing spin wave mode is minimized and therefore the in-plane propagating spin waves are robust against surface inhomogeneities and defects as long as their frequency lies inside the gap below the volume standing wave mode.  As a result, the nanoscale thickness of magnetic films can provide an intrinsic advantage of backscattering immunity, as studied theoretically by Mohseni~\textit{et al.}~\cite{Moh2019}.

When two ferromagnetic disks are overlapped, spin waves emitted from the vortex cores can generate interference patterns as studied by Behncke and coworkers~\cite{Beh2018}. The static magnetic state is shown in the bottom panel of Fig.~\ref{fig:vortex2}(d), where the yellow (pink) arrows represent in-plane magnetization with circularity of +1 (-1). With an in-plane excitation field $H_{\text{rf}}$, the spin-wave profile is characterized by TR-STXM as shown in the upper panel of Fig.~\ref{fig:vortex2}(d), which is in reasonably good agreement with the results of micromagnetic simulations (the middle panel of Fig.~\ref{fig:vortex2}). Spin-wave diffraction can also be clearly resolved from these results. Podbielski~\textit{et al.}~\cite{Pod2006} studied spin-wave interference in microscopic magnetic ring structures, but it is remarkable to be able to spatially image the spin-wave interference and diffraction pattern in the magnetic disks for coherent short-wavelength spin waves~\cite{Liu2018,Die2019}.

\section{Spin waves in antiferromagnetic spin textures}
\label{sec:afm}

\begin{figure}[hb]
  \centering
\includegraphics[width=0.98\textwidth]{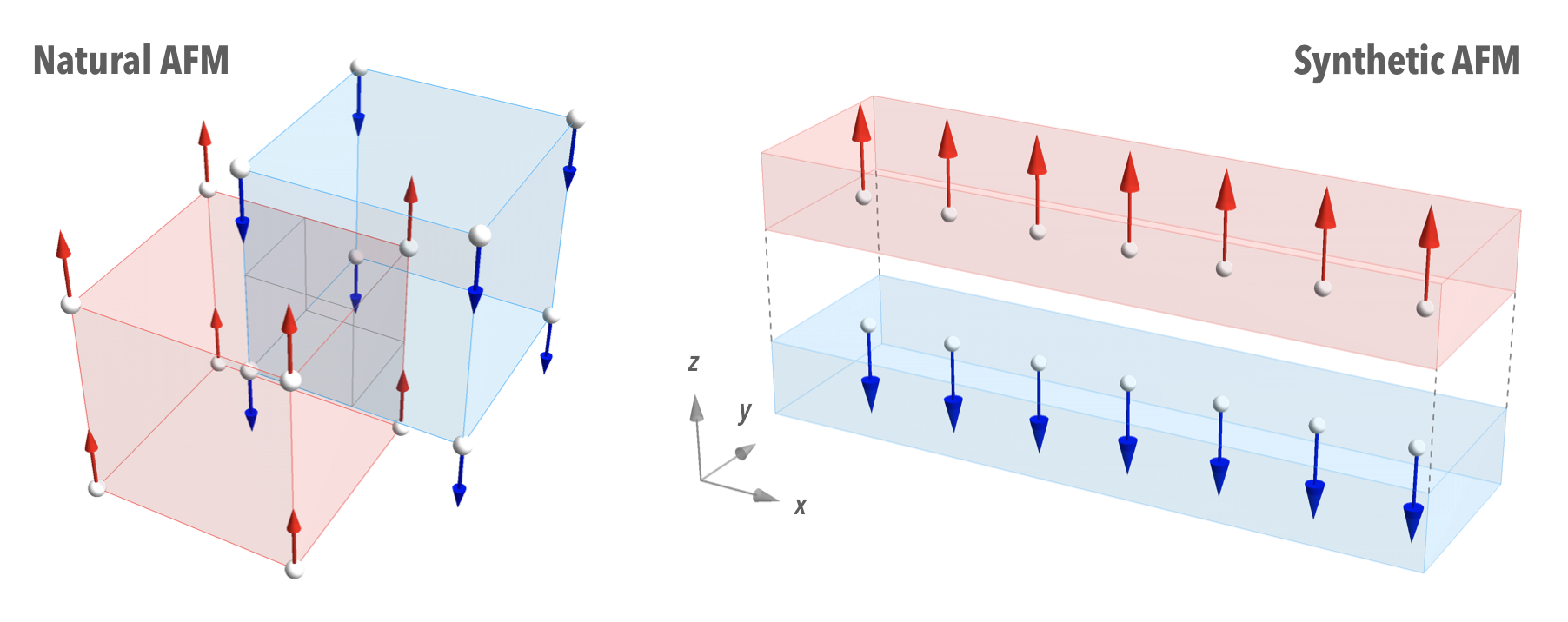}
\caption{A natural antiferromagnet with two interlocking magnetic sublattices and a synthetic antiferromagnet with two ferromagnetic sublayers coupled antiferromagnetically via RKKY coupling through a metallic spacer.}
\label{fig:AFM_SyAF}
\end{figure}

Because antiferromagnetic order is difficult to detect experimentally, and excitation of antiferromagnetic spin waves~\cite{Kampfrath2011,gitgeatpong_nonreciprocal_2017,Bialek2019,han_birefringence-like_2020} is difficult to achieve, the results presented in this section are mostly theoretical. In comparison with a natural antiferromagnet, a synthetic antiferromagnet~\cite{duine_synthetic_2018}, consisting of two (or many) antiferromagnetically coupled ferromagnetic layers (see \Figure{fig:AFM_SyAF}), can be fabricated and is relatively easier to measure in experiments. Most of the physics in synthetic antiferromagnets is rather similar to that in natural antiferromagnets. Therefore, synthetic antiferromagnets provide an ideal platform for studying the physics of antiferromagnetic spin waves.

%-----------------------------------
%-----------------------------------

Most ferromagnetic textures have their antiferromagnetic counterparts, such as antiferromagnetic domain walls or antiferromagnetic Skyrmions~\cite{diaz_topological_2019}.

\subsection{Spin waves interacting with antiferromagnetic domain walls}

{\bf Bound spin wave modes in antiferromagnetic domain walls}

\begin{figure}[t]
  \centering
\includegraphics[width=0.6\textwidth]{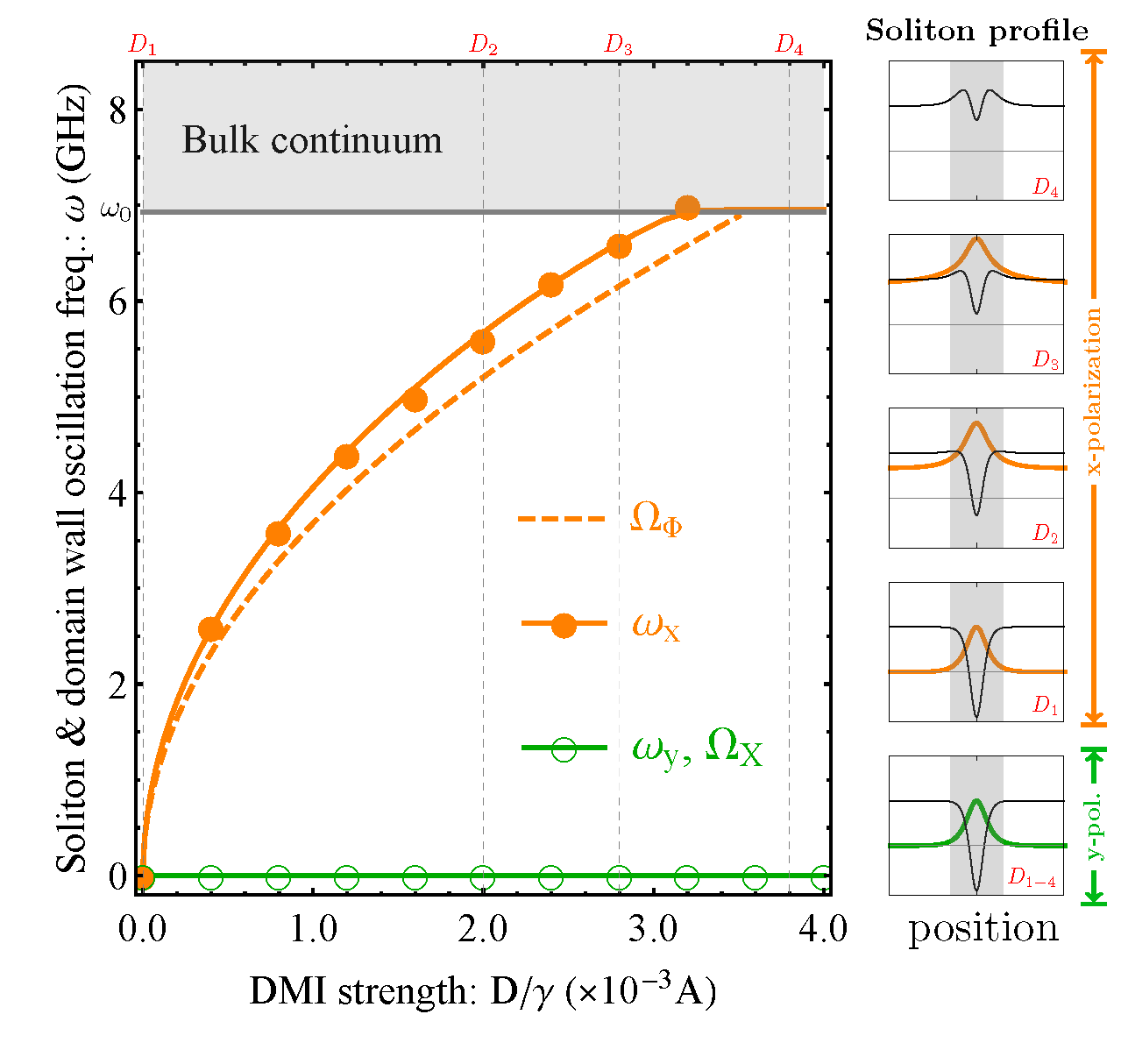}
\caption{The frequency of the bound spin wave modes as a function of the strength of DMI in an antiferromagnetic domain wall. Reproduced from Ref.~\cite{yu_polarization-selective_2018}.}
\label{fig:AFM_DW_Bound}
\end{figure}

Much of the spin wave physics discussed for the ferromagnetic textures can also be qualitatively applied to antiferromagntic (AFM) textures~\cite{gomonay_antiferromagnetic_2018}. For instance, similar to the bound spin wave mode confined in a ferromagnetic domain wall, an antiferromagnetic domain wall also accommodates bound spin wave modes~\cite{wieser_quantized_2009}. However, the difference lies in the fact that there are two bound spin wave modes in an antiferromagnetic domain wall, instead of one as in the ferromagnetic case. This is induced by the extra polarization degree of freedom in antiferromagnetic spin waves. The two bound states correspond to the sliding of the domain wall position and the tilting of the domain wall angle
\footnote{In ferromagnets, the sliding and the tilting of the domain wall profile are not independent, therefore there is only one bound state in ferromagnetic domain walls. However, in antiferromagnets, the domain walls can slide or tilt independently.}.
In the simplest scenario with only easy axis anisotropy, both the sliding mode and the tilting mode are the Goldstone modes with zero frequency, similar to the ferromagnetic case~\cite{yu_polarization-selective_2018}. In the presence of DMI or additional hard axis anisotropy, the sliding mode remains a Goldstone mode, but the tilting mode has finite frequencies. \Figure{fig:AFM_DW_Bound} shows the frequency of the sliding mode ($\Omega_\ssf{X}$) and the tilting mode ($\Omega_{\Phi}$) as a function of the DMI strength. It is seen that the sliding mode remains confined in the domain wall with zero frequency, but the tilting mode is squeezed out of the domain wall by the increasing DMI strength until it merges with the bulk continuum.

%-----------------------------------
%-----------------------------------

{\bf Polarization dependent spin wave scattering by an antiferromagnetic domain wall}\label{sssec:sw_pol_afm}

As shown in \Figure{fig:sw_pol} in Section~\ref{sssec:LLG}, antiferromagnetic spin waves, just like many other waves, also possesses polarization degrees of freedom, whereas ferromagnetic spin waves, being right circularly polarized, lack such degrees of freedom. Polarization is an extremely useful degree of freedom and can encode robustly both classical and quantum information~\cite{bennett_quantum_2000,goldstein_polarized_2010,sklan_phonon_2014,sklan_splash_2015,flebus_entangling_2019}. The antiferromagnetic spin wave polarization can be manipulated by tuning the DMI parameter via electric gating like in a field effect transistor~\cite{Cheng2016,ZQ_Liu2019}, which has been realized experimentally by Wimmer \etal \cite{wimmer_coherent_2020}.
More interestingly, antiferromagnetic textures such as domain walls could be used to realize more useful and simpler spin wave polarization manipulation functionality.

%-----------------------------------
\begin{figure}[t]
\centering
\includegraphics[width=\textwidth]{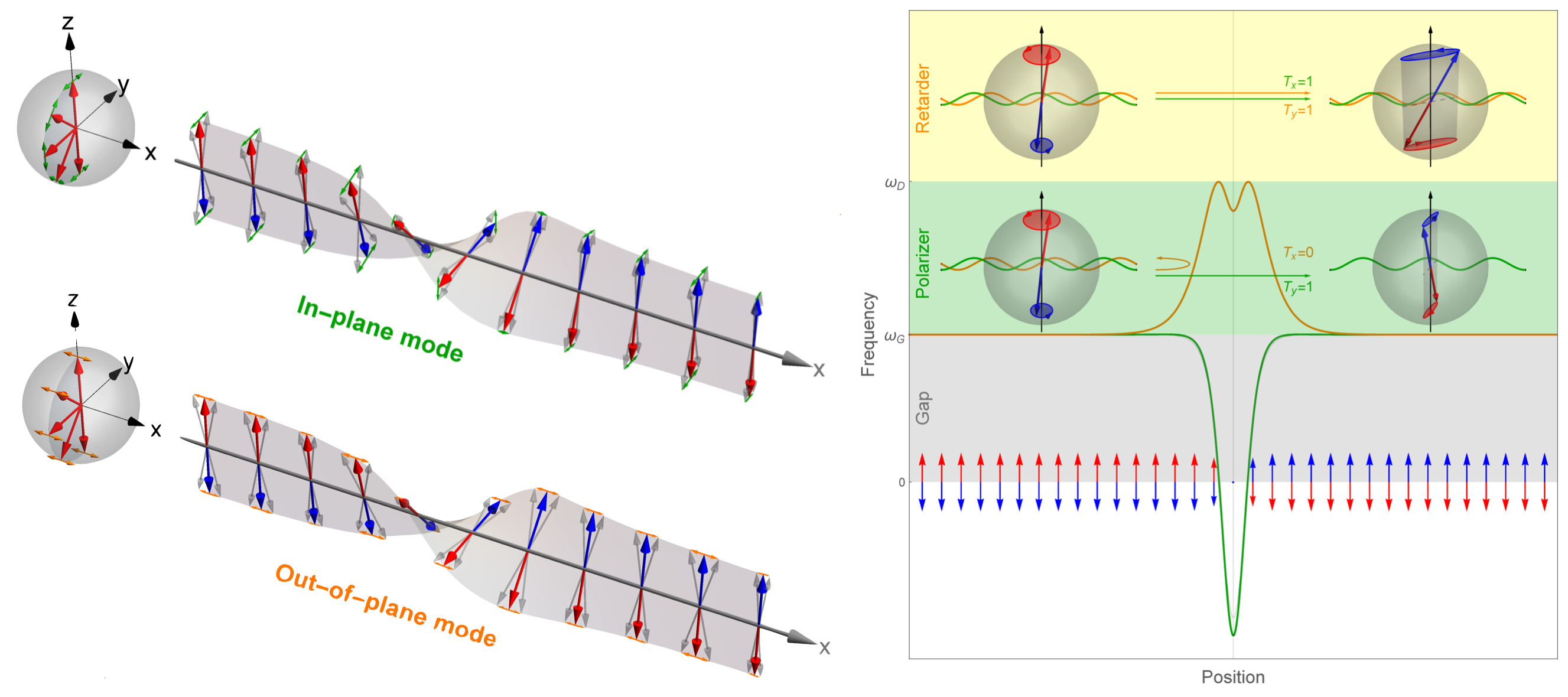}
\caption{Illustrations and simulations of antiferromagnetic domain wall polarizer. Reproduced from Ref.~\cite{Lan2017}.}
\label{fig:AFM_DW_SW}
\end{figure}
%-----------------------------------

Let us consider a Bloch-type \ang{180}-antiferromagnetic domain wall as shown in \Figure{fig:AFM_DW_SW}, which can be stabilized with a bulk-type DMI with a DM vector $\bD$ pointing in the $\hbx$ direction. The scattering of spin waves with such an antiferromagnetic domain wall is governed by \Eq{eqn:LLGnAFdw}. The two components of the N\'{e}el order $\bn$ transverse to the local static $\bn_0$ can be rephrased as~\cite{Lan2017}

%--------------
\begin{subequations}
\label{eqn:LLGnAFdw2}
\begin{align}
  J^{-1}\ddot{n}_\theta
  &= \qty(-{\hbar^2\ov 2\mu} \partial_x^2 + K\cos2\theta_0 ) n_\theta. \\
  J^{-1}\ddot{n}_\phi
  &= \qty(-{\hbar^2\ov 2\mu} \partial_x^2 + K\cos2\theta_0
  + {D\ov w}\sin\theta_0 ) n_\phi.
\end{align}
\end{subequations}
%--------------
Here we have neglected the Gilbert damping (i.e. $\alpha\ra 0$) for simplicity. The $n_\theta$ ($n_\phi$) component points along the longitude (latitude) lines on the Bloch sphere representing the magnetization direction (see \Figure{fig:AFM_DW_SW}).

The equations of motion shown in \Eq{eqn:LLGnAFdw2} suggest that spin waves polarized along the $\theta$ and $\phi$ directions are decoupled, and more importantly the effective scattering potential due to the antiferromagnetic domain wall is polarization dependent. For the in-plane polarization ($n_\theta$), the potential is purely due to the anisotropy $K$ and is a reflectionless Pöschl–Teller type potential well~\cite{poschl_bemerkungen_1933,dodd_solitons_1982,lekner_reflectionless_2007} just as in the ferromagnetic case. But for the out-of-plane polarization ($n_\phi$), there is an additional potential barrier due to the DMI, such that the total potential may become a barrier causing reflections of this polarization. \Figure{fig:AFM_DW_SW} shows the potential profile of an antiferromagnetic domain wall for the two types of linear polarization.

The origin of such interesting polarization dependent scattering at an antiferromagnetic domain wall is in fact caused by the DMI-induced potential barrier for the out-of-plane polarization, which can be understood as the following: the DMI has a preferred rotation axis ($\hbx$ axis in \Figure{fig:AFM_DW_SW}), and thus the magnetization tends to vary in the $y$-$z$ plane to minimize the DM energy. Any out-of-plane tilting will increase the DM energy, and consequently spin waves with out-of-plane polarization would experience increased energy potential inside the domain wall region. In other words, the DMI induces an effective hard axis ($\hbx$) in the antiferromagnetic domain wall, which prevents the magnetization from pointing away from the $y$-$z$ plane.

{\bf Antiferromagnetic domain wall motion driven by spin waves}\label{sec:AFM domain wall motion}

Similar to the situation in the ferromagnetic case, an antiferromagnetic domain wall can also be driven into motion by spin waves. However, attributed to the additional polarization degree of freedom in antiferromagnetic spin waves, the spin-wave driven domain wall motion in antiferromagnets has richer variants than its ferromagnetic counterpart.

As discussed in Section~\ref{sec:intro}, there are two physical mechanisms for domain wall motion, \ie via the angular momentum transfer and momentum transfer between the passing spin wave and the domain wall. In antiferromagnets, the angular momentum transfer gives rise to the precession of the antiferromagnetic texture in spin space, while the momentum transfer gives rise to the shift of the domain wall in real space. This behavior is more similar to that of conventional particles, but opposite to that of a ferromagnetic domain wall.

Tveten~\textit{et al.}~\cite{Tve2014} reported that AFM domain walls move in opposite directions for linearly and circularly polarized spin waves. The linearly polarized spin wave, carrying no angular momentum, is reflectionless at the AFM domain wall, and would cause the AFM domain wall to move towards the spin wave source. The circularly polarized spin wave carries angular momentum, which has opposite sign in opposite domains, therefore the transmission of circularly polarized spin wave would transfer angular momentum to the AFM domain wall. However, since an AFM domain wall cannot absorb angular momentum, the circularly polarized spin wave must be reflected at an AFM domain wall, such reflection transfers momentum to the domain wall, and drives the domain wall forward away from the spin wave source. Another spin wave reflection mechanism at an antiferromagnetic domain wall, proposed by Kim \textit{et al.}~\cite{Kim2015}, is induced by the precession of domain walls, and the magnon passing through the precessing domain wall would lose its frequency and thus reduce its momentum. This reduction in momentum is transferred to the domain wall and pushes the domain wall forward, away from the spin wave source. Another description of the same mechanism is provided by Qaiumzadeh \textit{et al.}~\cite{qaiumzadeh_controlling_2017}

%-----------------------------------
\begin{figure}[b]
\includegraphics[width=\textwidth]{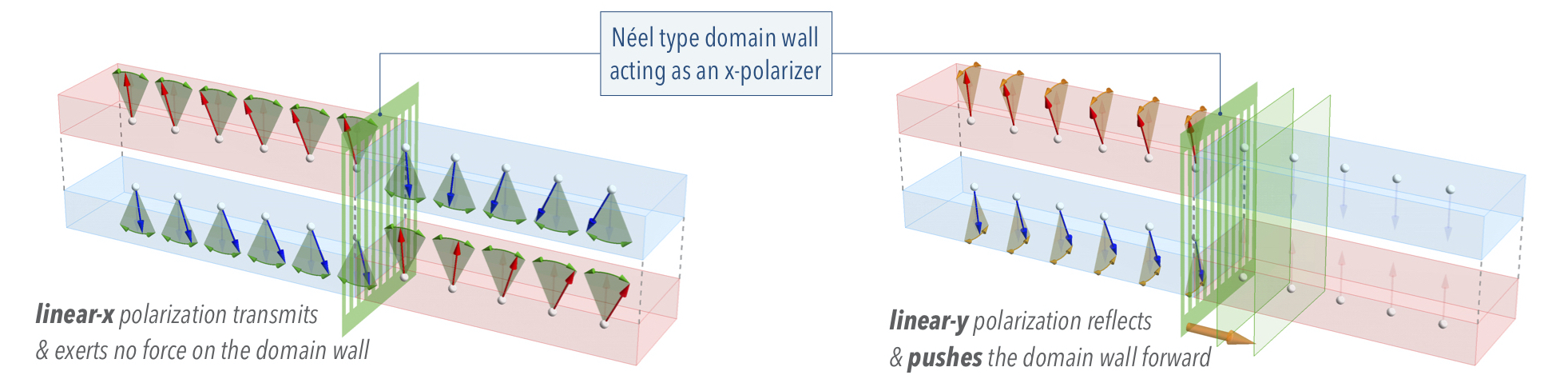}
\caption{Schematic diagrams of AFM domain wall motion driven by linearly polarized spin waves.}
\label{fig:AFM_DW_Move}
\end{figure}
%-----------------------------------

Yet another interesting case of linear momentum transfer mechanism is induced by DMI based on spin wave scattering by an antiferromagnetic domain wall in the presence of DMI. As shown in \Figure{fig:AFM_DW_SW}, the spin wave polarized perpendicular to the magnetization rotation plane (or parallel to the DMI vector) gets reflected at the domain wall, therefore the domain wall is pushed forward away from the spin wave source due to the transferred momentum (see \Figure{fig:AFM_DW_Move}). Depending on the frequency of the spin waves, this reflection can be as strong as 100\%, therefore the momentum transfer can be highly efficient.
This situation may be understood simply using its optical analogy (see \Figure{fig:AFM_DW_Move}): an optical grating polarizes light by reflecting the light polarized along the grating orientation, but the grating is also pushed forward due to such reflection. The difference between a magnonic polarizer and an optical polarizer is that the antiferromagnetic domain wall, functioning as a grating for spin waves, is not a material object such as an optical grating, but merely a spin texture, which makes it much easier to be controlled and manipulated. Antiferromagnetic domain wall motion driven by thermal magnon spin current has been studied by Selzer \textit{et al.}~\cite{selzer_inertia-free_2016} where the walls are found to move towards the hotter side, but when larger forces are driven, the Walker breakdown no longer holds and therefore antiferromagnetic domain walls exhibit higher effective mobility.

\subsection{Spin waves interacting with antiferromagnetic Skyrmions}

%-----------------------------------
\begin{figure}[t]
  \centering
\includegraphics[height=5.5cm]{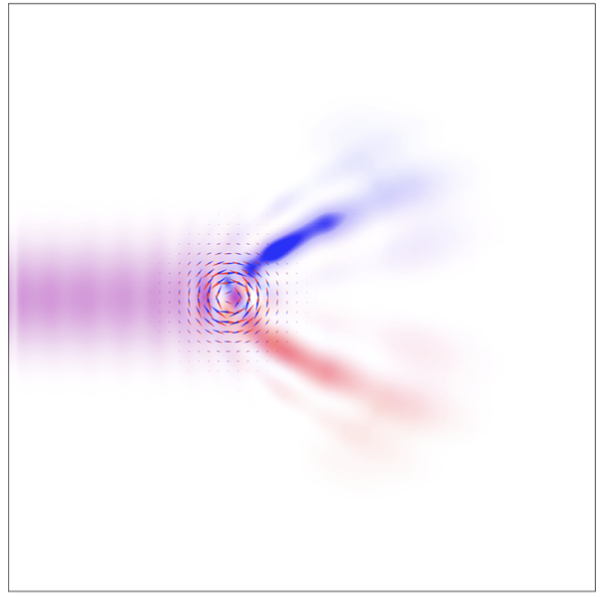}\hspace{0.5cm}
\includegraphics[height=5.5cm]{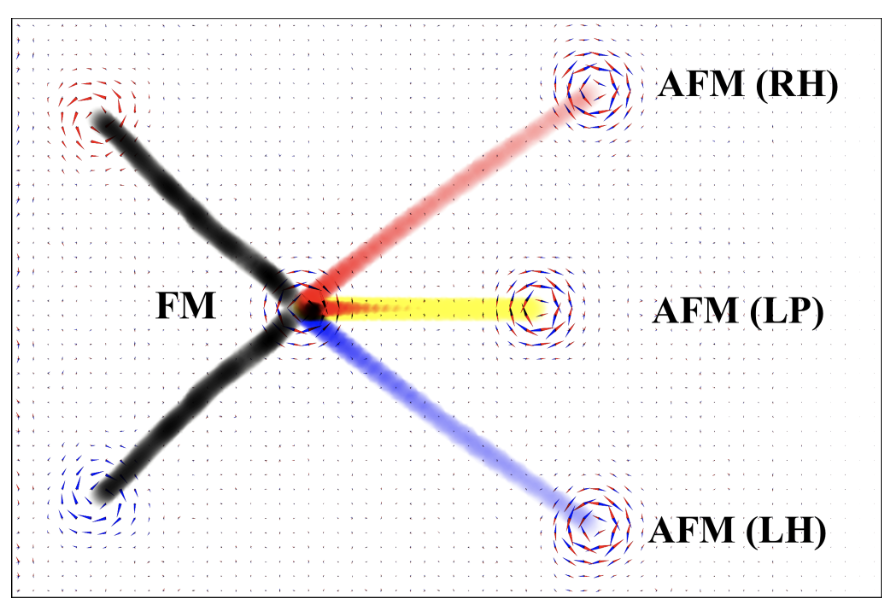}
\caption{Left panel: Micromagnetic simulations of the magnonic topological spin Hall effect in antiferromagnetic Skyrmions. Blue and red represent left- and right-handed spin wave precession of the local Néel order about equilbrium, plotted as the square of the spin wave amplitude integrated over the entire simulation time. A fluctuating magnetic field injects coherent spin waves ($\omega /2\pi = 30$ GHz) at the left of the frame. Linearly polarized spin waves, an equal superposition of the left- and right-handed magnon bands, are decomposed by the Skyrmion into its spin polarized eigenmodes. The sample size is 800 nm × 800 nm. Right panel: The trajectories for FM and AFM Skyrmions driven by (polarized) spin waves from the left. Reproduced from Ref.~\cite{daniels_topological_2019}.}
\label{fig:magnon_hall}
\end{figure}
%-----------------------------------

As mentioned previously in Section~\ref{sec:spin_wave_ferromagnetic_textures}, the spin wave dynamics upon a 2D magnetic texture can be described by an effective Schr\"{o}dinger equation with scalar and vector potentials associated with the texture. The potentials corresponding to a ferromagnetic Skyrmion give a ring-shaped potential well and an effective magnetic field perpendicular to the texture plane (see \Figure{fig:VB_Texture}(c)). For an antiferromagnetic Skyrmion, such an effective description still holds, but the two magnetic sublattices give rise to two sets of effective potentials acting on the left- and right-circularly polarization components of spin waves. The scalar potentials are the same for the two types of circular polarization, but the vector potentials and their corresponding effective magnetic fields are opposite because of the antiparallel alignment of the magnetization for two magnetic sublattices.

In an antiferromagnetic Skyrmion, the ring-shaped effective potential well as in \Figure{fig:VB_Texture} may confine one or more states to form localized spin wave modes. These modes appear as distortion of Skyrmions, such as breathing or squeezing of Skyrmions, similar to that for ferromagnetic Skyrmions as shown in \Figure{fig:Skyrmion}(a). On the other hand, propagating spin waves, when interacting with a Skyrmion, may be deflected preferentially in certain direction due to the effective Lorentz force caused by the effective magnetic field of Skyrmions.

More interestingly, in antiferromagnets, two degenerate circularly polarized spin wave modes have opposite effective charges~\cite{daniels_topological_2019}, and thus the resulting Lorentz forces are also in opposite directions. This means that an antiferromagnetic Skyrmion deflects two circularly polarized spin wave modes in opposite directions, causing an equivalent magnon spin Hall effect as shown in \Figure{fig:magnon_hall}(a).

As the Skyrmion deflects spin waves via an effective Lorentz force, the scattered spin waves also react back to the Skyrmion, \ie pushing the Skyrmion. The qualitative description of such back reaction behavior depends not only on its ferro- or antiferro-magnetism, but also relies on the polarization of spin waves.

\Figure{fig:magnon_hall} (right panel) shows how ferromagnetic (FM) and antiferromagnetic (AFM) Skyrmions respond when spin waves are injected from the left. The FM Skyrmion is dragged towards the spin wave source but in a sideways fashion. However, the AFM Skyrmion would be pushed forward by spin waves, and the sideways motion is determined by the polarization of the injected spin waves. More specifically, the AFM Skyrmion does not have transverse motion when the injected spin waves are linearly polarized. This situation is very similar to current driven AFM Skyrmion motion as demonstrated by Barker and Tretiakov~\cite{barker_static_2016}.

% !TEX root = review_magnonics_in_texture.tex
\section{Application \& Outlook}
The name \textit{magnonics} was inspired by the terms \textit{spintronics} and \textit{photonics} highlighting the potential of magnons to serve as information carriers in novel approaches beyond conventional CMOS information technology. The initial idea was based on wave-based computing meaning the logic operation is carried out by interference of multiple input magnons in which the information is encoded in the amplitude, frequency and phase. The advantages of this approach are twofold: First, information transport via magnons does not involve charge transport and, hence, does not suffer from Ohmic heating or voltage breakdowns. Second, wave-based computing falls in the category of non-boolean information processing schemes, which allow for non-destructive operations on the basic units of information~\cite{schneider_realization_2008,jamali_spin_2013,chumak_magnon_2014}. Therefore, magnonic devices are not subject to the fundamental Landauer limit describing the minimum energy consumption due to the increase of entropy in boolean logic gates, where the input information is lost after the computation~\cite{Landauer:wc,Balynskiy:2018km}.

There have been numerous experimental and micromagnetic demonstrations and prototypes of fundamental magnonic logic gates~\cite{ChumakNP15,Khi2010,Kru2010,Csa2017,Dav2015,Klingler:2014hk}. However, to be truly competitive with state of the art information technologies the above outlined original purpose of magnonics regarding the processing of information is facing two challenges: Firstly, the efficient generation of short-wavelength magnons and, secondly, the manipulation of the magnon's amplitude and phase on nanometer length scales and picosecond time scales. Spin textures are ideal candidates to tackle both challenges. We discussed in the previous chapters, that domain walls in magnonic waveguides are effective and nano-sized phase shifters operating in a broad frequency range~\cite{Bayer:2005ev,Han2019,Hama2018}. Domain walls are able to confine and channel propagating magnons~\cite{Gar2015,Wagner2016} and act as building blocks for larger magnonic networks~\cite{Lan2015} and magnetic vortices have been proven to be powerful sources for short-wavelength magnons without the need for nano-sized microwave antennas~\cite{Wintz2016}.

The original idea of magnonics to use magnons as information carriers for information processing in a non-boolean manner is still of great interest and is still being advanced by the international community. Nevertheless, developments in other key technologies such as vacancy based quantum metrology and quantum computing are beginning to recognize the potential of magnonics for delivering coherent microwave signals in a broad frequency range to nano-sized volumes for addressing single, solid state based quantum spins~\cite{Andrich:2017ey}. In a similar fashion researchers working in the field of magneto-photonics and magneto-plasmonics realized the advantages of magnons for enhanced coupling between photons and matter~\cite{Graf:2018im,Temnov:2012fn}. In this sense magnons are not only carriers of information, they act as a wavelength converter between microwave photons and photons in the visible range. Moreover, they also allow for a mutual coupling to charge currents~\cite{Vla2008}, mechanical oscillations~\cite{Duine2020,Klein2020} and superconducting circuits~\cite{LachanceQuirion:2019gm,Dobrovolskiy:2019ix}.

With the rise of antiferromagnetic spintronics~\cite{Jungwirth:2016dt}, an old class of materials was also rediscovered for magnonics. Antiferromagnets have several orders of magnitude higher frequency range compared to their ferromagnetic counterparts and they also bring in a new quantity which was not paid too much attention to up to this time: The antiferromagnetic spin system can be divided into two sub-lattices with different spin polarization which allows the introduction of the concept of magnon polarization (c.f. Section~\ref{sec:afm}). As we discuss in one of the following subsections spin textures play an important role in this concept since they can act as polarizing objects for spin waves.

Various prototypes of magnonic logic gates have already been reviewed in the literature and throughout the previous sections. At this point we want to highlight fundamental magnonic devices involving spin textures, which go beyond the scope of information processing and bring benefits to other research areas such as vacancy based solid state quantum spins and other hybrid systems. Fundamental refers to questions of how we can generate magnons inside spin textures and use them to steer the flow of magnons and how spin textures can polarize magnons in anti-ferromagnets or even store energy to release magnons pulses on demand.

\subsection{Spin-wave devices based on magnetic textures}

\subsubsection{Magnonic domain-wall circuit, fibers and spin wave diode}
\begin{figure}[t]
  \centering
\includegraphics[width=16cm]{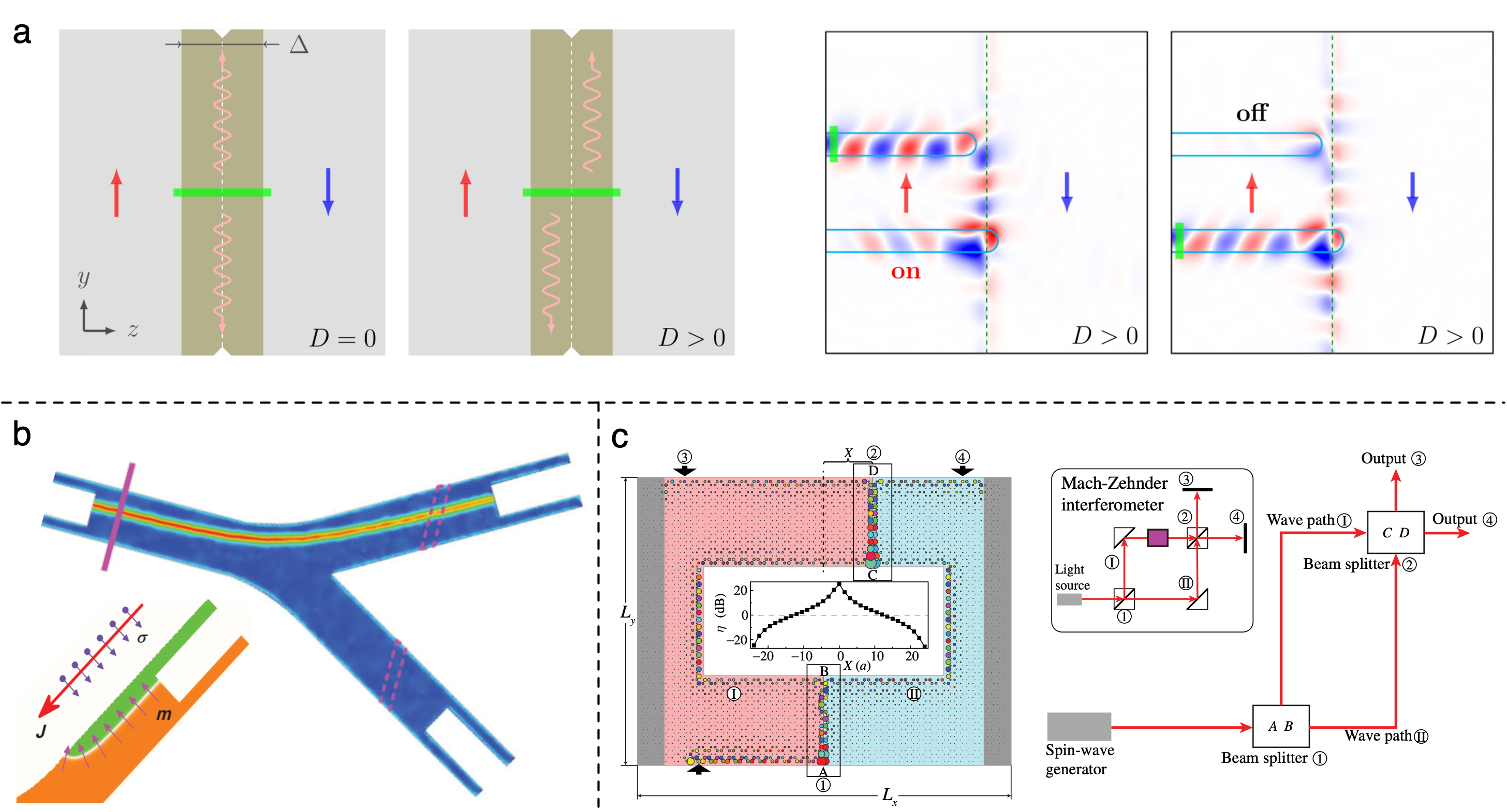}
\caption{(a) The left panels illustrate magnon propagation in domain walls with and without iDMI. The resulting break in symmetry is used in the right panels to achieve non-reciprocal spin wave propagation in complex domain wall networks for the realization of a spin wave diode. Images assembled from Ref.~\cite{Lan2015}. (b) Concept for a spin-wave fiber~\cite{xing_fiber_2016,Yu:2016bk} in a bilayer of a heavy metal and a perpendicularly magnetized ferromagnet. The spin-orbit torque resulting from a charge current $J$ in the heavy metal can inject a domain wall in the y-junction which acts as a spin-wave nanochannel. Images assembled from Ref.~\cite{Xing:2017ku}. (c) Concept of a spin-wave interferometer based on topological magnon surface states lying at the boundaries of magnetic domains shaded red (I) and blue (II). The inset in the left panel shows the ratio $\eta$ of the intensities arriving at the positions marked with $3$ and $4$ as a function of the domain wall position $X$ highlighted as region 1. The right panel shows the equivalent optical circuit. Images assembled from Ref.~\cite{Wang:2018eu}.}
\label{fig:APP1}
\end{figure}

The fundamental theory and experimental confirmation of spin waves propagating inside domain walls was discussed in Sections~\ref{Sec:spin waves in ferromagnetic thin films} and~\ref{sec:SW in domain wall}. Here, we want to discuss the opportunities which are given by implementing spin orbit torques, interfacial DMI (iDMI) and topological magnon states in magnonic devices based on domain walls. In the first study of spin-wave transport inside domain walls in a ferromagnet with perpendicular magnetic anisotropy and iDMI~\cite{Gar2015} the nonreciprocal character of spin-wave transport has already been described. Shortly after, the demonstration by Lan~\textit{et al.}~\cite{Lan2015} utilized the nonreciprocity to develop a theoretical and micromagnetic concept of a spin-wave diode (see Fig.~\ref{fig:APP1}(a)). As a result of the iDMI, the center of a spin wave beam propagating inside a domain wall is shifted off the domain wall center and the position depends on the direction of transport. Applying this to two dimensional networks of domain walls allows for building a magnonic circuit resembling the functionality of a diode where transport is possible only in a particular direction. Given the fact that domain walls are non-volatile but still reprogrammable, the idea of using domain walls for magnonic circuits is promising for the realization of reprogrammable devices~\cite{Vogel:2015ky,Heussner:2017ea,Wang:2018eg}. An important step in this direction was the discovery of Jiang and coworkers~\cite{Jiang2015}, where they demonstrated that divergent charge currents in a bilayer of a heavy metal and a perpendicularly magnetized ferromagnet with iDMI can inject spin textures (domain walls, bubbles~\cite{vukadinovic_spin-wave_2011} and skyrmions) with well defined chirality and spatial alignment with respect to the charge current direction. The combination of both creation of domain walls by spin-orbit torques and the channeling of spin waves inside the wall was used by Xing~\textit{et al.}~\cite{Xing:2017ku} as shown in Fig.~\ref{fig:APP1}(b). The path of a charge current in a y-junction traces out a trail for the injection of a Bloch domain wall, which was subsequently used in dynamic micromagnetic simulations to channel and steer spin waves in the desired direction. Even in more complex spin-wave waveguides the generation of the domain wall channel by a charge current takes only a few nanoseconds, showing the potential of spin textures for fast, reprogrammable spin-wave waveguides.

A more exotic concept for magnonic circuits is based on topological magnon edge states as introduced in Refs.~\cite{Mook:2014iu,Wang:2018eu,McClarty:2017im}, which are similar to the nonreciprocal electronic surface states in topological insulators. Wang~\textit{et al.}~\cite{Wang:2018eu} showed that these topological states do not only occur at the physical boundaries of magnonic crystals, but they also exist at the interface between magnetic domains. Combining the properties of edge modes and domain wall modes, they designed in micromagnetic simulations a magnon beam splitter on the nanometer length-scale and fused two of these beam splitters to a Mach-Zehnder interferometer as shown in Fig.~\ref{fig:APP1}(c). The output direction of this magnon interferometer is very sensitive to the position of the domain walls, which affects the phase of magnons propagating in the different arms of the interferometer. While this study is far from any experimental realization due to the complicated materials and small size, it nevertheless shows the potential of synergies between topological effects and spin textures.

\subsubsection{Domain wall based spin-Hall nano-oscillators}
The experiments and concepts relying on spin-wave transport in domain walls discussed so far used small antennas and the dynamic magnetic field generated by a microwave current for spin-wave excitation. For magnonic applications and also for using magnons as sources of microwave fields, \eg for driving solid state qubits in nitrogen vacancies, this approach suffers from the fact that microwave lines are difficult to miniaturize without crosstalk between neighbouring antennas. On the other hand, recent developments in the field of spin-Hall nano-oscillators showed that using spin-orbit torques is an efficient alternative for spin-wave excitation with the advantage of converting \textit{dc} into gigahertz oscillations of the magnetization. The idea of using intra-layer spin currents for driving magnons inside a transverse domain wall was already numerically studied in Ref.~\cite{Bisig:2009gq}. 

\begin{figure}[t]
  \centering
\includegraphics[width=16cm]{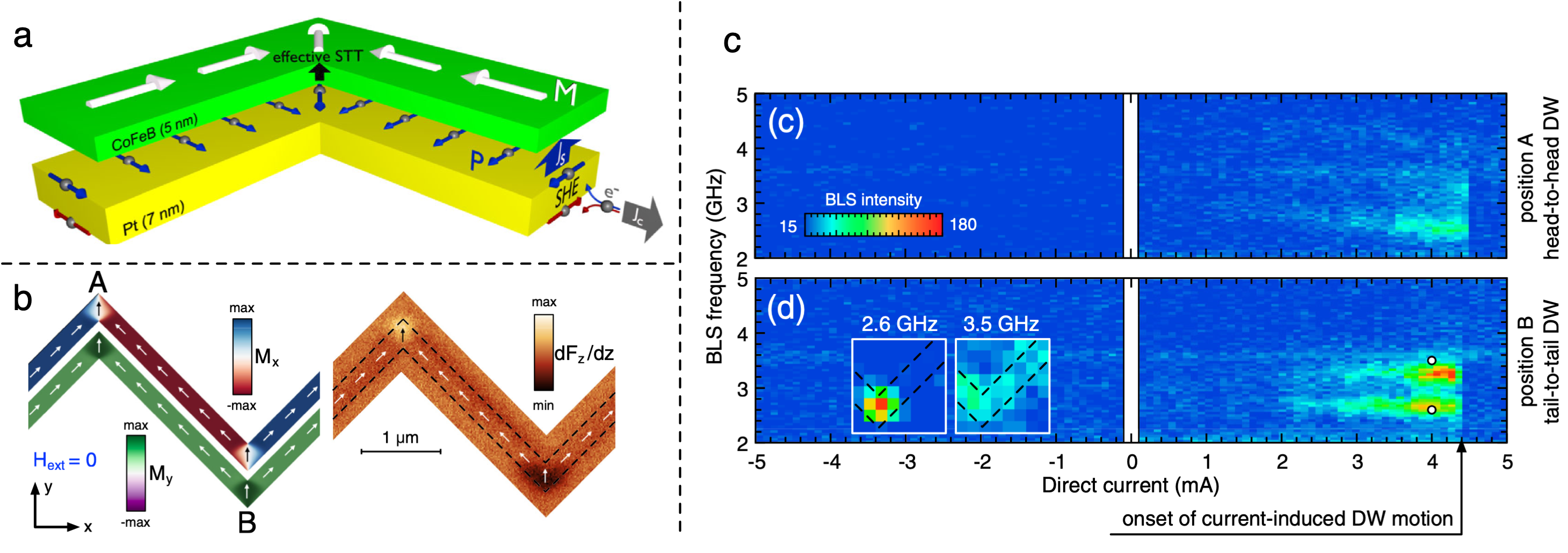}
\caption{Domain wall based spin Hall nano-oscillator. (a) Schematic device geometry: A bilayer of a heavy metal (Pt) and a metallic ferromagnet (CoFeB) is patterned into a nano-wire with a width of about 300~nm. A transverse magnetic domain wall can be stabilized at the corner by saturating the entire structure in the direction of the corner and allowing subsequent relaxation of the magnetic field. A charge current $J_c$ in the heavy metal is converted by the spin-Hall effect into a transverse spin current where the spin polarization meets the condition for driving auto-oscillations of the magnetization via spin transfer torque. (b) The transverse domain wall structure is confirmed by micromagnetic simulations (left) and magnetic force microscopy (right). (c) BLS microscopy as a function of the applied \textit{dc} current gives evidence for spin-torque driven magnons in the domain walls at the wire corners (position A and B) before the onset of current induced domain wall motion is reached. Images assembled from Ref.~\cite{Sato:2019cn}.}
\label{fig:APP2}
\end{figure}

The experimental realization of magnon auto-oscillations inside a domain wall~\cite{Sato:2019cn} was achieved using a pure spin current originating from the spin-Hall effect in a bilayer of a heavy metal and a metallic ferromagnet (see Fig.~\ref{fig:APP2}(a)). This geometry takes advantage of the fact that a compensation of the intrinsic magnetic damping is realized if the magnetization is aligned perpendicular to the charge current. Depending on the sign of the spin-Hall angle of the heavy metal and the sign of the charge current the effective torque on the magnetic moments either reduced or enhanced the damping. Since this device geometry requires only a charge current flowing in the heavy metal, the use of magnetic insulators is feasible as well. The internal structure of the transverse domain wall after saturating the bilayer with an external magnetic field along the y-direction and subsequent relaxation of the magnetic field to zero is shown by micromagnetic simulations and magnetic force microscopy in Fig.~\ref{fig:APP2}(b). Experimental evidence for magnon auto-oscillations driven by spin-orbit torque is given by BLS microscopy at the positions of the domain walls shown in Fig.~\ref{fig:APP2}(c). For a \textit{dc} current ranging from $-5$~mA to $+5$~mA, magnon spectra were recorded showing zero intensity for a negative \textit{dc} current but clear magnon intensities at $2.6$~GHz and $3.5$~GHz for a positive \textit{dc} current above a certain threshold. The abrupt disappearance of the BLS signal with a \textit{dc} current above $4.5$~mA is caused by current-induced domain wall motion~\cite{Beach2008}. These experiments show the feasibility of nanoscale spin textures to act as sources of magnons in magnonic circuits based on domain walls. However, they also open new opportunities for research on solid state quantum systems, where microwaves are needed in the GHz range and at low temperatures to address single qubits~\cite{Nakamura2015} with individual microwave pulses.

\subsubsection{Spin-wave polarizer}

The polarization dependent effective potential originating from an antiferromagnetic domain wall leads to very useful functionalities for manipulating the polarization of spin waves. An antiferromagnetic domain wall may function as a spin-wave polarizer for the in-plane polarization by reflecting the out-of-plane polarization at low frequencies, while at high frequencies the domain wall acts as a spin-wave retarder (wave-plate) above the barrier. Therefore, one simple antiferromagnetic texture realizes two essential device functionalities for polarization manipulation. With the above described polarization dependent scattering behaviour at an antiferromagnetic domain wall, it is straightforward to use such a domain wall as a spin-wave polarizer.

As discussed in Section \ref{sssec:sw_pol_afm}, because of the Dzyaloshinskii-Moriya interaction, the antiferromagnetic spin wave polarized along the DM vector is subject to an additional effective potential barrier within the domain wall. As a result, at low frequencies, the $y$-polarization experiences no (or little) reflection but the $x$-polarization hits a potential barrier and is strongly reflected. According to these principles, Lan~\textit{et al.} proposed that an antiferromagnetic domain wall works as a spin-wave polarizer \cite{Lan2017}. At higher frequencies, however, both polarizations transmit almost perfectly but with a relative phase delay, \ie the antiferromagnetic domain wall works as a spin-wave retarder~\cite{Lan2017}.

\subsubsection{Domain wall based universal logic gate}
\begin{figure}[t]
  \centering
\includegraphics[width=0.8\textwidth]{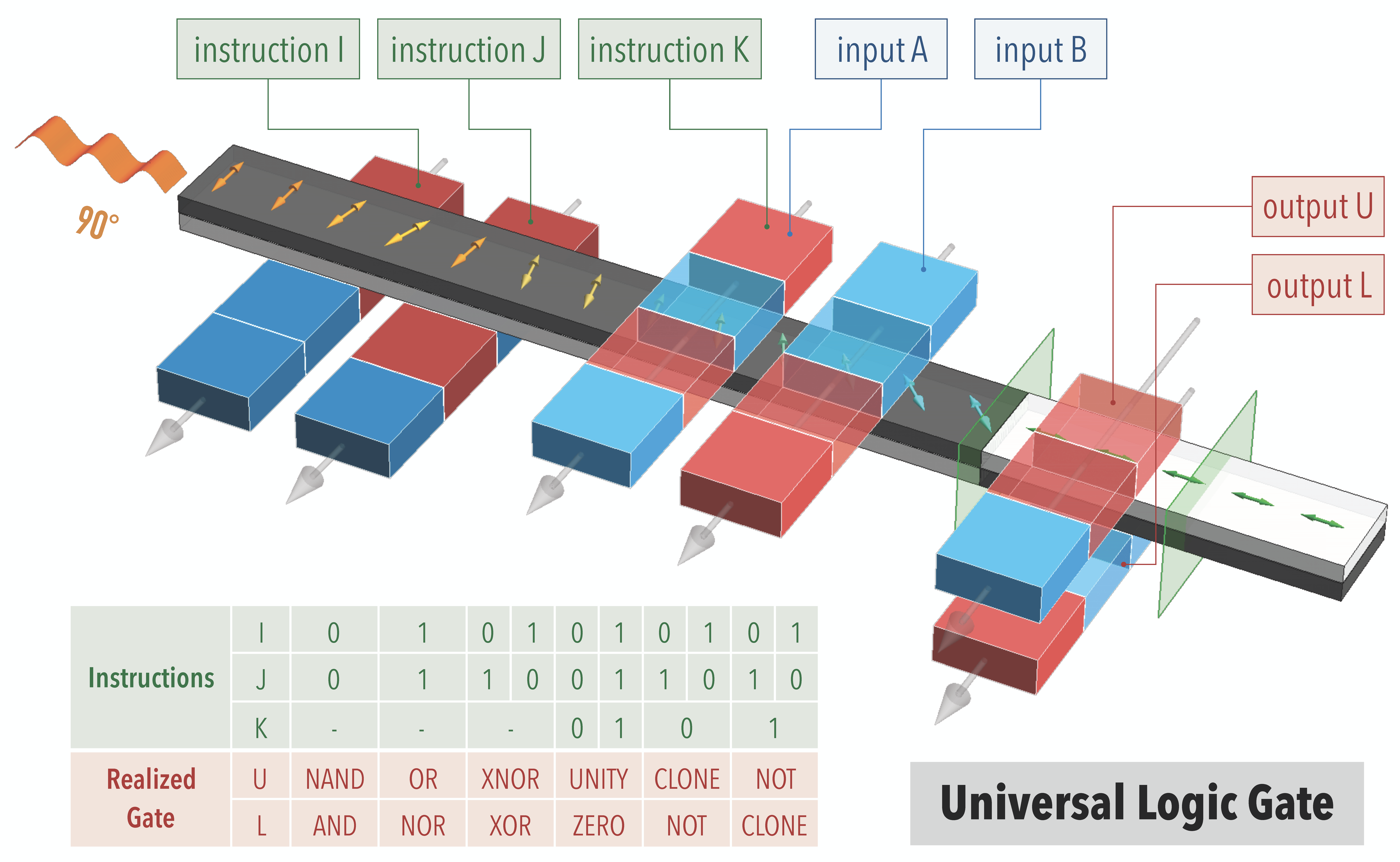}
\caption{A universal spin wave logic gate based on antiferromagnetic domain wall motion.}
\label{fig:UGate}
\end{figure}

As shown in Fig.~\ref{fig:UGate}, Yu~\textit{et al.}~\cite{W_Yu2020} proposed the realization of a universal logic gate based on the manipulation of antiferromagnetic domain walls by polarized spin waves. This computing scheme uses magnetic racetrack memories as inputs and outputs, and uses a synthetic antiferromagnetic wire as the logic track for computing~\cite{allwood_magnetic_2005}. The simplified working principle is the following: i) \ang{90} linearly polarized antiferromagnetic spin waves are injected into the logic track; ii) As linear spin waves pass through the input memories (the instructions and inputs), the magnetic state in the memory tracks alters the spin-wave polarization via the magnetic gating effect; iii) The overall rotation of the spin wave polarization relies on the combination of the magnetic states stored in the instructions and inputs; iv) The domain wall residing in the logic track may or may not be pushed forward according to the final spin-wave polarization, realizing a logic operation on the inputs. The exact logic function (OR, AND, XOR, \etc) is controlled by the memory bits stored in the instruction tracks.
%
%-----------------------------------
%-----------------------------------

\subsubsection{Energy storage in magnetic textures}

%-----------------------------------
\begin{figure}[t]
  \centering
\includegraphics[width=16cm]{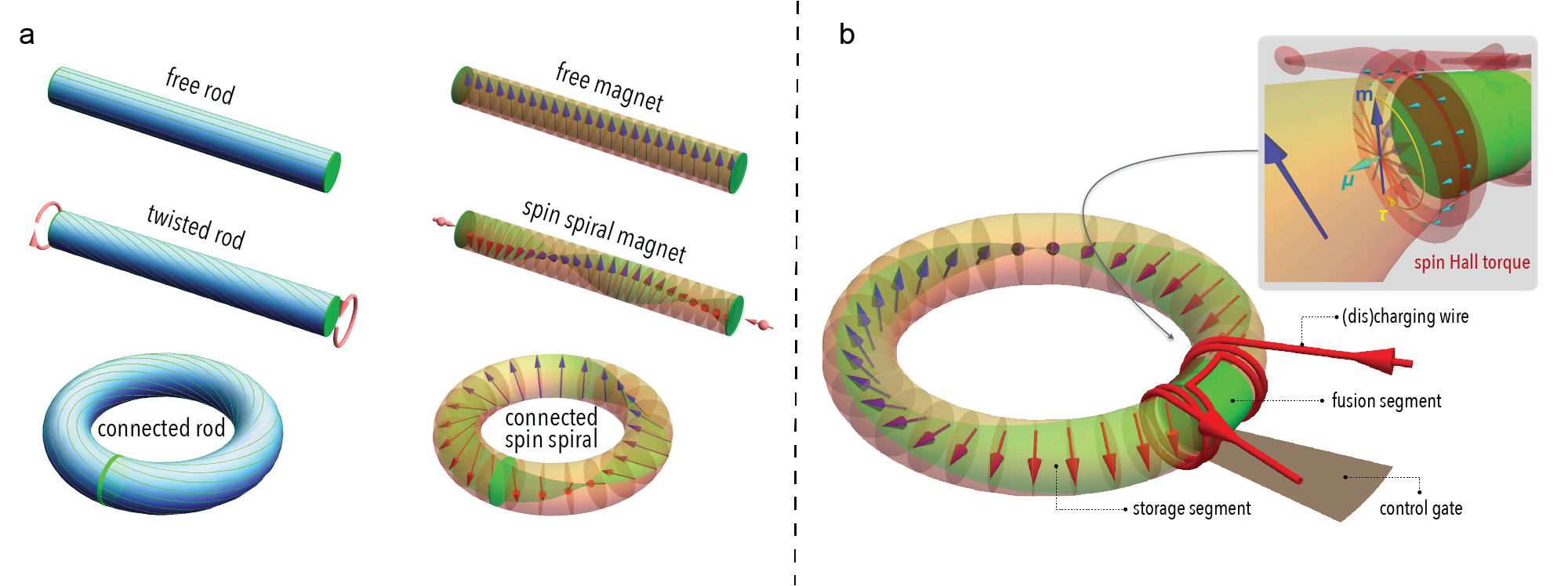}
\caption{(a) The comparison and analogy between a mechanically twisted rod as proposed by Sonin~\cite{Sonin2010} and a magnetically twisted spin texture as recently proposed by Tserkovnyak and Xiao~\cite{tserkovnyak_energy_2018}. (b) A conceptual spin battery based on twisted spin textures with a control gate at the fusion segment for changing and discharging of the stored energy.}
\label{fig:spin_bat}
\end{figure}
%-----------------------------------

The topological aspect of the magnetic texture can also be used for storing energy as proposed by Tserkovnyak and Xiao~\cite{tserkovnyak_energy_2018}. The idea can be illustrated in analogy with a twisted rod shown in \Figure{fig:spin_bat}(a), as proposed by Sonin~\cite{Sonin2010} in the context of a general super current. The energy is loaded into the rod by applying opposite torques at the ends, and then stored as the mechanical twisting. To prevent the rod from untwisting, the two ends are fused together.

A magnetic rod can store energy in a similar manner, but by the means of magnetic rather than mechanical twisting (\Figure{fig:spin_bat}(a)). The process is more or less the same as the twisted rod, except that the winding takes place in spin space, without any (mechanically) moving parts. By applying opposite spin torques (via a spin current injection) to the magnetization at two ends, the magnetization starts to twist in the easy plane and forms a spin spiral. The energy is stored in the spiral in the form of a magnetic exchange~\cite{Wang:2018ew}. To keep the spirals from unwinding, the magnetization at the ends has to be pinned or fused together. To release the stored magnetic energy, the spin-spiral has to be broken and then the stored energy is released in the form of magnetization dynamics, which can be harvested electrically via a spin-motive force~\cite{Maekawa2007}. A proposal of such a spin battery is illustrated in \Figure{fig:spin_bat}(b) with winding wires for spin injection and spin extraction, as well as a short segment for breaking and fusing the magnetic loop. In terms of energy density, competing with the traditional lithium ion technology with a spin battery appears challenging. However, the magnetic based energy storage has several unique advantages, such that it suffers no fatigue with infinite (dis)charging cycles, and for spintronic circuits such spin-texture energy storage can be naturally integrated with the nonvolatile logic and memory functionalities.

\subsection{Outlook and potential new directions}
The focus of this review article lies on the interaction of spin waves with non-collinear spin systems, namely spin textures, and it covers not only standard ferromagnets but highlights also recent developments in systems based on antiferromagnets. Since the term `interaction' covers transport along extended textures, displacement or manipulation of spatially confined spin textures and generation of magnons by dynamics of spin textures, the possible future directions are impossible to nail down. As outlined in this section, prototypes of fundamental building blocks for spin texture based magnonics have already been developed on the modelling side and experimental realization of complex networks of artificially created domain wall is progressing quickly~\cite{Alb2019,Alb2016,Alb2018a}. We already gave hints throughout this article that magnonics is going beyond its initial purpose of simply doing computation in a different, non-boolean manner. Magnons can not only be used as information carriers but also as a tool to deliver and focus microwave radiation to nano-sized volumes including phase manipulation on length-scales comparable to the magnon wavelength. This renders magnonics interesting in particular to the community working on the realization of quantum metrology and quantum computing~\cite{Dobrovolskiy:2019ix,Hoffmann2019,L_Liu2019,Nakamura2020} using \eg nitrogen-vacancy centers in diamond~\cite{Maz2008,Cas2018,Thi2019,Chau2019,H_Tu2020} or silicone vacancies in silicone carbide~\cite{Fuchs:2015ii}. In a similar fashion, the field of optomagnonics~\cite{Nakamura2016,H_Tang2016,JQ_You2017,CM_Hu2019} utilizes magnons as a wavelength converter between microwave photons and photonic cavities in the visible or near infrared regime. In this sense magnonics can serve as an interface technologies with all the benefits offered by magneto-optics~\cite{Ross2011} and spintronics~\cite{Fert2008}.

There is, however, a new, promising direction which makes use of the inherent nonlinear character of magnons. Grollier and coworkers applied the nonlinear magnetization dynamics in coupled spin torque oscillators to realize a brain inspired computing scheme~\cite{Torrejon:2017hj,Romera:2018dm} and recently the {\AA}kerman group succeeded at tranferring this concept to a two-dimensional array of spin-Hall nanooscillators~\cite{Zahedinejad:2019kd}. Especially in the times of big data and the Internet of Things, such new computational approaches mimicking the functionality of the human brain and its superior performance in classification and pattern recognition gives strong incentives for the field of magnonics. Spin textures play an important role since they bring features such as being non-volatile but still reprogrammable on a nanosecond timescale. Pushing these ideas forward could render magnonics the first alternative technology for a true, hardware based realization of neuromorphic computing.

\section{Acknowledgements}

The authors would like to thank J. Lan, O. A. Tretiakov, S. Granville, J. Chen and H. Wang for their helpful discussions. H.Y. thanks the financial support from NSF China under Grants 11674020, 12074026 and U1801661, the National Key Research and Development Program of China Grant 2016YFA0300802, and Shenzhen Institute for Quantum Science and Engineering Grant No. SIQSE202007. J.X. acknowledges the financial support from NSF China Grant 11722430. H.S. acknowledges the financial support by the Deutsche Forschungsgemeinschaft within programme SCHU2922/1-1 and SCHU2922/4-1.

\bibliography{Magnetic_texture_based_magnonics_v28.bib}

\begin{thebibliography}{100}
\expandafter\ifx\csname url\endcsname\relax
  \def\url#1{\texttt{#1}}\fi
\expandafter\ifx\csname urlprefix\endcsname\relax\def\urlprefix{URL }\fi
\expandafter\ifx\csname href\endcsname\relax
  \def\href#1#2{#2} \def\path#1{#1}\fi

\bibitem{bloch_zur_1930}
F.~Bloch, Zur {Theorie} des {Ferromagnetismus}, Z. Phys. A Hadr. and Nucl. 61
  (1930) 206--219.

\bibitem{Kittel1948}
C.~Kittel, {On the theory of ferromagnetic resonance absorption}, Phys. Rev. 73
  (1948) 155--161.

\bibitem{Kru2010}
V.~V. Kruglyak, S.~O. Demokritov, D.~Grundler, {Magnonics}, J. Phys. D Appl.
  Phys. 43 (2010) 264001.

\bibitem{Len2011}
B.~Lenk, H.~Ulrichs, F.~Garbs, M.~M{\"{u}}nzenberg, {The building blocks of
  magnonics}, Phys. Rep. 507 (2011) 107--136.

\bibitem{Stamps:2014en}
R.~L. Stamps, S.~Breitkreutz, J.~{\AA}kerman, A.~V. Chumak, Y.~Otani, G.~E.~W.
  Bauer, J.-U. Thiele, M.~Bowen, S.~A. Majetich, M.~Kl{\"a}ui, I.~L. Prejbeanu,
  B.~Dieny, N.~M. Dempsey, B.~Hillebrands, {The 2014 Magnetism Roadmap}, J.
  Phys. D Appl. Phys. 47 (2014) 333001.

\bibitem{ChumakNP15}
A.~V. Chumak, V.~I. Vasyuchka, A.~A. Serga, B.~Hillebrands, {Magnon
  spintronics}, Nat. Phys. 11 (2015) 453--461.

\bibitem{Yu_Xiao_Pirro_2018}
{Yu, H. and Xiao, J. and Pirro, P.}, {Magnon Spintronics}, J. Magn. Magn.
  Mater. 450 (2018) 1 -- 2.

\bibitem{Demi_Review_2020}
V.~E. Demidov, S.~Urazhdin, A.~Anane, V.~Cros, S.~O. Demokritov,
  {Spin–orbit-torque magnonics}, J. Appl. Phys. 127 (2020) 170901.

\bibitem{Khi2010}
A.~Khitun, M.~Bao, K.~L. Wang, {Magnonic logic circuits}, J. Phys. D Appl.
  Phys. 43 (2010) 264005.

\bibitem{Chu2017}
A.~Chumak, A.~Serga, B.~Hillebrands, {Magnonic crystals for data processing},
  J. Phys. D Appl. Phys. 50 (2017) 244001.

\bibitem{Csa2017}
G.~Csaba, {\'{A}}.~Papp, W.~Porod, {Perspectives of using spin waves for
  computing and signal processing}, Phys. Lett. A 381 (2017) 1471--1476.

\bibitem{Kra2014}
M.~Krawczyk, D.~Grundler, {Review and prospects of magnonic crystals and
  devices with reprogrammable band structure}, J. Phys. Condens. Matter 26
  (2014) 123202.

\bibitem{Neusser2009}
S.~Neusser, D.~Grundler, {Magnonics: Spin waves on the nanoscale}, Adv. Mater.
  21 (2009) 2927--2932.

\bibitem{Adeyeye_2010}
Z.~K. Wang, V.~L. Zhang, H.~S. Lim, S.~C. Ng, M.~H. Kuok, S.~Jain, A.~O.
  Adeyeye, {Nanostructured magnonic crystals with size-tunable bandgaps}, ACS
  Nano 4 (2010) 643--648.

\bibitem{Tacchi_BLS_2017}
S.~Tacchi, G.~Gubbiotti, M.~Madami, G.~Carlotti, {Brillouin light scattering
  studies of 2D magnonic crystals}, J. Phys. Condens. Matter 29 (2017) 073001.

\bibitem{demokritov_bose-einstein_2006}
S.~O. Demokritov, V.~E. Demidov, O.~Dzyapko, G.~A. Melkov, A.~A. Serga,
  B.~Hillebrands, A.~N. Slavin, Bose-{Einstein} condensation of
  quasi-equilibrium magnons at room temperature under pumping, Nature 443
  (2006) 430--433.

\bibitem{saitoh_current-induced_2004}
E.~Saitoh, H.~Miyajima, T.~Yamaoka, G.~Tatara, {Current-induced resonance and
  mass determination of a single magnetic domain wall}, Nature 432 (2004)
  203--206.

\bibitem{chou_direct_2007}
K.~W. Chou, A.~Puzic, H.~Stoll, D.~Dolgos, G.~Schütz, B.~Van~Waeyenberge,
  A.~Vansteenkiste, T.~Tyliszczak, G.~Woltersdorf, C.~H. Back, {Direct
  observation of the vortex core magnetization and its dynamics}, Appl. Phys.
  Lett. 90 (2007) 202505.

\bibitem{Wintz2016}
S.~Wintz, V.~Tiberkevich, M.~Weigand, J.~Raabe, J.~Lindner, A.~Erbe, A.~Slavin,
  J.~Fassbender, {Magnetic vortex cores as tunable spin-wave emitters}, Nat.
  Nanotechnol. 11 (2016) 948--953.

\bibitem{Ros2015}
H.~D. Rosales, D.~C. Cabra, P.~Pujol, {Three-sublattice skyrmion crystal in the
  antiferromagnetic triangular lattice}, Phys. Rev. B 92 (2015) 214439.

\bibitem{diaz_topological_2019}
S.~A. D\'{\i}az, J.~Klinovaja, D.~Loss, {Topological magnons and edge states in
  antiferromagnetic Skyrmion crystals}, Phys. Rev. Lett. 122 (2019) 187203.

\bibitem{Gao2020}
S.~Gao, H.~D. Rosales, F.~A. {G{\'{o}}mez Albarrac{\'{i}}n}, V.~Tsurkan,
  G.~Kaur, T.~Fennell, P.~Steffens, M.~Boehm, P.~{\v{C}}erm{\'{a}}k,
  A.~Schneidewind, E.~Ressouche, D.~C. Cabra, C.~R{\"{u}}egg, O.~Zaharko,
  {Fractional antiferromagnetic skyrmion lattice induced by anisotropic
  couplings}, Nature 586 (2020) 37--41.

\bibitem{Stancil}
D.~Stancil, A.~Prabhakar, {Spin waves: theory and applications, Appendix C},
  Springer, New York, 2009.

\bibitem{Daal1990}
G.~H.~O. Daalderop, P.~J. Kelly, M.~F.~H. Schuurmans, {First-principles
  calculation of the magnetocrystalline anisotropy energy of iron, cobalt, and
  nickel}, Phys. Rev. B 41 (1990) 11919--11937.

\bibitem{Guoqiang2015PMA}
G.~Yu, Z.~Wang, M.~Abolfath-Beygi, C.~He, X.~Li, K.~L. Wong, P.~Nordeen, H.~Wu,
  G.~P. Carman, X.~Han, I.~A. Alhomoudi, P.~K. Amiri, K.~L. Wang,
  {Strain-induced modulation of perpendicular magnetic anisotropy in
  Ta/CoFeB/MgO structures investigated by ferromagnetic resonance}, Appl. Phys.
  Lett. 106 (2015) 072402.

\bibitem{RMP_PMA_2017}
B.~Dieny, M.~Chshiev, {Perpendicular magnetic anisotropy at transition
  metal/oxide interfaces and applications}, Rev. Mod. Phys. 89 (2017) 025008.

\bibitem{Avci2017}
C.~O. Avci, A.~Quindeau, C.-F. Pai, M.~Mann, L.~Caretta, A.~S. Tang, M.~C.
  Onbasli, C.~A. Ross, G.~S.~D. Beach, {Current-induced switching in a magnetic
  insulator}, Nat. Mater. 16 (2017) 309--314.

\bibitem{Lat2018}
D.~M. Lattery, D.~Zhang, J.~Zhu, X.~Hang, J.~P. Wang, X.~Wang, {Low Gilbert
  damping constant in perpendicularly magnetized W/CoFeB/MgO films with high
  thermal stability}, Sci. Rep. 8 (2018) 13395.

\bibitem{Sou2018}
L.~Soumah, N.~Beaulieu, L.~Qassym, C.~Carr{\'{e}}tero, E.~Jacquet,
  R.~Lebourgeois, {Ultra-low damping insulating magnetic thin films get
  perpendicular}, Nat. Commun. 9 (2018) 3355.

\bibitem{Rana:2019ce}
B.~Rana, Y.~Otani, {Towards magnonic devices based on voltage-controlled
  magnetic anisotropy}, Commun. Phys. 2 (2019) 90.

\bibitem{Marko:2019bm}
D.~Mark{\'o}, F.~Vald{\'e}s-Bango, C.~Quir{\'o}s, A.~Hierro-Rodr{\'\i}guez,
  M.~V{\'e}lez, J.~I. Mart{\'\i}n, J.~M. Alameda, D.~S. Schmool, L.~M.
  {\'A}lvarez-Prado, {Tunable ferromagnetic resonance in coupled trilayers with
  crossed in-plane and perpendicular magnetic anisotropies}, Appl. Phys. Lett.
  115 (2019) 082401.

\bibitem{Chen2019PMA}
J.~Chen, C.~Wang, C.~Liu, S.~Tu, L.~Bi, H.~Yu, {Spin wave propagation in
  ultrathin magnetic insulators with perpendicular magnetic anisotropy}, Appl.
  Phys. Lett. 114 (2019) 212401.

\bibitem{X_Qiu2020}
H.~Chen, D.~Cheng, H.~Yang, D.~Wang, S.~Zhou, Z.~Shi, X.~Qiu, {Magnetization
  switching induced by magnetic field and electric current in perpendicular
  TbIG/Pt bilayers}, Appl. Phys. Lett. 116 (2020) 112401.

\bibitem{kittel_physical_1949}
C.~Kittel, Physical {theory} of {ferromagnetic} {domains}, Rev. Mod. Phys. 21
  (1949) 541--583.

\bibitem{dzyaloshinsky_thermodynamic_1958}
I.~Dzyaloshinsky, {A thermodynamic theory of “weak” ferromagnetism of
  antiferromagnetics}, J. Phys. Chem. Solids 4 (1958) 241--255.

\bibitem{Moriya1960}
T.~Moriya, {Anisotropic superexchange interaction and weak ferromagnetism},
  Phys. Rev. 120 (1960) 91--98.

\bibitem{fert_skyrmions_2013}
A.~Fert, V.~Cros, J.~Sampaio, {Skyrmions on the track}, Nat. Nanotechnol. 8
  (2013) 152--156.

\bibitem{Tok2010}
X.~Z. Yu, Y.~Onose, N.~Kanazawa, J.~H. Park, J.~H. Han, Y.~Matsui, N.~Nagaosa,
  Y.~Tokura, {Real-space observation of a two-dimensional skyrmion crystal},
  Nature 465 (2010) 901--904.

\bibitem{Abert2019}
C.~Abert, {Micromagnetics and spintronics: models and numerical methods}, Eur.
  Phys. J. B 92 (2019) 120.

\bibitem{Moon2013}
J.-H. Moon, S.-M. Seo, K.-J. Lee, K.-W. Kim, J.~Ryu, H.-W. Lee, R.~D.
  McMichael, M.~D. Stiles, {Spin-wave propagation in the presence of
  interfacial Dzyaloshinskii-Moriya interaction}, Phys. Rev. B 88 (2013)
  184404.

\bibitem{Cor2013}
D.~Cort{\'{e}}s-Ortu{\~{n}}o, P.~Landeros, {Influence of the
  Dzyaloshinskii-Moriya interaction on the spin-wave spectra of thin films}, J.
  Phys. Condens. Matter 25 (2013) 156001.

\bibitem{bogdanov_magnetic_2002}
A.~N. Bogdanov, U.~K. Rößler, M.~Wolf, K.-H. Müller, Magnetic structures and
  reorientation transitions in noncentrosymmetric uniaxial antiferromagnets,
  Phys. Rev. B 66 (2002) 214410.

\bibitem{landau_statistical_1980}
L.~D. Landau, E.~M. Lifsic, L.~P. Pitaevskii, J.~Sykes, M.~J. Kearsley,
  Statistical Physics {Vol}. 9, Butterworth-Heinemann, 1980.

\bibitem{Wegrowe1999}
J.-E. Wegrowe, D.~Kelly, Y.~Jaccard, P.~Guittienne, J.-P. Ansermet,
  {Current-induced magnetization reversal in magnetic nanowires}, EPL 45 (1999)
  626--632.

\bibitem{Ralph2000}
F.~J. Albert, J.~A. Katine, R.~A. Buhrman, D.~C. Ralph, Spin-polarized current
  switching of a co thin film nanomagnet, Appl. Phys. Lett. 77 (2000)
  3809--3811.

\bibitem{Beach2005}
G.~S.~D. Beach, C.~Nistor, C.~Knutson, M.~Tsoi, J.~L. Erskine, {Dynamics of
  field-driven domain-wall propagation in ferromagnetic nanowires}, Nat. Mater.
  4 (2005) 741--744.

\bibitem{Parkin2008}
S.~S. Parkin, M.~Hayashi, L.~Thomas, {Magnetic domain-wall racetrack memory},
  Science 320 (2008) 190.

\bibitem{Jung2008}
S.-W. Jung, W.~Kim, T.-D. Lee, K.-J. Lee, H.-W. Lee, {Current-induced domain
  wall motion in a nanowire with perpendicular magnetic anisotropy}, Appl.
  Phys. Lett. 92 (2008) 202508.

\bibitem{Sato:2019cn}
N.~Sato, K.~Schultheiss, L.~K{\"o}rber, N.~Puwenberg, T.~M{\"u}hl, A.~A. Awad,
  S.~S. P.~K. Arekapudi, O.~Hellwig, J.~Fassbender, H.~Schultheiss, {Domain
  wall based spin-Hall nano-oscillators}, Phys. Rev. Lett. 123 (2019) 057204.

\bibitem{hubert_magnetic_1998}
A.~Hubert, R.~Schäfer, {Magnetic Domains}, Springer, 1998.

\bibitem{lilley_lxxi_1950}
B.~Lilley, {LXXI}. {Energies} and widths of domain boundaries in
  ferromagnetics, Lond. Edinb. Dubl. Phil. Mag. 41 (1950) 792--813.

\bibitem{schoenherr_topological_2018}
P.~Schoenherr, J.~Müller, L.~Köhler, A.~Rosch, N.~Kanazawa, Y.~Tokura,
  M.~Garst, D.~Meier, Topological domain walls in helimagnets, Nat. Phys. 14
  (2018) 465--468.

\bibitem{chen_novel_2013}
G.~Chen, J.~Zhu, A.~Quesada, J.~Li, A.~T. N’Diaye, Y.~Huo, T.~P. Ma, Y.~Chen,
  H.~Y. Kwon, C.~Won, Z.~Q. Qiu, A.~K. Schmid, Y.~Z. Wu, Novel {chiral}
  {magnetic} {domain} {wall} {structure} in {Fe}/{Ni}/{Cu}(001) {films}, Phys.
  Rev. Lett. 110 (2013) 177204.

\bibitem{chen_tailoring_2013}
G.~Chen, T.~Ma, A.~T. N’Diaye, H.~Kwon, C.~Won, Y.~Wu, A.~K. Schmid,
  Tailoring the chirality of magnetic domain walls by interface engineering,
  Nat. Commun. 4 (2013) 2671.

\bibitem{dzyaloshinskii_theory_nodate}
I.~E. Dzyaloshinskii, Theory of {{Helicoidal Structures}} in
  {{Antiferromagnets}}. {{I}}. {{Nonmetals}}, JETP 19 (1963) 960.

\bibitem{bode_chiral_2007-2}
M.~Bode, M.~Heide, K.~{von Bergmann}, P.~Ferriani, S.~Heinze, G.~Bihlmayer,
  A.~Kubetzka, O.~Pietzsch, S.~Bl{\"u}gel, R.~Wiesendanger, Chiral magnetic
  order at surfaces driven by inversion asymmetry, Nature 447~(7141) (2007)
  190--193.

\bibitem{yoshimori_new_1959}
A.~Yoshimori, A {new} {type} of {antiferromagnetic} {structure} in the {rutile}
  {type} {crystal}, J. Phys. Soc. Jpn. 14 (1959) 807--821.

\bibitem{XSWang2018}
X.~S. Wang, H.~Y. Yuan, X.~R. Wang, {A theory on skyrmion size}, Commun. Phys.
  1 (2018) 31.

\bibitem{kezsmarki_ne-type_2015}
I.~Kézsmárki, S.~Bordács, P.~Milde, E.~Neuber, L.~M. Eng, J.~S. White, H.~M.
  Rønnow, C.~D. Dewhurst, M.~Mochizuki, K.~Yanai, H.~Nakamura, D.~Ehlers,
  V.~Tsurkan, A.~Loidl, Néel-type skyrmion lattice with confined orientation
  in the polar magnetic semiconductor {GaV4S8}, Nat. Mater. 14 (2015)
  1116--1122.

\bibitem{Muh2009}
S.~M{\"{u}}hlbauer, B.~Binz, F.~Jonietz, C.~Pfleiderer, A.~Rosch, A.~Neubauer,
  R.~Georgii, P.~B{\"{o}}ni, {Skyrmion lattice in a chiral magnet}, Science 323
  (2009) 915.

\bibitem{zheng_experimental_2018}
F.~Zheng, F.~N. Rybakov, A.~B. Borisov, D.~Song, S.~Wang, Z.-A. Li, H.~Du,
  N.~S. Kiselev, J.~Caron, A.~Kovács, M.~Tian, Y.~Zhang, S.~Blügel, R.~E.
  Dunin-Borkowski, Experimental observation of chiral magnetic bobbers in
  {B20}-type {FeGe}, Nat. Nanotechnol. 13 (2018) 451--455.

\bibitem{tveten_intrinsic_2016}
E.~G. Tveten, T.~Müller, J.~Linder, A.~Brataas, {Intrinsic magnetization of
  antiferromagnetic textures}, Phys. Rev. B 93 (2016) 104408.

\bibitem{bloch_zur_1932}
F.~Bloch, Zur {Theorie} des {Austauschproblems} und der {Remanenzerscheinung}
  der {Ferromagnetika}, Springer, Berlin, Heidelberg, 1932.

\bibitem{landau_3_1992}
L.~Landau, E.~Lifshitz, 3 - {On} the theory of the dispersion of magnetic
  permeability in ferromagnetic bodies, in: L.~Pitaevski (Ed.), Perspectives in
  {Theoretical} {Physics}, Pergamon, 1992, pp. 51--65.

\bibitem{gilbert_phenomenological_2004}
T.~Gilbert, {A phenomenological theory of damping in ferromagnetic materials},
  IEEE Trans. Magn. 40 (2004) 3443--3449.

\bibitem{keffer_theory_1952}
F.~Keffer, C.~Kittel, Theory of {antiferromagnetic} {resonance}, Phys. Rev. 85
  (1952) 329--337.

\bibitem{Serga2010}
A.~A. Serga, A.~V. Chumak, B.~Hillebrands, {YIG} magnonics, J. Phys. D Appl.
  Phys. 43 (2010) 264002.

\bibitem{Barker2016}
J.~Barker, G.~E.~W. Bauer, {Thermal spin dynamics of yttrium iron garnet},
  Phys. Rev. Lett. 117 (2016) 217201.

\bibitem{Nambu2020}
Y.~Nambu, J.~Barker, Y.~Okino, T.~Kikkawa, Y.~Shiomi, M.~Enderle, T.~Weber,
  B.~Winn, M.~Graves-Brook, J.~M. Tranquada, T.~Ziman, M.~Fujita, G.~E.~W.
  Bauer, E.~Saitoh, K.~Kakurai, {Observation of magnon polarization}, Phys.
  Rev. Lett. 125 (2020) 027201.

\bibitem{oh_coherent_2017}
S.-H. Oh, S.~K. Kim, D.-K. Lee, G.~Go, K.-J. Kim, T.~Ono, Y.~Tserkovnyak, K.-J.
  Lee, {Coherent terahertz spin-wave emission associated with ferrimagnetic
  domain wall dynamics}, Phys. Rev. B 96 (2017) 100407.

\bibitem{kim_self-focusing_2017}
S.~K. Kim, K.-J. Lee, Y.~Tserkovnyak, {Self-focusing skyrmion racetracks in
  ferrimagnets}, Phys. Rev. B 95 (2017) 140404.

\bibitem{kim_tunable_2019}
S.~K. Kim, K.~Nakata, D.~Loss, Y.~Tserkovnyak, Tunable {magnonic} {thermal}
  {Hall} {effect} in {Skyrmion} {crystal} {phases} of {ferrimagnets}, Phys.
  Rev. Lett. 122 (2019) 057204.

\bibitem{Demi_Nonlinear2001}
S.~Demokritov, B.~Hillebrands, A.~Slavin, {Brillouin light scattering studies
  of confined spin waves: linear and nonlinear confinement}, {Phys. Rep.} 348
  (2001) 441--489.

\bibitem{Wu_Nonlinear2007}
M.~Wu, C.~E. Patton, {Experimental observation of Fermi-Pasta-Ulam recurrence
  in a nonlinear feedback ring system}, Phys. Rev. Lett. 98 (2007) 047202.

\bibitem{Slavin_Nonlinear2009}
A.~Slavin, V.~Tiberkevich, {Nonlinear auto-oscillator theory of microwave
  generation by spin-polarized current}, IEEE Trans. Magn. 45 (2009)
  1875--1918.

\bibitem{Kam_Nonlinear2011}
M.~Kammerer, M.~Weigand, M.~Curcic, M.~Noske, M.~Sproll, A.~Vansteenkiste,
  B.~{Van Waeyenberge}, H.~Stoll, G.~Woltersdorf, C.~H. Back, G.~Schuetz,
  {Magnetic vortex core reversal by excitation of spin waves}, Nat. Commun. 2
  (2011) 279.

\bibitem{gurevich_magnetization_1996}
A.~G. Gurevich, G.~A. Melkov, {Magnetization oscillations and waves}, CRC
  Press, 1996.

\bibitem{Kal1986}
B.~Kalinikos, A.~Slavin, {Theory of dipole-exchange spin wave spectrum for
  ferromagnetic films with mixed exchange boundary conditions}, J. Phys. C
  Solid State Phys. 19 (1986) 7013--7033.

\bibitem{Demi2009}
V.~Demidov, M.~Kostylev, K.~Rott, P.~Krzysteczko, G.~Reiss, D.~S. O.,
  {Excitation of microwaveguide modes by a stripe antenna}, Appl. Phys. Lett.
  95 (2009) 112509.

\bibitem{Yu2014}
H.~Yu, O.~{D'Allivy Kelly}, V.~Cros, R.~Bernard, P.~Bortolotti, A.~Anane,
  F.~Brandl, R.~Huber, I.~Stasinopoulos, D.~Grundler, {Magnetic thin-film
  insulator with ultra-low spin wave damping for coherent nanomagnonics}, Sci.
  Rep. 4 (2014) 6848.

\bibitem{Yu2016}
H.~Yu, O.~d.~A. Kelly, C.~V., R.~Bernard, P.~Bortolotti, A.~Anane, F.~Brand,
  F.~Heimbach, D.~Grundler, {Approaching soft X-ray wavelengths in
  nanomagnet-based microwave technology}, Nat. Commun. 7 (2016) 11255.

\bibitem{Liu2018}
C.~Liu, J.~Chen, T.~Liu, F.~Heimbach, H.~Yu, Y.~Xiao, J.~Hu, M.~Liu, H.~Chang,
  T.~Stueckler, S.~Tu, Y.~Zhang, Y.~Zhang, P.~Gao, Z.~Liao, D.~Yu, K.~Xia,
  N.~Lei, W.~Zhao, M.~Wu, {Long-distance propagation of short-wavelength spin
  waves}, Nat. Commun. 9 (2018) 738.

\bibitem{Die2019}
G.~Dieterle, J.~F{\"{o}}rster, H.~Stoll, A.~S. Semisalova, S.~Finizio,
  A.~Gangwar, M.~Weigand, M.~Noske, M.~F{\"{a}}hnle, I.~Bykova, J.~Gr{\"{a}}fe,
  D.~A. Bozhko, H.~Y. Musiienko-Shmarova, V.~Tiberkevich, A.~N. Slavin, C.~H.
  Back, J.~Raabe, G.~Sch{\"{u}}tz, S.~Wintz, {Coherent excitation of
  heterosymmetric spin waves with ultrashort wavelengths}, Phys. Rev. Lett. 122
  (2019) 117202.

\bibitem{eshbach_surface_1960}
J.~R. Eshbach, R.~W. Damon, Surface {magnetostatic} {modes} and {surface}
  {spin} {waves}, Phys. Rev. 118 (1960) 1208.

\bibitem{damon_magnetostatic_1961}
R.~Damon, J.~Eshbach, Magnetostatic modes of a ferromagnet slab, J. Phys. Chem.
  Solids 19 (1961) 308--320.

\bibitem{camley_surface_1978}
R.~E. Camley, D.~L. Mills, Surface response of exchange- and dipolar-coupled
  ferromagnets: {Application} to light scattering from magnetic surfaces, Phys.
  Rev. B 18 (1978) 4821.

\bibitem{de_wames_dipole-exchange_1970}
R.~E. De~Wames, Dipole-{exchange} {spin} {waves} in {ferromagnetic} {films}, J.
  Appl. Phys. 41 (1970) 987.

\bibitem{Hil1990}
B.~Hillebrands, {Spin-wave calculations for multilayered structures}, Phys.
  Rev. B 41 (1990) 530.

\bibitem{hurben_theory_1995}
M.~J. Hurben, C.~E. Patton, {Theory of magnetostatic waves for in-plane
  magnetized isotropic films}, J. Magn. Magn. Mater. 139 (1995) 263--291.

\bibitem{xiao_chapter_2013}
J.~Xiao, Y.~Zhou, G.~E.~W. Bauer, Chapter {Two} - {Spin}-{Wave} {Excitation} in
  {Magnetic} {Insulator} {Thin} {Films} by {Spin}-{Transfer} {Torque}, in:
  M.~Wu, A.~Hoffmann (Eds.), Solid {State} {Physics}, Vol.~64 of Recent
  {Advances} in {Magnetic} {Insulators} – {From} {Spintronics} to {Microwave}
  {Applications}, Academic Press, 2013, pp. 29--51.

\bibitem{wu_book_2013}
M.~Wu, A.~Hoffmann, Recent {Advances} in {Magnetic} {Insulators} - {From}
  {Spintronics} to {Microwave} {Applications}, Academic Press, 2013.

\bibitem{Yan2011}
P.~Yan, X.~S. Wang, X.~R. Wang, {All-magnonic spin-transfer torque and domain
  wall propagation}, Phys. Rev. Lett. 107 (2011) 177207.

\bibitem{Lan2015}
J.~Lan, W.~Yu, R.~Wu, J.~Xiao, {Spin-wave diode}, Phys. Rev. X 5 (2015) 041049.

\bibitem{Gar2015}
F.~Garcia-sanchez, P.~Borys, R.~Soucaille, J.-p. Adam, R.~L. Stamps, J.-v. Kim,
  {Narrow magnonic waveguides based on domain walls}, Phys. Rev. Lett. 114
  (2015) 247206.

\bibitem{Wagner2016}
K.~Wagner, A.~K{\'{a}}kay, K.~Schultheiss, A.~Henschke, T.~Sebastian,
  H.~Schultheiss, {Magnetic domain walls as reconfigurable spin-wave
  nanochannels}, Nat. Nanotechnol. 11 (2016) 432--436.

\bibitem{Wintz2019}
V.~Sluka, T.~Schneider, R.~A. Gallardo, A.~K{\'{a}}kay, M.~Weigand, T.~Warnatz,
  R.~Mattheis, A.~Rold{\'{a}}n-molina, P.~Landeros, V.~Tiberkevich, A.~Slavin,
  G.~Sch{\"{u}}tz, A.~Erbe, A.~Deac, J.~Lindner, J.~Raabe, J.~Fassbender,
  S.~Wintz, {Emission and propagation of 1D and 2D spin waves with nanoscale
  wavelengths in anisotropic spin textures}, Nat. Nanotechnol. 14 (2019)
  328--333.

\bibitem{SF_Lee2020}
L.-J. Chang, J.~Chen, D.~Qu, L.-Z. Tsai, Y.-F. Liu, M.-Y. Kao, J.-Z. Liang,
  T.-S. Wu, T.-M. Chuang, H.~Yu, S.-F. Lee, {Spin wave injection and
  propagation in a magnetic nanochannel from a vortex core}, Nano Lett. 20
  (2020) 3140--3146.

\bibitem{Onose2010}
Y.~Onose, T.~Ideue, H.~Katsura, Y.~Shiomi, N.~Nagaosa, Y.~Tokura, {Observation
  of the magnon Hall effect}, Science 329 (2010) 297--299.

\bibitem{Cheng2016}
R.~Cheng, M.~W. Daniels, J.-g. Zhu, D.~Xiao, {Antiferromagnetic spin wave
  field-effect transistor}, Sci. Rep. 6 (2016) 24223.

\bibitem{Kam2017}
A.~Kamra, U.~Agrawal, W.~Belzig, {Noninteger-spin magnonic excitations in
  untextured magnets}, Phys. Rev. B 96 (2017) 020411.

\bibitem{shen_magnon_2020}
K.~Shen, Magnon {spin} {relaxation} and {spin} {Hall} {effect} {due} to the
  {dipolar} {interaction} in {antiferromagnetic} {insulators}, Phys. Rev. Lett.
  124 (2020) 077201.

\bibitem{Kawa2019}
M.~Kawano, C.~Hotta, {Thermal Hall effect and topological edge states in a
  square-lattice antiferromagnet}, Phys. Rev. B 99 (2019) 054422.

\bibitem{Woo2018}
S.~Woo, K.~M. Song, X.~Zhang, Y.~Zhou, M.~Ezawa, X.~Liu, S.~Finizio, J.~Raabe,
  N.~J. Lee, S.-i. Kim, S.-y. Park, Y.~Kim, J.-y. Kim, D.~Lee, O.~Lee, J.~W.
  Choi, B.-c. Min, H.~C. Koo, {Current-driven dynamics and inhibition of the
  skyrmion Hall effect of ferrimagnetic skyrmions in GdFeCo films}, Nat.
  Commun. 9 (2018) 959.

\bibitem{Velez2019}
S.~V{\'{e}}lez, J.~Schaab, M.~S. W{\"{o}}rnle, M.~M{\"{u}}ller,
  E.~Gradauskaite, P.~Welter, C.~Gutgsell, C.~Nistor, C.~L. Degen, M.~Trassin,
  M.~Fiebig, P.~Gambardella, {High-speed domain wall racetracks in a magnetic
  insulator}, Nat. Commun. 10 (2019) 4750.

\bibitem{Ding2019}
S.~Ding, A.~Ross, R.~Lebrun, S.~Becker, K.~Lee, I.~Boventer, S.~Das,
  Y.~Kurokawa, S.~Gupta, J.~Yang, G.~Jakob, M.~Kl{\"{a}}ui, {Interfacial
  Dzyaloshinskii-Moriya interaction and chiral magnetic textures in a
  ferrimagnetic insulator}, Phys. Rev. B 100 (2019) 100406.

\bibitem{Lukas2019}
L.~Liensberger, A.~Kamra, H.~Maier-Flaig, S.~Gepr\"ags, A.~Erb, S.~T.~B.
  Goennenwein, R.~Gross, W.~Belzig, H.~Huebl, M.~Weiler, Exchange-enhanced
  ultrastrong magnon-magnon coupling in a compensated ferrimagnet, Phys. Rev.
  Lett. 123 (2019) 117204.

\bibitem{tserkovnyak_enhanced_2002}
Y.~Tserkovnyak, A.~Brataas, G.~E.~W. Bauer, Enhanced {Gilbert} {damping} in
  {thin} {ferromagnetic} {films}, Phys. Rev. Lett. 88 (2002) 117601.

\bibitem{saitoh_conversion_2006}
E.~Saitoh, M.~Ueda, H.~Miyajima, G.~Tatara, Conversion of spin current into
  charge current at room temperature: {Inverse} spin-{Hall} effect, Appl. Phys.
  Lett. 88 (2006) 182509.

\bibitem{uchida_observation_2008}
K.~Uchida, S.~Takahashi, K.~Harii, J.~Ieda, W.~Koshibae, K.~Ando, S.~Maekawa,
  E.~Saitoh, Observation of the spin {Seebeck} effect, Nature 455 (2008)
  778--781.

\bibitem{uchida_spin_2010}
K.~Uchida, J.~Xiao, H.~Adachi, J.~Ohe, S.~Takahashi, J.~Ieda, T.~Ota,
  Y.~Kajiwara, H.~Umezawa, H.~Kawai, G.~E.~W. Bauer, S.~Maekawa, E.~Saitoh,
  Spin {Seebeck} insulator, Nat. Mater. 9 (2010) 894--897.

\bibitem{jaworski_observation_2010}
C.~M. Jaworski, J.~Yang, S.~Mack, D.~D. Awschalom, J.~P. Heremans, R.~C. Myers,
  Observation of the spin-{Seebeck} effect in a ferromagnetic semiconductor,
  Nat. Mater. 9 (2010) 898--903.

\bibitem{xiao_theory_2010}
J.~Xiao, G.~E.~W. Bauer, K.-c. Uchida, E.~Saitoh, S.~Maekawa, Theory of
  magnon-driven spin {Seebeck} effect, Phys. Rev. B 81 (2010) 214418.

\bibitem{adachi_linear-response_2011}
H.~Adachi, J.-i. Ohe, S.~Takahashi, S.~Maekawa, Linear-response theory of spin
  {Seebeck} effect in ferromagnetic insulators, Phys. Rev. B 83 (2011) 094410.

\bibitem{cheng_spin_2014}
R.~Cheng, J.~Xiao, Q.~Niu, A.~Brataas, Spin {pumping} and {spin}-{transfer}
  {torques} in {antiferromagnets}, Phys. Rev. Lett. 113 (2014) 057601.

\bibitem{li_spin_2020}
J.~Li, C.~B. Wilson, R.~Cheng, M.~Lohmann, M.~Kavand, W.~Yuan, M.~Aldosary,
  N.~Agladze, P.~Wei, M.~S. Sherwin, J.~Shi, {Spin current from
  sub-terahertz-generated antiferromagnetic magnons}, Nature 578 (2020) 70--74.

\bibitem{ohnuma_spin_2013}
Y.~Ohnuma, H.~Adachi, E.~Saitoh, S.~Maekawa, Spin {Seebeck} effect in
  antiferromagnets and compensated ferrimagnets, Phys. Rev. B 87 (2013) 014423.

\bibitem{Lan2017}
J.~Lan, W.~Yu, J.~Xiao, {Antiferromagnetic domain wall as spin wave polarizer
  and retarder}, Nat. Commun. 8 (2017) 178.

\bibitem{pierce_electromigration_1997}
D.~Pierce, P.~Brusius, Electromigration: {A} review, Microelectronics
  Reliability 37~(7) (1997) 1053--1072.

\bibitem{Thie1973}
A.~A. Thiele, {Steady-state motion of magnetic domains}, Phys. Rev. Lett. 30
  (1973) 230--233.

\bibitem{Jam2010}
M.~Jamali, H.~Yang, K.~J. Lee, {Spin wave assisted current induced magnetic
  domain wall motion}, Appl. Phys. Lett. 96 (2010) 98--101.

\bibitem{Thi2002}
J.~{A.Thiaville, J.M.Garcıa}, {Domain wall dynamics in nanowires}, J. Magn.
  Magn. Mater. 245 (2002) 1061--1063.

\bibitem{Han2009}
D.~S. Han, S.~K. Kim, J.~Y. Lee, S.~J. Hermsdoerfer, H.~Schultheiss, B.~Leven,
  B.~Hillebrands, {Magnetic domain-wall motion by propagating spin waves},
  Appl. Phys. Lett. 94 (2009) 22--25.

\bibitem{Wie2010}
R.~Wieser, E.~Y. Vedmedenko, R.~Wiesendanger, {Domain wall motion damped by the
  emission of spin waves}, Phys. Rev. B 81 (2010) 024405.

\bibitem{Seo2011}
S.~M. Seo, H.~W. Lee, H.~Kohno, K.~J. Lee, {Magnetic vortex wall motion driven
  by spin waves}, Appl. Phys. Lett. 98 (2011) 012514.

\bibitem{Wang2012}
X.~S. Wang, P.~Yan, Y.~H. Shen, G.~E.~W. Bauer, X.~R. Wang, {Domain wall
  propagation through spin wave emission}, Phys. Rev. Lett. 109 (2012) 167209.

\bibitem{GWang2012}
X.-g. Wang, G.-h. Guo, Y.-z. Nie, G.-f. Zhang, Z.-x. Li, {Domain wall motion
  induced by the magnonic spin current}, Phys. Rev. B 86 (2012) 054445.

\bibitem{Kim2012}
J.~S. Kim, M.~St{\"{a}}rk, M.~Kl{\"{a}}ui, J.~Yoon, C.~Y. You, L.~Lopez-Diaz,
  E.~Martinez, {Interaction between propagating spin waves and domain walls on
  a ferromagnetic nanowire}, Phys. Rev. B 85 (2012) 174428.

\bibitem{Wang2013}
X.~G. Wang, G.~H. Guo, G.~F. Zhang, Y.~Z. Nie, Q.~L. Xia, {Spin-wave resonance
  reflection and spin-wave induced domain wall displacement}, J. Appl. Phys.
  113 (2013) 213904.

\bibitem{Zhang2014}
S.~Zhang, C.~Mu, Q.~Zhu, Q.~Zheng, X.~Liu, J.~Wang, Q.~Liu, {Propagating and
  reflecting of spin wave in permalloy nanostrip with 360 domain wall}, J.
  Appl. Phys. 115 (2014) 013908.

\bibitem{Hata2014}
H.~Hata, T.~Taniguchi, H.~W. Lee, T.~Moriyama, T.~Ono, {Spin-wave-induced
  domain wall motion in perpendicularly magnetized system}, Appl. Phys. Express
  7 (2014) 033001.

\bibitem{Wang2015}
W.~Wang, M.~Albert, M.~Beg, M.-a. Bisotti, D.~Chernyshenko,
  D.~Cort{\'{e}}s-ortu{\~{n}}o, I.~Hawke, H.~Fangohr, {Magnon-driven
  domain-wall motion with the Dzyaloshinskii-Moriya interaction}, Phys. Rev.
  Lett. 114 (2015) 087203.

\bibitem{Yang2015}
J.~Yang, M.~W. Yoo, S.~K. Kim, {Spin-wave-driven high-speed domain-wall motions
  in soft magnetic nanotubes}, J. Appl. Phys. 118 (2015) 163902.

\bibitem{Nowak2011}
D.~Hinzke, U.~Nowak, {Domain wall motion by the magnonic spin seebeck effect},
  Phys. Rev. Lett. 107 (2011) 027205.

\bibitem{Jiang2013}
W.~Jiang, P.~Upadhyaya, Y.~Fan, J.~Zhao, M.~Wang, L.-T. Chang, M.~Lang, K.~L.
  Wong, M.~Lewis, Y.-t. Lin, J.~Tang, S.~Cherepov, X.~Zhou, Y.~Tserkovnyak,
  R.~N. Schwartz, K.~L. Wang, {Direct imaging of thermally driven domain wall
  motion in magnetic insulators}, Phys. Rev. Lett. 110 (2013) 177202.

\bibitem{Sch2014}
F.~Schlickeiser, U.~Ritzmann, D.~Hinzke, U.~Nowak, {Role of entropy in domain
  wall motion in thermal gradients}, Phys. Rev. Lett. 113 (2014) 097201.

\bibitem{Yan2015}
P.~Yan, Y.~Cao, J.~Sinova, {Thermodynamic magnon recoil for domain wall
  motion}, Phys. Rev. B 92 (2015) 097201.

\bibitem{Sel2016}
S.~Selzer, U.~Atxitia, U.~Ritzmann, D.~Hinzke, U.~Nowak, {Inertia-free
  thermally driven domain-wall motion in antiferromagnets}, Phys. Rev. Lett.
  117 (2016) 107201.

\bibitem{More2017}
S.~Moretti, V.~Raposo, E.~Martinez, L.~Lopez-Diaz, {Domain wall motion by
  localized temperature gradients}, Phys. Rev. B 95 (2017) 064419.

\bibitem{rodrigues_effective_2018}
D.~R. Rodrigues, A.~Abanov, J.~Sinova, K.~Everschor-Sitte, {Effective
  description of domain wall strings}, Phys. Rev. B 97 (2018) 134414.

\bibitem{Yan2013}
P.~Yan, A.~Kamra, Y.~Cao, G.~E. Bauer, {Angular and linear momentum of excited
  ferromagnets}, Phys. Rev. B 88 (2013) 144413.

\bibitem{Sch1974}
N.~L. Schryer, L.~R. Walker, {The motion of 180 domain walls in uniform dc
  magnetic fields}, J. Appl. Phys. 45 (1974) 5406--5421.

\bibitem{Kim2018}
S.~K. Kim, O.~Tchernyshyov, V.~Galitski, Y.~Tserkovnyak, {Magnon-induced
  non-Markovian friction of a domain wall in a ferromagnet}, Phys. Rev. B 97
  (2018) 174433.

\bibitem{Han2019}
J.~Han, P.~Zhang, J.~T. Hou, S.~A. Siddiqui, L.~Liu, {Mutual control of
  coherent spin waves and magnetic domain walls in a magnonic device.}, Science
  366 (2019) 1121--1125.

\bibitem{yu_polarization-selective_2018}
W.~Yu, J.~Lan, J.~Xiao, {Polarization-selective spin wave driven domain-wall
  motion in antiferromagnets}, Phys. Rev. B 98 (2018) 144422.

\bibitem{Zar2018}
R.~Zarzuela, S.~K. Kim, Y.~Tserkovnyak, {Magnetoelectric antiferromagnets as
  platforms for the manipulation of solitons}, Phys. Rev. B 97 (2018) 014418.

\bibitem{Rei2018}
D.~Reitz, A.~Ghosh, O.~Tchernyshyov, {Viscous dynamics of vortices in a
  ferromagnetic film}, Phys. Rev. B 97 (2018) 054424.

\bibitem{Kim2018a}
J.~Kim, J.~Yang, Y.~J. Cho, B.~Kim, S.~K. Kim, {Coupled breathing modes in
  one-dimensional Skyrmion lattices}, J. Appl. Phys. 123 (2018) 053903.

\bibitem{Chia2012}
H.-J. Chia, F.~Guo, L.~M. Belova, R.~D. McMichael, {Nanoscale spin wave
  localization using ferromagnetic resonance force microscopy}, Phys. Rev.
  Lett. 108 (2012) 087206.

\bibitem{Volo2018}
A.~Volodin, C.~Van~Haesendonck, E.~V. Skorokhodov, R.~V. Gorev, V.~L. Mironov,
  {Ferromagnetic resonance force microscopy of individual domain wall}, Appl.
  Phys. Lett. 113 (2018) 122407.

\bibitem{Hicken2020}
M.~D\k{a}browski, T.~Nakano, D.~M. Burn, A.~Frisk, D.~G. Newman, C.~Klewe,
  Q.~Li, M.~Yang, P.~Shafer, E.~Arenholz, T.~Hesjedal, G.~van~der Laan, Z.~Q.
  Qiu, R.~J. Hicken, Coherent transfer of spin angular momentum by evanescent
  spin waves within antiferromagnetic nio, Phys. Rev. Lett. 124 (2020) 217201.

\bibitem{Hart2003}
M.~Koblischka, U.~Hartmann, {Recent advances in magnetic force microscopy},
  Ultramicroscopy 97 (2003) 103--112.

\bibitem{Kimura2012}
S.~R. Bakaul, W.~Hu, T.~Wu, T.~Kimura, {Intrinsic domain-wall resistivity in
  half-metallic manganite thin films}, Phys. Rev. B 86 (2012) 184404.

\bibitem{JX_Zhang2018}
J.~Wang, S.~Wu, J.~Ma, L.~Xie, C.~Wang, I.~A. Malik, Y.~Zhang, K.~Xia, C.-W.
  Nan, J.~Zhang, {Nanoscale control of stripe-ordered magnetic domain walls by
  vertical spin transfer torque in La0.67Sr0.33MnO3 film}, Appl. Phys. Lett.
  112 (2018) 072408.

\bibitem{Claudia_Parkin_2020}
T.~Ma, A.~K. Sharma, R.~Saha, A.~K. Srivastava, P.~Werner, P.~Vir, V.~Kumar,
  C.~Felser, S.~S.~P. Parkin, {Tunable magnetic antiskyrmion size and helical
  period from nanometers to micrometers in a D-2d Heusler compound}, Adv.
  Mater. (2020) 2002043.

\bibitem{H_Du2018}
H.~Du, X.~Zhao, F.~N. Rybakov, A.~B. Borisov, S.~Wang, J.~Tang, C.~Jin,
  C.~Wang, W.~Wei, N.~S. Kiselev, Y.~Zhang, R.~Che, S.~Bl\"ugel, M.~Tian,
  {Interaction of individual Skyrmions in a nanostructured cubic chiral
  magnet}, Phys. Rev. Lett. 120 (2018) 197203.

\bibitem{W_Jiang2019}
W.~Jiang, S.~Zhang, X.~Wang, C.~Phatak, Q.~Wang, W.~Zhang, M.~B. Jungfleisch,
  J.~E. Pearson, Y.~Liu, J.~Zang, X.~Cheng, A.~Petford-Long, A.~Hoffmann,
  S.~G.~E. te~Velthuis, Quantifying chiral exchange interaction for n\'eel-type
  skyrmions via lorentz transmission electron microscopy, Phys. Rev. B 99
  (2019) 104402.

\bibitem{Back2020}
S.~P{\"o}llath, T.~Lin, N.~Lei, W.~Zhao, J.~Zweck, C.~Back, Spin structure
  relation to phase contrast imaging of isolated magnetic bloch and n{\'e}el
  skyrmions, Ultramicroscopy 212 (2020) 112973.

\bibitem{Y_Wu2019}
C.~Zhou, G.~Chen, J.~Xu, J.~Liang, K.~Liu, A.~K. Schmid, Y.~Wu, {Magnetic
  domain wall contrast under zero domain contrast conditions in spin polarized
  low energy electron microscopy}, Ultramicroscopy 200 (2019) 132--138.

\bibitem{Fle1966}
P.~Fleury, S.~Porto, L.~Cheesman, H.~Guggenheim, {Light scattering by spin
  waves in FeF$_{2}$}, Phys. Rev. Lett. 17 (1966) 84--87.

\bibitem{San1973}
J.~Sandercock, W.~W., {Light-scattering from thermal acoustic magnons in
  yttrium Iron-garnet}, Solid State Commun. 13 (1973) 1729--1732.

\bibitem{Seb2015}
T.~Sebastian, K.~Schultheiss, B.~Obry, B.~Hillebrands, H.~Schultheiss,
  {Micro-focused Brillouin light scattering: imaging spin waves at the
  nanoscale}, Front. Phys. 3 (2015) 1--23.

\bibitem{Ban2017}
C.~Banerjee, P.~Gruszecki, J.~W. Klos, O.~Hellwig, M.~Krawczyk, A.~Barman,
  {Magnonic band structure in a Co/Pd stripe domain system investigated by
  Brillouin light scattering and micromagnetic simulations}, Phys. Rev. B 96
  (2017) 024421.

\bibitem{Hama2018}
S.~J. H{\"{a}}m{\"{a}}l{\"{a}}inen, M.~Madami, H.~Qin, G.~Gubbiotti, S.~{van
  Dijken}, {Control of spin-wave transmission by a programmable domain wall},
  Nat. Commun. 9 (2018) 4853.

\bibitem{San2008}
C.~W. Sandweg, S.~J. Hermsdoerfer, H.~Schultheiss, S.~Sch{\"{a}}fer, B.~Leven,
  B.~Hillebrands, {Modification of the thermal spin-wave spectrum in a Ni81Fe19
  stripe by a domain wall}, J. Phys. D Appl. Phys. 41 (2008) 164008.

\bibitem{Sch2008}
H.~Schultheiss, S.~Sch{\"{a}}fer, P.~Candeloro, B.~Leven, B.~Hillebrands, A.~N.
  Slavin, {Observation of coherence and partial decoherence of quantized spin
  waves in nanoscaled magnetic ring structures}, Phys. Rev. Lett. 100 (2008)
  047204.

\bibitem{Per2005}
K.~Perzlmaier, M.~Buess, C.~H. Back, V.~E. Demidov, B.~Hillebrands, S.~O.
  Demokritov, {Spin-wave eigenmodes of permalloy squares with a closure domain
  structure}, Phys. Rev. Lett. 94 (2005) 057202.

\bibitem{Vogt2012}
K.~Vogt, H.~Schultheiss, S.~Jain, J.~E. Pearson, A.~Hoffmann, S.~D. Bader,
  B.~Hillebrands, K.~Vogt, H.~Schultheiss, S.~Jain, J.~E. Pearson, A.~Hoffmann,
  S.~D. Bader, {Spin waves turning a corner}, Appl. Phys. Lett. 101 (2012)
  042410.

\bibitem{Vogt2014}
K.~Vogt, F.~Y. Fradin, J.~E. Pearson, T.~Sebastian, S.~D. Bader,
  B.~Hillebrands, A.~Hoffmann, H.~Schultheiss, {Realization of a spin-wave
  multiplexer}, Nat. Commun. 5 (2014) 3727.

\bibitem{Pirro2015}
P.~Pirro, T.~Koyama, T.~Br{\"{a}}cher, T.~Sebastian, B.~Leven, B.~Hillebrands,
  {Experimental observation of the interaction of propagating spin waves with
  N{\'{e}}el domain walls in a Landau domain structure}, Appl. Phys. Lett. 106
  (2015) 232405.

\bibitem{Alb2017}
E.~Albisetti, D.~Petti, M.~Madami, S.~Tacchi, P.~Vavassori, E.~Riedo,
  R.~Bertacco, {Nanopatterning spin-textures: A route to reconfigurable
  magnonics}, AIP Adv. 7 (2017) 055601.

\bibitem{Sch2017}
{{K. Schultheiss, K. Wagner, A. Kakay, M. Althammer, R. Gross}, M. Weiler},
  {Steering magnons by noncollinear spin textures}, Spin Wave Confinement:
  Propagating Waves, Second Edition (2017) 261--294.

\bibitem{Vla2008}
V.~Vlaminck, M.~Bailleul, {Current-induced spin-wave Doppler shift}, Science
  322 (2008) 410--413.

\bibitem{Neu2010}
S.~Neusser, G.~Duerr, H.~G. Bauer, S.~Tacchi, M.~Madami, G.~Woltersdorf,
  G.~Gubbiotti, {Anisotropic propagation and damping of spin waves in a
  nanopatterned antidot lattice}, Phys. Rev. Lett. 105 (2010) 067208.

\bibitem{CPW1969}
C.~P. Wen, {Coplanar waveguide: a surface strip transmission line suitable for
  nonreciprocal gyromagnetic device applications}, IEEE Trans. Microwave Theory
  Tech. 17 (1969) 1087--1090.

\bibitem{Vla2010}
V.~Vlaminck, M.~Bailleul, {Spin-wave transduction at the submicrometer scale:
  Experiment and modeling}, Phys. Rev. B 81 (2010) 014425.

\bibitem{Liu2014}
T.~Liu, H.~Chang, V.~Vlaminck, Y.~Sun, M.~Kabatek, A.~Hoffmann, L.~Deng, M.~Wu,
  {Ferromagnetic resonance of sputtered yttrium iron garnet nanometer films},
  J. Appl. Phys. 115 (2014) 17A501.

\bibitem{Che2016}
P.~Che, Y.~Zhang, C.~Liu, S.~Tu, Z.~Liao, D.~Yu, F.~A. Vetro, J.~P. Ansermet,
  W.~Zhao, L.~Bi, H.~Yu, {Short-wavelength spin waves in yttrium iron garnet
  micro-channels on silicon}, IEEE Magn. Lett. 7 (2016) 3508404.

\bibitem{Kos2005}
M.~P. Kostylev, A.~A. Serga, T.~Schneider, B.~Leven, B.~Hillebrands, {Spin-wave
  logical gates}, Appl. Phys. Lett. 87 (2005) 153501.

\bibitem{Lee_Kim2008}
K.-S. Lee, S.-K. Kim, Conceptual design of spin wave logic gates based on a
  mach–zehnder-type spin wave interferometer for universal logic functions,
  J. Appl. Phys. 104 (2008) 053909.

\bibitem{Rousseau2015}
O.~Rousseau, B.~Rana, R.~Anami, M.~Yamada, K.~Miura, S.~Ogawa, Y.~Otani,
  {Realization of a micrometre-scale spin-wave interferometer}, Sci. Rep. 5
  (2015) 9873.

\bibitem{Her2004}
R.~Hertel, W.~Wulfhekel, J.~Kirschner, {Domain-wall induced phase shifts in
  spin waves}, Phys. Rev. Lett. 93 (2004) 257202.

\bibitem{Sam2013}
J.~Sampaio, V.~Cros, S.~Rohart, A.~Thiaville, A.~Fert, {Nucleation, stability
  and current-induced motion of isolated magnetic skyrmions in nanostructures},
  Nat. Nanotechnol. 8 (2013) 839--844.

\bibitem{Lee2016}
J.~M. Lee, C.~Jang, B.~C. Min, S.~W. Lee, K.~J. Lee, J.~Chang, {All-electrical
  measurement of interfacial Dzyaloshinskii-Moriya interaction using collective
  spin-wave dynamics}, Nano Lett. 16 (2016) 62--67.

\bibitem{DE1961}
R.~W. Damon, J.~R. Eshbach, {Magnetostatic modes of a ferromagnet slab}, J.
  Phys. Chem. Solids 19 (1961) 308--320.

\bibitem{Nem2015}
H.~T. Nembach, J.~M. Shaw, M.~Weiler, E.~Ju{\'{e}}, T.~J. Silva, {Linear
  relation between Heisenberg exchange and interfacial Dzyaloshinskii-Moriya
  interaction in metal films}, Nat. Phys. 11 (2015) 825--829.

\bibitem{Li2017}
X.~Ma, G.~Yu, S.~A. Razavi, S.~S. Sasaki, X.~Li, K.~Hao, S.~H. Tolbert, K.~L.
  Wang, X.~Li, {Dzyaloshinskii-Moriya interaction across an
  antiferromagnet-ferromagnet interface}, Phys. Rev. Lett. 119 (2017) 027202.

\bibitem{Saitoh2019}
K.~Yamamoto, G.~C. Thiang, P.~Pirro, K.-w. Kim, K.~Everschor-sitte, E.~Saitoh,
  Topological characterization of classical waves : the topological origin of
  magnetostatic surface spin waves, Phys. Rev. Lett. 122 (2019) 217201.

\bibitem{Liu2019}
C.~Liu, S.~Wu, J.~Zhang, J.~Chen, J.~Ding, J.~Ma, Y.~Zhang, Y.~Sun, S.~Tu,
  H.~Wang, P.~Liu, C.~Li, Y.~Jiang, P.~Gao, D.~Yu, J.~Xiao, R.~Duine, M.~Wu,
  C.~W. Nan, J.~Zhang, H.~Yu, {Current-controlled propagation of spin waves in
  antiparallel, coupled domains}, Nat. Nanotechnol. 14 (2019) 691--697.

\bibitem{Ciu2016}
F.~Ciubotaru, T.~Devolder, M.~Manfrini, C.~Adelmann, I.~P. Radu, {All
  electrical propagating spin wave spectroscopy with broadband wavevector},
  Appl. Phys. Lett. 109 (2016) 012403.

\bibitem{Tobias2017}
T.~St{\"{u}}ckler, C.~Liu, T.~Liu, H.~Yu, F.~Heimbach, J.~Chen, J.~Hu, S.~Tu,
  Y.~Zhang, S.~Granville, M.~Wu, Z.-M. Liao, D.~Yu, W.~Zhao, {Ultrabroadband
  spin-wave propagation in Co2(Mn0.6Fe0.4)Si thin films}, Phys. Rev. B 96
  (2017) 144430.

\bibitem{Yu2012}
H.~Yu, R.~Huber, T.~Schwarze, F.~Brandl, T.~Rapp, P.~Berberich, G.~Duerr, {High
  propagating velocity of spin waves and temperature dependent damping in a
  CoFeB thin film}, Appl. Phys. Lett. 100 (2012) 262412.

\bibitem{H_Wang2020}
H.~Wang, J.~Chen, T.~Liu, J.~Zhang, K.~Baumgaertl, C.~Guo, Y.~Li, C.~Liu,
  P.~Che, S.~Tu, S.~Liu, P.~Gao, X.~Han, D.~Yu, M.~Wu, D.~Grundler, H.~Yu,
  {Chiral spin-wave velocities induced by all-garnet interfacial
  Dzyaloshinskii-Moriya interaction in ultrathin yttrium iron garnet films},
  Phys. Rev. Lett. 124 (2020) 027203.

\bibitem{Luc2019}
J.~Lucassen, C.~F. Schippers, L.~Rutten, R.~A. Duine, H.~J. Swagten,
  B.~Koopmans, R.~Lavrijsen, {Optimizing propagating spin wave spectroscopy},
  Appl. Phys. Lett. 115 (2019) 012403.

\bibitem{Luc2020}
J.~Lucassen, C.~F. Schippers, M.~A. Verheijen, P.~Fritsch, E.~J. Geluk,
  B.~Barcones, R.~A. Duine, S.~Wurmehl, H.~J.~M. Swagten, B.~Koopmans,
  R.~Lavrijsen, Extraction of dzyaloshinskii-moriya interaction from
  propagating spin waves, Phys. Rev. B 101 (2020) 064432.

\bibitem{Che2020}
P.~Che, K.~Baumgaertl, A.~K{\'{u}}kol'ov{\'{a}}, C.~Dubs, D.~Grundler,
  {Efficient wavelength conversion of exchange magnons below 100 nm by magnetic
  coplanar waveguides}, Nat. Commun. 11 (2020) 1--9.

\bibitem{Nem2013}
H.~T. Nembach, J.~M. Shaw, C.~T. Boone, T.~J. Silva, {Mode- and size-dependent
  Landau-Lifshitz damping in magnetic nanostructures: Evidence for nonlocal
  damping}, Phys. Rev. Lett. 110 (2013) 117201.

\bibitem{Back1999}
C.~Back, R.~Allenspach, W.~Weber, S.~S.~P. Parkin, D.~Weller, E.~L. Garwin,
  H.~C. Siegmann, {Minimum field strength in precessional magnetization
  reversal}, Sience 285 (1999) 864--867.

\bibitem{Hubert_1998}
{A. Hubert, R. Sch\"afer}, {Magnetic domains: the analysis of magnetic
  microstructures}, Springer, 1998.

\bibitem{McCord2015}
J.~McCord, {Progress in magnetic domain observation by advanced magneto-optical
  microscopy}, J. Phys. D Appl. Phys. 48 (2015) 333001.

\bibitem{Kampen2002}
M.~van Kampen, C.~Jozsa, J.~T. Kohlhepp, P.~LeClair, L.~Lagae, W.~J. de~Jonge,
  B.~Koopmans, {All-optical probe of coherent spin waves}, Phys. Rev. Lett. 88
  (2002) 227201.

\bibitem{Hie1997}
W.~K. Hiebert, A.~Stankiewicz, M.~R. Freeman, {Direct observation of magnetic
  relaxation in a small permalloy disk by time-resolved scanning kerr
  microscopy}, Phys. Rev. Lett. 79 (1997) 1134--1137.

\bibitem{Ger2002}
T.~Gerrits, H.~A. {Van den Berg}, J.~Hohlfeld, L.~B{\"{a}}r, T.~Rasing,
  {Ultrafast precessional magnetization reversal by picosecond magnetic field
  pulse shaping}, Nature 418 (2002) 509--512.

\bibitem{Buess2005}
M.~Buess, T.~P. Knowles, R.~H{\"{o}}llinger, T.~Haug, U.~Krey, D.~Weiss,
  D.~Pescia, M.~R. Scheinfein, C.~H. Back, {Excitations with negative
  dispersion in a spin vortex}, Phys. Rev. B 71 (2005) 104415.

\bibitem{Neu2006}
I.~Neudecker, M.~Kl{\"{a}}ui, K.~Perzlmaier, D.~Backes, L.~J. Heyderman, C.~A.
  Vaz, J.~A. Bland, U.~R{\"{u}}diger, C.~H. Back, {Spatially resolved dynamic
  eigenmode spectrum of Co rings}, Phys. Rev. Lett. 96 (2006) 057207.

\bibitem{Au2011}
Y.~Au, T.~Davison, E.~Ahmad, P.~S. Keatley, R.~J. Hicken, V.~V. Kruglyak,
  {Excitation of propagating spin waves with global uniform microwave fields},
  Appl. Phys. Lett. 98 (2011) 122506.

\bibitem{Dav2015}
C.~S. Davies, A.~Francis, A.~V. Sadovnikov, S.~V. Chertopalov, M.~T. Bryan,
  S.~V. Grishin, D.~A. Allwood, Y.~P. Sharaevskii, S.~A. Nikitov, V.~V.
  Kruglyak, {Towards graded-index magnonics: Steering spin waves in magnonic
  networks}, Phys. Rev. B 92 (2015) 020408(R).

\bibitem{Holl2018}
R.~B. Holl{\"{a}}nder, C.~M{\"{u}}ller, J.~Schmalz, M.~Gerken, J.~McCord,
  {Magnetic domain walls as broadband spin wave and elastic magnetisation wave
  emitters}, Sci. Rep. 8 (2018) 13871.

\bibitem{mozooni_direct_2015}
B.~Mozooni, J.~McCord, {Direct observation of closure domain wall mediated spin
  waves}, Appl. Phys. Lett. 107 (2015) 042402.

\bibitem{Buess2004}
M.~Buess, R.~H{\"{o}}llinger, T.~Haug, K.~Perzlmaier, U.~Krey, D.~Pescia, M.~R.
  Scheinfein, D.~Weiss, C.~H. Back, {Fourier transform imaging of spin vortex
  eigenmodes}, Phys. Rev. Lett. 93 (2004) 077207.

\bibitem{Stig2016}
J.~Stigloher, M.~Decker, H.~S. K\"orner, K.~Tanabe, T.~Moriyama, T.~Taniguchi,
  H.~Hata, M.~Madami, G.~Gubbiotti, K.~Kobayashi, T.~Ono, C.~H. Back, {Snell's
  law for spin waves}, Phys. Rev. Lett. 117 (2016) 037204.

\bibitem{Farle2013}
M.~Farle, T.~Silva, G.~Woltersdorf, {Spin dynamics in the time and frequency
  domain}, Springer Tracts in Modern Physics 246 (2013) 37.

\bibitem{Dre2018}
R.~Dreyer, N.~Liebing, E.~R.~J. Edwards, G.~Woltersdorf, {Local spin-wave
  dispersion and damping in thin yttrium-iron-garnet films}.

\bibitem{Lie2019}
L.~Liensberger, L.~Flacke, D.~Rogerson, M.~Althammer, R.~Gross, M.~Weiler,
  {Spin-wave propagation in metallic Co25Fe75 films determined by microfocused
  frequency-resolved magneto-optic Kerr effect}, IEEE Magn. Lett. 10 (2019)
  5503905.

\bibitem{Ham2010}
J.~Hamrle, J.~Pi{\v{s}}tora, B.~Hillebrands, B.~Lenk, M.~M{\"{u}}nzenberg,
  {Analytical expression of the magneto-optical Kerr effect and Brillouin light
  scattering intensity arising from dynamic magnetization}, J. Phys. D Appl.
  Phys. 43 (2010) 325004.

\bibitem{Stig2018}
J.~Stigloher, T.~Taniguchi, H.~S. K{\"{o}}rner, M.~Decker, T.~Moriyama, T.~Ono,
  C.~H. Back, Observation of a goos-h{\"{a}}nchen-like phase shift for
  magnetostatic spin waves, Phys. Rev. Lett. 121 (2018) 137201.

\bibitem{Flac2019}
L.~Flacke, L.~Liensberger, M.~Althammer, H.~Huebl, S.~Gepr{\"{a}}gs,
  K.~Schultheiss, A.~Buzdakov, T.~Hula, H.~Schultheiss, E.~R. Edwards, H.~T.
  Nembach, J.~M. Shaw, R.~Gross, M.~Weiler, {High spin-wave propagation length
  consistent with low damping in a metallic ferromagnet}, Appl. Phys. Lett. 115
  (2019) 122402.

\bibitem{Sch2009}
H.~Schultheiss, X.~Janssens, M.~{Van Kampen}, F.~Ciubotaru, S.~J. Hermsdoerfer,
  B.~Obry, A.~Laraoui, A.~A. Serga, L.~Lagae, A.~N. Slavin, B.~Leven,
  B.~Hillebrands, {Direct current control of three magnon scattering processes
  in spin-valve nanocontacts}, Phys. Rev. Lett. 103 (2009) 157202.

\bibitem{Sch2012}
H.~Schultheiss, K.~Vogt, B.~Hillebrands, {Direct observation of nonlinear
  four-magnon scattering in spin-wave microconduits}, Phys. Rev. B 86 (2012)
  054414.

\bibitem{Sch2019}
K.~Schultheiss, R.~Verba, F.~Wehrmann, K.~Wagner, L.~K{\"{o}}rber, T.~Hula,
  T.~Hache, A.~K{\'{a}}kay, A.~A. Awad, V.~Tiberkevich, A.~N. Slavin,
  J.~Fassbender, H.~Schultheiss, {Excitation of whispering gallery magnons in a
  magnetic vortex}, Phys. Rev. Lett. 122 (2019) 097202.

\bibitem{Hache2019}
T.~Hache, T.~Weinhold, K.~Schultheiss, J.~Stigloher, F.~Vilsmeier, C.~Back,
  S.~S. Arekapudi, O.~Hellwig, J.~Fassbender, H.~Schultheiss, {Combined
  frequency and time domain measurements on injection-locked,
  constriction-based spin Hall nano-oscillators}, Appl. Phys. Lett. 114 (2019)
  102403.

\bibitem{Grundler2016}
D.~Grundler, {Nanomagnonics around the corner}, Nat. Nanotechnol. 11 (2016)
  407.

\bibitem{Ade2016}
A.~Haldar, D.~Kumar, A.~O. Adeyeye, {A reconfigurable waveguide for
  energy-efficient transmission and local manipulation of information in a
  nanomagnetic device}, Nat. Nanotechnol. 11 (2016) 437--443.

\bibitem{Demi2011}
V.~E. Demidov, M.~P. Kostylev, K.~Rott, J.~Mnchenberger, G.~Reiss, S.~O.
  Demokritov, {Excitation of short-wavelength spin waves in magnonic
  waveguides}, Appl. Phys. Lett. 99 (2011) 082507.

\bibitem{Ono2018}
J.~Stigloher, T.~Taniguchi, M.~Madami, M.~Decker, H.~S. K{\"{o}}rner,
  T.~Moriyama, G.~Gubbiotti, T.~Ono, C.~H. Back, {Spin-wave wavelength
  down-conversion at thickness steps}, Appl. Phys. Express 11 (2018) 053002.

\bibitem{Demi2010}
V.~E. Demidov, S.~Urazhdin, S.~O. Demokritov, {Direct observation and mapping
  of spin waves emitted by spin-torque nano-oscillators}, Nat. Mater. 9 (2010)
  984--988.

\bibitem{Madami2011}
M.~Madami, S.~Bonetti, G.~Consolo, S.~Tacchi, G.~Carlotti, G.~Gubbiotti, F.~B.
  Mancoff, M.~A. Yar, J.~{\AA}kerman, {Direct observation of a propagating spin
  wave induced by spin-transfer torque}, Nat. Nanotechnol. 6 (2011) 635--638.

\bibitem{Ura2014}
S.~Urazhdin, V.~E. Demidov, H.~Ulrichs, T.~Kendziorczyk, T.~Kuhn, J.~Leuthold,
  G.~Wilde, S.~O. Demokritov, {Nanomagnonic devices based on the spin-transfer
  torque}, Nat. Nanotechnol. 9 (2014) 509--513.

\bibitem{Ake2018}
A.~Houshang, R.~Khymyn, H.~Fulara, A.~Gangwar, M.~Haidar, S.~R. Etesami,
  R.~Ferreira, P.~P. Freitas, M.~Dvornik, R.~K. Dumas, J.~{\AA}kerman, {Spin
  transfer torque driven higher-order propagating spin waves in nano-contact
  magnetic tunnel junctions}, Nat. Commun. 9 (2018) 4374.

\bibitem{Yu2013}
H.~Yu, G.~Duerr, R.~Huber, M.~Bahr, T.~Schwarze, F.~Brandl, D.~Grundler,
  {Omnidirectional spin-wave nanograting coupler}, Nat. Commun. 4 (2013) 2702.

\bibitem{Yu2017}
H.~Yu, D.~Grundler, {Magnonic grating coupler effect and microwave-to-magnon
  transducers for exchange-dominated spin waves}, Spin Wave Confinement:
  Propagating Waves, Second Edition (2017) 197--218.

\bibitem{Hama2017}
S.~J. H{\"{a}}m{\"{a}}l{\"{a}}inen, F.~Brandl, K.~J.~A. Franke, D.~Grundler,
  S.~V. Dijken, {Tunable Short-Wavelength Spin-Wave Emission and Confinement in
  Anisotropy-Modulated Multiferroic Heterostructures}, Phys. Rev. Appl. 8
  (2017) 014020.

\bibitem{MAXYMUS}
{BESSY II—MAXYMUS BEAMLINE, Helmholtz-Zentrum Berlin}.

\bibitem{Kli2018}
S.~Klingler, V.~Amin, S.~Gepr{\"{a}}gs, K.~Ganzhorn, H.~Maier-Flaig,
  M.~Althammer, H.~Huebl, R.~Gross, R.~D. McMichael, M.~D. Stiles, S.~T.
  Goennenwein, M.~Weiler, {Spin-torque excitation of perpendicular standing
  spin waves in coupled YIG/Co heterostructures}, Phys. Rev. Lett. 120 (2018)
  127201.

\bibitem{Qin2018}
H.~Qin, S.~J. H{\"{a}}m{\"{a}}l{\"{a}}inen, S.~{Van Dijken},
  {Exchange-torque-induced excitation of perpendicular standing spin waves in
  nanometer-thick YIG films}, Sci. Rep. 8 (2018) 5755.

\bibitem{Pile2020}
S.~Pile, T.~Feggeler, T.~Schaffers, R.~Meckenstock, M.~Buchner, D.~Spoddig,
  B.~Zingsem, V.~Ney, M.~Farle, H.~Wende, H.~Ohldag, A.~Ney, K.~Ollefs,
  Non-standing spin-waves in confined micrometer-sized ferromagnetic structures
  under uniform excitation, Appl. Phys. Lett. 116 (2020) 072401.

\bibitem{Vogel:2018gr}
M.~Vogel, R.~A{\ss}mann, P.~Pirro, A.~V. Chumak, B.~Hillebrands, G.~von
  Freymann, {Control of spin-wave propagation using magnetisation gradients},
  Sci. Rep. 8 (2018) 1--10.

\bibitem{Alb2019}
E.~Albisetti, S.~Tacchi, R.~Silvani, G.~Scaramuzzi, S.~Finizio, S.~Wintz,
  C.~Rinaldi, M.~Cantoni, J.~Raabe, G.~Carlotti, R.~Bertacco, E.~Riedo,
  D.~Petti, Optically inspired nanomagnonics with nonreciprocal spin waves in
  synthetic antiferromagnets, Adv. Mater. 32 (2020) 1906439.

\bibitem{Alb2016}
E.~Albisetti, D.~Petti, M.~Pancaldi, M.~Madami, S.~Tacchi, J.~Curtis, W.~P.
  King, A.~Papp, G.~Csaba, W.~Porod, P.~Vavassori, E.~Riedo, R.~Bertacco,
  {Nanopatterning reconfigurable magnetic landscapes via thermally assisted
  scanning probe lithography}, Nat. Nanotechnol. 11 (2016) 545--551.

\bibitem{Alb2018}
E.~Albisetti, A.~Cal{\`{o}}, M.~Spieser, A.~W. Knoll, E.~Riedo, D.~Petti,
  {Stabilization and control of topological magnetic solitons via magnetic
  nanopatterning of exchange bias systems}, Appl. Phys. Lett. 113 (2018)
  162401.

\bibitem{Gru1981}
P.~Gr{\"{u}}nberg, {Magnetostatic spin-wave modes of a heterogeneous
  ferromagnetic double layer}, J. Appl. Phys. 52 (1981) 6824--6829.

\bibitem{duine_synthetic_2018}
R.~A. Duine, K.-J. Lee, S.~S.~P. Parkin, M.~D. Stiles, {Synthetic
  antiferromagnetic spintronics}, Nat. Phys. 14 (2018) 217--219.

\bibitem{Shiota2020}
Y.~Shiota, T.~Taniguchi, M.~Ishibashi, T.~Moriyama, T.~Ono, {Tunable
  magnon-magnon coupling mediated by dynamic dipolar interaction in synthetic
  antiferromagnets}, Phys. Rev. Lett. 125 (2020) 017203.

\bibitem{Di2015}
K.~Di, S.~X. Feng, S.~N. Piramanayagam, V.~L. Zhang, H.~S. Lim, S.~C. Ng, M.~H.
  Kuok, {Enhancement of spin-wave nonreciprocity in magnonic crystals via
  synthetic antiferromagnetic coupling}, Sci. Rep. 5 (2015) 10153.

\bibitem{Kwo2016}
J.~H. Kwon, J.~Yoon, P.~Deorani, J.~M. Lee, J.~Sinha, K.~J. Lee, M.~Hayashi,
  H.~Yang, {Giant nonreciprocal emission of spin waves in Ta/Py bilayers}, Sci.
  Adv. 2 (2016) e1501892.

\bibitem{Chen2019}
J.~Chen, T.~Yu, C.~Liu, T.~Liu, M.~Madami, K.~Shen, J.~Zhang, S.~Tu, M.~S.
  Alam, K.~Xia, M.~Wu, G.~Gubbiotti, Y.~M. Blanter, G.~E.~W. Bauer, H.~Yu,
  {Excitation of unidirectional exchange spin waves by a nanoscale magnetic
  grating}, Phys. Rev. B 100 (2019) 104427.

\bibitem{Wes2017}
D.~Wesenberg, T.~Liu, D.~Balzar, M.~Wu, B.~L. Zink, {Long-distance spin
  transport in a disordered magnetic insulator}, Nat. Phys. 13 (2017) 987--993.

\bibitem{Gomez2020}
J.~M. Gomez-Perez, K.~Oyanagi, R.~Yahiro, R.~Ramos, L.~E. Hueso, E.~Saitoh,
  F.~Casanova, {Absence of evidence of spin transport through amorphous
  Y3Fe5O12}, Appl. Phys. Lett. 116 (2020) 032401.

\bibitem{Forster2019}
J.~F\"orster, S.~Wintz, J.~Bailey, S.~Finizio, E.~Josten, C.~Dubs, D.~A.
  Bozhko, H.~Stoll, G.~Dieterle, N.~Tr\"ager, J.~Raabe, A.~N. Slavin,
  M.~Weigand, J.~Gräfe, G.~Sch\"utz, Nanoscale x-ray imaging of spin dynamics
  in yttrium iron garnet, J. Appl. Phys. 126 (2019) 173909.

\bibitem{Forster2019b}
J.~F\"orster, J.~Gr\"afe, J.~Bailey, S.~Finizio, N.~Tr\"ager, F.~Gross,
  S.~Mayr, H.~Stoll, C.~Dubs, O.~Surzhenko, N.~Liebing, G.~Woltersdorf,
  J.~Raabe, M.~Weigand, G.~Sch\"utz, S.~Wintz, Direct observation of coherent
  magnons with suboptical wavelengths in a single-crystalline ferrimagnetic
  insulator, Phys. Rev. B 100 (2019) 214416.

\bibitem{Saitoh2010}
Y.~Kajiwara, K.~Harii, S.~Takahashi, J.~Ohe, K.~Uchida, M.~Mizuguchi,
  H.~Umezawa, H.~Kawai, K.~Ando, K.~Takanashi, S.~Maekawa, E.~Saitoh,
  {Transmission of electrical signals by spin-wave interconversion in a
  magnetic insulator}, Nature 464 (2010) 262--266.

\bibitem{Chu2012}
A.~V. Chumak, A.~A. Serga, M.~B. Jungfleisch, R.~Neb, D.~A. Bozhko, V.~S.
  Tiberkevich, B.~Hillebrands, {Direct detection of magnon spin transport by
  the inverse spin Hall effect}, Appl. Phys. Lett. 100 (2012) 082405.

\bibitem{Madjid2014}
O.~{d'Allivy Kelly}, A.~Anane, R.~Bernard, J.~{Ben Youssef}, C.~Hahn, A.~H.
  Molpeceres, C.~Carr{\'{e}}t{\'{e}}ro, E.~Jacquet, C.~Deranlot, P.~Bortolotti,
  R.~Lebourgeois, J.~C. Mage, G.~{De Loubens}, O.~Klein, V.~Cros, A.~Fert,
  {Inverse spin Hall effect in nanometer-thick yttrium iron garnet/Pt system},
  Appl. Phys. Lett. 103 (2013) 082408.

\bibitem{Sai2006}
E.~Saitoh, M.~Ueda, H.~Miyajima, G.~Tatara, {Conversion of spin current into
  charge current at room temperature: Inverse spin-Hall effect}, Appl. Phys.
  Lett. 88 (2006) 182509.

\bibitem{Val2006}
S.~O. Valenzuela, M.~Tinkham, {Direct electronic measurement of the spin Hall
  effect}, Nature 442 (2006) 176--179.

\bibitem{Kim2007}
T.~Kimura, Y.~Otani, T.~Sato, S.~Takahashi, S.~Maekawa, {Room-temperature
  reversible spin Hall effect}, Phys. Rev. Lett. 98 (2007) 156601.

\bibitem{Hoffmann2015}
W.~Zhang, M.~B. Jungfleisch, W.~Jiang, J.~Sklenar, F.~Y. Fradin, J.~E. Pearson,
  J.~B. Ketterson, A.~Hoffmann, Spin pumping and inverse spin hall effects -
  insights for future spin-orbitronics, J. Appl. Phys. 117 (2015) 172610.

\bibitem{W_Han2019}
W.~Xing, L.~Qiu, X.~Wang, Y.~Yao, Y.~Ma, R.~Cai, S.~Jia, X.~C. Xie, W.~Han,
  Magnon transport in quasi-two-dimensional van der waals antiferromagnets,
  Phys. Rev. X 9 (2019) 011026.

\bibitem{Roma2020}
C.~Romanque-Albornoz, C.~Gonzalez-Fuentes, C.~Orellana, C.~Garcia, {Spin
  Seebeck effect detection by harmonic analysis}, Appl. Phys. Lett. 116 (2020)
  242402.

\bibitem{Maekawa_Spin_Current_2017}
S.~Maekawa, S.~O. Valenzuela, E.~Saitoh, T.~Kimura, {Spin Current (2nd
  edition)}, Oxford University Press, 2017.

\bibitem{W_Han2020}
W.~Han, S.~Maekawa, X.-C. Xie, {Spin current as a probe of quantum materials},
  Nat. Mater. 19 (2020) 139--152.

\bibitem{Heinrich2011}
B.~Heinrich, C.~Burrowes, E.~Montoya, B.~Kardasz, E.~Girt, Y.-Y. Song, Y.~Sun,
  M.~Wu, {Spin pumping at the magnetic insulator (YIG)/normal metal (Au)
  interfaces}, Phys. Rev. Lett. 107 (2011) 066604.

\bibitem{Y_Sun2013}
Y.~Sun, H.~Chang, M.~Kabatek, Y.-Y. Song, Z.~Wang, M.~Jantz, W.~Schneider,
  M.~Wu, E.~Montoya, B.~Kardasz, B.~Heinrich, S.~G.~E. te~Velthuis,
  H.~Schultheiss, A.~Hoffmann, {Damping in yttrium iron garnet nanoscale films
  capped by platinum}, Phys. Rev. Lett. 111 (2013) 106601.

\bibitem{Weiler2013}
M.~Weiler, M.~Althammer, M.~Schreier, J.~Lotze, M.~Pernpeintner, S.~Meyer,
  H.~Huebl, R.~Gross, A.~Kamra, J.~Xiao, Y.-T. Chen, H.~Jiao, G.~E.~W. Bauer,
  S.~T.~B. Goennenwein, {Experimental test of the spin mixing interface
  conductivity concept}, Phys. Rev. Lett. 111 (2013) 176601.

\bibitem{Mosendz2010PRB}
O.~Mosendz, V.~Vlaminck, J.~E. Pearson, F.~Y. Fradin, G.~E.~W. Bauer, S.~D.
  Bader, A.~Hoffmann, {Detection and quantification of inverse spin Hall effect
  from spin pumping in permalloy/normal metal bilayers}, Phys. Rev. B 82 (2010)
  214403.

\bibitem{Mosendz2010PRL}
O.~Mosendz, J.~E. Pearson, F.~Y. Fradin, G.~E.~W. Bauer, S.~D. Bader,
  A.~Hoffmann, {Quantifying spin Hall angles from spin pumping: Experiments and
  theory}, Phys. Rev. Lett. 104 (2010) 046601.

\bibitem{Ando2011}
K.~Ando, S.~Takahashi, J.~Ieda, Y.~Kajiwara, H.~Nakayama, T.~Yoshino, K.~Harii,
  Y.~Fujikawa, M.~Matsuo, S.~Maekawa, E.~Saitoh, {Inverse spin-Hall effect
  induced by spin pumping in metallic system}, J. Appl. Phys. 109 (2011)
  103913.

\bibitem{Col2016}
M.~Collet, X.~{De Milly}, O.~{D'Allivy Kelly}, V.~V. Naletov, R.~Bernard,
  P.~Bortolotti, J.~{Ben Youssef}, V.~E. Demidov, S.~O. Demokritov, J.~L.
  Prieto, M.~Mu{\~{n}}oz, V.~Cros, A.~Anane, G.~{De Loubens}, O.~Klein,
  {Generation of coherent spin-wave modes in yttrium iron garnet microdiscs by
  spin-orbit torque}, Nat. Commun. 7 (2016) 10377.

\bibitem{Chang2014}
H.~{Chang}, P.~{Li}, W.~{Zhang}, T.~{Liu}, A.~{Hoffmann}, L.~{Deng}, M.~{Wu},
  Nanometer-thick yttrium iron garnet films with extremely low damping, IEEE
  Magn. Lett. 5 (2014) 1--4.

\bibitem{Cor2015}
L.~J. Cornelissen, J.~Liu, R.~A. Duine, J.~B. Youssef, B.~J. {Van Wees},
  Long-distance transport of magnon spin information in a magnetic insulator at
  room temperature, Nat. Phys. 11 (2015) 1022--1026.

\bibitem{Men2019}
J.~Mendil, M.~Trassin, Q.~Bu, J.~Schaab, M.~Baumgartner, C.~Murer, P.~T. Dao,
  J.~Vijayakumar, D.~Bracher, C.~Bouillet, C.~A.~F. Vaz, M.~Fiebig,
  P.~Gambardella, {Magnetic properties and domain structure of ultrathin
  yttrium iron garnet/Pt bilayers}, Phys. Rev. Mater. 3 (2019) 034403.

\bibitem{Bai2013}
L.~Bai, P.~Hyde, Y.~S. Gui, C.-M. Hu, V.~Vlaminck, J.~E. Pearson, S.~D. Bader,
  A.~Hoffmann, {Universal method for separating spin pumping from spin
  rectification voltage of ferromagnetic resonance}, Phys. Rev. Lett. 111
  (2013) 217602.

\bibitem{C_Hu2016}
M.~Harder, Y.~Gui, C.-M. Hu, {Electrical detection of magnetization dynamics
  via spin rectification effects}, Phys. Rep. 661 (2016) 1--59.

\bibitem{Gui2007}
Y.~S. Gui, N.~Mecking, A.~Wirthmann, L.~H. Bai, C.-M. Hu, {Electrical detection
  of the ferromagnetic resonance: Spin-rectification versus bolometric effect},
  Appl. Phys. Lett. 91 (2007) 082503.

\bibitem{Hahn2013}
C.~Hahn, G.~de~Loubens, O.~Klein, M.~Viret, V.~V. Naletov, J.~Ben~Youssef,
  Comparative measurements of inverse spin hall effects and magnetoresistance
  in yig/pt and yig/ta, Phys. Rev. B 87 (2013) 174417.

\bibitem{Mal2012}
P.~Maletinsky, S.~Hong, M.~S. Grinolds, B.~Hausmann, M.~D. Lukin, R.~L.
  Walsworth, M.~Loncar, {A robust scanning diamond sensor for nanoscale imaging
  with single nitrogen-vacancy centres}, Nat. Nanotechnol. 7 (2012) 320--324.

\bibitem{App2016}
P.~Appel, E.~Neu, M.~Ganzhorn, A.~Barfuss, M.~Batzer, M.~Gratz,
  A.~Tsch{\"{o}}pe, P.~Maletinsky, Fabrication of all diamond scanning probes
  for nanoscale magnetometry, Rev. Sci. Instrum. 87 (2016) 063703.

\bibitem{Thi2019}
L.~Thiel, Z.~Wang, M.~A. Tschudin, D.~Rohner, I.~Guti{\'{e}}rrez-Lezama,
  N.~Ubrig, M.~Gibertini, E.~Giannini, A.~F. Morpurgo, P.~Maletinsky, {Probing
  magnetism in 2D materials at the nanoscale with single-spin microscopy},
  Science 364 (2019) 973--976.

\bibitem{Maz2008}
J.~R. Maze, P.~L. Stanwix, J.~S. Hodges, S.~Hong, J.~M. Taylor, P.~Cappellaro,
  L.~Jiang, M.~V. Dutt, E.~Togan, A.~S. Zibrov, A.~Yacoby, R.~L. Walsworth,
  M.~D. Lukin, {Nanoscale magnetic sensing with an individual electronic spin
  in diamond}, Nature 455 (2008) 644--647.

\bibitem{Cas2018}
F.~Casola, T.~{van der Sar}, A.~Yacoby, {Probing condensed matter physics with
  magnetometry based on nitrogen-vacancy centres in diamond}, Nat. Rev. Mater.
  3 (2018) 17088.

\bibitem{Dohe2013}
M.~W. Doherty, N.~B. Manson, P.~Delaney, F.~Jelezko, J.~Wrachtrup, L.~C.
  Hollenberg, {The nitrogen-vacancy colour centre in diamond}, Phys. Rep. 528
  (2013) 1--45.

\bibitem{Bala2009}
G.~Balasubramanian, P.~Neumann, D.~Twitchen, M.~Markham, R.~Kolesov,
  N.~Mizuochi, J.~Isoya, J.~Achard, J.~Beck, J.~Tissler, V.~Jacques, P.~R.
  Hemmer, F.~Jelezko, J.~Wrachtrup, {Ultralong spin coherence time in
  isotopically engineered diamond}, Nat. Mater. 8 (2009) 383--387.

\bibitem{Sar2015}
T.~{van der Sar}, F.~Casola, R.~Walsworth, A.~Yacoby, {Nanometre-scale probing
  of spin waves using single electron spins}, Nat. Commun. 6 (2015) 7886.

\bibitem{Du2017}
C.~Du, T.~{van der Sar}, T.~X. Zhou, P.~Upadhyaya, F.~Casola, H.~Zhang, M.~C.
  Onbasli, C.~A. Ross, R.~L. Walsworth, Y.~Tserkovnyak, A.~Yacoby, {Control and
  local measurement of the spin chemical potential in a magnetic insulator},
  Science 357 (2017) 195--198.

\bibitem{Zhang2019}
H.~Zhang, M.~J.~H. Ku, F.~Casola, C.~H. Du, T.~V.~D. Sar, M.~C. Onbasli, C.~A.
  Ross, {Spin-torque oscillation in a magnetic insulator probed by a
  single-spin sensor}, Phys. Rev. B 102 (2020) 024404.

\bibitem{App2019}
P.~Appel, B.~J. Shields, T.~Kosub, N.~Hedrich, R.~H{\"{u}}bner,
  J.~Fa{\ss}bender, D.~Makarov, P.~Maletinsky, {Nanomagnetism of
  magnetoelectric granular thin-film antiferromagnets}, Nano Lett. 19 (2019)
  1682--1687.

\bibitem{Tet2015}
J.~P. Tetienne, T.~Hingant, L.~J. Mart{\'{i}}nez, S.~Rohart, A.~Thiaville,
  L.~H. Diez, K.~Garcia, J.~P. Adam, J.~V. Kim, J.~F. Roch, I.~M. Miron,
  G.~Gaudin, L.~Vila, B.~Ocker, D.~Ravelosona, V.~Jacques, {The nature of
  domain walls in ultrathin ferromagnets revealed by scanning
  nanomagnetometry}, Nat. Commun. 6 (2015) 6733.

\bibitem{Jen2019}
A.~Jenkins, M.~Pelliccione, G.~Yu, X.~Ma, X.~Li, K.~L. Wang, A.~C. Jayich,
  Single-spin sensing of domain-wall structure and dynamics in a thin-film
  skyrmion host, Phys. Rev. Mater. 3 (2019) 083801.

\bibitem{Chau2019}
J.~Y. Chauleau, T.~Chirac, S.~Fusil, V.~Garcia, W.~Akhtar, J.~Tranchida,
  P.~Thibaudeau, I.~Gross, C.~Blouzon, A.~Finco, M.~Bibes, B.~Dkhil, D.~D.
  Khalyavin, P.~Manuel, V.~Jacques, N.~Jaouen, M.~Viret, Electric and
  antiferromagnetic chiral textures at multiferroic domain walls, Nat. Mater.
  19 (2020) 386--390.

\bibitem{Hay2020}
A.~Haykal, J.~Fischer, W.~Akhtar, J.-Y. Chauleau, D.~Sando, A.~Finco, F.~Godel,
  Y.~A. Birkh{\"{o}}lzer, C.~Carr{\'{e}}t{\'{e}}ro, N.~Jaouen, M.~Bibes,
  M.~Viret, S.~Fusil, V.~Jacques, V.~Garcia, {Antiferromagnetic textures in
  BiFeO3 controlled by strain and electric field}, Nat. Commun. 11 (2020) 1704.

\bibitem{Liu2012}
L.~Liu, C.-F. Pai, Y.~Li, H.~Tseng, D.~Ralph, R.~Buhrman, {Spin-torque
  switching with the giant spin hall effect of tantalum}, Science 336 (2012)
  555--558.

\bibitem{Du2020}
X.~Wang, Y.~Xiao, C.~Liu, E.~Lee-Wong, N.~J. McLaughlin, H.~Wang, M.~Wu,
  H.~Wang, E.~E. Fullerton, C.~R. Du, {Electrical control of coherent spin
  rotation of a single-spin qubit}, npj Quantum Inf. 6 (2020) 78.

\bibitem{Dirk2015}
T.~Schwarze, J.~Waizner, M.~Garst, A.~Bauer, I.~Stasinopoulos, H.~Berger,
  C.~Pfleiderer, D.~Grundler, {Universal helimagnon and skyrmion excitations in
  metallic, semiconducting and insulating chiral magnets}, Nat. Mater. 14
  (2015) 478--483.

\bibitem{Ding2013}
L.~Sun, R.~X. Cao, B.~F. Miao, Z.~Feng, B.~You, D.~Wu, W.~Zhang, A.~Hu, H.~F.
  Ding, {Creating an artificial two-dimensional Skyrmion crystal by
  nanopatterning}, Phys. Rev. Lett. 110 (2013) 167201.

\bibitem{Dustin2015}
D.~A. Gilbert, B.~B. Maranville, A.~L. Balk, B.~J. Kirby, P.~Fischer, D.~T.
  Pierce, J.~Unguris, J.~A. Borchers, K.~Liu, {Realization of ground-state
  artificial skyrmion lattices at room temperature}, Nat. Commun. 14 (2015)
  478--483.

\bibitem{Win1961}
J.~Winter, {Bloch wall excitation. Application to nuclear resonance in a Bloch
  wall}, Phys. Rev. 124 (1961) 452.

\bibitem{wieser_quantized_2009}
R.~Wieser, E.~Y. Vedmedenko, R.~Wiesendanger, Quantized spin waves in
  ferromagnetic and antiferromagnetic structures with domain walls, Phys. Rev.
  B 79 (2009) 144412.

\bibitem{umax}
$\mu$max.

\bibitem{Deb2019}
M.~Deb, E.~Popova, M.~Hehn, N.~Keller, S.~Petit-Watelot, M.~Bargheer,
  S.~Mangin, G.~Malinowski, {Damping of standing spin waves in
  bismuth-substituted yttrium iron garnet as seen via the time-resolved
  magneto-optical Kerr effect}, Phys. Rev. Appl. 12 (2019) 044006.

\bibitem{Keh2015}
A.~Kehlberger, K.~Richter, M.~C. Onbasli, G.~Jakob, D.~H. Kim, T.~Goto, C.~A.
  Ross, G.~G{\"{o}}tz, G.~Reiss, T.~Kuschel, M.~Kl{\"{a}}ui, {Enhanced
  magneto-optic Kerr effect and magnetic properties of CeY2Fe5O12 epitaxial
  thin films}, Phys. Rev. Appl. 4 (2015) 014008.

\bibitem{Bilei2019}
Y.~Zhang, Q.~Du, C.~Wang, W.~Yan, L.~Deng, J.~Hu, C.~A. Ross, L.~Bi,
  {Dysprosium substituted Ce:YIG thin films with perpendicular magnetic
  anisotropy for silicon integrated optical isolator applications}, {APL
  Mater.} {7} ({2019}) {081119}.

\bibitem{Wu2018}
C.~N. Wu, C.~C. Tseng, Y.~T. Fanchiang, C.~K. Cheng, K.~Y. Lin, S.~L. Yeh,
  S.~R. Yang, C.~T. Wu, T.~Liu, M.~Wu, M.~Hong, J.~Kwo, {High-quality thulium
  iron garnet films with tunable perpendicular magnetic anisotropy by off-axis
  sputtering – correlation between magnetic properties and film strain}, Sci.
  Rep. 8 (2018) 11087.

\bibitem{Beach2019}
C.~O. Avci, E.~Rosenberg, L.~Caretta, F.~B{\"{u}}ttner, M.~Mann, C.~Marcus,
  D.~Bono, C.~A. Ross, G.~S. Beach, Interface-driven chiral magnetism and
  current-driven domain walls in insulating magnetic garnets, Nat. Nanotechnol.
  14 (2019) 561--566.

\bibitem{Shao2019}
Q.~Shao, Y.~Liu, G.~Yu, S.~K. Kim, X.~Che, C.~Tang, Q.~L. He, Y.~Tserkovnyak,
  J.~Shi, K.~L. Wang, Topological hall effect at above room temperature in
  heterostructures composed of a magnetic insulator and a heavy metal, Nat.
  Electron. 2 (2019) 182--186.

\bibitem{F_Yang2020}
A.~J. Lee, A.~S. Ahmed, J.~Flores, S.~Guo, B.~Wang, N.~Bagu\'es, D.~W. McComb,
  F.~Yang, Probing the source of the interfacial dzyaloshinskii-moriya
  interaction responsible for the topological hall effect in
  $\text{metal}/{\mathrm{tm}}_{3}{\mathrm{fe}}_{5}{\mathrm{o}}_{12}$ systems,
  Phys. Rev. Lett. 124 (2020) 107201.

\bibitem{Beach2020}
L.~Caretta, E.~Rosenberg, F.~B{\"{u}}ttner, T.~Fakhrul, P.~Gargiani,
  M.~Valvidares, Z.~Chen, P.~Reddy, D.~A. Muller, C.~A. Ross, G.~S.~D. Beach,
  Interfacial dzyaloshinskii-moriya interaction arising from rare-earth orbital
  magnetism in insulating magnetic oxides, Nat. Commun. 11 (2020) 1090.

\bibitem{Sch2016}
M.~A.~W. Schoen, D.~Thonig, M.~L. Schneider, T.~J. Silva, H.~T. Nembach,
  O.~Eriksson, O.~Karis, J.~M. Shaw, Ultra-low magnetic damping of a metallic
  ferromagnet, Nat. Phys. 12 (2016) 839--842.

\bibitem{A_Lee2017}
A.~J. Lee, J.~T. Brangham, Y.~Cheng, S.~P. White, W.~T. Ruane, B.~D. Esser,
  D.~W. McComb, P.~C. Hammel, F.~Yang, Metallic ferromagnetic films with
  magnetic damping under $1.4\times10^{-3}$, Nat. Commun. 8 (2017) 234.

\bibitem{Y_Li2019}
Y.~Li, F.~Zeng, S.~S. Zhang, H.~Shin, H.~Saglam, V.~Karakas, O.~Ozatay, J.~E.
  Pearson, O.~G. Heinonen, Y.~Wu, A.~Hoffmann, W.~Zhang, {Giant anisotropy of
  Gilbert damping in epitaxial CoFe films}, Phys. Rev. Lett. 122 (2019) 117203.

\bibitem{J_Du2020}
R.~Li, F.~Kong, P.~Zhao, Z.~Cheng, Z.~Qin, M.~Wang, Q.~Zhang, P.~Wang, Y.~Wang,
  F.~Shi, J.~Du, {Nanoscale electrometry based on a magnetic-feld-resistant
  spin sensor}, Phys. Rev. Lett. 124 (2020) 247701.

\bibitem{Fel2009}
C.~Felser, B.~Hillebrands, Cluster issue on heusler compounds and devices, J.
  Phys. D Appl. Phys. 42 (2009) 080301.

\bibitem{Miz2009}
S.~Mizukami, D.~Watanabe, M.~Oogane, Y.~Ando, Y.~Miura, M.~Shirai, T.~Miyazaki,
  {Low damping constant for Co2FeAl Heusler alloy films and its correlation
  with density of states}, J. Appl. Phys. 105 (2009) 07D306.

\bibitem{Seb2012}
T.~Sebastian, Y.~Ohdaira, T.~Kubota, P.~Pirro, T.~Br{\"{a}}cher, K.~Vogt, A.~A.
  Serga, H.~Naganuma, M.~Oogane, Y.~Ando, B.~Hillebrands, Low-damping spin-wave
  propagation in a micro-structured {Co2Mn0.6Fe0.4Si} heusler waveguide, Appl.
  Phys. Lett. 100 (2012) 112402.

\bibitem{Bai2012}
L.~Bai, N.~Tezuka, M.~Kohda, J.~Nitta, Anisotropy and damping in
  {Co2FeAl0.5Si0.5} via electrical detection of ferromagnetic resonance, Jpn.
  J. Appl. Phys. 51 (2012) 083001.

\bibitem{Seb2013}
T.~Sebastian, T.~Br{\"{a}}cher, P.~Pirro, A.~A. Serga, B.~Hillebrands,
  T.~Kubota, H.~Naganuma, M.~Oogane, Y.~Ando, {Nonlinear emission of spin-wave
  caustics from an edge mode of a microstructured Co2Mn0.6Fe0.4Si waveguide},
  Phys. Rev. Lett. 110 (2013) 067201.

\bibitem{Loong2014}
L.~M. Loong, J.~H. Kwon, P.~Deorani, C.~N. {Tung Yu}, A.~Hirohata, H.~Yang,
  {Investigation of the temperature-dependence of ferromagnetic resonance and
  spin waves in Co2FeAl0.5Si0.5}, Appl. Phys. Lett. 104 (2014) 232409.

\bibitem{Qiao2016}
S.-Z. Qiao, Q.-N. Ren, R.-R. Hao, H.~Zhong, Y.~Kang, S.-S. Kang, Y.-F. Qin,
  S.-Y. Yu, G.-B. Han, S.-S. Yan, L.-M. Mei, {Broad-band FMR linewidth of
  Co2MnSi thin films with low damping factor: The role of two-magnon
  scattering}, {Chin. Phys. Lett.} {33} ({2016}) {047601}.

\bibitem{Ludbrook2017}
B.~M. Ludbrook, G.~Dubuis, A.-H. Puichaud, B.~J. Ruck, S.~Granville,
  {Nucleation and annihilation of skyrmions in Mn2CoAl observed through the
  topological Hall effect}, Sci. Rep. 7 (2017) 13620.

\bibitem{S_Wang2018}
S.~Wang, J.~Ding, X.~Guan, M.~B. Jungfleisch, Z.~Zhang, X.~Wang, W.~Gu, Y.~Zhu,
  J.~E. Pearson, X.~Cheng, A.~Hoffmann, X.~Miao, {Linear and nonlinear
  spin-wave dynamics in ultralow-damping microstructured Co2FeAl Heusler
  waveguide}, Appl. Phys. Lett. 113 (2018) 232404.

\bibitem{Z_Zhang2019}
W.~Zhu, Z.~Zhu, D.~Li, G.~Wu, L.~Xi, Q.~Y. Jin, Z.~Zhang, {Magnetization
  precession and damping in Co2FeSi Heusler alloy thin films}, J. Magn. Magn.
  Mater. {479} ({2019}) {179--184}.

\bibitem{Hen2019}
Y.~Henry, D.~Stoeffler, J.~Kim, M.~Bailleul, {Unidirectional spin-wave
  channeling along magnetic domain walls of Bloch type}, Phys. Rev. B 100
  (2019) 024416.

\bibitem{Pig2012}
B.~Pigeau, C.~Hahn, G.~D. Loubens, V.~V. Naletov, O.~Klein, {Measurement of the
  dynamical dipolar coupling in a pair of magnetic nanodisks using a
  ferromagnetic resonance force microscope}, Phys. Rev. Lett. 109 (2012)
  247602.

\bibitem{Tac2014}
S.~Tacchi, S.~Fin, G.~Carlotti, G.~Gubbiotti, M.~Madami, M.~Barturen,
  M.~Marangolo, M.~Eddrief, D.~Bisero, A.~Rettori, M.~G. Pini, {Rotatable
  magnetic anisotropy in a Fe0.8Ga0.2 thin film with stripe domains: Dynamics
  versus statics}, Phys. Rev. B 89 (2014) 024411.

\bibitem{Alb2018a}
E.~Albisetti, D.~Petti, G.~Sala, R.~Silvani, S.~Tacchi, S.~Finizio, S.~Wintz,
  A.~Cal{\`{o}}, X.~Zheng, J.~Raabe, E.~Riedo, R.~Bertacco, {Nanoscale
  spin-wave circuits based on engineered reconfigurable spin-textures}, Commun.
  Phys. 1 (2018) 56.

\bibitem{Bay2005}
C.~Bayer, H.~Schultheiss, B.~Hillebrands, R.~L. Stamps, {Phase shift of spin
  waves traveling through A 180$^{\circ}$ bloch domain wall}, IEEE Trans. Magn.
  41 (2005) 414.

\bibitem{macke_transmission_2010}
S.~Macke, D.~Goll, Transmission and reflection of spin waves in the presence of
  {Néel} walls, J. Phys. Conf. Ser. 200 (2010) 042015.

\bibitem{borys_spinwave_2016}
P.~Borys, F.~Garcia‐Sanchez, J.~Kim, R.~L. Stamps, Spin‐{wave} {eigenmodes}
  of {Dzyaloshinskii} {domain} {walls}, Adv. Electron. Mater. 2 (2016) 1500202.

\bibitem{Cha2018}
L.-J. Chang, Y.-F. Liu, M.-Y. Kao, L.-Z. Tsai, J.-Z. Liang, S.-F. Lee,
  {Ferromagnetic domain walls as spin wave filters and the interplay between
  domain walls and spin waves}, Sci. Rep. 8 (2018) 3910.

\bibitem{Woo2017}
S.~Woo, T.~Delaney, G.~S.~D. Beach, {Magnetic domain wall depinning assisted by
  spin wave bursts}, Nat. Phys. 13 (2017) 448--453.

\bibitem{Mah2009}
Y.~{Le Maho}, J.~V. Kim, G.~Tatara, {Spin-wave contributions to current-induced
  domain wall dynamics}, Phys. Rev. B 79 (2009) 174404.

\bibitem{Hatami2007}
M.~Hatami, G.~E. Bauer, Q.~Zhang, P.~J. Kelly, {Thermal spin-transfer torque in
  magnetoelectronic devices}, Phys. Rev. Lett. 99 (2007) 066603.

\bibitem{Yu2010}
H.~Yu, S.~Granville, D.~P. Yu, J.~P. Ansermet, {Evidence for thermal
  spin-transfer torque}, Phys. Rev. Lett. 104 (2010) 146601.

\bibitem{Slon2010}
J.~C. Slonczewski, {Initiation of spin-transfer torque by thermal transport
  from magnons}, Phys. Rev. B 82 (2010) 054403.

\bibitem{Lah2011}
T.~H. Lahtinen, J.~O. Tuomi, S.~{van Dijken}, {Pattern transfer and
  electric-field-induced magnetic domain formation in multiferroic
  heterostructures}, Adv. Mater. 23 (2011) 3187--3191.

\bibitem{Han2018}
H.~Wu, L.~Huang, C.~Fang, B.~S. Yang, C.~H. Wan, G.~Q. Yu, J.~F. Feng, H.~X.
  Wei, X.~F. Han, {Magnon valve effect between two magnetic insulators}, Phys.
  Rev. Lett. 120 (2018) 097205.

\bibitem{Cor2018}
L.~J. Cornelissen, J.~Liu, B.~J.~V. Wees, {Spin-current-controlled modulation
  of the magnon spin conductance in a three-terminal magnon transistor}, Phys.
  Rev. Lett. 120 (2018) 097702.

\bibitem{Cra2018}
J.~Cramer, F.~Fuhrmann, U.~Ritzmann, V.~Gall, T.~Niizeki, R.~Ramos, Z.~Qiu,
  D.~Hou, T.~Kikkawa, J.~Sinova, U.~Nowak, E.~Saitoh, M.~Kl{\"{a}}ui, {Magnon
  detection using a ferroic collinear multilayer spin valve}, Nat. Commun. 9
  (2018) 1089.

\bibitem{Gho2017}
A.~Ghosh, K.~S. Huang, O.~Tchernyshyov, {Annihilation of domain walls in a
  ferromagnetic wire}, Phys. Rev. B 95 (2017) 180408.

\bibitem{Ram2014}
M.~Ramu, I.~Purnama, S.~Goolaup, M.~{Chandra Sekhar}, W.~S. Lew, {Investigation
  of dominant spin wave modes by domain walls collision}, J. Appl. Phys. 115
  (2014) 243908.

\bibitem{ross_propagation_2020}
A.~Ross, R.~Lebrun, O.~Gomonay, D.~A. Grave, A.~Kay, L.~Baldrati, S.~Becker,
  A.~Qaiumzadeh, C.~Ulloa, G.~Jakob, F.~Kronast, J.~Sinova, R.~Duine,
  A.~Brataas, A.~Rothschild, M.~Kläui, Propagation {length} of
  {antiferromagnetic} {magnons} {governed} by {domain} {configurations}, Nano
  Lett. 20 (2020) 306--313.

\bibitem{Fis2017}
T.~Fischer, M.~Kewenig, D.~A. Bozhko, A.~A. Serga, I.~I. Syvorotka,
  F.~Ciubotaru, C.~Adelmann, B.~Hillebrands, A.~V. Chumak, {Experimental
  prototype of a spin-wave majority gate}, Appl. Phys. Lett. 110 (2017) 152401.

\bibitem{Khi2005}
A.~Khitun, K.~L. Wang, {Nano scale computational architectures with spin wave
  bus}, Superlattices Microstruct. 38 (2005) 184--200.

\bibitem{Bui2016}
F.~J. Buijnsters, Y.~Ferreiros, A.~Fasolino, M.~I. Katsnelson,
  {Chirality-dependent transmission of spin waves through domain walls}, Phys.
  Rev. Lett. 116 (2016) 147204.

\bibitem{Slon1996}
J.~Slonczewski, {Current-driven excitation of magnetic multilayers}, J. Magn.
  Magn. Mater. 195 (1996) L1--L7.

\bibitem{Ber1996}
L.~Berger, {Emission of spin waves by a magnetic multilayer traversed by a
  current}, Phys. Rev. B 54 (1996) 9353--9358.

\bibitem{Bra2012}
A.~Brataas, A.~D. Kent, H.~Ohno, {Current-induced torques in magnetic
  materials}, Nat. Mater. 11 (2012) 372--381.

\bibitem{Shi2011}
J.~Shibata, G.~Tatara, H.~Kohno, {A brief review of field- and current-driven
  domain-wall motion}, J. Phys. D Appl. Phys. 44 (2011) 384004.

\bibitem{Parkin2015}
S.~Parkin, S.~H. Yang, {Memory on the racetrack}, Nat. Nanotechnol. 10 (2015)
  195--198.

\bibitem{YWang2019}
Y.~Wang, D.~Zhu, Y.~Yang, K.~Lee, R.~Mishra, G.~Go, S.-h. Oh, D.-h. Kim,
  K.~Cai, E.~Liu, S.~D. Pollard, S.~Shi, J.~Lee, K.~L. Teo, Y.~Wu, K.-j. Lee,
  H.~Yang, {Magnetization switching by magnon-mediated spin torque through an
  antiferromagnetic insulator}, Science 366 (2019) 1125--1128.

\bibitem{Yama2004}
A.~Yamaguchi, T.~Ono, S.~Nasu, K.~Miyake, K.~Mibu, T.~Shinjo, {Real-space
  observation of current-driven domain wall motion in submicron magnetic
  wires}, Phys. Rev. Lett. 92 (2004) 077205.

\bibitem{Haya2006}
M.~Hayashi, L.~Thomas, Y.~B. Bazaliy, C.~Rettner, R.~Moriya, X.~Jiang, S.~S.
  Parkin, {Influence of current on field-driven domain wall motion in permalloy
  nanowires from time resolved measurements of anisotropic magnetoresistance},
  Phys. Rev. Lett. 96 (2006) 197207.

\bibitem{Rising2017}
V.~Risingg{\aa}rd, E.~G. Tveten, A.~Brataas, J.~Linder, Equations of motion and
  frequency dependence of magnon-induced domain wall motion, Phys. Rev. B 96
  (2017) 174441.

\bibitem{Sho2019}
Y.~A. Shokr, O.~Sandig, M.~Erkovan, B.~Zhang, M.~Bernien, A.~A. {\"{U}}nal,
  F.~Kronast, U.~Parlak, J.~Vogel, W.~Kuch, {Steering of magnetic domain walls
  by single ultrashort laser pulses}, Phys. Rev. B 99 (2019) 214404.

\bibitem{Mazo2016}
J.~Mazo-Zuluaga, E.~A. Velásquez, D.~Altbir, J.~Mejía-López, {Controlling
  domain wall nucleation and propagation with temperature gradients}, Appl.
  Phys. Lett. 109 (2016) 122408.

\bibitem{Kim2019}
K.~W. Kim, S.~W. Lee, J.~H. Moon, G.~Go, A.~Manchon, H.~W. Lee,
  K.~Everschor-Sitte, K.~J. Lee, {Unidirectional magnon-driven domain wall
  motion due to the interfacial dzyaloshinskii-moriya interaction}, Phys. Rev.
  Lett. 122 (2019) 147202.

\bibitem{bracher_creation_2017}
T.~Brächer, O.~Boulle, G.~Gaudin, P.~Pirro, Creation of unidirectional
  spin-wave emitters by utilizing interfacial {Dzyaloshinskii}-{Moriya}
  interaction, Phys. Rev. B 95 (2017) 064429.

\bibitem{qaiumzadeh_controlling_2017}
A.~Qaiumzadeh, L.~A. Kristiansen, A.~Brataas, {Controlling chiral domain walls
  in antiferromagnets using spin-wave helicity}, Phys. Rev. B 97 (2018) 020402.

\bibitem{Jan2010}
M.~Janoschek, P.~Link, {Helimagnon bands as universal excitations of chiral
  magnets}, Phys. Rev. B 81 (2010) 214436.

\bibitem{Kug2015}
M.~Kugler, G.~Brandl, J.~Waizner, M.~Janoschek, R.~Georgii, A.~Bauer,
  K.~Seemann, A.~Rosch, C.~Pfleiderer, P.~B{\"{o}}ni, M.~Garst, {Band structure
  of helimagnons in MnSi resolved by inelastic neutron scattering}, Phys. Rev.
  Lett. 122 (2015) 097203.

\bibitem{Wei2017}
M.~Weiler, A.~Aqeel, M.~Mostovoy, A.~Leonov, S.~Gepr{\"{a}}gs, R.~Gross,
  H.~Huebl, T.~T.~M. Palstra, S.~T.~B. Goennenwein, {Helimagnon resonances in
  an intrinsic chiral magnonic crystal}, Phys. Rev. Lett. 119 (2017) 237204.

\bibitem{Iwa2014}
J.~Iwasaki, A.~J. Beekman, N.~Nagaosa, {Theory of magnon-skyrmion scattering in
  chiral magnets}, Phys. Rev. B 89 (2014) 064412.

\bibitem{Seki2019}
S.~Seki, M.~Garst, J.~Waizner, R.~Takagi, Y.~Okamura, K.~Kondou, F.~Kagawa,
  Y.~Otani, Y.~Tokura, {Propagation dynamics of spin excitations along skyrmion
  strings}, Nat. Commun. 11 (2020) 256.

\bibitem{Nag2013}
N.~Nagaosa, Y.~Tokura, {Topological properties and dynamics of magnetic
  skyrmions}, Nat. Nanotechnol. 8 (2013) 899--911.

\bibitem{liu_skyrmion_2015}
Y.~Liu, G.~Yin, J.~Zang, J.~Shi, R.~K. Lake, {Skyrmion creation and
  annihilation by spin waves}, Appl. Phys. Lett. 107 (2015) 152411.

\bibitem{ye-hua_dynamics_2015}
L.~Ye-Hua, L.~You-Quan, Dynamics of magnetic skyrmions, Chin. Phys. B 24 (2015)
  017506.

\bibitem{Fert2017}
A.~Fert, N.~Reyren, V.~Cros, {Magnetic skyrmions: advances in physics and
  potential applications}, Nat. Rev. Mater. 2 (2017) 17031.

\bibitem{schulz_emergent_2012}
T.~Schulz, R.~Ritz, A.~Bauer, M.~Halder, M.~Wagner, C.~Franz, C.~Pfleiderer,
  K.~Everschor, M.~Garst, A.~Rosch, {Emergent electrodynamics of skyrmions in a
  chiral magnet}, Nat. Phys. 8 (2012) 301--304.

\bibitem{garst_collective_2017}
M.~Garst, J.~Waizner, D.~Grundler, Collective spin excitations of helices and
  magnetic skyrmions: review and perspectives of magnonics in
  non-centrosymmetric magnets, J. Phys. D Appl. Phys. 50 (2017) 293002.

\bibitem{C_Jin2017}
C.~Jin, Z.-A. Li, A.~Kov{\'{a}}cs, J.~Caron, F.~Zheng, F.~N. Rybakov, N.~S.
  Kiselev, H.~Du, S.~Bl{\"{u}}gel, M.~Tian, Y.~Zhang, M.~Farle, R.~E.
  Dunin-Borkowski, {Control of morphology and formation of highly geometrically
  confined magnetic skyrmions}, Nat. Commun. 8 (2017) 15569.

\bibitem{Jiang2015}
W.~Jiang, P.~Upadhyaya, W.~Zhang, G.~Yu, M.~B. Jungfleisch, F.~Y. Fradin, J.~E.
  Pearson, Y.~Tserkovnyak, K.~L. Wang, O.~Heinonen, S.~G.~E. Velthuis,
  A.~Hoffmann, {Blowing magnetic skyrmion bubbles}, Science 349 (2015)
  283--286.

\bibitem{Sou2017}
A.~Soumyanarayanan, M.~Raju, A.~L. {Gonzalez Oyarce}, A.~K.~C. Tan, M.-Y. Im,
  A.~P. Petrovi{\'{c}}, P.~Ho, K.~H. Khoo, M.~Tran, C.~K. Gan, F.~Ernult,
  C.~Panagopoulos, {Tunable room-temperature magnetic skyrmions in Ir/Fe/Co/Pt
  multilayers}, Nat. Mater. 16 (2017) 898--904.

\bibitem{Woo2016}
S.~Woo, K.~Litzius, B.~Kr{\"{u}}ger, M.~Y. Im, L.~Caretta, K.~Richter, M.~Mann,
  A.~Krone, R.~M. Reeve, M.~Weigand, P.~Agrawal, I.~Lemesh, M.~A. Mawass,
  P.~Fischer, M.~Kl{\"{a}}ui, G.~S. Beach, {Observation of room-temperature
  magnetic skyrmions and their current-driven dynamics in ultrathin metallic
  ferromagnets}, Nat. Mater. 15 (2016) 501--506.

\bibitem{Moc2012}
M.~Mochizuki, {Spin-wave modes and their intense excitation effects in Skyrmion
  crystals}, Phys. Rev. Lett. 108 (2012) 017601.

\bibitem{Kim2014}
J.~V. Kim, F.~Garcia-Sanchez, J.~Sampaio, C.~Moreau-Luchaire, V.~Cros, A.~Fert,
  {Breathing modes of confined skyrmions in ultrathin magnetic dots}, Phys.
  Rev. B 90 (2014) 064410.

\bibitem{Oka2013}
Y.~Okamura, F.~Kagawa, M.~Mochizuki, M.~Kubota, S.~Seki, S.~Ishiwata,
  M.~Kawasaki, Y.~Onose, Y.~Tokura, {Microwave magnetoelectric effect via
  skyrmion resonance modes in a helimagnetic multiferroic}, Nat. Commun. 4
  (2013) 2391.

\bibitem{Sta2017}
I.~Stasinopoulos, S.~Weichselbaumer, A.~Bauer, J.~Waizner, H.~Berger,
  S.~Maendl, M.~Garst, C.~Pfleiderer, I.~Stasinopoulos, S.~Weichselbaumer,
  A.~Bauer, J.~Waizner, H.~Berger, S.~Maendl, {Low spin wave damping in the
  insulating chiral magnet Cu2OSeO3}, Appl. Phys. Lett. 111 (2017) 032408.

\bibitem{Sat2018}
B.~Satywali, F.~Ma, S.~He, M.~Raju, P.~Volodymyr, M.~Garst, A.~Soumyanarayanan,
  C.~Panagopoulos, {Gyrotropic resonance of individual N{\'{e}}el skyrmions in
  Ir/Fe/Co/Pt multilayers}.

\bibitem{Lin2014}
S.-Z. Lin, C.~D. Batista, C.~Reichhardt, A.~Saxena, {AC current generation in
  chiral magnetic insulators and skyrmion motion induced by the spin Seebeck
  effect}, Phys. Rev. Lett. 112 (2014) 187203.

\bibitem{Moc2014}
M.~Mochizuki, X.~Z. Yu, S.~Seki, N.~Kanazawa, W.~Koshibae, J.~Zang,
  M.~Mostovoy, Y.~Tokura, N.~Nagaosa, {Thermally driven ratchet motion of a
  skyrmion microcrystal and topological magnon Hall effect}, Nat. Mater. 13
  (2014) 241--246.

\bibitem{X_Zhang2017}
X.~Zhang, J.~Muller, J.~Xia, M.~Garst, X.~Liu, Y.~Zhou, {Motion of skyrmions in
  nanowires driven by magnonic momentum-transfer forces}, New J. Phys. 9 (2017)
  065001.

\bibitem{Zhou2015}
X.~Zhang, M.~Ezawa, D.~Xiao, G.~P. Zhao, Y.~Liu, Y.~Zhou, {All-magnetic control
  of skyrmions in nanowires by a spin wave}, Nanotechnology 26 (2015) 225701.

\bibitem{li_dynamics_2018}
S.~Li, J.~Xia, X.~Zhang, M.~Ezawa, W.~Kang, X.~Liu, Y.~Zhou, W.~Zhao, {Dynamics
  of a magnetic skyrmionium driven by spin waves}, Appl. Phys. Lett. 112 (2018)
  142404.

\bibitem{Petrova2011}
O.~Petrova, O.~Tchernyshyov, {Spin waves in a skyrmion crystal}, Phys. Rev. B
  84 (2011) 214433.

\bibitem{F_Ma2015}
F.~Ma, Y.~Zhou, H.-b. Braun, W.~S. Lew, {Skyrmion-based dynamic magnonic
  crystal}, Nano Lett. 15 (2015) 4029--4036.

\bibitem{F_Ma2015b}
F.~{Ma}, Y.~{Zhou}, W.~S. {Lew}, {Magnonic band structure in a Skyrmion
  magnonic crystal}, IEEE Trans. Magn. 51 (2015) 1500404.

\bibitem{roldan-molina_topological_2016}
A.~Roldán-Molina, A.~S. Nunez, J.~Fernández-Rossier, Topological spin waves
  in the atomic-scale magnetic skyrmion crystal, New J. Phys. 18 (2016) 045015.

\bibitem{moon_control_2016}
K.-W. Moon, B.~S. Chun, W.~Kim, C.~Hwang, Control of {spin}-{wave} {refraction}
  {using} {arrays} of {Skyrmions}, Phys. Rev. Appl. 6 (2016) 064027.

\bibitem{ding_motion_2015}
J.~Ding, X.~Yang, T.~Zhu, The {motion} of {magnetic} {Skyrmions} {driven} by
  {propagating} {spin} {waves}, IEEE Trans. Magn. 51 (2015) 1500504.

\bibitem{Seki2016}
S.~Seki, Y.~Okamura, K.~Kondou, K.~Shibata, M.~Kubota, R.~Takagi, F.~Kagawa,
  M.~Kawasaki, G.~Tatara, Y.~Otani, Y.~Tokura, {Magnetochiral nonreciprocity of
  volume spin wave propagation in chiral-lattice ferromagnets}, Phys. Rev. B 93
  (2016) 235131.

\bibitem{Emo2013}
S.~Emori, U.~Bauer, S.-m. Ahn, E.~Martinez, G.~S.~D. Beach, {Current-driven
  dynamics of chiral ferromagnetic domain walls}, Nat. Mater. 12 (2013)
  611--616.

\bibitem{Z_Luo2020}
Z.~Luo, A.~Hrabec, T.~P. Dao, G.~Sala, S.~Finizio, J.~Feng, S.~Mayr, J.~Raabe,
  P.~Gambardella, L.~J. Heyderman, {Current-driven magnetic domain-wall logic},
  Nature 579 (2020) 214--218.

\bibitem{Heide2008}
M.~Heide, G.~Bihlmayer, {Dzyaloshinskii-Moriya interaction accounting for the
  orientation of magnetic domains in ultrathin films: Fe/W(110)}, Phys. Rev. B
  78 (2008) 140403.

\bibitem{Fran2014}
J.~H. Franken, M.~Herps, H.~J.~M. Swagten, B.~Koopmans, {Tunable chiral spin
  texture in magnetic domain-walls}, Sci. Rep. 4 (2014) 5248.

\bibitem{Z_luo2019}
Z.~Luo, T.~P. Dao, A.~Hrabec, J.~Vijayakumar, A.~Kleibert, M.~Baumgartner,
  E.~Kirk, J.~Cui, T.~Savchenko, G.~Krishnaswamy, L.~J. Heyderman,
  P.~Gambardella, {Chirally coupled nanomagnets}, Science 363 (2019)
  1435--1439.

\bibitem{F_Ma2014}
F.~Ma, Y.~Zhou, {Interfacial Dzialoshinskii–Moriya interaction induced
  nonreciprocity of spin waves in magnonic waveguides}, RSC Adv. 4 (2014)
  46454--46459.

\bibitem{Yang2015a}
K.~Di, V.~L. Zhang, H.~S. Lim, S.~C. Ng, M.~H. Kuok, J.~Yu, J.~Yoon, X.~Qiu,
  H.~Yang, {Direct observation of the Dzyaloshinskii-Moriya interaction in a
  Pt/Co/Ni film}, Phys. Rev. Lett. 114 (2015) 047201.

\bibitem{HYang2020}
H.~Bouloussa, Y.~Roussign\'e, M.~Belmeguenai, A.~Stashkevich, S.-M. Ch\'erif,
  S.~D. Pollard, H.~Yang, {Dzyaloshinskii-Moriya interaction induced asymmetry
  in dispersion of magnonic Bloch modes}, Phys. Rev. B 102 (2020) 014412.

\bibitem{Zak2010}
K.~Zakeri, Y.~Zhang, J.~Prokop, T.~Chuang, N.~Sakr, W.~X. Tang, J.~Kirschner,
  {Asymmetric spin-wave dispersion on Fe ( 110 ): Direct evidence of the
  Dzyaloshinskii-Moriya interaction}, Phys. Rev. Lett. 104 (2010) 137203.

\bibitem{Kor2015}
H.~K\"orner, J.~Stigloher, H.~Bauer, H.~Hata, T.~Taniguchi, T.~Moriyama,
  T.~Ono, C.~Back, {Interfacial Dzyaloshinskii-Moriya interaction studied by
  time-resolved scanning Kerr microscopy}, Phys. Rev. B 92 (2015) 220413.

\bibitem{Gla2016}
O.~Gladii, M.~Haidar, Y.~Henry, M.~Kostylev, M.~Bailleul, {Frequency
  nonreciprocity of surface spin wave in permalloy thin films}, Phys. Rev. B 93
  (2016) 054430.

\bibitem{Cos2018}
N.~P.~D. Costa, B.~G. Silva, R.~L. Sommer, F.~Bohn, M.~A. Correa, {Effects of
  second order surface anisotropy in YIG sputtered onto GGG ( 100 ) oriented
  substrate}, J. Magn. Magn. Mater. 469 (2018) 64--68.

\bibitem{Gal2019b}
R.~A. Gallardo, P.~Alvarado-Seguel, T.~Schneider, C.~Gonzalez-Fuentes,
  A.~Rold{\'{a}}n-Molina, K.~Lenz, J.~Lindner, P.~Landeros, Spin-wave
  non-reciprocity in magnetization-graded ferromagnetic films, New J. Phys. 21
  (2019) 033026.

\bibitem{Y_Jiang2020}
Q.~B. Liu, K.~K. Meng, Z.~D. Xu, T.~Zhu, X.~G. Xu, J.~Miao, Y.~Jiang, Unusual
  anomalous hall effect in perpendicularly magnetized yig films with a small
  gilbert damping constant, Phys. Rev. B 101 (2020) 174431.

\bibitem{F_Yang2020b}
A.~J. Lee, S.~Guo, J.~Flores, B.~Wang, N.~Bagués, D.~W. McComb, F.~Yang,
  {Investigation of the role of rare-earth elements in spin-Hall topological
  Hall effect in Pt/ferrimagnetic-garnet bilayers}, Nano Lett. 20 (2020)
  4667--4672.

\bibitem{J_Chen2020}
J.~Chen, J.~Hu, H.~Yu, Chiral magnonics: Reprogrammable nanoscale spin wave
  networks based on chiral domain walls, iScience 23 (2020) 101153.

\bibitem{Gal2019}
R.~A. Gallardo, D.~Cort\'es-Ortu\~no, T.~Schneider, A.~Rold\'an-Molina, F.~Ma,
  R.~E. Troncoso, K.~Lenz, H.~Fangohr, J.~Lindner, P.~Landeros, {Flat bands,
  indirect gaps, and unconventional spin-wave behavior induced by a periodic
  Dzyaloshinskii-Moriya interaction}, Phys. Rev. Lett. 122 (2019) 067204.

\bibitem{Tak2017}
R.~Takagi, D.~Morikawa, K.~Karube, N.~Kanazawa, K.~Shibata, G.~Tatara,
  Y.~Tokunaga, T.~Arima, Y.~Taguchi, Y.~Tokura, S.~Seki, {Spin-wave
  spectroscopy of the Dzyaloshinskii-Moriya interaction in room-temperature
  chiral magnets hosting skyrmions}, Phys. Rev. B 95 (2017) 220406.

\bibitem{Lee2017}
S.~J. Lee, J.~H. Moon, H.~W. Lee, K.~J. Lee, {Spin-wave propagation in the
  presence of inhomogeneous Dzyaloshinskii-Moriya interactions}, Phys. Rev. B
  96 (2017) 184433.

\bibitem{L_Zhang2013}
L.~Zhang, J.~Ren, J.-S. Wang, B.~Li, {Topological magnon insulator in
  insulating ferromagnet}, Phys. Rev. B 87 (2013) 144101.

\bibitem{Shindou:2013cp}
R.~Shindou, R.~Matsumoto, S.~Murakami, J.-i. Ohe, {Topological chiral magnonic
  edge mode in a magnonic crystal}, Phys. Rev. B 87 (2013) 174427.

\bibitem{K_Chang2018}
Y.-M. Li, J.~Xiao, K.~Chang, {Topological magnon modes in patterned
  ferrimagnetic insulator thin films}, Nano Lett. 18 (2018) 3032--3037.

\bibitem{Bussmann99}
K.~Bussmann, G.~A. Prinz, S.~F. Cheng, D.~Wang, {Switching of vertical giant
  magnetoresistance devices by current through the device}, Appl. Phys. Lett.
  75 (1999) 2476--2478.

\bibitem{Tretiakov07}
O.~Tretiakov, O.~Tchernyshyov, {Vortices in thin ferromagnetic films and the
  skyrmion number}, Phys. Rev. B 75 (2007) 012408.

\bibitem{hillebrands_spin-wave_1990}
B.~Hillebrands, {Spin-wave calculations for multilayered structures}, Phys.
  Rev. B 41 (1990) 530--540.

\bibitem{Chen2018}
J.~Chen, C.~Liu, T.~Liu, Y.~Xiao, K.~Xia, G.~E.~W. Bauer, M.~Wu, H.~Yu, {Strong
  interlayer magnon-magnon coupling in magnetic metal-insulator hybrid
  nanostructures}, Phys. Rev. Lett. 120 (2018) 217202.

\bibitem{Y_Fan2020}
Y.~Fan, P.~Quarterman, J.~Finley, J.~Han, P.~Zhang, J.~T. Hou, M.~D. Stiles,
  A.~J. Grutter, L.~Liu, {Manipulation of coupling and magnon transport in
  magnetic metal-insulator hybrid structures}, Phys. Rev. Appl. 13 (2020)
  061002.

\bibitem{Y_Li2020}
Y.~Li, W.~Cao, V.~P. Amin, Z.~Zhang, J.~Gibbons, J.~Sklenar, J.~Pearson, P.~M.
  Haney, M.~D. Stiles, W.~E. Bailey, V.~Novosad, A.~Hoffmann, W.~Zhang,
  Coherent spin pumping in a strongly coupled magnon-magnon hybrid system,
  Phys. Rev. Lett. 124 (2020) 117202.

\bibitem{Grunberg86}
P.~Gr$\ddot{\rm u}$nberg, R.~Schreiber, Y.~Pang, M.~Brodsky, H.~Sowers,
  {Layered magnetic structures: evidence for antiferromagnetic coupling of Fe
  layers across Cr interlayers}, Phys. Rev. Lett. 57 (1986) 2442--2445.

\bibitem{Gus2008}
M.~Alvarez, T.~L{\'{o}}pez, J.~A. Odriozola, R.~D. Gonzalez, {Magnetic vortex
  state stability, reversal and dynamics in restricted geometries}, J. Nanosci.
  Nanotech. 8 (2008) 2745--2760.

\bibitem{Choe2004}
S.~B. Choe, Y.~Acremann, A.~Scholl, A.~Bauer, A.~Doran, J.~St{\"{o}}hr, H.~A.
  Padmore, {Vortex core-driven magnetization dynamics}, Science 304 (2004)
  420--422.

\bibitem{lee_radiation_2005}
K.-S. Lee, S.~Choi, S.-K. Kim, {Radiation of spin waves from magnetic vortex
  cores by their dynamic motion and annihilation processes}, Appl. Phys. Lett.
  87 (2005) 192502.

\bibitem{park_interactions_2005}
J.~P. Park, P.~A. Crowell, {Interactions of spin waves with a magnetic vortex},
  Phys. Rev. Lett. 95 (2005) 167201.

\bibitem{Vansteenkiste09}
A.~Vansteenkiste, K.~W. Chou, M.~Weigand, M.~Curcic, V.~Sackmann, H.~Stoll,
  T.~Tyliszczak, G.~Woltersdorf, C.~H. Back, G.~Sch{\"{u}}tz, B.~{Van
  Waeyenberge}, {X-ray imaging of the dynamic magnetic vortex core
  deformation}, Nat. Phys. 5 (2009) 332--334.

\bibitem{Kammerer11}
M.~Kammerer, M.~Weigand, M.~Curcic, M.~Noske, M.~Sproll, A.~Vansteenkiste,
  B.~V. Waeyenberge, H.~Stoll, G.~Woltersdorf, C.~H. Back, G.~Schuetz,
  {Magnetic vortex core reversal by excitation of spin waves}, Nat. Commun. 2
  (2011) 276--279.

\bibitem{Shin2000}
T.~Shinjo, T.~Okuno, R.~Hassdorf, K.~Shigeto, T.~Ono, {Magnetic vortex core
  observation in circular dots of permalloy}, {Science} {289} ({2000})
  {930--932}.

\bibitem{Beh2018}
C.~Behncke, C.~F. Adolff, N.~Lenzing, M.~H{\"{a}}nze, B.~Schulte, M.~Weigand,
  G.~Sch{\"{u}}tz, G.~Meier, {Spin-wave interference in magnetic vortex
  stacks}, Commun. Phys. 1 (2018) 50.

\bibitem{Due2011}
G.~Duerr, M.~Madami, S.~Neusser, S.~Tacchi, G.~Gubbiotti, G.~Carlotti,
  D.~Grundler, {Spatial control of spin-wave modes in Ni80Fe20 antidot lattices
  by embedded Co nanodisks}, Appl. Phys. Lett. 99 (2011) 202502.

\bibitem{Ding2014}
J.~Ding, G.~N. Kakazei, X.~M. Liu, K.~Y. Guslienko, A.~O. Adeyeye, {Intensity
  inversion of vortex gyrotropic modes in thick ferromagnetic nanodots}, Appl.
  Phys. Lett. 104 (2014) 192405.

\bibitem{Lup2015}
P.~Lupo, D.~Kumar, A.~O. Adeyeye, {Size dependence of spin-wave modes in
  Ni80Fe20 nanodisks}, AIP Adv. 5 (2015) 077179.

\bibitem{Gub_book}
G.~Gubbiotti, {Three-dimensional magnonics: layered, micro- and
  nanostructures}, Jenny Stanford Publishing (2019).

\bibitem{Acre2000}
Y.~Acremann, C.~H. Back, M.~Buess, 0.Portmann, A.~Vaterlaus, D.~Pescia, {H.
  Melchior}, {Imaging precessional motion of the magnetization vector}, Science
  290 (2000) 492--495.

\bibitem{Wae2006}
B.~V. Waeyenberge, A.~Puzic, H.~Stoll, K.~W. Chou, T.~Tyliszczak, R.~Hertel,
  M.Fahnle, H.~Bruckl, K.~Rott, G.~Reiss, I.~Neudecker, D.~Weiss, C.~H. Back,
  G.~Schutz, {Magnetic vortex core reversal by excitation with short bursts of
  an alternating field}, Nature 444 (2006) 461--464.

\bibitem{Tac2019}
S.~Tacchi, R.~Silvani, G.~Carlotti, M.~Marangolo, M.~Eddrief, A.~Rettori, M.~G.
  Pini, {Strongly hybridized dipole-exchange spin waves in thin Fe-N
  ferromagnetic films}, Phys. Rev. B 100 (2019) 104406.

\bibitem{Moh2019}
M.~Mohseni, R.~Verba, T.~Br{\"{a}}cher, Q.~Wang, D.~A. Bozhko, B.~Hillebrands,
  P.~Pirro, {Backscattering immunity of dipole-exchange magnetostatic surface
  spin waves}, Phys. Rev. Lett. 122 (2019) 197201.

\bibitem{Pod2006}
J.~Podbielski, F.~Giesen, D.~Grundler, {Spin-wave interference in microscopic
  rings}, Phys. Rev. Lett. 96 (2006) 167207.

\bibitem{Kampfrath2011}
T.~Kampfrath, A.~Sell, G.~Klatt, A.~Pashkin, S.~M{\"{a}}hrlein, T.~Dekorsy,
  M.~Wolf, M.~Fiebig, A.~Leitenstorfer, R.~Huber, {Coherent terahertz control
  of antiferromagnetic spin waves}, Nat. Photon. 5 (2011) 31--34.

\bibitem{gitgeatpong_nonreciprocal_2017}
G.~Gitgeatpong, Y.~Zhao, P.~Piyawongwatthana, Y.~Qiu, L.~W. Harriger, N.~P.
  Butch, T.~J. Sato, K.~Matan, {Nonreciprocal magnons and symmetry-breaking in
  the noncentrosymmetric antiferromagnet}, Phys. Rev. Lett. 119 (2017) 047201.

\bibitem{Bialek2019}
M.~Bia\l{}ek, T.~Ito, H.~R\o{}nnow, J.-P. Ansermet, {Terahertz-optical
  properties of a bismuth ferrite single crystal}, Phys. Rev. B 99 (2019)
  064429.

\bibitem{han_birefringence-like_2020}
J.~Han, P.~Zhang, Z.~Bi, Y.~Fan, T.~S. Safi, J.~Xiang, J.~Finley, L.~Fu,
  R.~Cheng, L.~Liu, {Birefringence-like spin transport via linearly polarized
  antiferromagnetic magnons}, Nat. Nanotechnol. 15 (2020) 563--568.

\bibitem{gomonay_antiferromagnetic_2018}
O.~Gomonay, V.~Baltz, A.~Brataas, Y.~Tserkovnyak, {Antiferromagnetic spin
  textures and dynamics}, Nat. Phys. 14 (2018) 213--216.

\bibitem{bennett_quantum_2000}
C.~H. Bennett, D.~P. DiVincenzo, {Quantum information and computation}, Nature
  404 (2000) 247--255.

\bibitem{goldstein_polarized_2010}
D.~H. Goldstein, Polarized {Light}, 3rd Edition, Taylor and Francis, 2010.

\bibitem{sklan_phonon_2014}
S.~R. Sklan, J.~C. Grossman, {Phonon diodes and transistors from
  magneto-acoustics}, New J. Phys. 16 (2014) 053029.

\bibitem{sklan_splash_2015}
S.~R. Sklan, Splash, pop, sizzle: {Information} processing with phononic
  computing, AIP Adv. 5 (2015) 053302.

\bibitem{flebus_entangling_2019}
B.~Flebus, Y.~Tserkovnyak, {Entangling distant spin qubits via a magnetic
  domain wall}, Phys. Rev. B 99 (2019) 140403.

\bibitem{ZQ_Liu2019}
H.~Yan, Z.~Feng, S.~Shang, X.~Wang, Z.~Hu, J.~Wang, Z.~Zhu, H.~Wang, Z.~Chen,
  H.~Hua, W.~Lu, J.~Wang, P.~Qin, H.~Guo, X.~Zhou, Z.~Leng, Z.~Liu, C.~Jiang,
  M.~Coey, Z.~Liu, {A piezoelectric, strain-controlled antiferromagnetic memory
  insensitive to magnetic fields}, Nat. Nanotechnol. 14 (2019) 131--136.

\bibitem{wimmer_coherent_2020}
T.~Wimmer, A.~Kamra, J.~Gückelhorn, M.~Opel, S.~Geprägs, R.~Gross, H.~Huebl,
  M.~Althammer, Coherent {Control} of {Magnonic} {Spin} {Transport} in an
  {Antiferromagnetic} {Insulator}, arXiv:2008.00440 [cond-mat]ArXiv: 2008.00440
  (Aug. 2020).

\bibitem{poschl_bemerkungen_1933}
G.~Pöschl, E.~Teller, Bemerkungen zur {quantenmechanik} des anharmonischen
  {oszillators}, Z. Physik 83 (1933) 143--151.

\bibitem{dodd_solitons_1982}
R.~K. Dodd, {Solitons and Nonlinear Wave Equations}, Academic Press, 1982.

\bibitem{lekner_reflectionless_2007}
J.~Lekner, {Reflectionless eigenstates of the sech2 potential}, Am. J. Phys. 75
  (2007) 1151.

\bibitem{Tve2014}
E.~G. Tveten, A.~Qaiumzadeh, A.~Brataas, {Antiferromagnetic domain wall motion
  induced by spin waves}, Phys. Rev. Lett. 112 (2014) 147204.

\bibitem{Kim2015}
S.~K. Kim, Y.~Tserkovnyak, O.~Tchernyshyov, {Propulsion of a domain wall in an
  antiferromagnet by magnons}, Phys. Rev. B 90 (2014) 104406.

\bibitem{selzer_inertia-free_2016}
S.~Selzer, U.~Atxitia, U.~Ritzmann, D.~Hinzke, U.~Nowak, Inertia-{free}
  {thermally} {driven} {domain}-{wall} {motion} in {antiferromagnets}, Phys.
  Rev. Lett. 117 (2016) 107201.

\bibitem{daniels_topological_2019}
M.~W. Daniels, W.~Yu, R.~Cheng, J.~Xiao, D.~Xiao, Topological spin {Hall}
  effects and tunable skyrmion {Hall} effects in uniaxial antiferromagnetic
  insulators, Phys. Rev. B 99 (2019) 224433.

\bibitem{barker_static_2016}
J.~Barker, O.~A. Tretiakov, Static and {dynamical} {properties} of
  {antiferromagnetic} {Skyrmions} in the {presence} of {applied} {current} and
  {temperature}, Phys. Rev. Lett. 116 (2016) 147203.

\bibitem{schneider_realization_2008}
T.~Schneider, A.~A. Serga, B.~Leven, B.~Hillebrands, R.~L. Stamps, M.~P.
  Kostylev, {Realization of spin-wave logic gates}, Appl. Phys. Lett. 92 (2008)
  022505.

\bibitem{jamali_spin_2013}
M.~Jamali, J.~H. Kwon, S.-M. Seo, K.-J. Lee, H.~Yang, {Spin wave nonreciprocity
  for logic device applications}, Sci. Rep. 3 (2013) 3160.

\bibitem{chumak_magnon_2014}
A.~V. Chumak, A.~A. Serga, B.~Hillebrands, {Magnon transistor for all-magnon
  data processing}, Nat. Commun. 5 (2014) 4700.

\bibitem{Landauer:wc}
R.~Landauer, {Irreversibility and heat generation in the computing process},
  IBM J. Res. Dev. 5 (1961) 183--191.

\bibitem{Balynskiy:2018km}
M.~Balynskiy, H.~Chiang, D.~Gutierrez, A.~Kozhevnikov, Y.~Filimonov, A.~Khitun,
  {Reversible magnetic logic gates based on spin wave interference}, J. Appl.
  Phys. 123 (2018) 144501.

\bibitem{Klingler:2014hk}
S.~Klingler, P.~Pirro, T.~Br{\"a}cher, B.~leven, B.~Hillebrands, A.~V. Chumak,
  {Design of a spin-wave majority gate employing mode selection}, Appl. Phys.
  Lett. 105 (2014) 152410.

\bibitem{Bayer:2005ev}
C.~Bayer, H.~Schultheiss, B.~Hillebrands, R.~L. Stamps, {Phase shift of spin
  waves traveling through a 180 degrees Bloch-domain wall}, IEEE Trans. Magn.
  41 (2005) 3094--3096.

\bibitem{Andrich:2017ey}
P.~Andrich, C.~F. de~las Casas, X.~Liu, H.~L. Bretscher, J.~R. Berman, F.~J.
  Heremans, P.~F. Nealey, D.~D. Awschalom, {Long-range spin wave mediated
  control of defect qubits in nanodiamonds}, npj Quantum Inf. 3 (2017) 1--7.

\bibitem{Graf:2018im}
J.~Graf, H.~Pfeifer, F.~Marquardt, S.~V. Kusminskiy, {Cavity optomagnonics with
  magnetic textures: Coupling a magnetic vortex to light}, Phys. Rev. B 98
  (2018) 241406.

\bibitem{Temnov:2012fn}
V.~V. Temnov, {Ultrafast acousto-magneto-plasmonics}, Nat. Photonics 6 (2012)
  728--736.

\bibitem{Duine2020}
A.~R\"uckriegel, R.~A. Duine, {Long-range phonon spin transport in
  ferromagnet--nonmagnetic insulator heterostructures}, Phys. Rev. Lett. 124
  (2020) 117201.

\bibitem{Klein2020}
K.~An, A.~N. Litvinenko, R.~Kohno, A.~A. Fuad, V.~V. Naletov, L.~Vila,
  U.~Ebels, G.~de~Loubens, H.~Hurdequint, N.~Beaulieu, J.~Ben~Youssef,
  N.~Vukadinovic, G.~E.~W. Bauer, A.~N. Slavin, V.~S. Tiberkevich, O.~Klein,
  {Coherent long-range transfer of angular momentum between magnon Kittel modes
  by phonons}, Phys. Rev. B 101 (2020) 060407.

\bibitem{LachanceQuirion:2019gm}
D.~Lachance-Quirion, Y.~Tabuchi, A.~Gloppe, K.~Usami, Y.~Nakamura, {Hybrid
  quantum systems based on magnonics}, Appl. Phys. Express 12 (2019) 070101.

\bibitem{Dobrovolskiy:2019ix}
O.~V. Dobrovolskiy, R.~Sachser, T.~Br{\"a}cher, T.~B{\"o}ttcher, V.~V.
  Kruglyak, R.~V. Vovk, V.~A. Shklovskij, M.~Huth, B.~Hillebrands, A.~V.
  Chumak, {Magnon{\textendash}fluxon interaction in a
  ferromagnet/superconductor heterostructure}, Nat. Phys. 15 (2019) 477--482.

\bibitem{Jungwirth:2016dt}
T.~Jungwirth, X.~Marti, P.~Wadley, J.~Wunderlich, {Antiferromagnetic
  spintronics}, Nat. Nanotechnol. 11 (2016) 231--241.

\bibitem{xing_fiber_2016}
X.~Xing, Y.~Zhou, {Fiber optics for spin waves}, NPG Asia Mater. 8 (2016) e246.

\bibitem{Yu:2016bk}
W.~Yu, J.~Lan, R.~Wu, J.~Xiao, {Magnetic Snell's law and spin-wave fiber with
  Dzyaloshinskii-Moriya interaction}, Phys. Rev. B 94 (2016) 140410.

\bibitem{Xing:2017ku}
X.~Xing, P.~W.~T. Pong, J.~Akerman, Y.~Zhou, {Paving spin-wave fibers in
  magnonic nanocircuits using spin-orbit torque}, Phys. Rev. Appl. 7 (2017)
  054016.

\bibitem{Wang:2018eu}
X.~S. Wang, H.~W. Zhang, X.~R. Wang, {Topological magnonics: A paradigm for
  spin-wave manipulation and device design}, Phys. Rev. Appl. 9 (2018) 024029.

\bibitem{Vogel:2015ky}
M.~Vogel, A.~V. Chumak, E.~H. Waller, T.~Langner, V.~I. Vasyuchka,
  B.~Hillebrands, G.~von Freymann, {Optically reconfigurable magnetic
  materials}, Nat. Phys. 11 (2015) 487--491.

\bibitem{Heussner:2017ea}
F.~Heussner, A.~A. Serga, T.~Br{\"a}cher, B.~Hillebrands, P.~Pirro, {A
  switchable spin-wave signal splitter for magnonic networks}, Appl. Phys.
  Lett. 111 (2017) 122401.

\bibitem{Wang:2018eg}
Q.~Wang, P.~Pirro, R.~Verba, A.~Slavin, B.~Hillebrands, A.~V. Chumak,
  {Reconfigurable nanoscale spin-wave directional coupler}, Sci. Adv. 4 (2018)
  e1701517.

\bibitem{vukadinovic_spin-wave_2011}
N.~Vukadinovic, F.~Boust, {Spin-wave excitations of domain walls in
  bubble-state magnetic nanoelements}, Phys. Rev. B 84 (2011) 224425.

\bibitem{Mook:2014iu}
A.~Mook, J.~Henk, I.~Mertig, {Edge states in topological magnon insulators},
  Phys. Rev. B 90 (2014) 024412.

\bibitem{McClarty:2017im}
P.~A. McClarty, F.~Kr{\"u}ger, T.~Guidi, S.~F. Parker, K.~Refson, A.~W. Parker,
  D.~Prabhakaran, R.~Coldea, {Topological triplon modes and bound states in a
  Shastry{\textendash}Sutherland magnet}, Nat. Phys. 13 (2017) 736--741.

\bibitem{Bisig:2009gq}
A.~Bisig, L.~Heyne, O.~Boulle, M.~Klaui, {Tunable steady-state domain wall
  oscillator with perpendicular magnetic anisotropy}, Appl. Phys. Lett. 95
  (2009) 162504.

\bibitem{Beach2008}
G.~S.~D. Beach, M.~Tsoi, J.~L. Erskine, {Current-induced domain wall motion},
  J. Magn. Magn. Mater. 320 (2008) 1272--1281.

\bibitem{Nakamura2015}
Y.~Tabuchi, S.~Ishino, A.~Noguchi, T.~Ishikawa, R.~Yamazaki, K.~Usami,
  Y.~Nakamura, {Coherent coupling between a ferromagnetic magnon and a
  superconducting qubit}, Science 349 (2015) 405--408.

\bibitem{W_Yu2020}
W.~Yu, J.~Lan, J.~Xiao, {Magnetic logic gate based on polarized spin waves},
  Phys. Rev. Appl. 13 (2020) 024055.

\bibitem{allwood_magnetic_2005}
D.~A. Allwood, G.~Xiong, C.~C. Faulkner, D.~Atkinson, D.~Petit, R.~P. Cowburn,
  Magnetic {domain}-{wall} {logic}, Science 309 (2005) 1688--1692.

\bibitem{Sonin2010}
E.~B. Sonin, Spin currents and spin superfluidity, Adv. Phys. 59 (2010)
  181--255.

\bibitem{tserkovnyak_energy_2018}
Y.~Tserkovnyak, J.~Xiao, Energy {storage} via {topological} {spin} {textures},
  Phys. Rev. Lett. 121 (2018) 127701.

\bibitem{Wang:2018ew}
L.~Wang, L.~Gao, L.~Jin, Y.~Liao, T.~Wen, X.~Tang, H.~Zhang, Z.~Zhong,
  {Magnonic waveguide based on exchange-spring magnetic structure}, AIP Adv.
  8~(5) (2018) 055103.

\bibitem{Maekawa2007}
S.~E. Barnes, S.~Maekawa, {Generalization of Faraday's law to include
  nonconservative spin forces}, Phys. Rev. Lett. 98 (2007) 246601.

\bibitem{Hoffmann2019}
Y.~Li, T.~Polakovic, Y.-L. Wang, J.~Xu, S.~Lendinez, Z.~Zhang, J.~Ding,
  T.~Khaire, H.~Saglam, R.~Divan, J.~Pearson, W.-K. Kwok, Z.~Xiao, V.~Novosad,
  A.~Hoffmann, W.~Zhang, Strong coupling between magnons and microwave photons
  in on-chip ferromagnet-superconductor thin-film devices, Phys. Rev. Lett. 123
  (2019) 107701.

\bibitem{L_Liu2019}
J.~T. Hou, L.~Liu, Strong coupling between microwave photons and nanomagnet
  magnons, Phys. Rev. Lett. 123 (2019) 107702.

\bibitem{Nakamura2020}
D.~Lachance-Quirion, S.~P. Wolski, Y.~Tabuchi, S.~Kono, K.~Usami, Y.~Nakamura,
  {Entanglement-based single-shot detection of a single magnon with a
  superconducting qubit}, Science 367 (2020) 425--428.

\bibitem{H_Tu2020}
H.~Tu, J.~Wang, Z.~Huang, Y.~Zhai, Z.~Zhu, Z.~Zhang, J.~Qu, R.~Zheng, Y.~Yuan,
  R.~Liu, W.~Zhang, B.~You, J.~Du, Large anisotropy of magnetic damping in
  amorphous {CoFeB} films on {GaAs}(001), J. Phys Condens. Matter 32 (2020)
  335804.

\bibitem{Fuchs:2015ii}
F.~Fuchs, B.~Stender, M.~Trupke, D.~Simin, J.~Pflaum, V.~Dyakonov, G.~V.
  Astakhov, {Engineering near-infrared single-photon emitters with optically
  active spins in ultrapure silicon carbide}, Nat. Commun. 6 (2015) 7578.

\bibitem{Nakamura2016}
A.~Osada, R.~Hisatomi, A.~Noguchi, Y.~Tabuchi, R.~Yamazaki, K.~Usami,
  M.~Sadgrove, R.~Yalla, M.~Nomura, Y.~Nakamura, {Cavity optomagnonics with
  spin-orbit coupled photons}, Phys. Rev. Lett. 116 (2016) 223601.

\bibitem{H_Tang2016}
X.~Zhang, N.~Zhu, C.-L. Zou, H.~X. Tang, {Optomagnonic whispering gallery
  microresonators}, Phys. Rev. Lett. 117 (2016) 123605.

\bibitem{JQ_You2017}
D.~Zhang, X.-Q. Luo, Y.-P. Wang, T.-F. Li, J.~Q. You, {Observation of the
  exceptional point in cavity magnon-polaritons}, Nat. Commun. 8 (2017) 1368.

\bibitem{CM_Hu2019}
Y.-P. Wang, J.~W. Rao, Y.~Yang, P.-C. Xu, Y.~S. Gui, B.~M. Yao, J.~Q. You,
  C.-M. Hu, {Nonreciprocity and unidirectional invisibility in cavity
  magnonics}, Phys. Rev. Lett. 123 (2019) 127202.

\bibitem{Ross2011}
L.~Bi, J.~Hu, P.~Jiang, D.~H. Kim, G.~F. Dionne, L.~C. Kimerling, C.~A. Ross,
  {On-chip optical isolation in monolithically integrated non-reciprocal
  optical resonators}, Nat. Photon. 5 (2011) 758--762.

\bibitem{Fert2008}
A.~Fert, {Nobel Lecture: Origin, development, and future of spintronics}, Rev.
  Mod. Phys. 80 (2008) 1517--1530.

\bibitem{Torrejon:2017hj}
J.~Torrejon, M.~Riou, F.~A. Araujo, S.~Tsunegi, G.~Khalsa, D.~Querlioz,
  P.~Bortolotti, V.~Cros, K.~Yakushiji, A.~Fukushima, H.~Kubota, S.~Yuasa,
  M.~D. Stiles, J.~Grollier, {Neuromorphic computing with nanoscale spintronic
  oscillators}, Nature 547 (2017) 428--431.

\bibitem{Romera:2018dm}
M.~Romera, P.~Talatchian, S.~Tsunegi, F.~A. Araujo, V.~Cros, P.~Bortolotti,
  J.~Trastoy, K.~Yakushiji, A.~Fukushima, H.~Kubota, S.~Yuasa, M.~Ernoult,
  D.~Vodenicarevic, T.~Hirtzlin, N.~Locatelli, D.~Querlioz, J.~Grollier, {Vowel
  recognition with four coupled spin-torque nano-oscillators}, Nature 563
  (2018) 230--234.

\bibitem{Zahedinejad:2019kd}
M.~Zahedinejad, A.~A. Awad, S.~Muralidhar, R.~Khymyn, H.~Fulara, H.~Mazraati,
  M.~Dvornik, J.~{\AA}kerman, {Two-dimensional mutually synchronized spin Hall
  nano-oscillator arrays for neuromorphic computing}, Nat. Nanotechnol. 15
  (2020) 47--52.

\end{thebibliography}

\end{document}